\documentclass[numbook,envcountsame]{article}    
\usepackage[latin1]{inputenc}
\usepackage{latexsym}
\usepackage{amsmath}
\usepackage{amssymb}
\usepackage{amsfonts}
\usepackage{mathrsfs}               
\usepackage{bbm}                    
\usepackage{bbold}                  
\usepackage{textcomp}               
\usepackage{amstext}
\usepackage{enumerate}
\usepackage{units}
\usepackage{stmaryrd}
\usepackage{theorem}
\newcommand{\BB}{\mathbb{B}}
\newcommand{\CC}{\mathbb{C}}
\newcommand{\EE}{\mathbb{E}}

\newcommand{\NN}{\mathbb{N}}
\newcommand{\PP}{\mathbb{P}}
\newcommand{\QQ}{\mathbb{Q}}
\newcommand{\RR}{\mathbb{R}}

\newcommand{\Ran}{\mathrm{Ran}}

\newcommand{\pr}{\mathrm{pr}}

\newcommand{\loc}{\mathrm{loc}}
\newcommand{\Div}{{\mathrm{div}}}

\newcommand{\NE}{\mathrm{N}}
\newcommand{\Id}{\mathrm{d}}
\newcommand{\Ad}{\mathrm{ad}}
\newcommand{\scal}{\mathrm{sc}}
\newcommand{\SPn}[2]{\langle #1|#2 \rangle} 
\newcommand{\SPb}[2]{\big\langle #1\big|#2 \big\rangle} 

\newcommand{\ol}[1]{\overline{#1}} 
\newcommand{\ul}[1]{\underline{#1}} 

\newcommand{\wh}[1]{\widehat{#1}}
\newcommand{\wt}[1]{\widetilde{#1}}
\newcommand{\nf}[2]{\nicefrac{#1}{#2}}
\newcommand{\eh}{{\nf{1}{2}}}
\newcommand{\mh}{{-\nf{1}{2}}}
\newcommand{\cA}{\mathcal{A}}
\newcommand{\cB}{\mathcal{B}} 
\newcommand{\cC}{\mathcal{C}}
\newcommand{\cD}{\mathcal{D}}\newcommand{\cP}{\mathcal{P}} 
\newcommand{\cQ}{\mathcal{Q}}

\newcommand{\cS}{\mathcal{S}}
\newcommand{\cT}{\mathcal{T}}

\newcommand{\cM}{\mathcal{M}}       
 
\newcommand{\sA}{\mathscr{A}}\newcommand{\sN}{\mathscr{N}}
 
\newcommand{\sC}{\mathscr{C}}
\newcommand{\sD}{\mathscr{D}} 
\newcommand{\sE}{\mathscr{E}}\newcommand{\sQ}{\mathscr{Q}}
\newcommand{\sF}{\mathscr{F}}\newcommand{\sR}{\mathscr{R}}
\newcommand{\sG}{\mathscr{G}}
\newcommand{\sT}{\mathscr{T}}
\newcommand{\sI}{\mathscr{I}}\newcommand{\sU}{\mathscr{U}}
\newcommand{\sJ}{\mathscr{J}}
\newcommand{\sK}{\mathscr{K}}\newcommand{\sW}{\mathscr{W}}
\newcommand{\sL}{\mathscr{L}}         
\newcommand{\sM}{\mathscr{M}}\newcommand{\sY}{\mathscr{Y}}       
\newcommand{\sZ}{\mathscr{Z}} 
\newcommand{\fA}{\mathfrak{A}}\newcommand{\fN}{\mathfrak{N}}
\newcommand{\fB}{\mathfrak{B}}

\newcommand{\fF}{\mathfrak{F}}\newcommand{\fR}{\mathfrak{R}}
\newcommand{\fG}{\mathfrak{G}}
\newcommand{\fH}{\mathfrak{H}}

\newcommand{\fM}{\mathfrak{M}}       
 
\newcommand{\V}[1]{\boldsymbol{#1}}
\newcommand{\vsigma}{\boldsymbol{\sigma}}

\newcommand{\veps}{\boldsymbol{\varepsilon}}
\newcommand{\vXi}{\boldsymbol{\Xi}}
\newcommand{\vxi}{\boldsymbol{\xi}}
\newcommand{\vgamma}{\boldsymbol{\gamma}}
\newcommand{\ve}{\varepsilon}
\newcommand{\vp}{\varphi}

\newcommand{\vk}{\varkappa}
\newcommand{\vr}{\varrho}
\newcommand{\vt}{\vartheta}

\newcommand{\id}{\mathbbm{1}}                   
\newcommand{\dom}{\cD}                          
\newcommand{\fdom}{\cQ}                         
\newcommand{\HR}{\mathscr{H}}                   
\newcommand{\FHR}{\hat{\mathscr{H}}}            
\newcommand{\HP}{\mathfrak{h}}                  
\newcommand{\LO}{\mathscr{B}}                   
\newcommand{\ad}{a^\dagger}                     
\newcommand{\simplex}{{\triangle}}    
\newcommand{\w}[2]{w_{#1,#2}}         %
\newcommand{\olw}[2]{\overline{w}_{#1,#2}}         %
\newcommand{\filt}{{\mathfrak{F}}}    
\newcommand{\zeit}{I}                 
\newcommand{\const}{\mathfrak{c}}     
 
\newcommand{\W}[2]{W_{#1}^{#2}}
\newcommand{\WW}[2]{\mathbb{W}_{#1}^{#2}}
\newcommand{\AD}[2]{\sL^{\alpha_{#1}}_{#2}(t_{#1})}
\newcommand{\ADg}[2]{\sL^{\alpha_{#1}}_{#2}(t_{#1};g)}
\newcommand{\A}[1]{\sR_{\alpha_{#1}}(t_{#1})}

\newcommand{\Ah}[1]{\sR_{\alpha_{#1}}(t_{#1};h)}
\newcommand{\IN}[1]{\sI_{\alpha_{#1}}(t_{#1})}
\renewcommand{\Re}{\mathrm{Re}}
\renewcommand{\Im}{\mathrm{Im}}
\renewcommand{\le}{\leqslant}        
\renewcommand{\ge}{\geqslant}  
\newcommand{\proof}{\indent{{\it Proof.}}}
\newcommand{\qed}{${}$\hfill$\Box$}
\newtheorem{theorem}{Theorem}[section]
\newtheorem{lemma}[theorem]{Lemma}
\newtheorem{proposition}[theorem]{Proposition}
\newtheorem{corollary}[theorem]{Corollary}
\newtheorem{definition}[theorem]{Definition}
\newtheorem{remark}[theorem]{Remark}
\newtheorem{hyp}[theorem]{Hypothesis}
\newtheorem{example}[theorem]{Example}
\numberwithin{equation}{section}
\begin{document}

\title{Stochastic differential equations for models of non-relativistic matter interacting
   with quantized radiation fields}

\author{B. G\"uneysu, O. Matte, and J.S. M{\o}ller}

\maketitle

\begin{abstract}
\noindent We discuss Hilbert space-valued stochastic differential equations
associated with the heat semi-groups of the standard model of non-relativistic
quantum electrodynamics and of corresponding fiber Hamiltonians
for translation invariant systems. In particular, we prove the existence of a stochastic flow 
satisfying the strong Markov property and the Feller property. To this end we
employ an explicit solution ansatz. In the matrix-valued case, i.e., if the electron spin is taken into 
account, it is given by a series of operator-valued time-ordered integrals,
whose integrands are factorized into annihilation, preservation, creation, and scalar parts.
The Feynman-Kac formula implied by these results is new in the matrix-valued case.
Furthermore, we discuss stochastic differential equations
and Feynman-Kac representations for an operator-valued integral kernel of the semi-group. 
As a byproduct we obtain analogous results for Nelson's model.
\end{abstract}
%
%
%
%
%
%

\section{Introduction}\label{sec-intro}

\noindent
The present article is devoted to the stochastic analysis of certain models for non-relativistic quantum
mechanical matter interacting with quantized radiation fields.
While the time evolution of the matter particles alone would always be generated by Schr\"odinger operators in the models covered by our results, the radiation fields are described by relativistic 
quantum field theory. The fields obey Bose statistics and thus consist of an undetermined
number of bosons which may be created or annihilated along the time evolution.
In particular, the state space of the radiation field is the bosonic Fock space.
In the prime example, the {\em standard model of non-relativistic (NR)
quantum electrodynamics (QED)}, the matter particles are electrons and the bosons are photons
constituting the quantized electromagnetic field. 
In this model the electrons have internal spin degrees of freedom.
Another example is the {\em Nelson model} where the matter particles are (spinless)
nucleons and the bosons are mesons and thus have a mass.
The {\em massless} Nelson model can be used to describe the
interaction of electrons with acoustic phonons in solids, which are massless bosons.

After several decades of intensive studies of Schr\"odinger operators in classical electromagnetic 
fields, the mathematical analysis of NRQED became more and more popular in the late 90's. Since 
then various spectral theoretic aspects of NRQED have been investigated by new non-perturbative 
methods or sophisticated perturbative multi-scale methods; see, e.g.,
\cite{LHB2011,Spohn2004} for a general introduction and reference lists. 
In view of Feynman's famous article 
\cite{Feynman1950} where, in particular, the quantum mechanical time evolution of NR
matter particles coupled to the quantized electromagnetic field is discussed,
it is certainly most natural to generalize also path integral techniques 
developed in the mathematical study of Schr\"odinger operators to the case of
quantized radiation fields. In fact, Feynman-Kac formulas for the semi-group in the standard model 
of NRQED have already been derived earlier and exploited in spectral theoretic problems
mainly by F.~Hiroshima and his co-workers; 
see Subsect.~\ref{ssec-FK-intro} below for references and more remarks. 
These Feynman-Kac formulas have been obtained via a functional analytic 
approach based on Trotter product expansions. The aim of our work is 
to explore their relationship to corresponding stochastic differential equations (SDE) with the help
of the stochastic calculus in Hilbert spaces. 

In the first subsection below, we briefly describe the SDE analyzed in this paper and our
main results on it. In its full generality, our SDE escapes all frameworks we found in the 
literature; see Rem.~\ref{rem-abstractum} below. Therefore, we hope that readers interested in the 
theory of SDE in infinite dimensional Hilbert spaces will consider our analysis, which departs from an 
explicit solution ansatz, as an interesting case study.
In Subsect.~\ref{ssec-FK-intro} we comment on related Feynman-Kac formulas and future
applications of our main results. 

All notation used in the following two subsections will be re-introduced more carefully later on;
see in particular Sect.~\ref{sec-definitions}, where our basic hypotheses are formulated.
Concrete examples are given in App.~\ref{app-examples}. Another purpose of 
Sect.~\ref{sec-definitions} is to make this article accessible for readers who are experts in 
mathematical quantum field theory but might be less familiar with stochastic calculus in Hilbert 
spaces or vice versa. 
Hence, some basic information on Fock space calculus and stochastic calculus is collected
and suitably referenced.
The content of Sects.~\ref{sec-proc}--\ref{sec-ext} and Apps.~\ref{app-esa}--\ref{app-meas} will 
be indicated along the discussion in the following two subsections.

In App.~\ref{app-notation} we explain some general notation and provide a {\em list of symbols}.


\subsection{A class of stochastic differential equations and main results}\label{ssec-SDE-intro}

The present article provides a fairly comprehensive study of the type of Hilbert
space-valued SDE described in the following paragraphs:

Let $\zeit$ be a finite or infinite continuous time horizon,
$\BB:=(\Omega,\fF,(\fF_t)_{t\in I},\PP)$ be a filtered probability space satisfying the usual assumptions, 
and $\V{X}=(X_1,\ldots,X_\nu)$ be a continuous $\RR^\nu$-valued semi-martingale on $I$ with respect to $\BB$
whose quadratic covariation is equal to the identity matrix. 
In fact, $\V{X}$ will always be a solution of a suitable 
It\={o} equation. The two most important examples are Brownian motion and semi-martingale
realizations of Brownian bridges. Precise conditions on $\V{X}$ are formulated in Hyp.~\ref{hyp-B};
in App.~\ref{app-bridge} we verify that Brownian bridges
satisfy certain technical bounds appearing in it.

Let $\sF:=\Gamma_{\mathrm{s}}(\HP)$ denote the bosonic Fock space modeled over the
one-boson Hilbert space $\HP=L^2(\cM,\fA,\mu)$, which is assumed to be separable with a
$\sigma$-finite measure space $(\cM,\fA,\mu)$. As usual $\vp(f)$ is the field operator 
associated with $f\in\HP$ and $\Id\Gamma(\vk)$ denotes the differential second quantization
of the self-adjoint maximal multiplication operator in $\HP$ corresponding to some
measurable function $\vk:\cM\to\RR$. Then $\vp(f)$ and $\Id\Gamma(\vk)$
are unbounded self-adjoint operators in $\sF$ as soon as $f$ and $\vk$ are non-zero; 
they do not commute in general.  Suppose that
$G_{1,\V{x}},\ldots,G_{\nu,\V{x}},F_{1,\V{x}},\ldots,F_{S,\V{x}}\in\HP$, for every $\V{x}\in\RR^\nu$,
and $m_1,\ldots,m_\nu,\omega:\cM\to\RR$ are measurable with $\omega>0$ $\mu$-almost
everywhere ($\mu$-a.e.). In Hyp.~\ref{hyp-G} below we shall introduce 
appropriate assumptions on the latter functions. In particular, we shall require a certain regularity of the
maps $\V{x}\mapsto G_{\ell,\V{x}}$ and $\V{x}\mapsto F_{j,\V{x}}$ allowing for an application
of the stochastic calculus. Important from an algebraic point of view is the condition that
$G_{\ell,\V{x}}$ and $F_{j,\V{x}}$ belong to some fixed completely real subspace of $\HP$ which is
invariant under the multiplication operators induced by $\omega$ and $im_\ell$.

Finally, let $\sigma_1,\ldots,\sigma_S$ be hermitian $L$\texttimes$L$ matrices acting on 
(generalized) spin degrees of freedom and assume that the potential $V:\RR^\nu\to\RR$ is locally 
integrable. (The latter condition is Hyp.~\ref{hyp-V}.)

In the above situation we shall investigate the following SDE for an unknown
process $Y$ on $\zeit$ with values in the {\em fiber Hilbert space} $\FHR:=\CC^L\otimes\sF$,
\begin{align}\label{SDE-intro}
Y_\bullet&=\eta-\int_0^\bullet\wh{H}^V(\vxi,\V{X}_s)Y_s\Id s
-\sum_{\ell=1}^\nu\int_0^\bullet i\id_{\CC^L}\otimes v_\ell(\vxi,\V{X}_s)
Y_s\Id{X}_{\ell,s}.
\end{align}
The coefficients are unbounded operators defined, for fixed $\vxi,\V{x}\in\RR^\nu$, by
\begin{align}\label{v-intro}
v_\ell(\vxi,\V{x})&:=\xi_\ell-\Id\Gamma({m}_\ell)-\vp({G}_{\ell,\V{x}}),\quad\ell\in\{1,\ldots,\nu\},
\\\label{Hsc-intro}
\wh{H}^V_{\scal}(\vxi,\V{x})&:=
\frac{1}{2}\sum_{\ell=1}^\nu \big\{v_\ell(\vxi,\V{x})^2-i\vp(\partial_{x_\ell}G_{\ell,\V{x}})\big\}
+\Id\Gamma(\omega)+V(\V{x}),
\\\label{fHam-intro}
\wh{H}^V(\vxi,\V{x})&:=\id_{\CC^L}\otimes\wh{H}^V_{\scal}(\vxi,\V{x})
-\sum_{j=1}^S\sigma_j\otimes\vp({F}_{j,\V{x}}).
\end{align}
The $\fF_0$-measurable initial condition 
$\eta:\Omega\to\FHR$ attains its values in the $(\vxi,\V{x})$-ndependent domain $\wh{\dom}$
of the {\em generalized fiber Hamiltonians} $\wh{H}^V(\vxi,\V{x})$, which is explicitly given by
\begin{equation}\label{def-whD-intro}
\wh{\dom}:=\CC^L\otimes\dom({M}),\qquad
{M}:=\frac{1}{2}\sum_{\ell=1}^\nu\Id\Gamma(m_\ell)^2+\Id\Gamma(\omega).
\end{equation}
Here and henceforth, $\dom(\cdot)$ denotes the domain of a linear operator.
If the functions $G_\ell$ and $F_j$ are $\V{x}$-independent, then we denote
$\wh{H}^0(\vxi,\V{x})$ simply by $\wh{H}(\vxi)$ and call it a
{\em fiber Hamiltonian}. In this case $\wh{H}(\vxi)$ is self-adjoint
and has a direct physical interpretation: it generates the time-evolution of 
a combined particle-radiation system
moving at a fixed total momentum $\vxi$. Typically, the essential spectrum of $\wh{H}(\vxi)$
covers some half-line. Its $\V{x}$-dependent generalization $\wh{H}^V(\vxi,\V{x})$, which is closed but 
not self-adjoint in general, appears in the following formula for the
self-adjoint {\em total Hamiltonian} $H^V$ acting in $L^2(\RR^\nu,\FHR)$,
\begin{equation}\label{def-HV-intro}
(H^V\Psi)(\V{x}):=\sum_{\ell=1}^\nu\big\{\!-\tfrac{1}{2}\partial_{x_\ell}^2\Psi(\V{x})
+i\vp(G_{\ell,\V{x}})\partial_{x_\ell}\Psi(\V{x})\big\}
+\wh{H}^V(\V{0},\V{x})^*\Psi(\V{x}),
\end{equation}
for a.e. $\V{x}$ and $\Psi$ in the domain of $H^V$.
In App.~\ref{app-esa} we present a (partially well-known) elementary proof of the above
(essentially well-known) assertions on
self-adjointness/closedness and domains of the generalized fiber Hamiltonians.

Our main result is the following theorem. 
In its statement $\wh{\dom}$ is equipped with the graph norm of $\id_{\CC^L}\otimes M$.

\begin{theorem}\label{thm-SDE-intro}
Under our standing hypotheses Hyp.~\ref{hyp-G},~\ref{hyp-V}, and~\ref{hyp-B}
formulated below, the following assertions (1)--(4) hold where, in (2)--(4), we assume in addition that
$V$ is bounded and continuous. 
\begin{enumerate}
\item[{\rm (1)}]
Up to indistinguishability, there exists a unique
continuous $\FHR$-valued semi-martingale, whose paths belong $\PP$-a.s.
to $C(I,\wh{\dom})$ and which $\PP$-a.s. solves \eqref{SDE-intro} on $[0,\sup I)$. 
\item[{\rm (2)}] We can construct a {\em stochastic flow} for the system of SDE comprised of the It\={o} 
equation for $\V{X}$ and \eqref{SDE-intro}. 
\item[{\rm (3)}] The stochastic flow and the corresponding family of transition operators satisfy the 
{\em strong Markov} and {\em Feller} properties.
\item[{\rm (4)}] A {\em Blagove\v{s}\v{c}ensky-Freidlin} theorem holds, i.e., there exists 
a unique (probabilistically) strong solution to the SDE for $\V{X}$ and \eqref{SDE-intro}.
\end{enumerate}
\end{theorem}

Precise formulations of Statements (1)--(4) above are given in
Thm.~\ref{thm-Ito-spin}, Thm.~\ref{thm-flow}, Prop.~\ref{prop-Markov}, Thm.~\ref{thm-str-Markov},
and Thm.~\ref{thm-str-sol}.

With the help of some earlier ideas from mathematical quantum field
theory we prove Part~(1) by using an {\em explicit formula for the solution} as an ansatz.
We proceed in four steps:

\smallskip

\noindent{\em Step 1.}
In Sect.~\ref{sec-proc} we first analyze certain {\em basic processes}, 
namely a complex-valued 
semi-martingale $(u_{\vxi,t}^V)_{t\in I}$, an $\HP$-valued semi-martingale 
$U^+$, and a family of $\HP$-valued semi-martingales $(U_{\tau,t}^-)_{t\in I}$,
indexed by $\tau\in I$. These processes admit explicit stochastic integral
representations involving $\omega$, $m_\ell$, $G_\ell$, and $\V{X}$.

\smallskip

\noindent{\em Step 2.}
Next, we treat the {\em scalar case}, i.e. the case where $L=1$, $F_j=0$,
in Sect.~\ref{sec-Ito}. Here the ansatz is suggested by
Hiroshima's formula \cite{Hiroshima1997} for the Fock space operator-valued Feynman-Kac integrand 
in NRQED without spin. Applying it to an exponential vector we obtain an expression involving
the basic processes whose stochastic differential can be computed by means of the
stochastic calculus in Hilbert spaces \cite{daPrZa2014,Me1982,MePe1980}.

\smallskip

\noindent{\em Step 3.}
After that we turn to the general {\em matrix-valued case}, i.e., $L\ge1$ with non-zero 
$\sigma_j$ and $F_j$. It shall eventually turn out that the semi-martingale solving 
\eqref{SDE-intro} can be written as $(\WW{\vxi,t}{V}\eta)_{t\in I}$ with an operator-valued map
$$
\WW{\vxi}{V}:I\times\Omega\longmapsto\LO(\FHR)
$$ 
such that, with probability one, all operators $\WW{\vxi,t}{V}$, $t\in I$, are given by 
norm-convergent series of $\LO(\FHR)$-valued time-ordered strong integrals whose
integrands are factorized into an annihilation, a preservation, a creation,
and a scalar part. (This result is stated precisely in Sect.~\ref{sec-spin}
where all relevant definitions can be found as well.)
In the third step of the proof we choose again an
exponential vector as initial condition $\eta$ and apply It\={o}'s formula to the partial
sums of $\WW{\vxi,t}{V}\eta$. The corresponding algebraic manipulations 
are presented in Sect.~\ref{sec-alg}.
Two additional technical lemmas are deferred to App.~\ref{app-cont-tn}.

\smallskip

\noindent{\em Step 4.}
In the final step, carried out in Sect.~\ref{sec-weights}, 
we analyze the convergence of the time-ordered integral series, pass to general
initial conditions $\eta:\Omega\to\wh{\dom}$, and verify that $\WW{\vxi}{V}\eta$ has 
continuous paths in $\wh{\dom}$ and solves \eqref{SDE-intro}.
The analysis reveals in particular that
$\PP$-a.s. the following two bounds hold, for all $t\in I$,
\begin{align}\label{bd-intro1}
\|\WW{\vxi,t}{V}\|&\le e^{\const t-\int_0^tV(\V{X}_s)\Id s},
\\\label{bd-intro2}
\int_0^t\|\Id\Gamma(\omega)^\eh\WW{\vxi,s}{V}\psi\|^2\Id s
&\le \const'e^{\const''t-2\int_0^t(V\wedge0)(\V{X}_s)\Id s}\|\psi\|^2,\quad\psi\in\FHR.
\end{align}
Furthermore, the following weighted BDG type inequality holds, for all $p\in\NN$, $t\in I$, and
$\fF_0$-measurable $\eta:\Omega\to\wh{\dom}$ with $\|M\eta\|\in L^{4p}(\PP)$,
\begin{align}\label{bd-intro3}
\EE\big[\sup_{s\le t}\|M\WW{\vxi,s}{0}\eta\|^{2p}\big]&
\le\const_{p,t}\EE[\|M\eta\|^{4p}]^\eh.
\end{align}
Here the inclusion of the weight $M$ necessitates the operator norm bounds on commutators of
functions of second quantized multiplication operators and field operators derived in App.~\ref{app-comm}. The {\em pointwise} operator norm bound \eqref{bd-intro1} is owing to the 
{\em skew-symmetry} of $iv_\ell(\vxi,\V{x})$; it is crucially used to deal with the terms 
$\Id\Gamma(m_\ell)^2$ contained in the weight $M$ in \eqref{bd-intro3}.

Parts~(2)--(4) of Thm.~\ref{thm-SDE-intro} are proven in Sect.~\ref{sec-Markov} after we have 
discussed the continuous dependence on initial conditions in Sect.~\ref{sec-initial}.

 \begin{remark}{\rm \label{rem-for-intro}
(1)  Assume in addition that $|m_\ell|\le c\omega$,
for all $\ell$ and some $c>0$. Then $\WW{\vxi,t}{V}:\Omega\to\LO(\FHR)$ 
is $\fF_t$-$\fB(\LO(\FHR))$-measurable and
$\PP$-almost separably valued. In fact, $\WW{\vxi,t}{V}$ is $\PP$-a.s. given by a 
{norm}-convergent series of {\em$\LO(\FHR)$-valued} time-ordered Bochner-Lebesgue integrals. 
This is shown in App.~\ref{app-meas}.

\smallskip

\noindent(2) In Sect.~\ref{sec-C0} we verify that $\WW{\vxi,t}{V}$ goes over to its adjoint under a 
time-reversal.
 }\end{remark}

The computations in Sect.~\ref{sec-Ito} reveal the relation of some well-known
constructions in mathematical quantum field theory going back to Nelson \cite{Nelson1973} to
the stochastic calculus in Hilbert spaces, perhaps for the first time.
Working with explicit solution formulas certainly comes at the price of
lengthy expressions and complicated algebraic manipulations in the matrix-valued case. 
It is, however, nice to see that folkloric tools of quantum field theory
like time-ordered integration and normal ordering can be rigorously controlled
in our model by means of the stochastic calculus.

Next, we give some brief remarks on related abstract results.

 \begin{remark}{\rm \label{rem-abstractum}
(1) Under our general hypotheses, the SDE \eqref{SDE-intro} is not covered by any
of the results we encountered in the literature on the semi-group or variational approach
to the solution theory for Hilbert space valued SDE; see, e.g., 
\cite{Chow2007,daPrZa2014,PrevotRoeckner2007}. 
At the same time, Thm.~\ref{thm-SDE-intro}(1) together with the bounds 
\eqref{bd-intro1}--\eqref{bd-intro3} provides more information on the solutions than the usual
textbook theorems on the existence of unique mild, (analytically) weak/strong, or variational solutions,
even if one ignores our explicit solution formulas. The non-applicability of the abstract results 
is due to the fact that the operator-valued
coefficients $\wh{H}(\vxi,\V{X}_s)$, $v_1(\vxi,\V{X}_s),\ldots,v_\nu(\vxi,\V{X}_s)$ 
appearing in the finite variation and local martingale parts of our linear SDE are all unbounded, 
mutually non-commuting, random, and time-dependent in general. 
Alternatively, we could consider the SDE for $\V{X}$ together with \eqref{SDE-intro}, thus obtaining
a non-linear system comprising time-dependent vector fields and unbounded, non-commuting,
non-constant operator-valued coefficients. Recall also that the SDE for $\V{X}$ contains an 
unbounded drift vector field with a non-integrable singularity at $\sup I$ when $\V{X}$ is a 
Brownian bridge. Altogether, these features already rule out all general results we found.
In addition, we are in a critical situation with regards to coercivity estimates.
For, in general, it is impossible to replace $\Id\Gamma(\omega)$ by ${M}$ on the right hand side 
of the bound
\begin{align}\label{coe1}
\Re\wh{H}(\vxi,\V{x})-\frac{1}{2}\sum_{\ell=1}^\nu \id_{\CC^L}\otimes v_\ell(\vxi,\V{x})^2
\ge(1-\delta)\id_{\CC^L}\otimes\Id\Gamma(\omega)-c_\delta,
\end{align}
valid for arbitrary $\delta\in(0,1)$
in the sense of quadratic forms on the form domain of $\wh{H}(\vxi,\V{x})$.
Since the form domain of $\wh{H}(\vxi,\V{x})$ is equal to the form domain of $\id_{\CC^L}\otimes M$, 
this shows that the coercivity condition required in the variational approach 
(see \cite[(H3) on p.~56]{PrevotRoeckner2007} or \cite[(D.3) on p.~178]{Chow2007}) is not satisfied 
unless $m_1=\dots=m_\nu=0$. 
Likewise, it is impossible in general to have a constant $>1/2$ in front of the sum in 
\eqref{coe1}, which would correspond to assumptions one encounters in the semi-group approach to 
the study of mild or weak solvability; 
compare \cite[\textsection6.5., in particular, Thm.~6.26]{daPrZa2014}.
(At first sight it seems that the result on existence of (analytically) strong solutions in
\cite[\textsection6.6]{daPrZa2014} could apply to fiber Hamiltonians in the Nelson model,
where $G_\ell=0$, $L=S=1$, $\sigma_1=-1$, $F_1$ is constant, and the relevant choice for $\V{X}$ 
is Brownian motion. But also in this
situation a related problem arises: the operator on the left hand side of \eqref{coe1} is not
equal to $\Id\Gamma(\omega)+\vp(F_1)$, which is the negative generator of a $C_0$-semi-group.
Rather it is equal to its restriction to $\wh{\dom}$, so that the condition in Hyp.~6.5(iii)
of \cite{daPrZa2014} is violated.)

\smallskip

\noindent
(2) Assume that $I$ and $V$ are bounded, all $m_\ell$ are zero, $\V{X}$ is a Brownian motion or a 
diffusion with a bounded drift vector field, and $\eta$ is square-integrable. Then, without additional
elaboration, the variational approach implies the existence of a unique variational solution 
$Y^{\mathrm{var}}$ to \eqref{SDE-intro} satisfying 
$$
\EE\big[\sup_{s\in I}\|Y^{\mathrm{var}}_s\|^2\big]
+\EE\Big[\int_I\big\|\Id\Gamma(\omega)^\eh Y^{\mathrm{var}}_s \|^2\Id s\Big]<\infty,
$$
which should be compared with \eqref{bd-intro1}--\eqref{bd-intro3}; see
\cite[Def.~4.2.1\&Thm~4.2.2]{PrevotRoeckner2007}.
Moreover, Prop.~4.3.3. in \cite{PrevotRoeckner2007} implies a Markov property of the variational 
solutions which is weaker than our corresponding result as it is not formulated in terms of
a stochastic flow. If we were not interested in explicit solution formulas, then we could of course
start out from these abstract results and try to complement them by a discussion proceeding along 
parts of our Sect.~\ref{sec-weights} to arrive at Thm.~\ref{thm-SDE-intro}(1) in the present 
special case.

\smallskip

\noindent
(3) The measurability of the operator-valued map $\WW{\vxi,t}{V}$
claimed in Rem.~\ref{rem-for-intro}(1) is proved by means of our explicit representation formulas;
see Rem.~\ref{rem-FK-intro}(4) for its implications.
We did not find analogous results in the literature on Hilbert space-valued SDE.
 }\end{remark}


\subsection{Feynman-Kac formulas and applications to spectral theory}\label{ssec-FK-intro}

Let us add the argument $[\V{X}]$ to $\WW{\vxi}{V}$ in case we fix a special choice of $\V{X}$.
If $G_\ell$ and $F_j$ are constant and $\V{X}=\V{B}$ is a Brownian motion starting at zero, 
then the solution operator $\WW{\vxi,t}{0}[\V{B}]$ appears in the Feynman-Kac formula for the
semi-group of the fiber Hamiltonian,
\begin{align}\label{FK-fiber-intro}
e^{-t\wh{H}(\vxi)}\psi=\EE\big[\WW{\vxi,t}{0}[\V{B}]\psi\big],\quad\psi\in\FHR.
\end{align}
Furthermore, set $\V{B}^{\V{x}}:=\V{x}+\V{B}$, let $\V{b}^{t;\V{y},\V{x}}$ be a semi-martingale 
realization of a Brownian bridge from $\V{y}\in\RR^\nu$ to $\V{x}\in\RR^\nu$ in time $t>0$, and let 
\begin{equation}\label{gaussian}
p_t(\V{x},\V{y}):=(2\pi t)^{-\nf{\nu}{2}}e^{-|\V{x}-\V{y}|^2/2t}
\end{equation}
be the standard Gaussian.
Choose $m_1=\dots=m_\nu=0$. Then the Feynman-Kac formula for the total Hamiltonian reads
\begin{align}\nonumber
(e^{-tH^V}\!\Psi)(\V{x})&=\EE\big[\WW{\V{0},t}{V}[\V{B}^{\V{x}}]^*\Psi(\V{B}^{\V{x}}_t)\big]
\\\label{FK-intro}
&=\int_{\RR^\nu}p_t(\V{x},\V{y})\EE\big[\WW{\V{0},t}{V}[\V{b}^{t;\V{y},\V{x}}]\big]\Psi(\V{y})\Id\V{y},
\end{align}
for a.e. $\V{x}$ and all $\Psi\in L^2(\RR^\nu,\FHR)$; see Sect.~\ref{sec-ext} for precise formulations
and suitable assumptions on $V$. For the reader's convenience we present detailed proofs of
\eqref{FK-fiber-intro} and \eqref{FK-intro} in Sect.~\ref{sec-ext} after we have verified, in 
Sect.~\ref{sec-C0}, that the right hand sides of \eqref{FK-fiber-intro} and of the first line in 
\eqref{FK-intro} define symmetric $C_0$-semi-groups; recall Rem.~\ref{rem-for-intro}(2).

In the next remark we briefly discuss which features of the above formulas are well-known and 
which are new. An exhaustive presentation of the earlier results can be found in \cite{LHB2011}. 
This book also contains detailed discussions of Feynman-Kac formulas in 
{\em semi}-relativistic QED (see also the recent article \cite{Hiroshima2014}), as well as results
and references on path integral representations for related models with
paths running through the infinite-dimensional state space of the radiation field.

 \begin{remark}{\rm \label{rem-FK-intro}
(1) For the standard model of NRQED without spin, the first identity in \eqref{FK-intro}  
is due to \cite{Hiroshima1997}. The case of a single spinning electron has been
treated more recently in \cite{HiroshimaLorinczi2008}, where the
sesqui-linear form associated with the semi-group is represented
as a {\em limit} of expectations of certain {regularized} Feynman-Kac type integrands.
In \cite{HiroshimaLorinczi2008}, the discrete spin degrees of freedom are not put into the target
space, but accounted for by an additional Poisson jump process.
In both papers the Feynman-Kac formula is derived by means of 
repeated Trotter product expansions and Nelson's ideas on the free Markov field \cite{Nelson1973}. 
While this approach is constructive, it does not reveal the
relation of the Feynman-Kac integrand to a SDE, which is the aim of the present paper.

\smallskip

\noindent(2) In the earlier literature, the Feynman-Kac formula for the
fiber Hamiltonian \eqref{FK-fiber-intro} has been deduced from the one for
the total Hamiltonian by inserting suitable peak functions
localized at the corresponding total momenta of the system \cite{Hiroshima2007}.
By starting out with the SDE \eqref{SDE-intro} for the generalized fiber Hamiltonian
one can avoid this detour and unify the discussion of fiber and total Hamiltonians.

\smallskip

\noindent
(3) In the matrix-valued case, our representation of the Feynman-Kac integrands in
\eqref{FK-fiber-intro} and \eqref{FK-intro} as a time-ordered
integral series is new, and \eqref{FK-intro} also covers the case of several electrons. 
Since this representation is normal ordered, it immediately
gives fairly explicit formulas for vacuum expectation values of the semi-group
and, more generally, matrix elements of the semi-group in coherent states in terms
of the basic processes; cf. Rem.~\ref{rem-for-W-exp-vec}. 

\smallskip

\noindent(4)
The second relation in \eqref{FK-intro} is new in all cases. 
We also remark that, for the expectation 
$\EE[\WW{\V{0},t}{V}[\V{b}^{t;\V{y},\V{x}}]]$ to be a well-defined $\LO(\FHR)$-valued 
Bochner-Lebesgue integral, the (by no means obvious) measurability property of $\WW{\vxi}{V}$
asserted in Rem.~\ref{rem-for-intro}(1) is a necessary prerequisite. If the extra condition
in Rem.~\ref{rem-for-intro}(1) is fulfilled, which is mostly the case in applications, 
then we can actually drop the vector $\psi$ in \eqref{FK-fiber-intro} and represent the semi-group of 
the fiber Hamiltonian by means of $\LO(\FHR)$-valued expectations.

\smallskip

\noindent(5) The assumptions on $\omega$, $m_\ell$, $G_\ell$, $F_j$, and $V$ used here 
are more general than in earlier papers on Feynman-Kac formulas 
in NRQED \cite{HiHi2010,Hiroshima1997,HiroshimaLorinczi2008}.

\smallskip

\noindent(6) The formulas \eqref{FK-fiber-intro} and \eqref{FK-intro} cover
the Nelson model as well; see \cite[Thm.~6.3]{LHB2011} and the references given there
for earlier results. While the Nelson model is scalar, we shall read off the precise expression for the
corresponding Feynman-Kac integrand from our formula for the matrix-valued case
in order to illustrate the latter; see Ex.~\ref{ex-Nelson} and in particular the last remark in it.
 }\end{remark}

Feynman-Kac formulas in NRQED and related models have various applications in their spectral 
theory. For instance, in NRQED, the existence of invariant
domains under semi-groups, diamagnetic inequalities, and the (essential) self-adjointness of the 
total Hamiltonian have been analyzed in \cite{Hiroshima2000esa,Hiroshima2002}; 
see \cite{Hiroshima2007} for similar results on fiber Hamiltonians.
In the scalar case, ergodic properties of the semi-group and Perron-Frobenius type 
theorems have been studied in \cite{Hiroshima2000}. Further
properties of ground state eigenvectors like, for instance, their spatial exponential decay are 
investigated in \cite{HiHi2010,Hiroshima2003}. Starting from Feynman-Kac representations, 
Gibbs measures associated with ground state 
eigenvectors have been constructed in \cite{BetzHiroshima2009,BHLMS2002}.
Again we refer to \cite{LHB2011} for a textbook presentation and numerous references.

By means of our results on the SDE \eqref{SDE-intro} one can add many more results to the list.
In fact, under suitable assumptions on $G_\ell$ and $F_j$, weighted BDG type estimations like
\eqref{bd-intro3} can be substantially pushed forward: we can consider higher powers of 
$t\Id\Gamma(\omega)$ instead of $M$ on the left hand side of \eqref{bd-intro3} and {\em drop}
$M$ on its right hand side at the same time, by properly exploiting the regularizing effect of
the term $e^{-t\Id\Gamma(\omega)}$ contained in $\WW{\vxi,t}{V}$. Using this one of us worked out
a semi-group theory for NRQED in the spirit of \cite{BHL2000,Carmona1979,Simon1982} which, in 
addition to the regularizing effects known from Schr\"odinger semi-groups with Kato decomposable 
potentials, takes into account the smoothing effect of $e^{-t\Id\Gamma(\omega)}$ on the position 
coordinates of the bosons; see \cite{Matte2015}.
In a second companion paper \cite{Matte2016} the second-named author discusses differentiability 
properties of the stochastic flow in weighted spaces, by employing our SDE and adapting 
strategies from \cite{Kunita1990}. Under suitable assumptions he infers smoothing properties 
of the semi-group, a Bismut-Elworthy-Li type formula, and smoothness of the operator-valued 
integral kernel.


\section{Definitions, assumptions, and examples}\label{sec-definitions}

\subsection{Operators in Fock space}\label{ssec-Fock}

In this subsection we introduce the bosonic Fock space $\sF$, which is the 
state space of the radiation field, and recall the definition of certain operators acting in it. 
$\sF$ is modeled over the one-boson Hilbert space 
\begin{align}\label{def-one-boson-space}
\HP:={\sF}^{(1)}:=L^2(\cM,\fA,\mu).
\end{align}
We assume that $\fA$ is generated
by a countable semi-ring $\fR$ such that $\mu\!\!\upharpoonright_{\fR}$ is $\sigma$-finite, 
which entails separability of $\HP$.
Let $n\in\NN$ with $n>1$, and let $\mu^n$ denote the $n$-fold product of $\mu$ defined on the
$n$-fold product $\sigma$-algebra $\fA^{n}$. Then
the $n$-boson subspace of $\sF$, denoted by ${\sF}^{(n)}$, is equal to the closed subspace 
in $L^2(\cM^n,\fA^{n},\mu^{n})$ of all elements $\psi^{(n)}$ satisfying
$\psi^{(n)}(k_1,\ldots,k_n)=\psi^{(n)}(k_{\pi(1)},\ldots,k_{\pi(n)})$, $\mu^n$-a.e., for every
permutation $\pi$ of $\{1,\ldots,n\}$. Finally,
\begin{align}\label{def-Fockspace}
{\sF}:=\CC\oplus\bigoplus_{n=1}^\infty{\sF}^{(n)}\ni\psi=(\psi^{(0)},\psi^{(1)},\ldots,\psi^{(n)},\ldots\:).
\end{align}
We shall make extensive use of the exponential vectors 
\begin{equation}\label{def-exp-vec}
\zeta(h):=\big(1,ih,\ldots,(n!)^\mh i^n h^{\otimes_n},\ldots\:\big)\in\sF,
\qquad h\in\HP,
\end{equation}
where as usual we identify $h^{\otimes_n}(k_1,\ldots,k_n)=h(k_1)\dots h(k_n)$, $\mu^n$-a.e. Let
\begin{equation}\label{def-E}
\sE[\mathfrak{v}]:=\big\{\zeta(h):\,h\in\mathfrak{v}\big\},
\qquad\sC[\mathfrak{v}]:=\mathrm{span}_{\CC}(\sE[\mathfrak{v}]),
\end{equation}
be the set of exponential vectors corresponding to one-boson states
in some subset $\mathfrak{v}\subset\HP$ and its complex linear hull, 
respectively. The set $\sE[\mathfrak{h}]$ is linearly independent
and $\sC[\mathfrak{v}]$ is dense in $\sF$, if $\mathfrak{v}$ is dense in $\HP$; see, e.g.,
\cite[Prop.~19.4 and Cor.~19.5]{Parthasarathy1992}.

Let $\tilde{\HP}$ be another $L^2$-space satisfying the same assumptions as $\HP$ and
$\tilde{\sF}$ the corresponding bosonic Fock space.
If $f\in\tilde{\HP}$ and $J:\HP\to\tilde{\HP}$ is an isometry, then we may define an isometry
$\sW(f,J):\sF\to\tilde{\sF}$ first on $\sE[\HP]$ by
\begin{align}\label{def-Weyl}
\sW(f,J)\zeta(h)&:=e^{-\|f\|^2/2-\SPn{f}{Jh}}\zeta(f+Jh),\quad h\in\HP,
\end{align}
then on $\sC[\HP]$ by linear extension, and finally on $\sF$ by isometric extension;
compare, e.g., \cite[\textsection20]{Parthasarathy1992}.
If $J$ is unitary, then $\sW(f,J)$ is unitary as well.
Writing $\Gamma(J):=\sW(0,J)$ and, in the case $\HP=\tilde{\HP}$, $\sW(f):=\sW(f,\id)$, we have
\begin{align}\label{exp-vec2}
\Gamma(J)\zeta(h)=\zeta(Jh),\quad
\sW(f)\zeta(h)&=e^{-\|f\|^2/2-\SPn{f}{h}}\zeta(f+h),\quad h\in\HP.
\end{align}
If $J:\HP\to\tilde{\HP}$ is a conjugate linear isometry,
then we obtain a conjugate linear isometry $\Gamma(J):\sF\to\tilde{\sF}$ by 
the first relation in \eqref{exp-vec2} and conjugate linear and isometric extension.
If the set $\sU(\HP)$ of unitary operators on $\HP$ is equipped with the strong topology, then the 
correspondence $\HP\times\sU(\HP)\ni(f,J)\mapsto\sW(f,J)$ is strongly continuous.
In particular, for $f\in\HP$ and every self-adjoint operator $T$ in $\HP$, there exist unique self-adjoint
operators $\vp(f)$ and $\Id\Gamma(T)$ in $\sF$ such that
\begin{align}
\sW(tf)&=e^{it\vp(f)},\quad\Gamma(e^{itT})=e^{it\Id\Gamma(T)},\quad t\in\RR.
\end{align}
More generally, for every $J\in\LO(\HP,\tilde{\HP})$ with $\|J\|\le1$, there is a unique operator
$\Gamma(J)\in\LO(\sF,\tilde{\sF})$ with $\|\Gamma(J)\|\le1$ satisfying the first relation in 
\eqref{exp-vec2}. If $A\in\LO(\tilde{\HP},\HP)$ with $\|A\|\le1$, then
$\Gamma(A)\Gamma(J)=\Gamma(AJ)$. If $T$ is a self-adjoint non-negative operator in
$\HP$, then $\Gamma(e^{-tT})=e^{-t\Id\Gamma(T)}$, $t\ge0$.

Let $f\in\HP$. Then the symbols $\ad(f)$ and $a(f)$ denote the usual (smeared) creation and 
annihilation operators in $\sF$ given by
\begin{align*}
(\ad(f)\psi)^{(n)}(k_1,\ldots,k_n)
&=n^\mh\sum_{\ell=1}^{n}f(k_\ell)\,\psi^{(n-1)}(\ldots,k_{\ell-1},k_{\ell+1},\ldots),
\\
(a(f)\psi)^{(n)}(k_1,\ldots,k_n)
&=(n+1)^\eh\int_{\cM}\ol{f(k)}\,\psi^{(n+1)}(k_1,\ldots,k_n,k)\,\Id\mu(k),
\end{align*}
$\mu^n$-a.e., for $n\in\NN$, and $(\ad(f)\psi)^{(0)}=0$ and $a(f)\zeta(0)=0$.
They are defined on their maximal domains and mutually adjoint to each other, 
$a(f)^*=\ad(f)$, $\ad(f)^*=a(f)$. For all $f,g\in\HP$, we have the following relations,
\begin{align}\label{for-vp}
&\vp(f)=\ad(f)+a(f),&[\vp(f),\vp(g)]=2i\Im\SPn{f}{g}\id,
\\\label{CCR}
&[a(f),a(g)]=[\ad(f),\ad(g)]=0,
&[a(f),\ad(g)]=\SPn{f}{g}\id,
\end{align}
on, e.g., $\dom(\Id\Gamma(\id))\supset\sC[\HP]$. For a self-adjoint operator $T$ in $\HP$,
we further have
\begin{align}\label{comm-dGamma-aad}
[a(f),\Id\Gamma(T)]&=a(Tf),\quad[\ad(f),\Id\Gamma(T)]=-\ad(Tf),
\\\label{CR-Segal3}
[\vp(f),\Id\Gamma(T)]&=i\vp(iTf),
\end{align}
on $\sC[\dom(T)]$, where $f\in\dom(T)$. For $f,h\in\HP$ and $g\in\dom(T)$, 
\begin{align}\label{exp-vec1}
&a(f)\zeta(h)=i\SPn{f}{h}\zeta(h),\quad\Id\Gamma(T)\zeta(g)=i\ad(Tg)\zeta(g).
\end{align}

Exponential vectors are analytic, as we shall see in the following lemma. 
We recall that a map $F:\sK\to\sK'$
from one complex Hilbert space $\sK$ into another $\sK'$ is analytic, if and only if it is Fr\'{e}chet
differentiable. In this case the Taylor series
$F(y+h)=\sum_{n=0}^\infty(n!)^{-1}F^{(n)}(y)(h^{\otimes_n})$, where
$F^{(n)}(y)$ is the $n$-th Fr\'{e}chet derivative of $F$ at $y$ interpreted as a linear map from 
$\sK^{\otimes_n}$ to $\sK'$, converges absolutely, for all $y,h\in\sK$;
see, e.g., \cite[\textsection III.3.3]{HillePhillips1957}
for more information on analytic maps. 

\begin{lemma}\label{lem-Taylor-exp}
The map $\HP\ni h\mapsto\zeta(h)\in\sF$ is analytic and
\begin{align}\label{exp-vec1b}
&\zeta^{(n)}(h)(f_1\otimes\dots\otimes f_n)=i^n\ad(f_1)\dots\ad(f_n)\zeta(h).
\end{align}
for all $h,f_1,\ldots,f_n\in\HP$.  For all $n\in\NN_0$ and $f,h\in\HP$, we have the error bound
\begin{align}\label{Taylor-exp1}
\Big\|\zeta(h+f)-\sum_{\ell=0}^n\frac{i^\ell}{\ell!}\ad(f)^\ell\zeta(h)\Big\|
&\le e^{\|h\|^2}\sum_{\ell=n+1}^\infty\frac{2^{\nf{\ell}{2}}\|f\|^\ell}{(\ell!)^\eh}.
\end{align}
\end{lemma}

{\proof}
The proof is a straightforward exercise starting from the observation that
$\ad(f)^\ell h^{\otimes_{n-\ell}}=(\ell!)^\eh{n\choose\ell}^\eh
\cS_n(f^{\otimes_\ell}\otimes h^{\otimes_{n-\ell}})$, where $\cS_n$ is the orthogonal projection onto 
$\sF^{(n)}$ in $L^2(\cM^n,\fA^n,\mu^n)$.
\qed

\begin{lemma}\label{lem-APC-fact}
For all $f\in\HP$ and $z\in\CC$, the series $\exp\{z\ad(f)\}$, $\exp\{za(f)\}$, and $\exp\{z\vp(f)\}$ are 
strongly convergent on the normed space $\sC[\HP]$ and map it into itself.
For $A,B\in\LO(\HP,\tilde{\HP})$ with $\|A\|,\|B\|\le1$, $g\in\tilde{\HP}$, $h\in\HP$, and $z\in\CC$, 
\begin{align}\nonumber
\Gamma(B^*)\exp\{z\vp(g)\}\Gamma(A)\zeta(h)
=e^{z^2\|g\|^2/2+iz\SPn{g}{Ah}}\zeta(B^*Ah-izB^*g)\,
\\\label{APC-fact}
=e^{z^2\|g\|^2/2}\exp\{z\ad(B^*g)\}\Gamma(B^*A)\exp\{za(A^*g)\}\zeta(h).
\end{align}
\end{lemma}

{\proof}
The first statement follows from \eqref{exp-vec1}, \eqref{Taylor-exp1}, and 
the following consequence of \eqref{exp-vec2},
$i^n\vp(f)^n\zeta(h)=\frac{\Id^n}{\Id t^n}\big|_{t=0}e^{-t^2\|f\|^2/2-t\SPn{f}{h}}\zeta(h+tf)$, $h\in\HP$.
Together with \eqref{exp-vec2}, \eqref{exp-vec1}, and \eqref{Taylor-exp1} it implies
the second equality in \eqref{APC-fact}. 
Let $z=it$ with $t\in\RR$. Then $\exp\{it\vp(g)\}=\sW(tg)$ on $\sC[\tilde{\HP}]$ and 
the first equality in \eqref{APC-fact} follows from \eqref{exp-vec2}.
For general $z\in\CC$, the first equality in \eqref{APC-fact} is obtained by analytic continuation.
(See \cite[Thm.~3.11.5]{HillePhillips1957} for analytic continuation of vector-valued functions.)
\qed

It is helpful to keep in mind that, if $\vk$ is a real-valued measurable function on $\cM$
and if the maximal operator of multiplication with $\vk$ is denoted
by the same symbol, then $\Id\Gamma(\vk)$ is again a
self-adjoint maximal multiplication operator in ${\sF}$
given by $\Id\Gamma(\vk)\zeta(0)=0$ and, for $n\in\NN$,
\begin{align*}
(\Id\Gamma(\vk)\,\psi)^{(n)}(k_1,\ldots,k_n)
&=\sum_{\ell=1}^n\vk(k_\ell)\,\psi^{(n)}(k_1,\ldots,k_n),\quad\psi\in\dom(\Id\Gamma(\vk)).
\end{align*}
For instance, this remark is useful in order to derive the basic relative bounds
\begin{align}\label{rb-a}
\|a(f)^n\,\psi\|&\le\|\vk^\mh f\|^n\,\|\Id\Gamma(\vk)^{\nf{n}{2}}\,\psi\|,\quad
n\in\NN,
\\\label{rb-ad}
\|\ad(f)\|&\le\|(1+\vk^{-1})^\eh f\|\,\|(\Id\Gamma(\vk)+1)^\eh\,\psi\|,
\\\label{rb-vp1}
\|\vp(f)\,\psi\|
&\le2^\eh\|(1+\vk^{-1})^\eh f\|\,\|(\Id\Gamma(\vk)+1)^\eh\,\psi\|,
\\\label{rb-vp2}
\|\vp(f)^2\,\psi\|&\le6\|(1+\vk^{-1})^\eh f\|^2\,\|(\Id\Gamma(\vk)+1)\,\psi\|,
\end{align}
where we assume that $\vk>0$, $\mu$-a.e., and, in each line, $f$ and $\psi$ are chosen
such that all norms on its right hand side are well-defined.
The bound in \eqref{rb-a} follows from a standard
exercise using a weighted Cauchy-Schwarz inequality, Fubini's theorem,
and a little combinatorics. The other bounds are consequences
of \eqref{CCR} and \eqref{rb-a}. Another consequence of \eqref{for-vp} and \eqref{rb-a} is 
\begin{align}\label{qfb-vp}
\Id\Gamma(\vk)+\vp(f)\ge-\|\vk^\mh f\|^2\quad\text{on}\;\,\fdom(\Id\Gamma(\vk)).
\end{align}

Given a row vector of boson wave functions,
$\V{f}=(f_1,\ldots,f_\nu)$, we set $\vp(\V{f}):=(\vp(f_1),\ldots,\vp(f_\nu))$, and we shall
employ an analogous convention for the creation and annihilation operators. 


\subsection{Generalized fiber Hamiltonians}\label{ssec-models}

\noindent
Next, we add (generalized) spin degrees of freedom to our
model by tensoring the Fock space with ${\CC^L}$, for some fixed $L\in\NN$. We call 
\begin{equation}\label{def-FHR}
\FHR:=\CC^L\otimes\sF
\end{equation}
the \emph{fiber Hilbert space}, a notion motivated by Ex.~\ref{ex-G}(4). 
We assume that, for some $S\in\NN$,
$$
\sigma_1,\ldots,\sigma_S\in\LO({\CC^L})
$$
are hermitian matrices with $\|\sigma_j\|\le1$. 
Most of the time we regard them as operators on $\FHR$
by identifying $\sigma_j\equiv\sigma_j\otimes\id_{\sF}$.
We shall write $\vsigma:=(\sigma_1,\ldots,\sigma_S)$ and
$\vsigma\cdot\V{v}:=\sigma_1\,v_1+\dots+\sigma_S\,v_S$, where
$\V{v}=(v_1,\ldots,v_{S})$ is a vector of complex numbers or suitable operators.

Furthermore, we fix some $\nu\in\NN$ and collect the coefficient functions appearing in the
SDE \eqref{SDE-intro} in row vectors,
$\V{G}_{\V{x}}=(G_{1,\V{x}},\ldots,G_{\nu,\V{x}})\in\HP^\nu$ and
$\V{F}_{\V{x}}=(F_{1,\V{x}},\ldots,F_{S,\V{x}})\in\HP^S$,
parametrized by $\V{x}=(x_1,\ldots,x_\nu)\in\RR^\nu$.
We will exclusively work under the following standing hypothesis:

\begin{hyp}{\rm \label{hyp-G}
{\rm(1)} $\omega:\cM\to\RR$ and $\V{m}:\cM\to\RR^\nu$ 
are measurable such that $\omega$ is $\mu$-a.e. strictly positive. 
We introduce the following dense subspace of $\HP$,
\begin{equation}\label{def-fd}
\mathfrak{d}:=\dom(\omega+\tfrac{1}{2}\V{m}^2).
\end{equation}
{\rm(2)} The map 
$\V{x}\mapsto\V{G}_{\V{x}}$ is in $C^2(\RR^\nu,\HP^{\nu})$,
and $\V{x}\mapsto \V{F}_{\V{x}}\in\HP^S$ is globally Lipschitz continuous on $\RR^\nu$.
The components of $\V{G}_{\V{x}}$, $\partial_{x_\ell}\V{G}_{\V{x}}$, $\V{F}_{\V{x}}$,  
and $i\V{m}\cdot\V{G}_{\V{x}}$ belong to 
\begin{equation}\label{def-fk}
\mathfrak{k}:=L^2\big(\cM,\fA,[\omega^{-1}+(\omega+\tfrac{1}{2}\V{m}^2)^2]\mu\big),
\end{equation}
and the following map is continuous and bounded,
$$
\RR^\nu\ni\V{x}\longmapsto
(\V{G}_{\V{x}},\partial_{x_1}\V{G}_{\V{x}},\ldots,\partial_{x_\nu}\V{G}_{\V{x}},\V{F}_{\V{x}},
i\V{m}\cdot\V{G}_{\V{x}})\in\mathfrak{k}^{(\nu+1)\nu+S+1}.
$$
\noindent{\rm(3)}
There exists a conjugation $C\colon\HP\to\HP$, i.e., an anti-linear isometry with $C^2=\id_{\HP}$,
such that, for all $t\ge0$, $\V{x}\in\RR^\nu$, $\ell=1,\ldots,\nu$, and $j=1,\ldots,S$,
\begin{align}\label{hyp-sym2}
[C,e^{-t\omega+i\V{m}\cdot\V{x}}]=0,\quad G_{\ell,\V{x}},F_{j,\V{x}}\in\HP_C:=\{f\in\HP:\,Cf=f\}.
\end{align}
 }\end{hyp}

As a consequence of \eqref{hyp-sym2} we also have
\begin{align}\label{sym-q}
q_{\V{x}}:=\Div_{\V{x}}\V{G}_{\V{x}}\in\HP_C,\quad i\V{m}\cdot\V{G}_{\V{x}}\in\HP_C,\quad
\breve{q}_{\V{x}}:=\tfrac{1}{2}q_{\V{x}}
-\tfrac{i}{2}\V{m}\cdot\V{G}_{\V{x}}\in\HP_C,
\end{align}
for all $\V{x}\in\RR^\nu$.
In view of \eqref{hyp-sym2} we observe that $C$ is isometric on $\mathfrak{k}$ as well,
and we introduce the completely real subspaces
\begin{equation}\label{def-dCkC}
\mathfrak{d}_C:=\big\{f\in\mathfrak{d}:\,Cf=f\big\},\qquad
\mathfrak{k}_C:=\big\{f\in\mathfrak{k}:\,Cf=f\big\},
\end{equation}
and, noticing that $\Gamma(-C)$ is a conjugation on $\sF$,
\begin{align}\label{conj-F}
\sF_{C}:=\big\{\psi\in\sF:\,\Gamma(-C)\,\psi=\psi\big\}.
\end{align}
Then the real linear hull
$\mathrm{span}_\RR\sE[\mathfrak{d}_C]=\sC[\mathfrak{d}_C]\cap\sF_C$
is dense in  $\sF_C$; see, e.g., \cite[Cor.~19.5]{Parthasarathy1992}. Since $\sF=\sF_C+i\sF_C$, 
we see that $\CC^L\otimes\sC[\mathfrak{d}_C]$ is dense in $\FHR$.

Concerning the electrostatic potential $V$, we introduce the following standing hypothesis:

\begin{hyp}{\rm \label{hyp-V}
$V:\RR^{\nu}\to \RR$ is locally integrable.
 }\end{hyp}

To treat the total Hamiltonian and the fiber Hamiltonians in a unified way, we
introduce a mathematical model {Hamiltonian} in the next definition.
We again use the notation introduced in \eqref{def-whD-intro}.
Henceforth, we shall also employ a common, self-explanatory notation involving tuples of
operators and formal scalar products between them; simply compare the formulas
in Def.~\ref{defn-gen-Ham} with \eqref{v-intro}--\eqref{def-whD-intro} to interprete them correctly.
The various terms in \eqref{def-fib-Ham-scalar} are well-defined on the
given domains in view of \eqref{rb-vp1}, \eqref{rb-vp2}, and Hyp.~\ref{hyp-G}.

\begin{definition}[{\bf Generalized fiber Hamiltonian}]\label{defn-gen-Ham}
Let $\vxi,\V{x}\in\RR^\nu$ and 
\begin{equation}\label{def-vecv}
\V{v}(\vxi,\V{x}):=\vxi-\Id\Gamma(\V{m})-\vp(\V{G}_{\V{x}}).
\end{equation}
We introduce a \emph{generalized fiber Hamiltonian} $\wh{H}^V(\vxi,\V{x})$ in $\FHR$, 
defined on the domain of definition $\wh{\dom}$ by
\begin{align}\label{def-fib-Ham-spin}
\wh{H}^V(\vxi,\V{x})
&:=\id_{\CC^L}\otimes\wh{H}^V_{\scal}(\vxi,\V{x})-\vsigma\cdot\vp(\V{F}_{\V{x}}),
\end{align}
whose scalar part is defined on the domain of definition $\dom(M)$ by
\begin{align}
\wh{H}^V_{\scal}(\vxi,\V{x})\nonumber
&:=\tfrac{1}{2}\V{v}(\vxi,\V{x})^2-\tfrac{i}{2}\,\vp(q_{\V{x}})+\Id\Gamma(\omega)+V(\V{x})
\\\nonumber
&=
\tfrac{1}{2}(\vxi-\Id\Gamma(\V{m}))^2-\vp(\V{G}_{\V{x}})\cdot(\vxi-\Id\Gamma(\V{m}))
+\tfrac{1}{2}\vp(\V{G}_{\V{x}})^2
\\\label{def-fib-Ham-scalar}
&\quad-\tfrac{i}{2}\,\vp(q_{\V{x}})-\tfrac{i}{2}\vp(i\V{m}\cdot\V{G}_{\V{x}})
+\Id\Gamma(\omega)+V(\V{x}).
\end{align}
If $\V{G}$ and $\V{F}$ are $\V{x}$-independent, then we denote
$\wh{H}^0(\vxi,\V{x})$ simply by $\wh{H}(\vxi)$.
\end{definition}

To get the equality in \eqref{def-fib-Ham-scalar} we used 
the following consequence of \eqref{comm-dGamma-aad} and a simple approximation argument,
$$
[\Id\Gamma(\V{m}),\vp(\V{g})]=\ad(\V{m}\cdot\V{g})-a(\V{m}\cdot\V{g})
=-i\vp(i\V{m}\cdot\V{g})\quad\text{on $\dom(M)$}.
$$

In the next proposition we collect some essentially well-known basic properties
of the generalized fiber Hamiltonians; see \cite{Hiroshima2007} where the essential self-adjointness
of fiber Hamiltonians is proved via the method of invariant domains and Feynman-Kac formulas.
The existence of invariant domains is, however, a deeper result
than the essential self-adjointness itself which turns out to be a consequence of the relations and 
bounds recalled in Subsect.~\ref{ssec-Fock}. To illustrate this we present a short proof of 
Prop.~\ref{prop-esa} in App.~\ref{app-esa}. We abbreviate
\begin{equation}\label{def-Maxi}
{M}_{a}(\vxi):=\id_{\CC^L}\otimes\big(\tfrac{1}{2}(\vxi-\Id\Gamma(\V{m}))^2+a\Id\Gamma(\omega)\big),
\quad\vxi\in\RR^\nu,\,a\ge1.
\end{equation}
Obviously, ${M}_{a}(\vxi)$ is self-adjoint on $\wh{\dom}$.
Let $\mathfrak{a}_C$ denote a dense set of analytic vectors for $\tfrac{1}{2}\V{m}^2+\omega$
in $\HP_C$. With the help of the semi-analytic vector theorem one can easily show that $M_a(\vxi)$ is
essentially self-adjoint on $\CC^L\otimes\sC[\mathfrak{a}_C]$ and, hence, 
on $\CC^L\otimes\sC[\mathfrak{d}_C]$. 

\begin{proposition}\label{prop-esa} 
Let $\vxi,\V{x}\in\RR^\nu$. Then $\wh{H}^0(\vxi,\V{x})$ is well-defined and closed on its 
domain $\wh{\dom}$ and, for all $\ve>0$, there exists $a\ge1$ such that, for all $\psi\in\wh{\dom}$,
\begin{align}\label{rb-whH}
\big\|\big(\wh{H}^0(\vxi,\V{x})-{M}_{1}(\vxi)\big)\psi\big\|
&\le\ve\|{M}_{a}(\vxi)\psi\|+\const(\ve)\|\psi\|,
\\\label{ub-whH}
\|\wh{H}^0(\vxi,\V{x})\psi\|&\le\const(\|{M}_{1}(\vxi)\psi\|+\|\psi\|).
\end{align}
The subspace $\CC^L\otimes\sC[\mathfrak{d}_C]$ and, more generally, every core of $M_1(\V{0})$
is a core of $\wh{H}^0(\vxi,\V{x})$.
If $q_{\V{x}}=0$, then $\wh{H}^0(\vxi,\V{x})$ is self-adjoint on $\wh{\dom}$.
\end{proposition}


\subsection{Probabilistic objects and assumptions on the driving process}\label{sec-prel}

\noindent
In the whole article, $\zeit$ denotes a time horizon, which is either equal to $[0,\infty)$ or
to $[0,\mathcal{T}]$ with $\mathcal{T}>0$, and $\BB=(\Omega,\filt,(\filt_t)_{t\in I},\mathbb{P})$ 
is some stochastic basis satisfying the usual assumptions.
This means that $(\Omega,\fF,\PP)$ is a complete probability space, 
the filtration $(\filt_t)_{t\in I}$ is right-continuous, and $\fF_0$ contains all $\PP$-zero sets.
The letter $\EE$ denotes expectation with respect to $\PP$ and, for any sub-$\sigma$-algebra
$\fH$ of $\fF$, the symbol $\EE^{\fH}$ denotes the corresponding conditional expectation.
For $s\in I$, we shall sometimes consider the time-shifted basis
\begin{equation}\label{shifted-stoch-basis}
\BB_s:=\big(\Omega,\filt,(\filt_{s+t})_{t\in I^s},\PP\big),\qquad I^s:=\{t\ge0:s+t\in I\},
\end{equation}
so that $I=I^0$. If $\sK$ is a real separable Hilbert space, then we denote the space of
all {\em continuous} $\sK$-valued semi-martingales defined on $I^s$ by $\mathsf{S}_{I^s}(\sK)$.
The bold letter $\V{B}\in\mathsf{S}_{I}(\RR^{\nu})$ always denotes a $\nu$-dimensional
$\BB$-Brownian motion (with covariance matrix $\id_{\RR^\nu}$) defined on $I$ and,
for all $0\le s<t\in I$, the $\sigma$-algebra $\fF_{s,t}$ is the completion of the $\sigma$-algebra
generated by all increments $\V{B}_r-\V{B}_s$ with $r\in[s,t]$.
If $X$ is any process on $I^s$ with values in a separable Hilbert space $\sK$, 
then $X_\bullet:\Omega\to\sK^{I^s}$ denotes the corresponding path map
given by $(X_{\bullet}(\vgamma))(t):=X_t(\vgamma)$, $t\in I^s$, $\vgamma\in\Omega$.

With this we introduce a third (and last) standing hypothesis on a
$\RR^\nu$-valued process $\V{X}$ which will enter into all our constructions
and play the role of the driving process in the SDE studied in this paper.

\begin{hyp}{\rm \label{hyp-B}
The bold letter $\V{X}\in\mathsf{S}_{I}(\RR^{\nu})$ denotes a
semi-martingale with respect to $\BB$ solving the It\={o} equation
\begin{align}\label{Ito-eq-X}
\V{X}_t=\V{q}+\V{B}_t+\int_0^t\V{\beta}(s,\V{X}_s)\,\Id s\,,\quad t\in
[0,\sup I)\,,
\end{align}
for some $\fF_0$-measurable $\V{q}:\Omega\to\RR^\nu$.
When it becomes relevant, we shall indicate the dependence of $\V{X}$
on $\V{q}$ by writing $\V{X}^{\V{q}}$ for the solution of \eqref{Ito-eq-X}.
We assume that the drift vector field $\V{\beta}\in C([0,\sup I)\times\RR^\nu,\RR^\nu)$
in \eqref{Ito-eq-X} is such that the following holds:
\begin{enumerate}
\item[{\rm(1)}] For all $s\in[0,\sup I)$ and every 
$\fF_s$-measurable $\V{q}:\Omega\to\RR^\nu$ the SDE (with underlying basis $\BB_s$)
\begin{align}\label{Ito-eq-sX}
{}^s\!\V{X}_{t}=\V{q}+\V{B}_{s+t}-\V{B}_s+\int_0^t\V{\beta}(s+r,{}^s\!\V{X}_{r})\,\Id r,
\quad t\in[0,\sup I^s),
\end{align} 
has a global solution ${}^s\!\V{X}^{\V{q}}\in\mathsf{S}_{I^s}(\RR^\nu)$
which is unique up to indistinguishability.
\item[{\rm(2)}] 
\eqref{Ito-eq-X} admits a stochastic flow, i.e., there is a family $(\vXi_{s,t})_{0\le s\le t\in I}$ 
of maps $\vXi_{s,t}:\RR^\nu\times\Omega\to\RR^\nu$, such that
\begin{enumerate}
\item[{\rm(a)}]
$\V{x}\mapsto\vXi_{s,s+\bullet}(\V{x},\vgamma)$ is continuous from
$\RR^\nu$ into $C(I^s,\RR^\nu)$, for all $s\in I$ and $\vgamma\in\Omega$;
\item[{\rm(b)}] 
$(\tau,\V{x},\vgamma)\mapsto\vXi_{s,\tau}(\V{x},\vgamma)$
is $\fB([s,t])\otimes\fB(\RR^\nu)\otimes\fF_{s,t}$-measurable for fixed $0\le s<t\in I$;
\item[{\rm(c)}] if $s\in I$, then $\vXi_{s,s}(\V{x},\vgamma)=\V{x}$, for all
$(\V{x},\vgamma)\in \RR^\nu\times\Omega$, and, 
if $\V{q}:\Omega\to\RR^\nu$ is $\fF_s$-measurable, then 
\begin{equation*}
\vXi_{s,s+\bullet}(\V{q}(\vgamma),\vgamma)={}^s\!\V{X}_{s+\bullet}^{\V{q}}(\vgamma)
\quad\text{on $I^s$, for $\PP$-a.e. $\vgamma$.}
\end{equation*}
\end{enumerate}
\item[{\rm(3)}]
For all $\kappa\ge1$ and all {\em bounded} $\fF_0$-measurable
$\mathfrak{q}:\Omega\to[0,\infty)$, it holds
\begin{align}\label{hyp-Y2}
\forall\;t\in I:\quad&\int_{0}^t(1\wedge(\sup I-s)^\kappa)\,
\EE\big[\sup_{|\V{q}|\le\mathfrak{q}}|\V{\beta}(s,\V{X}_s^{{\V{q}}})|^{2\kappa}\big]\,\Id s<\infty,
\end{align} 
where the supremum under the expectation is taken over all $\fF_0$-measurable functions
${\V{q}}:\Omega\to\RR^\nu$ with $|\V{q}|\le\mathfrak{q}$.
\item[{\rm(4)}] 
There exist $p\ge2$ with $p>\nu$ and an increasing function $L:I\to[0,\infty)$ such that
\begin{equation}\label{Lip-EX}
\EE\big[|\vXi_{0,t}(\V{x},\cdot)-\vXi_{0,t}(\V{y},\cdot)|^p\big]
\le L(t)^p\,|\V{x}-\V{y}|^p,\quad\V{x},\V{y}\in\RR^\nu,\;t\in I.
\end{equation}
\end{enumerate}
Finally, we assume that
\begin{align}\label{hyp-XV}
\PP\{V(\V{X}_{\bullet})\in L^1_{\mathrm{loc}}(\zeit)\}=1.
\end{align}
 }\end{hyp}

 \begin{remark}{\rm 
(1) Of course, \eqref{hyp-XV} imposes no restriction on $\V{X}$, if $V\in C(\RR^\nu,\RR)$.

\smallskip

\noindent(2) Notice that the time-dependent vector field $\V{\beta}$
may be unbounded at $\cT$, if $I$ is finite, and that the validity of
the integral equations \eqref{Ito-eq-X} and \eqref{Ito-eq-sX} 
is required only strictly before $\cT$ and $\cT-s$, respectively.
Then the technical condition \eqref{hyp-Y2} says that the possible
singularity of $\V{\beta}$ at $\cT$ is not too strong in a certain sense.
Nevertheless the paths of $\V{X}$ and ${}^s\!\V{X}$ are assumed 
to be continuous {\em on all of} $I$ and $I^s$, respectively.
 
\smallskip

\noindent(3) In many parts of the paper
we won't use all properties of $\V{X}$ imposed in Hyp.~\ref{hyp-B}. In fact,
the arguments of Sects.~\ref{sec-proc} and~\ref{sec-Ito}
hold true as soon as $\V{X}$ is a continuous $\RR^\nu$-valued semi-martingale on $I$
with quadratic covariation $\id_{\RR^\nu}$ satisfying \eqref{hyp-XV}.
The technical extra condition \eqref{hyp-Y2} will be used in Sects.~\ref{sec-alg} and~\ref{sec-weights}
to prove the statements in Sect.~\ref{sec-spin}, which in turn are used to
derive the results of Sects.~\ref{sec-initial}--\ref{sec-ext}.
The continuity properties of the flow $\vXi$ and in particular the
$L^p$-Lipschitz condition \eqref{Lip-EX} are exploited in
Sect.~\ref{sec-initial} whose results are used in Sects.~\ref{sec-Markov}--\ref{sec-ext}.
 }\end{remark}

\begin{example}\label{ex-hyp-B}{\rm 
(1) The most important example of a process satisfying Hyp.~\ref{hyp-B}
with an infinite time horizon $I=[0,\infty)$ is the trivial choice $\V{X}=\V{q}+\V{B}$.

\smallskip

\noindent(2)
If $I=[0,\infty)$ and $\V{\beta}\in C(I\times\RR^\nu,\RR^\nu)$ is such that
$|\V{\beta}(s,\V{x})-\V{\beta}(s,\V{y})|\le \ell(t)|\V{x}-\V{y}|$,
$0\le s\le t$, $\V{x},\V{y}\in\RR^\nu$, with some increasing
function $\ell:I\to(0,\infty)$, then the validity
of all conditions imposed in Hyp.~\ref{hyp-B} follows from standard textbook results;
see, e.g., \cite[Chap.~6]{HackenbrochThalmaier1994}.

\smallskip

\noindent(3) The most important example with a finite time horizon $I=[0,\cT]$
is a semi-martingale realization of a Brownian bridge from an $\fF_0$-measurable 
$\V{q}:\Omega\to\RR^\nu$ to $\V{y}\in\RR^\nu$ in time $\cT$. 
The  definition of such a process is recalled in Sect.~\ref{sec-C0}. 
In App.~\ref{app-bridge} we shall verify that Brownian bridges actually fulfill Hyp.~\ref{hyp-B}.
 }\end{example}

For later reference, we state an It\={o} formula suitable for our applications
in the next proposition. The construction of the Hilbert space-valued stochastic integrals with 
integrator $\V{X}$ appearing in its statement and in the following sections is standard and we refer 
readers who wish to recall that construction to the textbooks \cite{daPrZa2014,Me1982,MePe1980}.


\begin{proposition}\label{prop-Ito}
Let $\sY$ be a real separable Hilbert space and $\sK$ be a real or complex separable Hilbert space.
Let $\V{A}:I\times\Omega\to\LO(\RR^\nu,\sY)$ and $\wt{A}:I\times\Omega\to\sY$ be predictable such
that, for every $t\in I$, $\|\V{A}_\bullet\|$ is $\PP$-a.s. square-integrable on $[0,t]$
and $\wt{A}_{\bullet}$ is $\PP$-a.s. Bochner-Lebesgue integrable on $[0,t]$. 
Finally, let $\eta:\Omega\to\sY$ be $\fF_0$-measurable. Set
\begin{align}\label{Z-Ito}
Z_\bullet&:=\eta+\int_0^\bullet\V{A}_s\Id\V{X}_s+\int_0^\bullet \wt{A}_{s}\Id s.
\end{align}
Assume that the partial derivatives $\partial_sf$, $d_yf$, and $d_y^2f$ of $f:I\times\sY\to\sK$ exist
and are uniformly continuous on every bounded subset of $I\times\sY$. Then $(f(t,Z_t))_{t\in I}$ is a
$\sK$-valued continuous semi-martingale and $\PP$-a.s. satisfies
\begin{align}\label{Ito}
f(t,Z_t)&=f(0,\eta)+\int_0^t\partial_sf(s,Z_s)\Id s+\int_0^td_yf(s,Z_s)\wt{A}_{s}\Id s
\\\nonumber
&\quad+\int_0^td_yf(s,Z_s)\V{A}_s\Id\V{X}_s
+\frac{1}{2}\int_0^td_y^2f(s,Z_s)\V{A}_s^{\otimes_2}\Id s,\;\;t\in[0,\sup I).
\end{align} 
\end{proposition}

{\proof}
If $\phi\in\sK$ and we replace $f$ by $\Re\SPn{\phi}{f}$ or $\Im\SPn{\phi}{f}$, then the claim follows 
from \cite[Thm.~4.32]{daPrZa2014}. The general case follows by applying this observation for every 
$\phi$ in a countable dense subset of $\sK$ and using that all 
so-obtained derivatives and (stochastic) integrals commute 
(up to indistinguishability) with $\Re$, $\Im$, and $\SPn{\phi}{\cdot}$.
\qed 

The next example will be applied with $\sY=\FHR=\sF_C^\nu+i\sF_C^\nu$.

\begin{example}{\rm \label{ex-Ito-SP}
Assume that $Z$ is given as in Prop.~\ref{prop-Ito} with the only exception that $\sY$ is now a
complex Hilbert space which can be written as $\sY=\sY_\RR+i\sY_\RR$, for some completely
real subspace $\sY_\RR\subset\sY$.
Then $\|Z\|^2$ is a continuous real semi-martingale and $\PP$-a.s. satisfies
\begin{align*}
\|Z_t\|^2&=\|\eta\|^2+\int_0^t2\Re\SPn{Z_s}{\wt{A}_{s}}\Id s+\int_0^t2\Re\SPn{Z_s}{\V{A}_s}\Id\V{X}_s
+\int_0^t\|\V{A}_s\|^2\Id s,
\end{align*}
for all $t\in [0,\sup I)$. In fact, $Z$ can be uniquely written as $Z=Z_1+iZ_2$, with $\sY_\RR$-valued
processes $Z_j$, $j=1,2$, given by formulas analogous to \eqref{Z-Ito}. 
Since $\|\phi+i\psi\|^2=\|\phi\|^2+\|\psi\|^2$, for all $\phi,\psi\in\sY_\RR$, we may apply
Prop.~\ref{prop-Ito} to $\|Z_1\|^2+\|Z_2\|^2$ and obtain the asserted formula after some trivial
rearrangements.
 }\end{example}

For later reference, we also recall a substitution rule sufficient for our purposes.
For remarks on its proof see, e.g., \cite[\textsection26.4]{Me1982}.

\begin{proposition}\label{prop-stoch-calc}
Assume that $Z$ is given as in Prop.~\ref{prop-Ito} with a finite dimensional $\sY$ and
let $D$ be a uniformly bounded, predictable $\LO(\sY,\CC)$-valued process on $I$. Then
\begin{align*}
\int_0^\bullet D_s\Id Z_s=\int_0^\bullet D_s\V{A}_s\Id\V{X}_s+\int_0^\bullet D_s\wt{A}_s\Id s,
\quad\text{$\PP$-a.s.}
\end{align*}
\end{proposition}

The following dominated convergence theorem for stochastic integrals shall be used repeatedly:

\begin{theorem}\label{thm-stoch-dom-conv} 
Let $\sK\!$ be a real or complex separable Hilbert space, $Z\in\mathsf{S}_{\zeit}(\RR^\nu)$, and
$A,A^{(n)}$, $n\in\NN$, be left continuous adapted $\LO(\RR^\nu,\sK)$-valued processes.
Let $R:\zeit\times\Omega\to \RR$ be a predictable process with locally
bounded paths and assume that, $\PP$-a.s., the following relations hold on $I$,
\begin{align}\label{w1}
A^{(n)}\rightarrow A\,\>\>\text{as}\>\> n\rightarrow\infty,\qquad\|A^{(n)}\|\le R,\;\;n\in\NN. 
\end{align}
Then 
\begin{align}\label{gabi}
\underset{n\to\infty}{\mathrm{lim\,prob}}
\sup_{t\in[0,\tau]}\Big\|\int_{0}^t
A^{(n)}_s\Id Z_s-\int_{0}^t A_s\Id Z_s\Big\|=0,\quad\tau\in\zeit,
\end{align}
and there is a subsequence $(A^{(n_k)})_{k\in\NN}$ of
$(A^{(n)})_{n\in\NN}$ such that, $\PP$-a.s., one has
\begin{align}
\lim_{k\to \infty} \sup_{t\in [0,\tau]}
\Big\|\int^{t}_{0} A^{(n_k)}_s\Id Z_s-\int^{t}_{0} A_s\Id Z_s\Big\|=0,\quad\tau\in I.\label{konn}
\end{align}
\end{theorem}

\begin{example}{\rm \label{ex-Ito-limprob}
Let $\sK$, $A$, and $Z$ be as in Thm.~\ref{thm-stoch-dom-conv}
and let $\tau\in\zeit$. For every $n\in\NN$, let $(\sigma_\ell^{(n)})_{\ell\in\NN}$
be an increasing sequence of stopping times such that
$\sup_{\ell}(\sigma^{(n)}_{\ell+1}-\sigma^{(n)}_\ell)\to0$, $n\to\infty$, $\PP$-a.s.,
and such that $\PP\{\sigma^{(n)}_\ell<t\}\to0$, $\ell\to\infty$, for all $n\in\NN$ and $t\in I$. Then
\begin{align}\label{limprob-Ito}
\underset{n\to\infty}{\mathrm{lim\,prob}}\sup_{t\in[0,\tau]}\Big\|\int_{0}^tA_s\Id Z_s
&-\sum_{\ell\in\NN_0}A_{\sigma_{\ell}^{(n)}}\,\big(Z_{\sigma_{\ell+1}^{(n)}\wedge t}
-Z_{\sigma_{\ell}^{(n)}\wedge t}\big)\Big\|=0.
\end{align}
In fact, the sum appearing under the norm in \eqref{limprob-Ito} equals 
$\int_{0}^tA^{(n)}_s\Id Z_s$ with
$A^{(n)}=\sum_{\ell\in\NN_0}1_{(\sigma_{\ell}^{(n)},\sigma_{\ell+1}^{(n)}]}A_{\sigma_{\ell}^{(n)}}$. 
Since $A$ has left-continuous paths we see that \eqref{w1} holds with $R_t:=\sup_{0\le s<t}\|A_s\|$.
 }\end{example}

{\it Proof of Thm.~\ref{thm-stoch-dom-conv}.}
We refer to \cite[\textsection26.1]{Me1982} for a construction of the stochastic
integral which, under the assumptions of the theorem,
implies the existence of an increasing sequence of stopping times $\tau_m$, $m\in\NN$,
with $\PP\{\sup_m\tau_m<t\}=0$, $t\in I$, and
$\EE[\vr_{\tau_m}^*(n)^2]\to0$, $n\to\infty$,
where $\vr_\tau^*(n):=\sup_{t\le \tau}\vr_t(n)$
with $\vr_t(n)$ denoting the norm $\|\cdots\|$ on the left hand side of \eqref{gabi};
cf. the proofs of \cite[Thm.~24.2 and Thm.~26.3]{Me1982}.
Now let $\tau\in\zeit$ and $\ve,\ve_1>0$.
Choose some $m\in\NN$ with $\PP\{\tau_m<\tau\}<\ve_1$.
Then the above remarks imply
$\limsup_n\PP\{\vr_{\tau}^*(n)\ge\ve\}
\le\limsup_n\PP\{\vr_{\tau_m}^*(n)\ge\ve\}+\PP\{\tau_m<\tau\}<\ve_1$,
which proves \eqref{gabi}. The remaining statements 
follow as in the proof of \cite[Thm.~24.2]{Me1982}.
\qed

 \begin{remark}{\rm \label{cons}
Let us recall that the mutual variation of two real-valued continuous semi-martingales 
$Z_1$ and $Z_2$ on $I$ is defined (up to indistinguishability) by
\begin{align}\label{def-covar}
\llbracket Z_1,Z_2\rrbracket_\bullet&:=Z_{1,t}Z_{2,t}-Z_{1,0}Z_{2,0}
-\int_0^\bullet Z_{1,s}\Id Z_{2,s}-\int_0^\bullet Z_{2,s}\Id Z_{1,s}.
\end{align}
If both semi-martingales are of the form
$Z_{j,\bullet}=\int_0^\bullet\V{A}_{j,s}\Id\V{X}_s+\int_0^\bullet\wt{A}_{j,s}\Id s$, $j=1,2$,
with processes $\V{A}_{j}$ and $\wt{A}_j$ as in Prop.~\ref{prop-Ito} (with $\sY=\RR$), then
\begin{align}\label{for-covar}
\llbracket Z_1,Z_2\rrbracket_\bullet=\int_0^\bullet\V{A}_{1,s}\cdot\V{A}_{2,s}\Id s,\quad\text{$\PP$-a.s.}
\end{align}
 }\end{remark}

We end this summary of results from stochastic analysis with a standard criterion for a
stochastic integral with respect to Brownian motion to be a martingale (where $\lambda$
denotes the one-dimensional Lebesgue-Borel measure):

\begin{proposition}\label{prop-mart} 
Let $\sK$ be a real or complex separable Hilbert space and $\V{A}$ be an
adapted, left continuous, $\LO(\RR^\nu,\sK)$-valued process on $I$ such that 
$\EE[\|\V{A}_\bullet\|^2]\in L^1_\loc(I,\lambda)$.
Then $(\int^{t}_0 \V{A}_s\Id \V{B}_s)_{t\in I}$ is a martingale.
\end{proposition}


\section{Some basic Hilbert space-valued processes}\label{sec-proc}

\noindent
In this section, we define and discuss the basic processes
appearing in our ansatz for the solution of \eqref{SDE-intro}; recall the remarks on Step 1 of
the proof of Thm.~\ref{thm-SDE-intro} given below its statement.

To this end we first recall the definition of Nelson's isometries $j_t$ \cite{Nelson1973} mapping
$\HP$ and $\mathfrak{k}$ into $\HP_{+1}$ and $\mathfrak{k}_{+1}$, respectively, where
\begin{align*}
\HP_{+1}&:=L^2(\RR\times\cM,\lambda\otimes\mu),\quad\mathfrak{k}_{+1}:=
L^2\big(\RR\times\cM,[\omega^{-1}+(\omega+\tfrac{1}{2}\V{m}^2)^2]\lambda\otimes\mu\big),
\end{align*}
with $\lambda$ denoting the Lebesgue-Borel measure on $\RR$. They are defined by
\begin{align}\label{def-jt}
j_tf(k_0,k):=\pi^\mh e^{-itk_0}\omega(k)^\eh(\omega(k)^2+k_0^2)^\mh f(k),
\end{align}
for all $t\in\RR$ and a.e. $(k_0,k)\in\RR\times\cM$. (Usually, $j_t$ is defined in
the position representation for a single boson in a -- sometimes weighted -- $L^2$-space 
over $\RR^3$, which explains the discrepancy between \eqref{def-jt} and the formulas in 
\cite{Hiroshima1997,LHB2011,Nelson1973,Simon1974}.)
The isometry of the maps $j_t:\HP\to\HP_{+1}$ and 
$j_t\!\!\upharpoonright_{\mathfrak{k}}:\mathfrak{k}\to\mathfrak{k}_{+1}$ follows from 
\begin{align}\label{j-omega}
j_s^*j_tf&=\frac{\omega}{\pi}\int_\RR\frac{e^{ik_0(s-t)}\,\Id k_0}{\omega^2+k_0^2}\,f
=e^{-|s-t|\omega}f,\quad s,t\in\RR,\,f\in\HP,
\end{align}
which is easily verified by contour deformation. 
The maps $t\mapsto j_t\in\LO(\HP,\HP_{+1})$ and 
$t\mapsto j_t\!\!\upharpoonright_{\mathfrak{k}}\in\LO(\mathfrak{k},\mathfrak{k}_{+1})$ are strongly 
continuous. A direct inspection reveals that $t\mapsto j_t^*\in\LO(\HP_{+1},\HP)$ and 
$t\mapsto(j_t\!\!\upharpoonright_{\mathfrak{k}})^*\in\LO(\mathfrak{k}_{+1},\mathfrak{k})$
are strongly continuous as well. It is convenient to introduce the random isometries
\begin{equation}\label{def-iota}
\iota_t:=j_te^{-i\V{m}\cdot(\V{X}_t-\V{X}_0)},\qquad t\in\zeit.
\end{equation}
Obviously, if $A$ is an adapted process with values in $\HP$ or $\mathfrak{k}$, then
$\iota A=(\iota_t A_t)_{t\in I}$ is an adapted process with values in $\HP_{+1}$ or $\mathfrak{k}_{+1}$,
respectively. If $A$ is continuous, then $\iota A$ is continuous as well.
Analogous remarks hold for $\iota^*$.
 
\begin{definition}[{\bf Basic processes}]\label{defn-basic-proc}
We define ${K},(K_{\tau,t})_{t\in I}\in\mathsf{S}_{\zeit}(\mathfrak{k}_{+1})$ by
\begin{align}\label{def-ulK}
K_{\tau,\bullet}&:=\int_0^\bullet1_{(\tau,\infty)}(s)\iota_s\V{G}_{\V{X}_s}{\Id}\V{X}_s
+\int_0^\bullet1_{(\tau,\infty)}(s)\iota_s\breve{q}_{\V{X}_s}\Id s,\quad K_\bullet:=K_{0,\bullet},
\end{align}
for every $\tau\in I$. With this we further define $\mathfrak{k}$-valued processes on $\zeit$ by
\begin{align}\label{def-Uminus}
U_{\tau,t}^-:=(\iota_t^\tau)^*K_{\tau,t},
\qquad
U_t^{-}&:=U^-_{0,t}=j_0^*K_t,
\qquad
U^{+}_t:=\iota_t^*K_t,
\end{align}
for $t\in I$, where $\iota^\tau$ is $\iota$ stopped at $\tau$.
For every $\vxi\in\RR^\nu$, we finally set
\begin{equation}\label{def-u}
u^V_{\vxi,\bullet}:=\frac{1}{2}\,\|K_\bullet\|^2_{\HP_{+1}}+\int_0^{\bullet}V(\V{X}_s)\Id s
-i\vxi\cdot(\V{X}_\bullet-\V{X}_0).
\end{equation}
\end{definition}

All processes introduced in Def.~\ref{defn-basic-proc} are well-defined 
up to indistinguishability. In NRQED (using slightly stronger assumptions) the process $K$ has 
been introduced in \cite{Hiroshima1997}. Lem.~\ref{lem-def-W} below motivates the
definitions in \eqref{def-Uminus} and \eqref{def-u}. 
The parameter $\tau$ is needed only in the matrix-valued case.

The reader might have noticed that $K_{\tau,t}$ looks formally like a Stratonovich integral.
According to the following technical lemma it can indeed be approximated by the usual average 
of left and right Riemann sums; this result will become important in Sect.~\ref{sec-C0} where
we discuss time-reversals. The only reason why Lem.~\ref{rem-K-strat} might not immediately follow 
from the textbook literature is that the embeddings 
$j_s$ are not strongly differentiable with respect to $s$.

\begin{lemma}\label{rem-K-strat}
Fix $\tau,t\in I$ with $\tau\le t$. Then 
\begin{equation}\label{approx-KR}
\big\|{K}_{\tau,t}-\Sigma_{\tau,t}^{n}\big\|_{\mathfrak{h}_{+1}}\xrightarrow{\;\;n\to\infty\;\;}0
\quad\text{in probability,}
\end{equation} 
where the sum corresponds to the sample points
$\sigma_\ell^n=\sigma_\ell^n(\tau,t):=\tau+\ell(t-\tau)/n$,
\begin{align}\label{ulk}
\Sigma_{\tau,t}^{n}:=\frac{1}{2}\sum_{\ell=0}^{n-1}
\big(\iota_{\sigma_{\ell+1}^n}\V{G}_{\V{X}_{\sigma_{\ell+1}^n}}\!\!
+\iota_{\sigma_{\ell}^n}\V{G}_{\V{X}_{\sigma_{\ell}^n}}\big)\cdot
\big(\V{X}_{\sigma_{\ell+1}^n}\!\!-\V{X}_{\sigma_{\ell}^n}\big),\quad n\in\NN.
\end{align}
\end{lemma}

{\proof}
We set $\V{D}(s,\V{x}):=j_se^{-i\V{m}\cdot(\V{x}-\V{X}_0)}\V{G}_{\V{x}}$, $s\in I$,  $\V{x}\in\RR^\nu$. 
Then Taylor's formula yields 
$\Sigma_{\tau,t}^{n}=\frac{1}{2}(I_1^n+I_2^n+J_n+R_n)$, for every $n\in\NN$, with
\begin{align}\nonumber
I_{1+\alpha}^n&:=\int_0^t\sum_{\ell=0}^{n-1}1_{(\sigma_\ell^n,\sigma_{\ell+1}^n]}(s)
\V{D}(\sigma_{\ell+\alpha}^n,\V{X}_{\sigma_\ell^n})\Id\V{X}_s,\quad\alpha=0,1,
\\\nonumber
J_n&:=\sum_{a,b=1}^\nu\sum_{\ell=0}^{n-1}\partial_{x_a}
{D}_b(\sigma_{\ell+1}^n,\V{X}_{\sigma_\ell^n})
(X_{a,\sigma_{\ell+1}^n}-X_{a,\sigma_\ell^n})(X_{b,\sigma_{\ell+1}^n}-X_{b,\sigma_\ell^n}),
\\\label{ruth1}
\|R_n\|&\le \max_{\tilde{\ell}=1,...,n-1}
\frac{r_{\tilde{\ell}}^n}{2}\sum_{\ell=1}^{n-1}\|\V{X}_{\sigma_{\ell+1}^n}-\V{X}_{\sigma_\ell^n}\|^2.
\end{align}
Here we further abbreviate
\begin{align*}
r_\ell^n:=\!\sum_{a,b=1}^\nu\int_0^1\!
\big\|\partial_{{x_a}}{D}_b\big(\sigma_{\ell+1}^n,(1-s)\V{X}_{\sigma_\ell^n}
+s\V{X}_{\sigma_{\ell+1}^n}\big)-\partial_{{x_a}}
{D}_b(\sigma_{\ell+1}^n,\V{X}_{\sigma_\ell^n})\big\|\Id s.
\end{align*}
Since the integrands in $I_{1}^n$ and $I_2^n$ are adapted, left continuous, uniformly bounded, and 
converge both to the process $\big(1_{(\tau,t]}(s)\V{D}(s,\V{X}_s)\big)_{s\in I}$ 
pointwise on $I\times\Omega$ as $n$ goes to infinity, it follows from 
Thm.~\ref{thm-stoch-dom-conv} that $I_{1}^n$ and $I_2^n$ converge both to
$\int_0^t1_{(\tau,\infty)}(s)\V{D}(s,\V{X}_s)\Id\V{X}_s$ in probability. 
Expressions similar to $J_n$ are well-known from the proof of the It\={o} formula. In fact, writing
\begin{align*}
X_{a,s}^{(n)}&:=\sum_{\ell=0}^{n-1}1_{(\sigma_\ell^n,\sigma_{\ell+1}^n]}(s)X_{a,\sigma_\ell^n},
\quad
Z_{ab,s}^{(n)}:=\sum_{\ell=0}^{n-1}1_{(\sigma_\ell^n,\sigma_{\ell+1}^n]}(s)
\partial_{x_a}{D}_b(\sigma_{\ell+1}^n,\V{X}_{\sigma_\ell^n}),
\end{align*}
for all $s\in I$, $a,b\in\{1,\ldots,\nu\}$, and $n\in\NN$, we find
\begin{align*}
J_n&:=\sum_{a,b=1}^\nu\Big(\int_0^tZ_{ab,s}^{(n)}\Id(X_aX_b)_s
-\int_0^tZ_{ab,s}^{(n)}X_{a,s}^{(n)}\Id X_{b,s}-\int_0^tZ_{ab,s}^{(n)}X_{b,s}^{(n)}\Id X_{a,s}\Big).
\end{align*}
Here the uniformly bounded, left continuous, and adapted processes $Z_{ab}^{(n)}$, $n\in\NN$,
converge pointwise on $I\times\Omega$ to 
$\big(1_{(\tau,t]}(s)\partial_{x_a}{D}_b(s,\V{X}_s)\big)_{s\in I}$.
Applying successively Thm.~\ref{thm-stoch-dom-conv}, Prop.~\ref{prop-stoch-calc}, \eqref{def-covar},
and $\llbracket X_a,X_b\rrbracket_s=s\delta_{a,b}$, we readily verify that $J_n$ converges in 
probability to
$$
\int_0^t\mathrm{div}_{\V{x}}\V{D}(s,\V{X}_{s})\Id s
=2\int_0^t1_{(\tau,\infty)}(s)\iota_s\breve{q}_{\V{X}_s}\Id s.
$$
Finally, fix $\vgamma\in\Omega$ and let $P(\vgamma)$ be the compact convex hull of
the path $\{\V{X}_s(\vgamma):s\in[0,t]\}$. Since the maps $[0,t]\ni s\mapsto\V{X}_s(\vgamma)$ and
$[0,t]\times P(\vgamma)\ni(s,\V{x})\mapsto\partial_{x_a}{D}_b(s,\V{x})$ are uniformly continuous,
the sequence of random variables $(\max_{\ell}r_\ell^n)_{n\in\NN}$ converges to $0$
pointwise on $\Omega$, as $n$ goes to infinity. Thanks to Hyp.~\ref{hyp-G} we further
find some constant $c>0$ such that $0\le r_\ell^n\le c$ on $\Omega$, for all $\ell$ and $n$.
At the same time we know that the sequence
$\big(\sum_{\ell=1}^{n-1}\|\V{X}_{\sigma_{\ell+1}^n}-\V{X}_{\sigma_\ell^n}\|^2\big)_{n\in\NN}$
converges in probability to $\sum_{a=1}^\nu\big(\llbracket X_a,X_a\rrbracket_t
-\llbracket X_a,X_a\rrbracket_\tau\big)=\nu(t-\tau)$.
Employing these remarks, it is easy to show that $\|R_n\|\to0$, $n\to\infty$, in probability.
In fact, let $\ve,\ve_1>0$. Then we find some $n_0\in\NN$ such that
$$
\PP\Big\{\Big|\nu(t-\tau)-\sum_{\ell=1}^{n-1}\|\V{X}_{\sigma_{\ell+1}^n}-\V{X}_{\sigma_\ell^n}\|^2
\Big|\ge1\Big\}<\ve_1,\quad n\ge n_0.
$$
Set $A_n:=\big\{\big|\nu(t-\tau)-\sum_{\ell=1}^{n-1}\|\V{X}_{\sigma_{\ell+1}^n}
-\V{X}_{\sigma_\ell^n}\|^2 \big|< 1\big\}$. Then the previous bound and \eqref{ruth1}
permit to get, for all $n\ge n_0$,
\begin{align*}
\PP\big\{\|R_n\|\ge\ve\big\}
&\le\ve_1+\EE\big[1_{A_n}1_{\{\|R_n\|\ge\ve\}}\big]
\le\ve_1+\frac{1}{\ve}\EE\big[1_{A_n}\|R_n\|\big]
\\
&\le\ve_1+\frac{1+\nu(t-\tau)}{2\ve}\EE\big[\max_{\ell}r_\ell^n\big]\xrightarrow{\;\;n\to\infty\;\;}
\ve_1,
\end{align*}
where we also applied the dominated convergence theorem in the last step.
Since $\ve_1>0$ was arbitrary, this proves that $\|R_n\|$ goes to $0$ in probability.
\qed 

To derive stochastic integral representations for $U_{\tau,\bullet}^-$ and $U^+$ we set
\begin{align}\label{def-w}
\w{\tau}{t}:=\olw{\tau}{t}^*,\quad\olw{\tau}{t}:=(\iota_t^\tau)^*\iota_t=
\left\{\begin{array}{ll}
e^{-(t-\tau)\omega-i\V{m}\cdot(\V{X}_t-\V{X}_\tau)},\;\:&{t>\tau},
\\
1,&{t\le\tau},
\end{array}
\right.
\end{align}
for all $\tau,t\in I$. Depending on the circumstances, we consider $\w{\tau}{t}$ and $\olw{\tau}{t}$ as 
maps from $\Omega$ into $\LO(\HP)$ or $\LO(\mathfrak{k})$, which should cause no confusion. 
They leave the real space $\HP_C$
(resp. $\mathfrak{k}_C$) invariant; recall \eqref{hyp-sym2} and \eqref{def-dCkC}.
If $A$ is an adapted continuous process with values in $\HP$ or $\mathfrak{k}$, then so are
$(\w{\tau}{t}A_t)_{t\in I}$ and $(\olw{\tau}{t}A_t)_{t\in I}$.

\begin{lemma}\label{cor-strat}
Let $R\in\NN$ and set $\chi_R:=1_{\{\frac{1}{2}\V{m}^2+\omega\le R\}}$. Then, $\PP$-a.s.,
\begin{equation}\label{tracy2}
\chi_R\,U_t^+=\w{0}{t}\int_0^t
\chi_R\,e^{s\omega-i\V{m}\cdot(\V{X}_s-\V{X}_0)}\{\V{G}_{\V{X}_s}\Id\V{X}_s
+\breve{q}_{\V{X}_s}\Id s\},\quad t\in I.
\end{equation}
\end{lemma}

{\proof}
Fix $t\in I$ and set $\sigma_\ell^n:=t\ell/n$, $\ell\in\NN_0$. Then Lem.~\ref{rem-K-strat} implies
\begin{equation*}
\big\|\chi_R\iota_t^*{K}_{0,t}-\chi_R\iota_t^*\Sigma_{0,t}^{n}
\big\|_{\mathfrak{h}}\xrightarrow{\;\;n\to\infty\;\;}0
\quad\text{in probability,}
\end{equation*} 
with $\Sigma_{0,t}^n$ as in \eqref{ulk}. Next, we observe that 
$\chi_R\iota_t^*\Sigma_{0,t}^{n}=\w{0}{t}\wt{\Sigma}_{t}^{n}$ with
\begin{align*}
\wt{\Sigma}_{t}^{n}:=\frac{1}{2}\sum_{\ell=0}^{n-1}
\big(J_{\sigma_{\ell+1}^n}\V{G}_{\V{X}_{\sigma_{\ell+1}^n}}\!\!
+J_{\sigma_{\ell}^n}\V{G}_{\V{X}_{\sigma_{\ell}^n}}\big)\cdot
\big(\V{X}_{\sigma_{\ell+1}^n}\!\!-\V{X}_{\sigma_{\ell}^n}\big),\quad n\in\NN,
\end{align*}
where $J_s:=\chi_Re^{s\omega-i\V{m}\cdot(\V{X}_{s}-\V{X}_0)}$, $s\in I$.
Replacing $\V{D}$ by the function $\wt{\V{D}}$ defined by
$\wt{\V{D}}(s,\V{x}):=\chi_Re^{s\omega-i\V{m}\cdot(\V{x}-\V{X}_0)}\V{G}_{\V{x}}$, 
$s\in I$, $\V{x}\in\RR^\nu$, in the proof of Lem.~\ref{rem-K-strat}, we may further verify that
\begin{equation*}
\underset{n\to\infty}{\mathrm{lim\,prob}}\,\wt{\Sigma}_{t}^{n}=
\int_0^t\chi_R\,e^{s\omega-i\V{m}\cdot(\V{X}_s-\V{X}_0)}\{\V{G}_{\V{X}_s}\Id\V{X}_s
+\breve{q}_{\V{X}_s}\Id s\}.
\end{equation*} 
Together with \eqref{def-Uminus}, these remarks prove the equality in \eqref{tracy2}, a priori outside 
some $t$-dependent $\PP$-zero set. We conclude by noting that the processes on both sides of 
\eqref{tracy2} are continuous.
\qed 

\begin{lemma}\label{lem-U-K}
{\rm(1)}  Let $\tau\in I$. Then 
$(U_{\tau,t}^{-})_{t\in I}\in\mathsf{S}_I(\mathfrak{k})\subset\mathsf{S}_I(\HP)$ and, $\PP$-a.s.,
\begin{align}\label{laura2}
U_{\tau,\bullet}^{-}&=\int_0^\bullet1_{(\tau,\infty)}(s)\olw{\tau}{s}\,
\V{G}_{\V{X}_s}\Id\V{X}_s+\int_0^\bullet1_{(\tau,\infty)}(s)\olw{\tau}{s}\,\breve{q}_{\V{X}_s}\Id s.
\end{align}
{\rm(2)} $U^+$ is adapted and continuous with values in $\mathfrak{k}$.
Moreover, $\omega\,U^{+}$, $\V{m}^2\,U^{+}$, 
and the components of $\V{m}\,U^{+}$ are adapted and continuous as $\HP$-valued processes.

\smallskip

\noindent{\rm(3)} $U^{+}\in\mathsf{S}_{\zeit }(\HP)$ with
\begin{align}\nonumber
U^{+}_\bullet&=\int_0^\bullet\big(\V{G}_{\V{X}_s}+i\V{m}\,U_s^{+}\big)\Id\V{X}_s
-\int_0^\bullet(\omega+\tfrac{1}{2}\V{m}^2)\,U_s^{+}\Id s
\\\label{laura3}
&\quad+
\int_0^\bullet\big(\tfrac{i}{2}\V{m}\cdot\V{G}_{\V{X}_s}+\tfrac{1}{2}q_{\V{X}_s}\big)
\Id s,\quad\PP\text{-a.s.}
\end{align}
{\rm(4)} $U_t^+$ and $U_{\tau,t}^-$ attain their values in the {\em real} space $\HP_C$. 

\smallskip

\noindent{\rm(5)} By passing to suitable modifications of $(K_{\tau,t})_{t\in I}$ and 
$(U_{\tau,t}^-)_{t\in I}$, for each $\tau\in I$, we may assume that, for all $\vgamma\in \Omega$, 
the maps $(\tau,t)\mapsto K_{\tau,t}(\vgamma)\in\mathfrak{k}_{+1}$ 
and $(\tau,t)\mapsto U^-_{\tau,t}(\vgamma)\in\mathfrak{k}$
are continuous on $I\times I$ with $K_{s,s}(\vgamma)=0$ and $U^-_{s,s}(\vgamma)=0$,
for every $s\in I$.
\end{lemma}

{\proof} 
(1) follows by definition of $U_{\tau,t}^-$, \eqref{j-omega}, and
the fact that, if $t\ge\tau$, then the integrals defining $K_{\tau,t}$ 
commute with $(\iota_t^\tau)^*=\iota_\tau^*$, $\PP$-a.s.

(2): By the remarks preceding Def.~\ref{defn-basic-proc},
$U^+$ is adapted and continuous. The remaining statements are clear since
$\tfrac{1}{2}\V{m}^2+\omega\in\LO(\mathfrak{k},\HP)$. 

(3): Let $R\in\NN$ and consider the function 
$f_R:[0,\infty)\times\RR^\nu\times\mathfrak{k}_C\to\HP_C$ given by
$f_R(t,\V{x},y):=e^{-t\omega+i\V{m}\cdot\V{x}}\chi_Ry$, where $\chi_R$ is
the same as in Lem.~\ref{cor-strat}. Thanks to the cut-off function, $f_R$ satisfies the assumptions of 
Prop.~\ref{prop-Ito}. According to Lem.~\ref{cor-strat} we $\PP$-a.s. have
$\chi_RU_t^+=f_R(t,\V{X}_t-\V{X}_0,Y_t)$, $t\in I$, where $Y_t$ abbreviates the integral on the
right hand side of \eqref{tracy2}. 
Notice that Hyp.~\ref{hyp-G} and the presence of $\chi_R$ ensure that $Y$ is in fact a 
$\mathfrak{k}_C$-valued semi-martingale.
Applying Prop.~\ref{prop-Ito} and using \eqref{tracy2} to
simplify the result, we see that $\chi_RU_t^+\in\mathsf{S}_I(\HP_C)$ with
\begin{align}\nonumber
\chi_RU_t^+&=\int_0^t\chi_R(\V{G}_{\V{X}_s}+i\V{m}\,U_s^+)\Id\V{X}_s
-\int_0^t\chi_R\big(\omega+\tfrac{1}{2}\V{m}^2\big)\,U_s^+\Id s
\\\label{mille1}
&\quad
+\int_0^t\chi_R\big(\tfrac{i}{2}\V{m}\cdot\V{G}_{\V{X}_s}+\tfrac{1}{2}q_{\V{X}_s}\big)\Id s,
\quad t\in I,\;\;\PP\text{-a.s.}
\end{align}
By Part~(2), $\omega U^+$, $\V{m}U^+$, and $\V{m}^2U^+$
are adapted, continuous processes, whence all integrals in \eqref{mille1} are still well-defined
$\HP_C$-valued (stochastic) integrals, if the cut-off function $\chi_R$ is dropped. 
In particular, we may (up to indistinguishability) commute all integration signs
in \eqref{mille1} with $\chi_R$, regarding the latter as a bounded operator on $\HP_C$.
This finally leads to \eqref{laura3}.

(4) follows from \eqref{hyp-sym2}, \eqref{sym-q}, \eqref{laura2}, and \eqref{laura3}.

(5):  A suitable modification of $(K_{\tau,t})_{t\in I}$ is simply given by
$1_{(\tau,\infty)}(t)(K_{t}-K_\tau)$. Applying $(\iota^\tau)^*$, 
with $\V{X}_t(\vgamma)=\V{X}^{\V{q}}_t(\vgamma)$ replaced by 
$\vXi_{0,t}(\V{q}(\vgamma),\vgamma)$  in its definition (see Hyp.~\ref{hyp-B}(2)),
to the latter modification we may produce a suitable modification of $(U^-_{\tau,t})_{t\in I}$.
\qed

\begin{lemma}\label{lem-u-k} It holds
$u_{\vxi}^V\in\mathsf{S}_I(\CC)$ and one $\PP$-a.s. has
\begin{align}\nonumber
u^{V}_{\vxi,\bullet}&=\int_{0}^\bullet\SPn{U_s^{+}}{\V{G}_{\V{X}_s}}\,\Id\V{X}_s+
\int_0^\bullet\SPn{U_s^{+}}{\breve{q}_{\V{X}_s}}\,\Id s
+\frac{1}{2}\int_{0}^\bullet\|\V{G}_{\V{X}_s}\|^2\Id s
\\\label{for-u}
&\quad
+\int_{0}^\bullet V(\V{X}_s)\,\Id s-i\vxi\cdot(\V{X}_\bullet-\V{X}_0).
\end{align}
\end{lemma}

{\proof} 
The fact that $u_{\vxi}^V$ is a continuous semi-martingale follows from \eqref{def-u} 
and Ex.~\ref{ex-Ito-SP}. By means of Ex.~\ref{ex-Ito-SP} and the isometry of $\iota_s$ 
we $\PP$-a.s. obtain
\begin{align}\nonumber
\|K_\bullet\|^2&=
\int^\bullet_02\Re\SPn{K_s}{\iota_s\V{G}_{\V{X}_s}}\,\Id\V{X}_s
+\int^\bullet_0\big(2\Re\SPn{K_s}{\iota_s\breve{q}_{\V{X}_s}}+\|\iota_s\V{G}_{\V{X}_s}\|^2\big)\,\Id s
\\\nonumber
&=
\int^\bullet_02\SPn{U_s^+}{\V{G}_{\V{X}_s}}\,\Id\V{X}_s
+\int^\bullet_0\big(2\SPn{U_s^+}{\breve{q}_{\V{X}_s}}+\|\V{G}_{\V{X}_s}\|^2\big)\,\Id s.
\end{align}
Here we also used \eqref{hyp-sym2}, \eqref{sym-q}, and $U^+=CU^+$ in the second step.
\qed


\section{Stochastic calculus in the scalar case}\label{sec-Ito}

\noindent
In this section we verify that the ansatz \eqref{W-Hi} suggested by Hiroshima's expression for
the Feynman-Kac integrand \cite{Hiroshima1997} gives rise to solutions of the SDE
 \eqref{SDE-intro} in the scalar case. 
We consider only deterministic exponential vectors
as initial conditions, which effectively simplifies computations.
A proper existence and uniqueness result with a natural class
of initial conditions for the SDE \eqref{SDE} will be contained in Thm.~\ref{thm-Ito-spin}  
as a special case ($L=1$, $\V{F}=\V{0}$). 

In the following definition we use the notation introduced in \eqref{def-iota}, \eqref{def-ulK},
and the discussion of the Weyl representation $\sW$ following \eqref{def-Weyl}.

\begin{definition} For all $\vxi\in\RR^\nu$, we define $\W{\vxi}{V}:I\times\Omega\to\LO(\sF)$ by
\begin{align}\label{W-Hi}
\W{\vxi,t}{V}&:=e^{-i\vxi\cdot(\V{X}_t-\V{X}_0)-\int_0^tV(\V{X}_s)\Id s}
\Gamma(\iota_t^*)\sW(K_t)\Gamma(\iota_0),\quad t\in I.
\end{align}
\end{definition}

 \begin{remark}{\rm \label{rem-def-W-a}
It is the {\em adjoint} of $\W{\vxi,t}{V}$ which appears in the Feynman-Kac formula in the scalar case.
It is advantageous to study $\W{\vxi}{V}$, instead of its adjoint, because it yields solutions to a 
backward SDE.
 }\end{remark}

\begin{lemma}\label{lem-def-W}
$\W{\vxi,t}{V}$ maps $\sC[\HP]$ into itself and
\begin{align}
\W{\vxi,t}{V}\zeta(h)
&=e^{-u_{-\vxi,t}^V-\SPn{U_t^-}{h}}\zeta\big(\w{0}{t}h+U_t^+\big),\quad h\in\HP.\label{def-W}
\end{align}
\end{lemma}

{\proof}
Combine \eqref{APC-fact}, \eqref{j-omega}, \eqref{def-Uminus}, \eqref{def-u}, and \eqref{W-Hi}.
\qed

 \begin{remark}{\rm \label{rem-def-W}
(1) In view of \eqref{def-exp-vec}, \eqref{exp-vec2}, \eqref{def-u}, \eqref{def-W},
and Lem.~\ref{lem-U-K}(4), the operator $\W{\V{0},t}{V}$ is manifestly real, i.e., it maps $\sF_C$ 
into itself. Just recall that $\sF_C$ is the closure of $\mathrm{span}_\RR(\sE[\mathfrak{d}_C])$.

\smallskip

\noindent(2) Another formula for $\W{\vxi,t}{V}$ is given in Rem.~\ref{rem-for-W-meas}(1).

\smallskip

\noindent(3) From \eqref{def-u} and \eqref{W-Hi} it is obvious that
\begin{equation}\label{norm-W-scalar}
\ln\|\W{\vxi,t}{V}\|\le-\int_0^tV({\V{X}}_s)\,\Id s,\quad t\in I.
\end{equation}
 }\end{remark}

To prepare for an application of It\={o}'s formula, we compute a few derivatives in the next lemma,
where the real Hilbert space $\RR^2\times\HP_C\times\HP_C$ will play the role of $\sY$ 
in Prop.~\ref{prop-Ito}.

\begin{lemma}\label{lem-Frechet}
Let $h\in\HP_C$ and define the function $f\colon\RR^2\times\HP_C\times\HP_C\to\sF$ by
\begin{equation}\label{def-fuvw}
f[u,v,w]:=e^{-u_1-iu_2-\SPn{w}{h}}\zeta(v),\qquad(u,v,w)\in\RR^2\times\HP_C\times\HP_C.
\end{equation}
Then $f$ is smooth and satisfies all conditions of Prop.~\ref{prop-Ito}.
Given any $g\in\HP_C$ and any self-adjoint operator, $T$, in $\HP_C$,
the diagonal parts of its first two Fr\'{e}chet derivatives at $(u,v,w)$ 
applied to tangent vectors $(x,y,z)\in\RR^2\times\HP_C\times\HP_C$ can be written as
\begin{align}
f'[&u,v,w](x,y,z)\label{Frechet1}
\\\nonumber
&=\big(\SPn{g}{v}-\SPn{z}{h}-x_1-ix_2+\Id\Gamma(T)+i\ad(y-Tv)+ia(g)\big)\,f[u,v,w],
\\\nonumber
f''[&u,v,w](x,y,z)^{\otimes_2}
\\\nonumber
&=\big(\SPn{g}{v}-\SPn{z}{h}-x_1-ix_2+\Id\Gamma(T)+i\ad(y-Tv)+ia(g)\big)^2\,f[u,v,w]
\\\label{Frechet2}
&\quad
+\big(\SPn{g}{y}-i\ad(Ty)\big)\,f[u,v,w],
\end{align}
provided that $v\in\dom(T)$ in \eqref{Frechet1} 
(resp. $y\in\dom(T)$, $v\in\dom(T^2)$ in \eqref{Frechet2}).
\end{lemma}

{\proof} 
With the help of Lem.~\ref{lem-Taylor-exp} it is
elementary to check that $f$ satisfies the condition in Prop.~\ref{prop-Ito} 
and that the diagonal parts of its $n$-th Fr\'{e}chet derivatives are given by
\begin{align}\label{Frechet-n}
f^{(n)}[u,v,w](x,y,z)^{\otimes_n}
&=\big(-\SPn{z}{h}-x_1-ix_2+i\ad(y)\big)^nf[u,v,w].
\end{align}
Finally, we use \eqref{exp-vec1}, \eqref{exp-vec2}, and \eqref{CCR} to 
include $a(g)$ and $\Id\Gamma(T)$.
\qed 

 \begin{remark}{\rm \label{rem-smooth}
As another consequence of Lem.~\ref{lem-Taylor-exp}, the
function $f_n:\RR^2\times\HP_C^{2+n}\to\sF$ defined by
\begin{align*}
f_n[u,v,w,y_1,\ldots,y_n]&:=(d_v^nf)[u,v,w](y_1,\ldots,y_n)
\\
&=e^{-u_1-iu_2-\SPn{w}{h}}i^n\ad(y_1)\dots\ad(y_n)\,\zeta(v),
\end{align*}
for $u\in\RR^2$ and $v,w,y_j\in\HP_C$, $j=1,\ldots,n$, is smooth as well.
 }\end{remark}

\begin{theorem}\label{thm-Ito1}
Let $h\in\mathfrak{d}_C$. Then the process $\W{\vxi}{V}\zeta(h)$ belongs to ${\sf S}_{\zeit}(\sF)$
and, $\PP$-a.s., we have, for all $t\in[0,\sup I)$,
\begin{align}\nonumber
&\W{\vxi,t}{V}\zeta(h)-\zeta(h)
\\\label{SDE}
&=-\int_0^ti\V{v}(\vxi,\V{X}_s)\,\W{\vxi,s}{V}\zeta(h)\,\Id\V{X}_s
-\int_0^t\wh{H}^V_{\scal}(\vxi,\V{X}_s)\,\W{\vxi,s}{V}\zeta(h)\,\Id s,
\end{align}
where $\V{v}(\vxi,\V{x})$ and $\wh{H}^V_{\scal}(\vxi,\V{x})$ are defined by 
\eqref{def-vecv} and \eqref{def-fib-Ham-scalar}, respectively.
\end{theorem}

{\proof} 
By definition, ${\W{\vxi,s}{V}}\,\zeta(h)=f[u,v,w]$ with $f$ as in
\eqref{def-fuvw} and with
\begin{equation}\label{yvonne}
u=u_{-\vxi,s}^V,\qquad v=\w{0}{s}\,h+U_s^+,\qquad w=U_s^-.
\end{equation}
(Here and in what follows, we consider the complex-valued quantities 
$u$ and $x$ as $\RR^2$-valued objects when we plug
them into the formulas of Lem.~\ref{lem-Frechet} and apply Prop.~\ref{prop-Ito}.)
Applying Prop.~\ref{prop-Ito} (with $f(s,\V{x})=e^{-s\omega+i\V{m}\cdot\V{x}}h$) and
Lem.~\ref{lem-U-K}(3), we see that, with the above choice of $v$,
\begin{align}
\Id_s v&=-\big((\omega+\tfrac{1}{2}\V{m}^2)v
+\tfrac{1}{2}q_{\V{X}_s}+\tfrac{i}{2}\V{m}\cdot\V{G}_{\V{X}_s}\big)
\Id s\label{Id-v}
+(i\V{m}v+\V{G}_{\V{X}_s})\Id\V{X}_s.
\end{align}
On account of Lem.~\ref{lem-Frechet} we may now apply the It\={o} formula of Prop.~\ref{prop-Ito}.
In combination with \eqref{def-u}, \eqref{laura2}, and \eqref{Id-v} this results $\PP$-a.s. in
\begin{align*}
{\W{\vxi,t}{V}}\zeta(h)-\zeta(h)&=
\int_0^t \V{I}_{\V{X}_s}\Id\V{X}_s+\int_0^t I_{0,\V{X}_s}\Id s
+\frac{1}{2}\int_0^t II_{\V{X}_s}\Id \llbracket\V{X}\rrbracket_s,\;\; t<\sup I,
\end{align*}
where $I_0$ and the components of $\V{I}$ are equal to 
$f'[u,v,w](x,y,z)$ in \eqref{Frechet1} with $(u,v,w)$ substituted according to
\eqref{yvonne} and $(x,y,z,g,M)$ substituted according to the table below. Likewise,
$II$ equals $f''[u,v,w](x,y,z)^{\otimes_2}$ in \eqref{Frechet2} with $(u,v,w)$
as in \eqref{yvonne} and $(x,y,z,g,M)$ given by the following table
(where we drop all subscripts and arguments $\V{X}_s$):
\begin{center}
\begin{tabular}{c||l|l|l|l|l}
&$x$&$y$&$z$&$g$&$M$\\
\hline
\hline
$\V{I}\phantom{\Big|}$ \& $II$&$\SPn{U_s^+}{\V{G}}+i\vxi$&$i\V{m}\,v+\V{G}$
&$\olw{0}{s}\,\V{G}$&$\V{G}$&$i\V{m}$\\
\hline
$I_0\phantom{\Big|}$&$\SPn{U_s^+}{\breve{q}}+\tfrac{1}{2}\|\V{G}\|^2+V$&
$-(\omega+\tfrac{1}{2}\,\V{m}^2)\,v$
&$\olw{0}{s}\,\breve{q}$&$\breve{q}$&$-\omega$\\
& &$+\tfrac{1}{2}q+\tfrac{i}{2}\V{m}\cdot\V{G}$
& & &\\
\end{tabular}
\end{center}
Using that $\SPn{U_s^+}{\V{G}_{\V{X}_s}}$ and $\SPn{U_s^+}{\breve{q}_{\V{X}_s}}$
are real, we see that we have equalities according to the next table:
\begin{center}
\begin{tabular}{c||l|l}
&$\SPn{g}{v}-\SPn{z}{h}-x$&$y-M\,v$\\
\hline
\hline
$\V{I}\phantom{\Big|}$ \& $II$&$-i\vxi$&$\V{G}$\\
\hline
$I_0\phantom{\Big|}$&$-\tfrac{1}{2}\|\V{G}\|^2-V$&
$-\tfrac{1}{2}\V{m}^2\,v+\tfrac{1}{2}q+\tfrac{i}{2}\V{m}\cdot\V{G}$\\
\end{tabular}
\end{center}
Putting these remarks together we obtain
\begin{align*}
I_0+\tfrac{1}{2}II
&=\Big\{-\tfrac{1}{2}\|\V{G}\|^2-V+\tfrac{1}{2}\SPn{\V{G}}{i\V{m}\,v+\V{G}}-\Id\Gamma(\omega)
\\
&\qquad+i\ad(-\tfrac{1}{2}\V{m}^2\,v+\tfrac{1}{2}q+\tfrac{i}{2}\V{m}\cdot\V{G})
+ia(\breve{q})
-\tfrac{i}{2}\ad\big(i\V{m}(i\V{m}\,v+\V{G})\big)
\\
&\qquad
+\tfrac{1}{2}\big(-i\vxi+i\Id\Gamma(\V{m})+i\ad(\V{G})+ia(\V{G})\big)^2\Big\}W^V_{\vxi}\zeta(h)
\\
&=\Big\{-V-\Id\Gamma(\omega)+\tfrac{i}{2}\vp(q)
+ia(-\tfrac{i}{2}\V{m}\cdot\V{G})+\tfrac{1}{2}\SPn{\V{G}}{i\V{m}\,v}
\\
&\qquad
-\tfrac{1}{2}\big(\vxi-\Id\Gamma(\V{m})-\vp(\V{G})\big)^2\Big\}W^V_{\vxi}\zeta(h).
\end{align*}
On account of \eqref{exp-vec1} and since $\W{\vxi}{V}\,\zeta(h)$
is proportional to $\zeta(v)$ the eigenvalue equation
$a(\V{m}\cdot\V{G})\,\W{\vxi}{V}\zeta(h)=\SPn{\V{m}\cdot\V{G}}{iv}\,\W{\vxi}{V}\zeta(h)$ holds, whence
$$
I_{0,\V{X}_s}+\tfrac{1}{2}\,II_{\V{X}_s}=-\wh{H}_{\scal}(\vxi,\V{X}_s)\,\W{\vxi,s}{V}\,\zeta(h).
$$
Moreover, by \eqref{Frechet1}, $f[u,v,w]={\W{\vxi,s}{V}}\,\zeta(h)$, and the above tables,
$$
\V{I}_{\V{X}_s}=\big(-i\vxi+\Id\Gamma(i\V{m})+i\ad(\V{G}_{\V{X}_s})+ia(\V{G}_{\V{X}_s})\big)
{\W{\vxi,s}{V}}\,\zeta(h).
$$
We thus arrive at  \eqref{SDE}.
\qed


\section{The matrix-valued case: Definitions and results}\label{sec-spin}

\noindent
Our main existence and uniqueness theorem for solutions of the SDE \eqref{SDE-intro}
associated with the generalized fiber Hamiltonian in the general matrix-valued case will 
be formulated at the end of the present section. 
For this purpose, we shall first introduce and discuss the required notation.
In Ex.~\ref{ex-Nelson} the somewhat involved formulas below will be illustrated by showing how they
simplify in the special case of the Nelson model.

In what follows we shall use the symbol
$\sum\limits_{{\cA\cup\cB\cup\cC=[n]\atop\#\cC\in2\NN_0}}$
for the sum over all  {\em disjoint} partitions of $[n]:=\{1,\ldots,n\}$ into three sets,
where each set $\cA$, $\cB$, or $\cC$ may be empty
and the cardinality of $\cC$ is always even.  It appears in the following instance of Wick's
theorem saying that, on a suitable dense domain like $\sC[\HP]$,
$$
\vp(f_1)\ldots\vp(f_n)=\!\!\sum_{{\cA\cup\cB\cup\cC=[n]\atop\#\cC\in2\NN_0}}\!\!
\Big\{\sum_{{\cC=\cup\{c_p,c_p'\}\atop c_p<c_p'}}\!\!
\Big(\prod_{p=1}^{\#\cC/2}\SPn{f_{c_p}}{f_{c_{p}'}}\Big)\Big\}
\Big(\prod_{a\in\cA}\ad(f_a)\Big)\prod_{b\in\cB}a(f_b).
$$
Here the sum in the curly brackets runs over all possibilities to split $\cC$
into disjoint subsets $\{c_p,c_p'\}\subset\cC$ with $c_p<c_p'$, $p=1,\ldots,\#\cC/2$.
If $\cC$ is empty, then the whole term $\{\cdots\}$ should be read as $1$, of course.
We shall further write
$$
t\simplex_n:=\big\{(s_1,\ldots,s_n)\in\RR^n:\;0\le s_1\le\ldots\le s_n\le t\big\},\quad t\ge0.
$$
If $t_1,\ldots,t_n\in\RR$ and $\cA\subset[n]$, then we set
$t_{\cA}:=(t_{a_1},\ldots,t_{a_{m}})$ where $\cA=\{a_1,\ldots,a_m\}$ with $a_1<\dots<a_m$.
For a multi-index $\alpha\in[S]^n$ with $[S]:=\{1,\ldots,S\}$,
the notation $\alpha_{\cA}$ is defined in the same way.

\begin{definition}[{\bf Time-ordered integral series}]\label{defn-TOE}
Let $\tau,t_1,\ldots,t_n\in I$, $\alpha\in[S]^n$, and $\cA,\cB\subset[n]$. We define
$\AD{\varnothing}{\tau}:=\A{\varnothing}:=\id$ and, in case $\cA$ (resp. $\cB$) is non-empty, 
\begin{align}\nonumber
\AD{\cA}{\tau}&:=\prod_{a\in\cA}\{\ad(\w{t_a}{\tau}
{F}_{\alpha_a,\V{X}_{t_a}})+i\SPn{U_{t_a,\tau}^{-}}{{F}_{\alpha_a,\V{X}_{t_a}}}\},
\\\nonumber 
\A{\cB}&:=\prod_{b\in\cB}\{a(\olw{0}{t_b}{F}_{\alpha_b,\V{X}_{t_b}})
+i\SPn{{F}_{\alpha_b,\V{X}_{t_b}}}{U_{t_b}^{+}}\},
\end{align}
on the domain 
${\CC^L}\otimes\sC[\mathfrak{d}_C]$,
noticing that, by \eqref{CCR}, the order of factors is immaterial.
If $\cC\subset[n]$ with $\#\cC\in2\NN_0$, then we further set $\IN{\varnothing}:=1$ and
\begin{align}\nonumber
\IN{\cC}&:=\sum_{{\cC=\cup\{c_p,c_p'\}\atop c_p<c_p'}}\prod_{p=1}^{\#\cC/2}
\SPn{{F}_{\alpha_{c_p'},\V{X}_{t_{c_p'}}}}{\w{t_{c_p}}{t_{c_p'}}\,{F}_{\alpha_{c_p},\V{X}_{t_{c_p}}}},
\end{align}
if $\cC$ is non-empty. Writing $\Id t_{[n]}:=\Id t_1\ldots\Id t_n$, we finally define 
\begin{align*}
&\WW{\vxi,t}{V,(n)}\psi
\\
&:=\sum_{\alpha\in[S]^n}
\sigma_{\alpha_n}\dots\sigma_{\alpha_1}\!\!\!\sum_{{\cA\cup\cB\cup\cC=[n]\atop\#\cC\in2\NN_0}}
\int_{t\simplex_n}\!\!\IN{\cC}\,\AD{\cA}{t}\W{\vxi,t}{V}\,\A{\cB}\,\psi\,\Id t_{[n]},
\end{align*}
for $\psi\in\sC[\mathfrak{d}_C]$ and $t\in\zeit$, and, using the convention
$\WW{\vxi,t}{V,(0)}:=W^{V}_{\vxi,t}$, 
\begin{align*}
\WW{\vxi,t}{V,(N,M)}\psi&:=\sum_{n=N}^M \WW{\vxi,t}{V,(n)}\psi,\quad
\psi\in\CC^L\otimes\sC[\mathfrak{d}_C],\;N,M\in\NN_0,\;N\le M.
\end{align*}
\end{definition}

For later reference, we shall collect a few relations in the following remark.
It also shows that $\WW{\vxi}{V,(n)}\psi$ with $\psi\in\CC^L\otimes\sC[\mathfrak{d}_C]$
is a well-defined adapted continuous process
given by a manageable formula when $\psi$ is an exponential vector.

 \begin{remark}{\rm \label{rem-sec6}
(1) Let $g,h\in\mathfrak{d}_C$. Then we set
\begin{align}\label{def-ADg}
\ADg{\cA}{\tau}&:=\prod_{a\in\cA}\SPn{i\olw{t_a}{\tau}g-iU_{t_a,\tau}^{-}}{{F}_{\alpha_a,\V{X}_{t_a}}},
\\\label{def-Ah}
\Ah{\cB}&:=\prod_{b\in\cB}
\SPn{{F}_{\alpha_b,\V{X}_{t_b}}}{i\w{0}{t_b}h+iU_{t_b}^{+}},
\end{align}
and we shall repeatedly use the following consequences of \eqref{exp-vec1},
\begin{align}\label{eq-ADg}
\ADg{\cA}{\tau}\SPn{\zeta(g)}{\psi}
&=\SPn{\zeta(g)}{\AD{\cA}{\tau}\,\psi},\quad\psi\in\sC[\HP],
\\\label{eq-Ah}
\Ah{\cB}\,\zeta(h)&=\A{\cB}\,\zeta(h).
\end{align}
For instance, we see that, for an exponential vector $\zeta(h)\in\sE[\mathfrak{d}_C]$, 
\begin{align}\label{WWzeta(h)}
\WW{\vxi,t}{V,(n)}\zeta(h)&=\int_{t\simplex_n}\!
\sQ^{(n)}_{t}(h;t_{[n]})\,\W{\vxi,t}{V}\,\zeta(h)\,\Id t_{[n]},
\end{align}
considered as an identity in $\LO(\CC^L)\otimes\sF$, with 
\begin{align}\label{def-Qh}
&\sQ^{(n)}_{\tau}(h;t_{[n]}):=\sum_{\alpha\in[S]^n}
\sigma_{\alpha_n}\dots\sigma_{\alpha_1}\!\!\sum_{{\cA\cup\cB\cup\cC=[n]\atop\#\cC\in2\NN_0}}
\!\!\IN{\cC}\,\Ah{\cB}\,\AD{\cA}{\tau}.
\end{align}
In our computations below it shall also be convenient to use the relation
\begin{align}\label{Qh-Qgh}
\SPb{\zeta(g)}{\sQ_\tau^{(n)}(h;t_{[n]})\,W_{\vxi,t}^{V}\,\zeta(h)}
&=\SPn{\zeta(g)}{W_{\vxi,t}^{V}\,\zeta(h)}\,\sQ_\tau^{(n)}(g,h;t_{[n]}),
\end{align}
which is an identity in $\LO({\CC^L})$, with
\begin{align}\label{spin2}
&\sQ^{(n)}_{\tau}(g,h;t_{[n]}):=\!\!\sum_{\alpha\in[S]^n}
\sigma_{\alpha_n}...\sigma_{\alpha_1}\!\!\!\sum_{{\cA\cup\cB\cup\cC=[n]\atop\#\cC\in2\NN_0}}
\!\!\!\IN{\cC}\ADg{\cA}{\tau}\Ah{\cB}.
\end{align}
The matrix element of $\WW{\vxi,t}{V,(n)}$ for
two exponential vectors $\zeta(g),\zeta(h)\in\sE[\mathfrak{d}_C]$ reads
\begin{align}
\SPn{\zeta(g)}{\WW{\vxi,t}{V,(n)}\,\zeta(h)}\label{spin2b}
&=\SPn{\zeta(g)}{\W{\vxi,t}{V}\,\zeta(h)}\int_{t\simplex_n}\!\sQ^{(n)}_{t}(g,h;t_{[n]})\,\Id t_{[n]}.
\end{align}
(2) We shall consider the domain $\dom(M)$ defined in \eqref{def-whD-intro} as a
Hilbert space equipped with the graph norm of 
$M=\tfrac{1}{2}\Id\Gamma(\V{m})^2+\Id\Gamma(\omega)$.
Then, for each $\vgamma\in\Omega$, the following map is continuous,
\begin{align}\label{sarah0}
I^{n+1}\times\mathfrak{d}_C\ni(t_{[n]},t,h)&\longmapsto
\big(\sQ_t^{(n)}(h;t_{[n]})\,W_{\vxi,t}^V\zeta(h)\big)(\vgamma)\in\LO(\CC^L)\otimes{\dom(M)}.
\end{align}
In particular, the Bochner integral in \eqref{WWzeta(h)} exists and
defines an adapted $\LO(\CC^L)\otimes{\dom(M)}$-valued process such that 
$I\times\mathfrak{d}_C\ni(t,h)\mapsto
\WW{\vxi,t}{V,(n)}(\vgamma)\zeta(h)\in\LO(\CC^L)\otimes{\dom(M)}$
is continuous, for every $\vgamma\in\Omega$, and such that
\begin{align}\label{sarah1}
\V{v}(\vxi,\V{X}_t)\,\WW{\vxi,t}{V,(n)}\,\zeta(h)
&=\int_{t\simplex_n}\V{v}(\vxi,\V{X}_t)\,\sQ_t(h;t_{[n]})\,W_{\vxi,t}^V\zeta(h)\,\Id t_{[n]},
\\\label{sarah2}
\wh{H}^V(\vxi,\V{X}_t)\,\WW{\vxi,t}{V,(n)}\,\zeta(h)
&=\int_{t\simplex_n}\wh{H}^V(\vxi,\V{X}_t)\,\sQ_t(h;t_{[n]})\,W_{\vxi,t}^V\zeta(h)\,\Id t_{[n]},
\end{align}
on $\Omega$ for all $t\in I$ and $h\in\mathfrak{d}_C$.

In fact, recall that, by Hyp.~\ref{hyp-G}, Hyp.~\ref{hyp-B}, and Lem.~\ref{lem-U-K}(5), 
the maps $(s,t)\mapsto U_{s,t}^-\in\mathfrak{k}_C$
and $(s,t)\mapsto \w{s}{t}\,\V{F}_{\V{X}_s}\in\mathfrak{k}_C^S$ are
jointly continuous on $I\times I$, at every $\vgamma\in\Omega$.
Since $u_{\vxi}^V$ (resp. $U^\pm$) are continuous 
complex-valued (resp. $\mathfrak{k}_C$-valued) processes as well,
it is straightforward to infer the continuity of \eqref{sarah0} from
\eqref{exp-vec2}, \eqref{comm-dGamma-aad}, \eqref{def-W}, and \eqref{def-Qh}
in combination with Hyp.~\ref{hyp-G} and Rem.~\ref{rem-smooth}. 
The relations \eqref{sarah1} and \eqref{sarah2} hold true
since $\wh{H}^V(\vxi,\V{x})$ and the components of $\V{v}(\vxi,\V{x})$ can be considered as
bounded operators from $\wh{\dom}$ into $\HR$, whose norms are bounded uniformly in $\V{x}$;
recall \eqref{rb-vp1} and \eqref{ub-whH}.
 }\end{remark}

In Thm.~\ref{thm-Ito-spin} below, we collect our main results on the objects introduced above. 
Recall our standing Hypotheses~\ref{hyp-G},~\ref{hyp-V}, and~\ref{hyp-B}.
Recall also that $\wh{H}^V(\vxi,\V{x})$ in \eqref{SDE-spin} is defined by \eqref{def-fib-Ham-spin}
on the domain $\wh{\dom}$ defined in \eqref{def-whD-intro}. Since we shall consider measurable
functions with values in $\wh{\dom}$, it might make sense to recall that the $\sigma$-algebra
on $\wh{\dom}$ corresponding to the graph norm of 
$M_1(\V{0})$ (defined in \eqref{def-Maxi}) coincides with the trace $\sigma$-algebra
$\wh{\dom}\cap\fB(\FHR)$ of the Borel $\sigma$-algebra on $\FHR$.

\begin{theorem}\label{thm-Ito-spin}
{\rm(1)}
For all $N\in\NN$ and $t\in\zeit$, the operator $\WW{\vxi,t}{V,(0,N)}$, 
defined a priori on $\CC^L\otimes\sC[\mathfrak{d}_C]$, extends uniquely to an element of
$\LO(\FHR)$, which is henceforth again denoted by the same symbol. 
Furthermore, the limit
\begin{equation}\label{limit-WW}
\WW{\vxi,t}{V}:=\WW{\vxi,t}{V,(0,\infty)}:=\lim_{N\to\infty}\WW{\vxi,t}{V,(0,N)}
\end{equation}
exists in $\LO({\FHR})$ $\mathbb{P}$-a.s. and locally uniformly in $t\in \zeit$,
and it $\mathbb{P}$-a.s. satisfies 
\begin{equation}\label{norm-W}
\ln\|\WW{\vxi,t}{V}\|\le\int_0^t\big(\Lambda(\V{X}_s)^2-V(\V{X}_s)\big)\Id s,
\quad t\in I,
\end{equation}
where $\Lambda(\V{x})$ denotes the operator norm of the matrix
$\big(\|\omega^\mh(\vsigma\cdot\V{F}_{\V{x}})_{ij}\|\big)_{i,j=1}^L$.

\smallskip

\noindent{\rm(2)} 
Let $\eta:\Omega\to\wh{\dom}$ be $\fF_0$-measurable. Then 
$\WW{\vxi}{V}\,\eta\in {\sf S}_{\zeit}(\FHR)$
and, up to indistinguishability, $\WW{\vxi}{V}\,\eta$ is the unique
element of ${\sf S}_{\zeit}(\FHR)$ whose paths belong $\PP$-a.s. to  
$C(I,\wh{\dom})$ and which $\PP$-a.s. solves
\begin{align}\label{SDE-spin}
X_\bullet&=\eta-\int_0^\bullet i\V{v}(\vxi,\V{X}_s)X_s\Id\V{X}_s
-\int_0^\bullet\wh{H}^V(\vxi,\V{X}_s)X_s\Id s\,\>\>\text{on $[0,\sup I)$}.
\end{align}
\end{theorem}

{\proof} 
The proof of this theorem can be found at the end of Sect.~\ref{sec-weights};
the rest of Sect.~\ref{sec-weights} and the whole Sect.~\ref{sec-alg} serve as a preparation for it.
\qed

 \begin{remark}{\rm \label{rem-for-W-exp-vec}
In view of \eqref{def-exp-vec}, \eqref{def-W}, and \eqref{spin2b} the matrix element of 
$\WW{\vxi,t}{V}$ for two exponential vectors $\zeta(g),\zeta(h)\in\sE[\mathfrak{d}_C]$ reads
\begin{align}\nonumber
\SPn{\zeta(g)}{\WW{\vxi,t}{V}\zeta(h)}
&=\SPn{\zeta(g)}{W^V_{\vxi,t}\zeta(h)}\,Q_{t}(g,h)
\\\label{spin2bW}
&=e^{-u_{-\vxi,t}^V-\SPn{U_t^-}{h}+\SPn{g}{U^+_t}+\SPn{g}{\w{0}{t}h}}\,Q_{t}(g,h),
\end{align}
which are identities in $\LO({\CC^L})$ with 
\begin{align}\label{spin2a}
Q_{t}(g,h)&:=\id+\sum_{n=1}^\infty\int_{t\simplex_n}
\sQ^{(n)}_{t}(g,h;t_{[n]})\,\Id t_{[n]}\in\LO({\CC^L});
\end{align}
see \eqref{spin2} for a formula for $\sQ^{(n)}_{t}$.
The $\PP$-a.s. locally uniform convergence of the series in \eqref{spin2a}
follows from Thm.~\ref{thm-Ito-spin}; the exceptional subset of $\Omega$
where the series might not converge neither depends on $g$, $h$, nor $t\in I$.
 }\end{remark}


\section{Stochastic calculus in the matrix-valued case}\label{sec-alg}

\noindent
The first step towards the proof of
Thm.~\ref{thm-Ito-spin} essentially comprises applications of It\={o}'s formula
and algebraic manipulations. These are carried through in the present section.
The final result of this section is formulated in
Lem.~\ref{lem-spin1} below, whose derivation is split into three preparatory lemmas
and a concluding proof at the end of the section. The latter proof requires two 
additional technical lemmas which are deferred to App.~\ref{app-cont-tn}.

As the potential $V$ does not influence the convergence properties of
the time ordered integral series, we set it equal to zero in this and in the most part
of the next section; it will be re-introduced only at the very end of the proof of Thm.~\ref{thm-Ito-spin}.

By a simple function we shall always mean a function on $\Omega$ attaining only finitely many values.

\begin{lemma}\label{lem-spin1}
Let $M,N\in\NN_0$ with $N\le M$, and let $\eta$ be a
$\CC^L\otimes\sC[\mathfrak{d}_C]$-valued $\fF_0$-measurable simple function.
Then $\WW{\vxi}{V,(N,M)}\,\eta\in\mathsf{S}_I(\FHR)$ and we $\PP$-a.s. have
\begin{align}\label{spin78}
\WW{\vxi,\bullet}{0,(N,M)}&\eta=\delta_{0,N}\,\eta
-\int_{0}^\bullet\wh{H}^0_{\scal}(\vxi,\V{X}_s)\WW{\vxi,s}{0,(N,M)}\eta\,\Id s
\\\nonumber
&-\int_{0}^\bullet i\V{v}(\vxi,\V{X}_s)\WW{\vxi,s}{0,(N,M)}\eta\,\Id\V{X}_s
+\int_0^\bullet\vsigma\cdot\vp(\V{F}_{\V{X}_{s}})\WW{\vxi,s}{0,(N-1,M-1)}\eta\,\Id s
\end{align}
on $[0,\sup I)$, with $\WW{\vxi,t}{0,(-1,n)}:=\WW{\vxi,t}{0,(0,n)}$, $n\in\NN_0$, 
and $\WW{\vxi,t}{0,(-1,-1)}:=0$.
\end{lemma}

In the rest of this section we fix $g,h\in\mathfrak{d}_C$; recall \eqref{def-fd} and \eqref{def-dCkC}. 

\begin{lemma}\label{lem-spin-adam}
For all $n\in\NN$ and $0\le t_1\le \dots\le t_n\in I$, we have
\begin{align}\nonumber
\SPn{\zeta(g)}{\W{\vxi,t_n}{0}&\zeta(h)}\,\sQ^{(n)}_{t_n}(g,h;t_{[n]})
\\\label{spin9}
&=
\SPb{\zeta(g)}{\vsigma\cdot\vp(\V{F}_{\V{X}_{t_n}})\,
\sQ^{(n-1)}_{t_n}(h;t_{[n-1]})\,\W{\vxi,t_n}{0}\zeta(h)},
\end{align}
where we use the convention $\sQ_{t_1}^{(0)}(h;t_{[0]}):=\id$.
\end{lemma}

{\proof}
Setting $\tau=t_n$ in \eqref{spin2} and taking into account that
$U_{t_n,t_n}^-=0$ on $\Omega$ (see Lem.~\ref{lem-U-K}(5))
and $\w{t_n}{t_n}=1$ on $\Omega$, and we obtain, since $t_n$ is contained in precisely 
one of the sets $\cA$, $\cB$, or $\cC$,
\begin{align}\label{spin5}
\sQ^{(n)}_{t_n}(g,h;t_{[n]})
&=-i\SPn{g}{\vsigma\cdot\V{F}_{\V{X}_{t_n}}}\,\sQ^{(n-1)}_{t_n}(g,h;t_{[n-1]})
\\\nonumber
&\quad
+i\SPn{\vsigma\cdot\V{F}_{\V{X}_{t_n}}}{\w{0}{t_n}h+U_{t_n}^+}\,
\sQ^{(n-1)}_{t_n}(g,h;t_{[n-1]})+J_n(g,h),
\end{align}
where $t_{[n]}=(t_{[n-1]},t_n)=(t_1,\ldots,t_{n-1},t_n)$, $J_1(g,h):=0$, and 
\begin{align}\nonumber
J_n(g,h)&:=\vsigma\cdot\sum_{\alpha\in[S]^{n-1}}
\sigma_{\alpha_{n-1}}\dots\sigma_{\alpha_1}\sum_{c\in[n-1]}
\SPn{\V{F}_{\V{X}_{t_n}}}{\w{t_c}{t_n}{F}_{\alpha_c,\V{X}_{t_c}}}
\\\nonumber
&\qquad\cdot\sum_{{\cA\cup\cB\cup\cC=[n-1]\setminus\{c\}\atop\#\cC\in2\NN_0}}
\IN{\cC}\,\ADg{\cA}{t_n}\,\Ah{\cB},\quad n\ge2.
\end{align}
Next, we observe that, by \eqref{exp-vec1} and \eqref{def-W},
\begin{align}\label{spin6}
i\SPn{\V{F}_{\V{X}_{t_n}}}{\w{0}{t_n}\,h+U_{t_n}^+}\,\W{\vxi,t_n}{0}\zeta(h)
&=a(\V{F}_{\V{X}_{t_n}})\,\W{\vxi,t_n}{0}\zeta(h),
\\\label{spin7}
-i\SPn{\zeta(g)}{\psi}\SPn{g}{\V{F}_{\V{X}_{t_n}}}
&=\SPn{\zeta(g)}{\ad(\V{F}_{\V{X}_{t_n}})\,\psi},\quad\psi\in\sC[\mathfrak{d}_C].
\end{align}
Hence, using \eqref{Qh-Qgh} first and \eqref{spin6} and \eqref{spin7} afterwards we see that
\begin{align}\nonumber
\SPn{\zeta(g)}{\W{\vxi,t_n}{0}&\zeta(h)}\sQ^{(n)}_{t_n}(g,h;t_{[n]})
=\SPn{\zeta(g)}{\W{\vxi,t_n}{0}\zeta(h)}\,J_n(g,h)
\\\nonumber
&+
\SPb{\zeta(g)}{\vsigma\cdot\ad(\V{F}_{\V{X}_{t_n}})\sQ^{(n-1)}_{t_n}(h;t_{[n-1]})
\W{\vxi,t_n}{0}\zeta(h)}
\\\label{spin1234}
&+\vsigma\cdot\SPn{\zeta(g)}{\sQ_{t_n}^{(n-1)}(h;t_{[n-1]})a(\V{F}_{\V{X}_{t_n}})
\W{\vxi,t_n}{0}\zeta(h)}.
\end{align}
Moreover, for $n\ge2$ and every subset $\cA\subset[n-1]$, \eqref{CCR} implies
\begin{align}\nonumber
\big[a(\V{F}_{\V{X}_{t_n}}),\AD{\cA}{t_n}\big]
&=\sum_{c\in\cA} \SPn{\V{F}_{\V{X}_{t_n}}}{\w{t_c}{t_n}{F}_{\alpha_c,\V{X}_{t_c}}}
\,\AD{\cA\setminus{\{c\}}}{t_n},
\end{align}
which together with \eqref{eq-ADg}, \eqref{def-Qh}, and a rearrangement of summations yields
\begin{align*}
\SPn{\zeta(g)}{\W{\vxi,t_n}{0}\zeta(h)}J_n(g,h)
&=\vsigma\cdot\SPb{\zeta(g)}{\big[a(\V{F}_{\V{X}_{t_n}}),\sQ_{t_n}^{(n-1)}(h;t_{[n-1]})
\big]\W{\vxi,t_n}{0}\zeta(h)}.
\end{align*}
Combining the previous identity with \eqref{spin1234} we arrive at \eqref{spin9}.
\qed

In the next lemmas we shall apply the formulas of the stochastic calculus with
respect to the time-shifted stochastic basis $\BB_{t_n}$. For this purpose, we shall
first introduce some convenient notation. 

As usual stochastic integrals
starting at $t_n\in I$ are defined as follows: If $\sK$ is a separable Hilbert space,
$(A_t)_{t\in I}$ a family of $\LO(\RR^m,\sK)$-valued random variables such that
$(A_{t_n+t})_{t\in I^{t_n}}$ is left-continuous and  $\BB_{t_n}$-adapted,
and if $(Z_t)_{t\in I}$ is a family of $\RR^m$-valued random variables such that
$Z^{(t_n)}:=(Z_{t_n+t})_{t\in I^{t_n}}$ is a continuous $\BB_{t_n}$-semi-martingale, then we set
\begin{align*}
\int_{t_n}^tA_{s}\,\Id Z_s:=\int_{0}^{t-t_n}A_{t_n+s}\,\Id Z_s^{(t_n)},\quad t_n\le t<\sup I.
\end{align*}
For instance,  if $t_a\in[0,t_n]$, then by using It\={o}'s formula with $\BB_{t_n}$ as underlying 
stochastic basis we obtain the formulas
\begin{align}\nonumber
&\w{t_a}{\tau}{F}_{\alpha_a,\V{X}_{t_a}}-\w{t_a}{t_n}{F}_{\alpha_a,\V{X}_{t_a}}
\\\label{spin55}
&\quad=\int_{t_n}^\tau i\V{m}\,\w{t_a}{s}{F}_{\alpha_a,\V{X}_{t_a}}\Id\V{X}_s
-\int_{t_n}^\tau(\tfrac{1}{2}\V{m}^2+\omega)\w{t_a}{s}{F}_{\alpha_a,\V{X}_{t_a}}\Id s,
\\\nonumber
&\SPn{U_{t_a,\tau}^-}{F_{\alpha_a,\V{X}_{t_a}}}
-\SPn{U_{t_a,t_n}^-}{F_{\alpha_a,\V{X}_{t_a}}}
\\\label{spin56}
&\quad=
\int_{t_n}^\tau\SPb{\V{G}_{\V{X}_s}}{\w{t_a}{s}{F}_{\alpha_a,\V{X}_{t_a}}}\Id\V{X}_s
+\int_{t_n}^\tau\SPb{\breve{q}_{\V{X}_s}}{\w{t_a}{s}{F}_{\alpha_a,\V{X}_{t_a}}}\Id s,
\end{align}
$\PP$-a.s. for all $\tau\in[t_n,\sup I)$. 
(If one wishes to prove the first one by means of Prop.~\ref{prop-Ito}, then one should
apply this proposition to $f_R(s+t_n-t_a,\V{X}_{t_n+s}-\V{X}_{t_a},{F}_{\alpha_a,\V{X}_{t_a}})$, 
where $f_R$ is the same as in the proof of Lem.~\ref{lem-U-K}(3), consider 
${F}_{\alpha_a,\V{X}_{t_a}}$ as a time-independent process, and remove the cut-off
afterwards.)

\begin{lemma}\label{lem-spin-eva}
For all $n\in\NN$, $0\le t_1\le\dots\le t_n<\sup I$, and $\sA\subset[n]$,
we $\PP$-a.s. have, for all $t\in[t_n,\sup I)$,
\begin{align}\nonumber
&\int_{t_n}^t\SPn{\zeta(g)}{\W{\vxi,\tau}{0}\zeta(h)}\,\Id_\tau\ADg{\cA}{\tau}
\\\nonumber
&\quad+\int_{t_n}^t\SPb{\zeta(g)}{[\AD{\cA}{\tau},\V{v}(\vxi,\V{X}_\tau)]
\V{v}(\vxi,\V{X}_\tau)\W{\vxi,\tau}{0}\zeta(h)}\Id\tau
\\\nonumber
&\quad\quad=\int_{t_n}^t\!\!
\SPb{\zeta(g)}{[\AD{\cA}{\tau},\wh{H}_{\scal}^0(\vxi,\V{X}_\tau)]\W{\vxi,\tau}{0}\zeta(h)}\Id\tau
\\\label{spin1000}
&\quad\quad\quad+i\int_{t_n}^t\SPb{\zeta(g)}{[\AD{\cA}{\tau},\V{v}(\vxi,\V{X}_\tau)]\,
\W{\vxi,\tau}{0}\zeta(h)}\Id\V{X}_\tau.
\end{align}
\end{lemma}

{\proof}
We may assume that $\cA$ is non-empty, for otherwise all terms in \eqref{spin1000} are zero.
First, we compute the stochastic differential of the process $[t_n,t]\ni\tau\mapsto\ADg{\cA}{\tau}$ 
given by \eqref{def-ADg}. Employing the conventions introduced in the paragraph preceding 
the lemma and \eqref{spin55} and \eqref{spin56}
we find by straightforward computations and It\={o}'s product rule for $\#\cA$ factors,
\begin{align}\nonumber
&\ADg{\cA}{t}-\ADg{\cA}{t_n}
\\\nonumber
&=i\sum_{c\in\cA}\int_{t_n}^t\ADg{\cA\setminus{c}}{\tau}
\SPb{i\V{m}g+\V{G}_{\V{X}_\tau}}{\w{t_c}{\tau}{F}_{\alpha_c,\V{X}_{t_c}}}\Id\V{X}_\tau
\\\nonumber
&+i\sum_{c\in\cA}\int_{t_n}^t\ADg{\cA\setminus{c}}{\tau}
\SPb{(\omega+\tfrac{1}{2}\V{m}^2)g+\breve{q}_{\V{X}_\tau}}{\w{t_c}{\tau}
{F}_{\alpha_c,\V{X}_{t_c}}}\Id\tau
\\\label{spin11}
&-\sum_{{\cP\subset\cA\atop\#\cP=2}}\int_{t_n}^t\ADg{\cA\setminus\cP}{\tau}
\Big(\prod_{b\in\cP}\SPb{i\V{m}g+\V{G}_{\V{X}_\tau}}{\w{t_b}{\tau}
{F}_{\alpha_b,\V{X}_{t_b}}}\Big)\,d\tau,
\end{align}
$\PP$-a.s. for all $t\in[t_n,\sup I)$.
Here and henceforth we write $\cA\setminus c:=\cA\setminus\{c\}$ for short;
if $\#\cA=1$, then the last line of \eqref{spin11} should be ignored.
It will shortly turn out that all terms on the right hand side of \eqref{spin11} 
can be related to certain commutators involving $\V{v}(\vxi,\V{X}_\tau)$ or the scalar fiber Hamiltonian.
In fact, if the multiplication operator $\vk$ in $\HP$ is either equal to $1$ or
equal to one of the components of $\V{m}$, then we first observe that 
\eqref{CCR} and \eqref{comm-dGamma-aad} entail
\begin{align}\nonumber
\big[\ad(\vk\,\w{t_c}{\tau}&{F}_{\alpha_c,\V{X}_{t_c}})\,,-\Id\Gamma(\V{m})-\vp(\V{G}_{\V{X}_\tau})\big]
\\\label{spin57}
&=\ad(\vk\,\V{m}\,\w{t_c}{\tau}{F}_{\alpha_c,\V{X}_{t_c}})
+\SPn{\vk\,\V{G}_{\V{X}_\tau}}{\w{t_c}{\tau}{F}_{\alpha_c,\V{X}_{t_c}}}.
\end{align}
By virtue of the Leibnitz rule for commutators and \eqref{CCR}, this shows that
\begin{align}\nonumber
&\big[\AD{\cA}{\tau},\V{v}(\vxi,\V{X}_\tau)\big]
\\\label{spin50}
&=
\sum_{c\in \cA}\big(\ad(\V{m}\,\w{t_c}{\tau}{F}_{\alpha_c,\V{X}_{t_c}})
+\SPn{\V{G}_{\V{X}_\tau}}{\w{t_c}{\tau}{F}_{\alpha_c,\V{X}_{t_c}}}\big)\AD{\cA\setminus{c}}{\tau}.
\end{align}
Applying \eqref{spin57} and the Leibnitz rule and \eqref{CCR} once more we deduce that
\begin{align}\label{spin51}
\big[&\AD{\cA}{\tau},\,
\tfrac{1}{2}\V{v}(\vxi,\V{X}_\tau)^2\big]-[\AD{\cA}{\tau},\V{v}(\vxi,\V{X}_\tau)]\,\V{v}(\vxi,\V{X}_\tau)
\\\nonumber
&=-\frac{1}{2}\,\big[[\AD{\cA}{\tau},\V{v}(\vxi,\V{X}_\tau)],\,\V{v}(\vxi,\V{X}_\tau)\big]
\\\nonumber
&=
-\sum_{c\in \cA}\big\{
\ad\big(\tfrac{1}{2}\V{m}^2
\w{t_c}{\tau}{F}_{\alpha_c,\V{X}_{t_c}}\big)
+\tfrac{1}{2}\SPn{\V{m}\cdot\V{G}_{\V{X}_\tau}}{\w{t_c}{\tau}{F}_{\alpha_c,\V{X}_{t_c}}}\big\}
\AD{\cA\setminus{c}}{\tau}
\\\nonumber
&\quad-\sum_{{\cP\subset \cA\atop\#\cP=2}}\!\!\AD{\cA\setminus\cP}{\tau}\prod_{b\in\cP}
\{\ad(\V{m}\,\w{t_{b}}{\tau}{F}_{\alpha_b,\V{X}_{t_{b}}})
+\SPn{\V{G}_{\V{X}_\tau}}{\w{t_{b}}{\tau}{F}_{\alpha_b,\V{X}_{t_{b}}}}\},
\end{align}
where the last line should again be ignored in the case $\#\cA=1$.
Likewise, we obtain the following relation for the remaining term in $\wh{H}_{\scal}^0(\vxi,\V{X}_\tau)$,
\begin{align}\label{spin52}
&\big[\AD{\cA}{\tau},\,\Id\Gamma(\omega)-\tfrac{i}{2}\vp(q_{\V{X}_\tau})\big]
\\\nonumber
&=\sum_{c\in \cA}\big\{-\ad(\omega
\,\w{t_c}{\tau}{F}_{\alpha_c,\V{X}_{t_c}})
+\tfrac{i}{2}\SPn{q_{\V{X}_\tau}}{\w{t_c}{\tau}{F}_{\alpha_c,\V{X}_{t_c}}}\big\}
\AD{\cA\setminus{c}}{\tau}.
\end{align}
Next, we observe that, if we apply the operators on the right hand sides of 
\eqref{spin50}--\eqref{spin52} to any vector in $\sC[\mathfrak{d}_C]$ and scalar-multiply
the results with $\zeta(g)$, then the creation operators $\ad(f)$ on the right hand sides
can be replaced by $\SPn{ig}{f}$ and $\AD{\cB}{\tau}$ 
can be replaced by $\ADg{\cB}{\tau}$; see \eqref{exp-vec1} and \eqref{eq-ADg}.
We conclude by comparing the so-obtained identities with 
\eqref{sym-q} and \eqref{spin11}, and by employing the substitution rule of
Prop.~\ref{prop-stoch-calc} (w.r.t. the basis $\BB_{t_n}$) to compute the first line of \eqref{spin1000}.
\qed

\begin{lemma}
For all $n\in\NN$, $0\le t_1\le\dots \le t_n<\sup I$, and $\cA\subset[n]$, we $\PP$-a.s. have
\begin{align}\nonumber
&\SPn{\zeta(g)}{\W{\vxi,t}{0}\zeta(h)}\,\ADg{\cA}{t}
-\SPn{\zeta(g)}{\W{\vxi,t_n}{0}\zeta(h)}\,\ADg{\cA}{t_n}
\\\nonumber
&=-\int_{t_n}^t\SPb{\zeta(g)}{\wh{H}_{\scal}^0(\vxi,\V{X}_\tau)
\,\AD{\cA}{\tau}\,\W{\vxi,\tau}{0}\zeta(h)}\Id\tau
\\\label{spin1001}
&\quad
-i\int_{t_n}^t\SPb{\zeta(g)}{\V{v}(\vxi,\V{X}_\tau)\,\AD{\cA}{\tau}\,\W{\vxi,\tau}{0}\zeta(h)}\Id\V{X}_\tau,
\quad t\in[t_n,\sup I).
\end{align}
\end{lemma}

{\proof}
If $\cA=\varnothing$, then \eqref{spin1001} follows directly from \eqref{SDE}.
Hence, we may assume in the following that $\cA$ is non-empty.

We shall denote the mutual variation, defined by means of the
time-shifted stochastic basis $\BB_{t_n}$, of 
$(\SPn{\zeta(g)}{\W{\vxi,t_n+s}{0}\zeta(h)})_{s\in I^{t_n}}$
and $(\ADg{\cA}{t_n+s})_{s\in I^{t_n}}$  by
$(\llbracket\SPn{\zeta(g)}{\W{\vxi}{0}\zeta(h)},\ADg{\sA}{}\rrbracket_{t_n,t_n+s})_{s\in I^{t_n}}$.
Then, on the one hand, by the definition \eqref{def-covar} and by \eqref{SDE}, we $\PP$-a.s. have
\begin{align}\label{spin1002}
\llbracket&\SPn{\zeta(g)}{\W{\vxi}{0}\zeta(h)},\ADg{\sA}{}\rrbracket_{t_n,t}
+\int_{t_n}^t\SPn{\zeta(g)}{\W{\vxi,\tau}{0}\zeta(h)}\,\Id_\tau\ADg{\cA}{\tau}
\\\nonumber
&=
\SPn{\zeta(g)}{\W{\vxi,t}{0}\zeta(h)}\,\ADg{\cA}{t}
-\SPn{\zeta(g)}{\W{\vxi,t_n}{0}\zeta(h)}\,\ADg{\cA}{t_n}
\\\nonumber
&\quad+
\int_{t_n}^t\ADg{\cA}{\tau}\,
\SPb{\zeta(g)}{\wh{H}_{\scal}^0(\vxi,\V{X}_\tau)\,\W{\vxi,\tau}{0}\zeta(h)}\,\Id\tau
\\\nonumber
&\quad+\int_{t_n}^t\ADg{\cA}{\tau}\,
\SPb{\zeta(g)}{i\V{v}(\vxi,{\V{X}_\tau})\,\W{\vxi,\tau}{0}\zeta(h)}\,\Id\V{X}_\tau,\quad t\in[t_n,\sup I).
\end{align}
On the other hand we may compute the mutual variation defined above
by applying \eqref{for-covar}
in combination with \eqref{SDE} and \eqref{spin11}. In this way we obtain
\begin{align}\nonumber
&\llbracket\SPn{\zeta(g)}{\W{\vxi}{0}\zeta(h)},\ADg{\sA}{}\rrbracket_{t_n,t}
\\\nonumber
&=\sum_{c\in\cA}\int_{t_n}^t\SPn{\zeta(g)}{\V{v}(\vxi,\V{X}_\tau)\W{\vxi,\tau}{0}\zeta(h)}
\SPn{i\V{m}g+\V{G}_{\V{X}_\tau}}{\w{t_c}{\tau}\V{F}^{(c)}_{\V{X}_{t_c}}}
\ADg{\cA\setminus{c}}{\tau}\Id\tau
\\\label{spin1003}
&=\int_{t_n}^t\SPb{\zeta(g)}{[\AD{\cA}{\tau},\V{v}(\vxi,\V{X}_\tau)]\V{v}(\vxi,\V{X}_\tau)
\W{\vxi,\tau}{0}\zeta(h)}\Id\tau,
\end{align}
where we also used \eqref{exp-vec1}, \eqref{eq-ADg}, and \eqref{spin50} in the second step.
By virtue of \eqref{spin1003} we see that the left hand sides of
\eqref{spin1000} and \eqref{spin1002} are equal, $\PP$-a.s. for all $t\in[t_n,\sup I)$.
Equating the right hand sides of the latter identities 
and applying \eqref{eq-ADg} we arrive at \eqref{spin1001}.
\qed 

\smallskip

{\it Proof of Lem.~\ref{lem-spin1}.}
Let $n\in \NN$ and $0\le t_1\le\dots\le t_n$.
Multiplying both sides of the identity \eqref{spin1001}
with the $\fF_{t_n}$-measurable, $\LO(\CC^L)$-valued random variable
$\sigma_{\alpha_n}\dots\sigma_{\alpha_1}\IN{\cC}\Ah{\cB}$
(which commutes $\PP$-a.s. with the stochastic integral in \eqref{spin1001})
and summing over all partitions of sets and components of the multi-index $\alpha$
afterwards, we $\PP$-a.s. obtain
\begin{align}\nonumber
\SPn{\zeta(g)&}{\W{\vxi,t}{0}\zeta(h)}\,\sQ_t^{(n)}(g,h;t_{[n]})
=\SPn{\zeta(g)}{\W{\vxi,t_n}{0}\zeta(h)}\,\sQ_{t_n}^{(n)}(g,h;t_{[n]})
\\\nonumber
&-\int_{t_n}^t\SPb{\zeta(g)}{\wh{H}_{\scal}^0(\vxi,\V{X}_\tau)\,
\sQ_\tau^{(n)}(h;t_{[n]})\,\W{\vxi,\tau}{0}\zeta(h)}\,\Id\tau
\\\label{spin-vera}
&-\int_{t_n}^t\SPb{\zeta(g)}{i\V{v}(\vxi,\V{X}_\tau)\,
\sQ_\tau^{(n)}(h;t_{[n]})\,\W{\vxi,\tau}{0}\zeta(h)}\,\Id\V{X}_\tau,
\end{align}
for all $t\in[t_n,\sup I)$. In Lem.~\ref{lem-cont-tn} below we shall verify that the exceptional 
$\PP$-zero set where \eqref{spin-vera} might not hold can actually be chosen independently
of $0\le t_1\le\dots\le t_n\le t<\sup I$.
Hence, we may integrate \eqref{spin-vera} over the simplex $t\simplex_n$, for every $t<\sup I$.
Rewriting the second member of the first line of the above identity
by means of \eqref{spin9} we thus obtain (recall that $\Id t_{[n]}:=\Id t_1\ldots \Id t_n$)
\begin{align*}
&\int_{t\simplex_n}\SPn{\zeta(g)}{\sQ_t^{(n)}(h;t_{[n]})\,\W{\vxi,t}{0}\zeta(h)}\,\Id t_{[n]}
\\
&=\int_{t\simplex_n}\SPn{\zeta(g)}{\vsigma\cdot\vp(\V{F}_{\V{X}_{t_n}})\,
\sQ_{t_n}^{(n-1)}(h;t_{[n-1]})\,\W{\vxi,t_n}{0}\zeta(h)}\,\Id t_1\ldots \Id t_n
\\
&\quad-\int_{0}^t\!\!\int_{t_n}^t\!\int_{t_{n}\simplex_{n-1}}
\!\!\!\SPb{\zeta(g)}{\wh{H}_{\scal}^0(\vxi,\V{X}_\tau)
\sQ_\tau^{(n)}(h;t_{[n]})\W{\vxi,\tau}{0}\zeta(h)}
\,\Id t_1\ldots \Id t_{n-1}\,\Id\tau\,\Id t_n
\\
&\quad-\int_{I^n}\int_{0}^t1_{\{t_1\le\cdots\le t_n\le\tau\le t\}}\times
\\
&\qquad\qquad\qquad\;\;\times\SPb{\zeta(g)}{i\V{v}(\vxi,\V{X}_\tau)\,
\sQ_\tau^{(n)}(h;t_{[n]})\,\W{\vxi,\tau}{0}\zeta(h)}\,\Id\V{X}_\tau\,\Id t_{[n]}.
\end{align*}
Next, we apply the rule 
\begin{equation}\label{Fub}
\int_0^t\int_{t_n}^t f(\tau,t_n)\,\Id\tau\,\Id t_n=\int_0^t\int_0^{\tau}f(\tau,t_n)\,\Id t_n\,\Id \tau
\end{equation}
to the integral in the third line and change the name of the
integration variable of the most exterior integral in the second line from $t_n$ to $\tau$.
In the last integral we write $\Id\V{X}_\tau=\Id\V{B}_\tau+\V{\beta}(\tau,\V{X}_\tau)\Id\tau$,
employ \eqref{Fub} once more to deal with the term containing $\V{\beta}$,
and use the stochastic Fubini theorem 
to interchange the $\Id\V{B}_\tau$- and $\Id t_{[n]}$-integration;
see, e.g., \cite[\textsection4.5]{daPrZa2014} for a suitable version
of the stochastic Fubini theorem and Lem.~\ref{lem-stoch-Fub} for justification.
After that we apply the relations \eqref{sarah1} and \eqref{sarah2}.
Finally, we use that the (stochastic)
integrals commute with the scalar product and that 
$\{\zeta(g):\,g\in\mathfrak{a}_C\}$ is a countable total set in $\sF$, 
if $\mathfrak{a}_C\subset\mathfrak{d}_C$ is countable and dense in $\HP_C$.
Taking these remarks into account we $\PP$-a.s. arrive at
\begin{align}\nonumber
\WW{\vxi,t}{0,(n)}\,\psi
&=\int_0^t\vsigma\cdot\vp(\V{F}_{\V{X}_{\tau}})\,\WW{\vxi,\tau}{0,(n-1)}\,\psi\,\Id\tau
-\int_{0}^t\wh{H}_{\scal}^{0}(\vxi,\V{X}_\tau)\,\WW{\vxi,\tau}{0,(n)}\,\psi\,\Id\tau
\\\label{spin77}
&\quad-\int_{0}^ti\V{v}(\vxi,\V{X}_\tau)\,
\WW{\vxi,\tau}{0,(n)}\,\psi\,\Id\V{X}_\tau,\quad t\in[0,\sup I),\;n\in\NN_0,
\end{align}
for a given $\psi\in\CC^L\otimes\sC[\mathfrak{d}_C]$.
Here we introduced the convention $\WW{\vxi,t}{0,(-1)}:=0$, so that \eqref{spin77}
follows immediately from \eqref{SDE} in the case $n=0$.
Adding the identities \eqref{spin77} for $n=N,\ldots,M$ we arrive at
\eqref{spin78} with constant $\eta=\psi\in\CC^L\otimes\sC[\mathfrak{d}_C]$. We
conclude by noting that the integrals in \eqref{spin78} commute $\PP$-a.s. with
multiplications by characteristic functions of sets in $\fF_0$.
\qed


\section{Weighted estimates}\label{sec-weights}

\noindent
It is not hard to infer the existence of the limit \eqref{limit-WW} from the results of 
Sect.~\ref{sec-alg}, which is done in Lem.~\ref{lem-spin2} below by an iterative application of 
Gronwall inequalities. What is more involved is to prove that the limiting objects $\WW{\vxi}{V}$ 
give rise to solutions of the SDE \eqref{SDE-spin}, for every $\fF_0$-measurable initial condition 
$\eta:\Omega\to\wh{\dom}$. In particular, we first have to study some mapping properties of the
operators $\WW{\vxi}{0}$ ensuring that $\WW{\vxi}{0}\eta$ again attains its values in $\wh{\dom}$, 
i.e., in the domain of the generalized fiber Hamiltonians, and that it is continuous as a 
$\wh{\dom}$-valued process (so that the (stochastic) integrals in \eqref{SDE-spin} actually exist).
This is the purpose of the weighted estimates derived in Lem.~\ref{lem-spin4} and 
Lem.~\ref{lem-spin-k}, which require some more preparations themselves. The latter two lemmas are 
obtained for bounded initial conditions $\V{q}$ in \eqref{Ito-eq-X} only, as we shall use the bound 
\eqref{hyp-Y2} in their proofs. To get rid of this restriction on $\V{q}$ we shall invoke the pathwise 
uniqueness properties discussed in Rem.~\ref{rem-pw-unique}.
When we pass to more general $\eta$ and to the limit $M\to\infty$ in the
SDE \eqref{spin78}, then our analysis will again rest on Lem.~\ref{lem-spin4} and 
Lem.~\ref{lem-spin-k}. Everything will be put together in the proof of Thm.~\ref{thm-Ito-spin} at the end 
of this section.

The following corollary will be applied with various choices for the weights $\Theta$
later on, namely the trivial choice $\Theta=\id$ and the ones defined in \eqref{def-Xi} and
\eqref{def-Upsilon} below. We shall use the convenient notation $\Ad_ST:=[S,T]$, and
the symbols $\const(a,\ldots),\const'(a,\ldots)$, etc., denote positive constants which depend 
only on the objects appearing in Hyp.~\ref{hyp-G} and the quantities
displayed in their arguments (if any) as long as nothing else
is stated explicitly. Their values might change from one estimate to another.

\begin{corollary}\label{cor-spin1}
Let $\Theta$ be a bounded, strictly positive measurable function of a second quantized
multiplication operator. Let $\vt$ denote one of the operators 
$\id$ or $1+\Id\Gamma(\omega)$ and abbreviate
\begin{align}\nonumber
T_1(s)
&:=\vt^\mh\Theta^{-1}\big[[\Theta^2,\vp(\V{G}_{\V{X}_s})]\,,\,
\Id\Gamma(\V{m})+\vp(\V{G}_{\V{X}_s})\big]\,\Theta^{-1}\vt^\mh
\\\nonumber
&\quad-\vt^\mh\,\Re\big(i[\Theta,\vp({q}_{\V{X}_s})]\Theta^{-1}\big)\,\vt^\mh
\\\nonumber
&=
2i\vt^\mh\Theta^{-1}(\Ad_{\vp(\breve{q}_{\V{X}_s})}\Theta)\vt^\mh
+2i\vt^\mh(\Ad_{\vp(\breve{q}_{\V{X}_s})}\Theta)\Theta^{-1}\vt^\mh
\\\label{eva1}
&\quad
+\vt^\mh\Theta^{-1}(\Ad_{\vp(\V{G}_{\V{X}_s})}^2\Theta^2)\Theta^{-1}\vt^\mh,
\\\label{eva3}
T_2(s)
&:=(1+\Id\Gamma(\omega))^\mh\,\Theta\,\vsigma\cdot\vp(\V{F}_{\V{X}_{s}})\,\Theta^{-1},
\\\label{eva4}
\V{T}(s)&:=-(\Ad_{\vp(\V{G}_{\V{X}_s})}\Theta)\,\Theta^{-1}\vt^\mh,
\end{align}
assuming that the operators in \eqref{eva1} and \eqref{eva4}, 
which are well-defined a priori on $\CC^L\otimes\sC[\mathfrak{d}_C]$, extend to
bounded operators on ${\FHR}$ whose norms are locally uniformly bounded in $s\in I$. 
(For the one in \eqref{eva3} this is clear in view of \eqref{rb-vp1}.) 
Let $p\in\NN$, $\delta>0$, $M,N\in\NN_0$ with $N\le M$, and let
$\eta$ be a $\CC^L\otimes\sC[\mathfrak{d}_C]$-valued $\fF_0$-measurable simple function. 
Outside some $\PP$-zero set we then have, for all $t\in[0,\sup I)$,
\begin{align}\label{spin81n}
&\|\Theta\WW{\vxi,t}{0,(N,M)}\,\eta\|^{2p}-\|\Theta\WW{\vxi,0}{0,(N,M)}\,\eta\|^{2p}
\\\nonumber
&\le-p(2-\delta)\int_0^t\|\Theta \WW{\vxi,s}{0,(N,M)}\eta\|^{2p-2}
\,\big\|\Id\Gamma(\omega)^\eh\Theta\WW{\vxi,s}{0,(N,M)}\eta\big\|^2\Id s
\\\nonumber
&\;\,+\const(p)\int_0^t\|\Theta\WW{\vxi,s}{0,(N,M)}\eta\|^{2p-2}
\big(\|T_1(s)\|+\|\V{T}(s)\|^2\big)\big\|\vt^\eh\Theta \WW{\vxi,s}{0,(N,M)}\eta\big\|^2\Id s
\\\nonumber
&\;\,+{\textstyle{p(\delta+\frac{1}{\delta})}}\int_0^t
\|\Theta \WW{\vxi,s}{0,(N,M)}\eta\|^{2p}\Id s
+\int_0^t \frac{\|T_2(s)\|^{2p}}{\delta}\,\|\Theta \WW{\vxi,s}{0,(N-1,M-1)}\eta\|^{2p}\Id s
\\\nonumber
&\;\,+\int_0^t
2p\|\Theta \WW{\vxi,s}{0,(N,M)}\eta\|^{2p-2}\Re\SPb{\Theta \WW{\vxi,s}{0,(N,M)}\eta}{
i\V{T}(s)\,\vt^\eh\Theta \WW{\vxi,s}{0,(N,M)}\eta}\Id \V{X}_s.
\end{align}
\end{corollary}

{\proof} 
By virtue of the integral representation \eqref{spin78}
we know that $\psi^{(N,M)}:=\Theta\,\WW{\vxi}{0,(N,M)}\eta\in{\sf S}_I({\FHR})$.
Applying Ex.~\ref{ex-Ito-SP}, we $\PP$-a.s. find
\begin{align}\nonumber
\|\psi_t^{(N,M)}\|^2&=\|\psi_0^{(N,M)}\|^2
-\int_0^t2\SPb{\psi_s^{(N,M)}}{\Id\Gamma(\omega)\,\psi_s^{(N,M)}}\,\Id s
\\\nonumber
&\quad-\int_0^t2\Re\SPb{\WW{\vxi,s}{0,(N,M)}\eta}{\Theta^2\,
\tfrac{1}{2}\V{v}(\vxi,\V{X}_s)^2\,\WW{\vxi,s}{0,(N,M)}\eta}\,\Id s
\\\nonumber
&\quad+\int_0^t2\Re\SPb{\psi_s^{(N,M)}}{\Theta\,\tfrac{i}{2}\vp(q_{\V{X}_s})\,
\WW{\vxi,s}{0,(N,M)}\eta}\,\Id s
\\\nonumber
&\quad+
\int_0^t2\Re\SPb{\psi_s^{(N,M)}}{\Theta\,\vsigma\cdot\vp(\V{F}_{\V{X}_{s}})
\,\WW{\vxi,s}{0,(N-1,M-1)}\eta}\,\Id s
\\\nonumber
&\quad-\int_0^t
2\Re\SPb{\psi_s^{(N,M)}}{\Theta\,i\V{v}(\vxi,\V{X}_s)
\,\WW{\vxi,s}{0,(N,M)}\eta}\,\Id \V{X}_s
\\\label{spin79}
&\quad+\int_0^t
\big\|\Theta\,\V{v}(\vxi,\V{X}_s)
\,\WW{\vxi,s}{0,(N,M)}\eta\big\|^2\Id s,\quad t\in[0,\sup I).
\end{align}
Next, we commute $\Theta^2$ with one of the factors $\V{v}(\vxi,\V{X}_s)$
in the second line and take the cancellation with the term in the last
line into account. Furthermore, 
we use that $2\Re\{[A,B]C\}=[[A,B],C]$, if $A$, $B$, and $C$ are symmetric, 
and $\Re\SPn{\psi_s^{(N,M)}}{\tfrac{i}{2}\vp(q_{\V{X}_s})\psi_s^{(N,M)}}=0$,
to see that the sum of the terms in the second, third, and last lines
above is equal to the term in the second line of \eqref{spin80} below.
Likewise, we commute $\Theta$
with $\V{v}(\vxi,\V{X}_s)=\vxi-\Id\Gamma(\V{m})-\vp(\V{G}_{\V{X}_s})$ 
in the penultimate line above and observe that
$
\Re\SPn{\psi_s^{(N,M)}}{i\V{v}(\vxi,\V{X}_s)\,\psi_s^{(N,M)}}=0
$
to see that the $\Id\V{X}_s$-integrals in \eqref{spin79} and \eqref{spin80} are identical.
Altogether we $\PP$-a.s. arrive at 
\begin{align}\nonumber
\|&\psi_{t}^{(N,M)}\|^2=\|\psi_{0}^{(N,M)}\|^2
-\int_0^t2\SPb{\psi_s^{(N,M)}}{\Id\Gamma(\omega)\,\psi_s^{(N,M)}}\,\Id s
\\\nonumber
&\quad-\frac{1}{2}\int_0^t\SPb{\vt^\eh\psi_s^{(N,M)}}{T_1(s)\,
\vt^\eh\psi_s^{(N,M)}}\,\Id s
\\\nonumber
&\quad+
\int_0^t2\Re\SPb{(1+\Id\Gamma(\omega))^\eh\psi_s^{(N,M)}}{
T_2(s)\,\psi_{s}^{(N-1,M-1)}}\,\Id s
\\\label{spin80}
&\quad+\int_0^t2\Re\SPb{\psi_s^{(N,M)}}{i\V{T}(s)
\,\vt^\eh\psi_s^{(N,M)}}\,\Id \V{X}_s,\quad t\in[0,\sup I).
\end{align}
For every $p\in\NN$, $p\ge 2$, another application of It\={o}'s formula 
(to the function $f(t)=t^p$, using \eqref{spin79}) yields
\begin{align}\nonumber
&\|\psi_t^{(N,M)}\|^{2p}=\|\psi_0^{(N,M)}\|^{2p}
-\int_0^t2p\,\|\psi_s^{(N,M)}\|^{2p-2}\,\SPb{\psi_s^{(N,M)}}{\Id\Gamma(\omega)\,
\psi_s^{(N,M)}}\,\Id s
\\\nonumber
&\;\;-\frac{p}{2}\int_0^t\|\psi_s^{(N,M)}\|^{2p-2}\,
\SPb{\vt^\eh\psi_s^{(N,M)}}{T_1(s)\,\vt^\eh\psi_s^{(N,M)}}\,\Id s
\\\nonumber
&\;\;+
\int_0^t2p\,\|\psi_s^{(N,M)}\|^{2p-2}\,\Re\SPb{(1+\Id\Gamma(\omega))^\eh\psi_s^{(N,M)}}{
T_2(s)\,\,\psi_s^{(N-1,M-1)}}\,\Id s
\\\nonumber
&\;\;+\int_0^t2p\,\|\psi_s^{(N,M)}\|^{2p-2}\,\Re\SPb{\psi_s^{(N,M)}}{
i\V{T}(s)\,\vt^\eh\psi_s^{(N,M)}}\,\Id \V{X}_s
\\\label{spin80n}
&\;\;
+\frac{p(p-1)}{2}\int_0^t\|\psi_s^{(N,M)}\|^{2p-4}\big(2\Re\SPb{\psi_s^{(N,M)}}{
i\V{T}(s)\,\vt^\eh\psi_s^{(N,M)}}\big)^2\Id s,
\end{align}
$\PP$-a.s. for all $t\in[0,\sup I)$. Finally, we apply the bounds
\begin{align*}
&2p\,\|\phi\|^{2p-2}\,|\SPn{(1+\Id\Gamma(\omega))^\eh\phi}{T_2(s)\,\phi'}|
\\
&\le \delta\,p\,\|\phi\|^{2p-2}\SPn{\phi}{(1+\Id\Gamma(\omega))\,\phi}
+p\,\|T_2(s)\|^2\,\|\phi\|^{2p-2}\,\|\phi'\|^2/\delta
\\
&\le \delta\,p\,\|\phi\|^{2p-2}\SPn{\phi}{\Id\Gamma(\omega)\,\phi}
+p(\delta+1/\delta)\,\|\phi\|^{2p}+\|T_2(s)\|^{2p}\,\|\phi'\|^{2p}/\delta,
\end{align*}
with $\phi:=\psi_s^{(N,M)}$ and $\phi':=\psi_s^{(N-1,M-1)}$, to arrive at the asserted estimate.
\qed

We recall that Gronwall's lemma states that, for all
non-negative, continuous functions $a$, $\beta$, and $\rho$ on $I$, we have the implication
\begin{equation}\label{Gronwall1}
\rho\le a+\int_0^\bullet (\beta\rho)(s)\Id s\;\;\Rightarrow\;\;
\rho(t)\le a(t)+\int_0^t (a\,\beta)(s)\,e^{\int_s^t \beta(\tau)\Id\tau}\Id s,\;t\in I.
\end{equation}
If $a$ is the integral of another continuous function, $c$, on $I$, then
\begin{equation}\label{Gronwall2}
\rho\le \int_0^\bullet (c+\beta\rho)(s)\,\Id s\;\;\Rightarrow\;\;
\rho(t)\le \int_0^t c(s)\,e^{\int_s^t \beta(\tau)\Id\tau}\Id s,\;t\in I.
\end{equation}

\begin{lemma}\label{lem-spin2}
There is a $\PP$-zero set $\sN$
such that, for all $(t,\vgamma)\in I\times(\Omega\setminus\sN)$ and $0\le N\le M<\infty$, 
the operators $\WW{\vxi,t}{V,(N,M)}(\vgamma)$, defined
a priori on $\CC^L\otimes\sC[\mathfrak{d}_C]$, extend uniquely to continuous operators
on ${\FHR}$ (which are denoted by the same symbols). The limits 
$\WW{\vxi}{V,(N,\infty)}:=\lim_{M\to\infty}\WW{\vxi}{V,(N,M)}$
converge in $\LO({\FHR})$, pointwise on $\Omega\setminus\sN$ and locally uniformly on $I$.
Moreover, for all $0\le N\le M\le\infty$ and $\psi\in\FHR$, the $\FHR$-valued process
$\WW{\vxi}{V,(N,M)}\,\psi$
is adapted and has continuous paths on $\Omega\setminus\sN$. For every $p\in\NN$,
we finally have the following bound on $\Omega\setminus\sN$,
\begin{align}\label{norm-WQ1}
\|\WW{\vxi,t}{V,(N,M)}\|^{2p}&\le e^{2pt-2p\int_0^tV(\V{X}_s)\Id s}
\sum_{n=N}^M\frac{1}{n!}\Big(\int_0^t\gamma_p(s)\,\Id s\Big)^n,\;\;t\in I,
\end{align}
with $\gamma_p(s):=\const(p)\,\|(1+\omega^{-1})^\eh\V{F}_{\V{X}_s}\|^{2p}$.
\end{lemma}

{\proof} 
Obviously, it is sufficient to prove the lemma for $V=0$; recall \eqref{hyp-XV}.
Let $\psi\in\CC^L\otimes\sC[\mathfrak{d}_C]$ and suppose that $0\le N\le M<\infty$.
We apply \eqref{spin81n} with $\Theta=\vt=\id$.
Then the term in the last line of \eqref{spin81n} vanishes,
$\|T_2(s)\|^{2p}\le\gamma_p(s)$ by \eqref{rb-vp1},
and $T_1=0$, $\V{T}=\V{0}$. Moreover, we choose $\delta=1$ and abbreviate
$$
\rho_{N,M}:=\|\WW{\vxi}{0,(N,M)}\psi\|^{2p},\qquad b(\tau,t):=e^{2p(t-\tau)},
\quad0\le\tau\le t,
$$ 
so that $b(r,s)\,b(s,t)=b(r,t)$, for $0\le r\le s\le t$. Taking also the initial values
$\rho_{N,M}(0)=\delta_{N,0}\,\|\psi\|^{2p}$, $0\le N\le M<\infty$,
into account in \eqref{spin81n} and applying \eqref{Gronwall2}
we $\PP$-a.s. arrive at the following recursive system of inequalities,
\begin{align*}
\rho_{N,M}(t)
&\le \int_0^tb(\tau,t)\,\gamma_p(\tau)\,\rho_{N-1,M-1}(\tau)\,\Id
\tau,\hspace{1.95cm} N\in\NN,\;M>N,
\\
\rho_{0,N}(t)
&\le b(0,t)\,\|\psi\|^2+\int_0^tb(\tau,t)\,\gamma_p(\tau)\,\rho_{0,N-1}(\tau)\,\Id\tau,\quad N\in\NN,
\\
\rho_{0,0}(t)&\le\|\psi\|^2\,b(0,t),
\end{align*}
the last one of which is following from \eqref{norm-W-scalar}. From this we readily infer that
\begin{align*}
\rho_{0,N}(t)&\le\|\psi\|^2\,b(0,t)\,\Big(1+
\sum_{n=1}^N\int_{t\simplex_n}\gamma_p(t_1)\dots\gamma_p(t_n)\,\Id t_1\dots\Id t_n\Big)
\end{align*}
and, hence,
\begin{align}\nonumber
\rho_{N,M}(t)&\le
\int_{t\simplex_N}b(t_1,t_2)\dots b(t_N,t)\,\gamma_p(t_1)\dots
\gamma_p(t_N)\,\rho_{0,M-N}(t_1)\,\Id t_1\dots\Id t_N
\\\nonumber
&\le\|\psi\|^2\,b(0,t)
\sum_{n=N}^M\int_{t\simplex_n}\gamma_p(t_1)\dots\gamma_p(t_n)\,\Id t_1\dots\Id t_n
\\\label{norbert}
&= \|\psi\|^2\,b(0,t)\sum_{n=N}^M\frac{1}{n!}\Big(\int_0^t\gamma_p(s)\,\Id s\Big)^n.
\end{align}
Here we find a $\PP$-zero set $\sN$ such that \eqref{norbert} holds
on $\Omega\setminus\sN$, for all $t\in I$, $N\le M<\infty$, and all
$\psi$ contained in the following countable subset of
$\CC^L\otimes\sC[\mathfrak{d}_C]$,
\begin{align*}
\sA:=\Big\{\sum_{\ell=1}^nv_\ell\otimes\zeta(h_\ell)\,:\;v_\ell\in(\QQ+i\QQ)^L,
\;h_\ell\in\mathfrak{a}_C,\;\ell=1,\ldots,n,\;n\in\NN\Big\},
\end{align*}
where $\mathfrak{a}_C$ is some countable dense subset of $\mathfrak{d}_C$.
Employing Rem.~\ref{rem-sec6}(2) we then deduce that \eqref{norbert}
actually holds on $\Omega\setminus\sN$, for all $t\in I$, $N\le
M<\infty$, and {\em every} $\psi\in\CC^L\otimes\sC[\mathfrak{d}_C]$.
This shows that all $\WW{\vxi,t}{0,(N,M)}$, $t\in I$, have unique extensions to elements
of $\LO(\FHR)$ on $\Omega\setminus\sN$, and we see that
\eqref{norm-WQ1} holds on $\Omega\setminus\sN$ as well. 
If $\psi\in\FHR$ and $\psi_n\in\CC^L\otimes\sC[\mathfrak{d}_C]$, $n\in\NN$,
with $\psi_n\to\psi$, then we also see that, on $\Omega\setminus\sN$, the convergence
$\WW{\vxi,t}{0,(N,M)}\,\psi_n\to \WW{\vxi,t}{0,(N,M)}\,\psi$ is locally uniform
in $t$. Since, by Rem.~\ref{rem-sec6}(2), each process $\WW{\vxi}{0,(N,M)}\,\wt{\psi}$ with
$\wt{\psi}\in\CC^L\otimes\sC[\mathfrak{d}_C]$ and $M<\infty$
is adapted and has continuous paths, we conclude that 
$\WW{\vxi}{0,(N,M)}\,\psi$ is adapted and has continuous paths on
$\Omega\setminus\sN$, for every $\psi\in\FHR$.
The assertions on the limiting objects with $M=\infty$ are now clear as well.
\qed 

 \begin{remark}{\rm \label{rem-pw-unique}
Recall that $\WW{\vxi}{V,(N,M)}$ depends on $\V{X}$.
Let $\wt{\V{X}}$ be another process fulfilling Hyp.~\ref{hyp-B} with the same stochastic basis $\BB$,
and denote the corresponding processes constructed in Lem.~\ref{lem-spin2}
by $\wt{\mathbb{W}}_{\vxi}^{V,(N,M)}$. Then, for all $0\le N\le M\le\infty$, 
\begin{align}\label{pwu-WW}
\WW{\vxi,\bullet}{V,(N,M)}=\wt{\mathbb{W}}_{\vxi,\bullet}^{V,(N,M)},\;
\;\text{$\PP$-a.s. on $\{\V{X}_\bullet=\wt{\V{X}}_{\bullet}\}$.}
\end{align}
For a start, it is clear that $\WW{\vxi,\bullet}{V,(N,M)}\psi=\wt{\mathbb{W}}_{\vxi,\bullet}^{V,(N,M)}\psi$
holds on some $(\psi,N,M)$-independent set $A\in\fF$ with
$\PP(\{\V{X}_\bullet=\wt{\V{X}}_\bullet\}\setminus A)=0$,
if $\psi\in\CC^L\otimes\sC[\mathfrak{d}_C]$ and $0\le N\le M<\infty$.
This can be read off from \eqref{def-W}, \eqref{WWzeta(h)}, and \eqref{def-Qh},
if one keeps in mind that the (stochastic) integrals defining the
basic processes in Def.~\ref{defn-basic-proc} remain $\PP$-a.s.
unchanged on $\{\V{X}_\bullet=\wt{\V{X}}_\bullet\}$, when $\V{X}$ is replaced by $\wt{\V{X}}$.
Since \eqref{norm-WQ1} holds, however, on some $(N,M)$-independent
set of full $\PP$-measure, these observations are sufficient to verify \eqref{pwu-WW}.
 }\end{remark}

 \begin{remark}{\rm \label{rem-spin}
The previous lemma and its proof imply
\begin{align}\label{apriori-omega}
\int_0^t\big\|\Id\Gamma(\omega)^\eh\WW{\vxi,s}{0,(N,M)}\eta\big\|^2\Id s
\le\const'e^{\const t}\|\eta\|^2,\;\; t\in I,\;0\le N\le M<\infty,
\end{align}
$\PP$-a.s. for all $\CC^L\otimes\sC[\mathfrak{d}_C]$-valued $\fF_0$-measurable
simple functions $\eta$.
In fact, choose $p=1$ and, as before, $\Theta=\vt=\id$ and $\delta=1$ in
\eqref{spin81n}. Then solve \eqref{spin81n} for the left hand side of
\eqref{apriori-omega} (instead of just throwing it away as in the proof of the lemma).
Combining the result with \eqref{norm-WQ1} we obtain
\eqref{apriori-omega}. Taking the expectation of \eqref{apriori-omega} and using the trival bound 
$\|\Id\Gamma(\omega)^\eh(1+\ve\,\Id\Gamma(\omega))^\mh\phi\|\le\|\Id\Gamma(\omega)^\eh\phi\|$,
$\ve>0$, we further infer from \eqref{apriori-omega} and the dominated
convergence theorem that, in the limit $\ve\downarrow0$, 
$\eta_\ve':=\Id\Gamma(\omega)^\eh(1+\ve\,\Id\Gamma(\omega))^\mh\WW{\vxi}{0,(N,M)}\eta$
converges to $\Id\Gamma(\omega)^\eh\WW{\vxi}{0,(N,M)}\eta$ in 
$L^2_{\FHR}([0,t]\times\Omega,\lambda\otimes\PP)$.
Since each $\eta_\ve'$ is predictable, $\Id\Gamma(\omega)^\eh\WW{\vxi}{0,(N,M)}\eta$
is predictable as well.
 }\end{remark}

\begin{lemma}\label{lem-spin3}
We consider the process on $I$ defined by
\begin{equation*}
\sM_\bullet:=\int_0^\bullet
\|\Theta\WW{\vxi,s}{0,(N,M)}\eta\|^{2p-2}\Re\SPb{\Theta\WW{\vxi,s}{0,(N,M)}\eta}{
i\V{T}(s)\vt^\eh\Theta\WW{\vxi,s}{0,(N,M)}\eta}\Id \V{B}_s,
\end{equation*}
where $\vt$ is $\id$ or $1+\Id\Gamma(\omega)$;
compare it with the last line of \eqref{spin81n} and with \eqref{Ito-eq-X}. Then, under the 
assumptions of Cor.~\ref{cor-spin1} and for all $\CC^L\otimes\sC[\mathfrak{d}_C]$-valued 
$\fF_0$-measurable simple functions $\eta$, $\sM$ is a martingale with
\begin{align}\label{mone1}
\EE\big[&\sup_{s\le t}|\sM_s|\big]
\le\epsilon\,\EE\big[\sup_{s\le t}\|\Theta\WW{\vxi,s}{0,(N,M)}\eta\|^{2p}\big]
\\\nonumber
&+\frac{\const}{\epsilon}\,\EE\Big[\int_0^t\|\Theta\WW{\vxi,s}{0,(N,M)}\eta\|^{2p-2}
\big\|\V{T}(s)\vt^\eh\Theta\WW{\vxi,s}{0,(N,M)}\eta\big\|^2\Id s\Big],\quad t\in I,\;\epsilon>0.
\end{align}
\end{lemma}

{\proof} 
First, let $\vt=\id$. On account of \eqref{norm-WQ1} and the boundedness
of $\Theta$ the criterion given in Prop.~\ref{prop-mart}
can be applied to show that $\sM$ is a martingale in this case.
Notice also that the integrand in the definition of $\sM$ is predictable
because $\WW{\vxi}{0,(N,M)}\eta$ is adapted and continuous.
If $\vt=1+\Id\Gamma(\omega)$, then we apply Prop.~\ref{prop-mart} using \eqref{norm-WQ1}
and Rem.~\ref{rem-spin} in addition.
The estimate \eqref{mone1} is an easy consequence of Davis' inequality 
(see, e.g., \cite[Thm.~3.28 in Chap.~3]{KaratzasShreve})
$\EE[\sup_{s\le t}|\sM_s|]\le\const\,\EE[\llbracket\sM,\sM\rrbracket_t^\eh]$,
Prop.~\ref{prop-stoch-calc}(2), and Cauchy-Schwarz inequalities.
\qed 

In the statement of the next lemma and henceforth we abbreviate
\begin{equation}\label{def-Yt}
\V{Y}_t:=\V{\beta}(t,\V{X}_t),\quad t\in[0,\sup I),
\end{equation}
so that $\Id\V{X}_t=\Id\V{B}_t+\V{Y}_t\Id t$.

\begin{lemma}\label{lem-spin4}
Assume that $\V{q}$ in \eqref{Ito-eq-X} is bounded so that
\eqref{hyp-Y2} is available, and set $\theta:=1+\Id\Gamma(\omega)$.
Then there is a $\PP$-zero set $\sN$ such that, on $\Omega\setminus\sN$, every
$\WW{\vxi,t}{0,(N,M)}$, $t\in I$,  maps the domain of $\Id\Gamma(\omega)$ into itself.
Moreover, for all $p\in\NN$, $t\in I$, $N\in\NN_0$, $M\in\NN_0\cup\{\infty\}$ with $N\le M$, 
and $\fF_0$-measurable $\eta:\Omega\to\dom(\Id\Gamma(\omega))$ 
with $\EE[\|\theta\eta\|^{4p}]<\infty$,
\begin{align}\label{bound-spin4}
\EE\big[\sup_{s\le t}\|\theta\,\WW{\vxi,s}{0,(N,M)}\eta\|^{2p}\big]
&\le c_{p,\V{Y},I}(t)\,\EE[\|\theta\eta\|^{4p}]^\eh\sum_{\ell=N}^M\frac{(\const(p)t)^\ell}{\ell!}.
\end{align}
Here $c_{p,\V{Y},I}:I\to(0,\infty)$ is continuous and monotonically increasing.
\end{lemma}

{\proof} 
Let us first treat the case $I=[0,\cT]$ with $0<\cT<\infty$.
We assume that $0\le N\le M<\infty$ and that $\eta$ is a
$\CC^L\otimes\sC[\mathfrak{d}_C]$-valued $\fF_0$-measurable simple function to begin with. 
We apply \eqref{spin81n} with $\delta=1$, $\vt=1$, and
$\Theta=\theta_\ve$, where
\begin{equation}\label{def-Xi}
\theta_\ve:=(1+\Id\Gamma(\omega))(1+\ve\Id\Gamma(\omega))^{-1},\qquad\ve\in(0,1].
\end{equation}
As a consequence of Hyp.~\ref{hyp-G} and Lem.~\ref{lem-comm} we then know that
$T_1(s)$ and the components of $\V{T}(s)$ extend to bounded operators on ${\FHR}$
and that $\|T_1(s)\|$, $\|T_2(s)\|$, and $\|\V{T}(s)\|^2$ 
are bounded by deterministic constants, {uniformly in $\ve\in(0,1]$} and $s\in I$.
In fact, $\|\V{T}(s)\theta^\eh_\ve\|$ is bounded uniformly in
$\ve$ and $s$ as well; see \eqref{nina1}. We set
\begin{align*}
\psi_{\ve,s}^{(N,M)}&:=\theta_\ve\WW{\vxi,s}{0,(N,M)}\eta,\qquad
\rho_{N,M}^\ve(t):=\EE\big[\sup_{s\le t}\|\psi_{\ve,s}^{(N,M)}\|^{2p}\big].
\end{align*}
According to \eqref{Ito-eq-X}, \eqref{spin81n}, Lem.~\ref{lem-spin3} 
(where we choose $\epsilon=1/4p$), and the above remarks we then obtain 
\begin{align}\nonumber
&\rho_{N,M}^\ve(t) \le\rho_{N,M}^\ve(0)
+\const(p)\int_0^t\big(\rho_{N,M}^\ve(s))+\rho_{N-1,M-1}^\ve(s)\big)\,\Id s+\frac{1}{2}\,\rho_{N,M}^\ve(t)
\\\nonumber
&+\const(p)\,\EE\Big[\int_0^t\|\psi_{\ve,s}^{(N,M)}\|^{2p-2}\big\|\{\V{T}(s)\theta_\ve^\eh\}
\theta_\ve^\eh\WW{\vxi,s}{0,(N,M)}\eta\big\|^2\Id s\Big]
\\\label{spin1974}
&+2p\int_0^t\EE\Big[\|\psi_{\ve,s}^{(N,M)}\|^{2p-1}
\big\|\{\V{T}(s)\theta_\ve^\eh\}\theta_\ve^\eh\WW{\vxi,s}{0,(N,M)}\eta\big\|\,|\V{Y}_s|\Big]\,\Id s,
\end{align}
for all $t\in[0,\cT)$. The Cauchy-Schwarz inequality implies
\begin{equation}\label{david1}
\|\theta_\ve^\eh
\,\WW{\vxi,s}{0,(N,M)}\eta\|\le\|\psi_{\ve,s}^{(N,M)}\|^\eh\|\WW{\vxi,s}{0,(N,M)}\eta\|^\eh,
\end{equation}
and combining this with a weighted H\"older inequality (w.r.t. the measure $\lambda\otimes\PP$)
and the bounds $\|\V{T}(s)\theta_\ve^\eh\|\le\const$ and \eqref{norm-WQ1},
we see that the term in last line of \eqref{spin1974} is bounded
by some $p$-dependent constant times
\begin{align}\label{david3}
&\int_0^t\EE\big[\|\psi_{\ve,s}^{(N,M)}\|^{2p-\eh}\|\WW{\vxi,s}{0,(N,M)}\eta\|^\eh|\V{Y}_s|\big]\Id s
\\\nonumber
&\le\Big(\int_0^t\frac{\EE[\|\psi_{\ve,s}^{(N,M)}\|^{2p}]}{(\cT-s)^{{2p}/({4p-1})}}\Id s\Big)^{1-\frac{1}{4p}}
\Big(\int_0^t\Sigma_{N}^{M}(s)(\cT-s)^{2p}\,
\EE\big[\|\eta\|^{2p}|\V{Y}_s|^{4p}\big]\Id s\Big)^{\frac{1}{4p}}.
\end{align}
Here we abbreviate $\Sigma_{N}^{M}(s):=e^{cs}\sum_{n=0\vee N}^M(cs)^n/n!$,
for integers $N\le M$, where $c>0$ is chosen such that 
$\|\WW{\vxi,s}{0,(N,M)}\|^{2p}\le\Sigma_{N}^{M}(s)$, which is possible thanks to \eqref{norm-WQ1}.
It is now also clear that
\begin{align*}
\big(\text{second line of
\eqref{spin1974}}\big)
&\le\const(p)\int_0^t\rho_{N,M}^\ve(s)\,\Id
s+\const(p)\int_0^t\Sigma_N^M(s)\,\Id s\,\EE[\|\eta\|^{4p}]^\eh.
\end{align*}
Applying Young's and H\"older's inequalities to \eqref{david3}, using \eqref{hyp-Y2},
and applying the obtained estimates to \eqref{spin1974} we obtain
\begin{align}\nonumber
\rho_{N,M}^\ve(t)
&\le2\rho_{N,M}^\ve(0)+\const(p)\int_0^t\alpha_N^M(s)\,\Id s\,\EE[\|\eta\|^{4p}]^\eh
\\\label{david3b}
&\quad+\int_0^t\beta_{p,\cT}(s)\,\rho^\ve_{N,M}(s)\,\Id s
+\const(p)\int_0^t\rho_{N-1,M-1}^\ve(s)\,\Id s,
\end{align}
for $t\in[0,\cT)$, where
\begin{align*}
\alpha_N^M(s)&:=\Sigma_N^M(s)\big(1+(\cT-s)^{2p}\EE[|\V{Y}_s|^{8p}]^\eh\big),
\\
\beta_{p,\cT}(s)&:=
\const(p)\big(1+(\cT-s)^{\frac{-2p}{(4p-1)}}\big).
\end{align*}
Finally, an application of \eqref{Gronwall1} and an integration by
parts using $\rho_{N,M}^\ve(0)=\delta_{N,0}\EE[\|\theta_\ve\eta\|^{2p}]$ yield
\begin{align}\nonumber
\rho_{N,M}^\ve(t)
&\le2\delta_{N,0}\,b_{p,\cT}(0,t)\,\EE[\|\theta_\ve\eta\|^{2p}]
+\const(p)\int_0^tb_{p,\cT}(s,t)\,\alpha_N^M(s)\,\Id s\,\EE[\|\eta\|^{4p}]^\eh
\\\label{spin1999}
&\quad+\const(p)\int_0^tb_{p,\cT}(s,t)\,\rho_{N-1,M-1}^\ve(s)\,\Id s,\quad t\in[0,\cT)\,,
\end{align}
with $b_{p,\cT}(s,t):=e^{\int_s^t\beta_{p,\cT}(r)\Id r}$.
Observe that $b_{p,\cT}(r,s)\,b_{p,\cT}(s,t)= b_{p,\cT}(r,t)$, $0\le r\le s\le t<\cT$.
We may now argue similarly as in the proof of Lem.~\ref{lem-spin2} to see that
the following inequalities hold, for $0\le t<\cT$,
\begin{align}\nonumber
\rho_{N,M}^\ve(t)
&\le2\EE[\|\theta_\ve\eta\|^{2p}]\,b_{p,\cT}(0,t)
\sum_{n=N}^M\int_{t\simplex_n}\!\const(p)^n\Id t_{[n]}
\\\label{spin200}
&\qquad+J_{N,M}(t)\,\EE[\|\eta\|^{4p}]^\eh,
\\\label{david2}
J_{N,M}(t)
&:=\sum_{m=0}^{M}\int_{t\simplex_{m+1}}\!\const(p)^{m+1}\,
b_{p,\cT}(t_1,t)\,\alpha_{N-m}^{M-m}(t_1)\,\Id t_{[m+1]}.
\end{align}
Since $\alpha_L^K\ge0$ we may replace $b_{p,\cT}(t_1,t)$
by $b_{p,\cT}(0,t)$ in \eqref{david2}. After that we estimate
$\int_0^{t_2}\alpha_L^K(t_1)\Id t_1\le\Sigma_L^K(t)(t+\int_0^t(\cT-s)^{2p}\EE[|\V{Y}_s|^{8p}]^\eh\Id s)$ 
and evaluate the remaining integrals over the simplices in $J_{N,M}(t)$, which yields
\begin{align*}
J_{N,M}(t)
&\le\const(p)\,e^{ct}\,b_{p,\cT}(0,t)\Big(t+\int_0^t(\cT-s)^{2p}\EE[|\V{Y}_s|^{8p}]^\eh\Id s\Big)\,S_{N,M},
\\
S_{N,M}
&:=\sum_{m=0}^{M}
\sum_{n=0\vee(N-m)}^{M-m}\frac{(\const(p)t)^{m}}{m!}\,\frac{(ct)^n}{n!}
\\
&=\sum_{m=0}^N\frac{(\const(p)t)^m}{m!}\sum_{j=N}^M\frac{(ct)^{j-m}}{(j-m)!}
+\sum_{m=N+1}^M\frac{(\const(p)t)^m}{m!}\sum_{j=m}^M\frac{(ct)^{j-m}}{(j-m)!}
\\
&\le\sum_{j=N}^{M}\frac{(\const(p)\vee c)^jt^j}{j!}\sum_{m=0}^j{j\choose m}
=\sum_{j=N}^M\frac{(\const'(p)t)^j}{j!}.
\end{align*}
Thanks to Lem.~\ref{lem-spin2} and since $\theta_\ve$ is bounded we may
use Lebesgue's dominated convergence theorem, first to extend \eqref{spin200} to all 
$\fF_0$-measurable $\eta:\Omega\to\FHR$ with $\EE[\|\eta\|^{4p}]<\infty$, and then
to pass to the limit $M\to\infty$. Combining this with the bounds on $J_{N,M}(t)$ we obtain
\begin{align}
\rho_{N,\infty}^\ve(t)\label{spin201}
&\le c_{p,\V{Y},I}(t)\,\EE[\|\theta_\ve\eta\|^{4p}]^\eh\,b_{p,\cT}(0,t)
\sum_{n=N}^\infty\frac{(\const''(p)t)^n}{n!},
\end{align}
for $t\in[0,\cT)$ and $N\in\NN_0$. Since $\theta_\ve$
is merely a multiplication operator in each Fock space sector $\sF^{(m)}$, $m\in\NN$,
we may now pass to the limit $\ve\downarrow0$ in \eqref{spin200}
and \eqref{spin201} by means of the monotone convergence theorem,
for all $\eta$ as in the statement of this lemma. Finally, we observe that $p\ge1$ entails
$2p/{(4p-1)}\in(\nf{1}{2},\nf{2}{3}]$. Therefore, $b_{p,\cT}(0,\cT)$ is finite and we may extend our
estimates to the case $t=\cT$ again by monotone convergence.

The same proof works in the case $I=[0,\infty)$, provided that all
factors $(\cT-s)^a$, which were used to control a possible singularity
of $\V{Y}$ at $\cT$, are replaced by $1$.
\qed 

In what follows we again consider $\wh{\dom}$, i.e., the domain of
the generalized fiber Hamiltonians, as a Hilbert space equipped with the graph norm of
$M_1(\V{0})$ (defined in \eqref{def-Maxi}).

\begin{lemma}\label{lem-spin-k}
Assume that $\V{q}$ in \eqref{Ito-eq-X} is bounded and set $\Upsilon:=1+\Id\Gamma(\V{m})^2$.
Then there is a $\PP$-zero set $\sN_0$ such that, on $\Omega\setminus\sN_0$, every
$\WW{\vxi,t}{0,(N,M)}$, $t\in I$,  maps $\wh{\dom}$ into $\dom(\Id\Gamma(\V{m})^2)$.
Furthermore, for all $p\in\NN$, $t\in I$, $N\in\NN_0$, $M\in\NN_0\cup\{\infty\}$ with
$N\le M$, and all $\fF_0$-measurable $\eta:\Omega\to\wh{\dom}$
with $\EE[\|\eta\|_{\wh{\dom}}^{4p}]<\infty$,
\begin{align}\label{bound-spin-k}
\EE\big[\sup_{s\le t}\|\Upsilon\,\WW{\vxi,s}{0,(N,M)}\,\eta\|^{2p}\big]
&\le\tilde{c}_{p,\V{Y},I}(t)\,\EE[\|\eta\|_{\wh{\dom}}^{4p}]^\eh
\sum_{\ell=N}^M\frac{(\const(p)t)^\ell}{\ell!}.
\end{align}
Here $\tilde{c}_{p,\V{Y},I}:I\to(0,\infty)$ is continuous and monotonically increasing.
\end{lemma}

{\proof} 
Again we start with the case $I=[0,\cT]$, $0<\cT<\infty$.
Let $0\le N\le M<\infty$ and suppose that $\eta$ is a
$\CC^L\otimes\sC[\mathfrak{d}_C]$-valued $\fF_0$-measurable simple function to begin with.
We apply \eqref{spin81n} with $\Theta:=\Upsilon_{\ve}$, where
\begin{equation}\label{def-Upsilon}
\Upsilon_\ve:=(E+\Id\Gamma(\V{m})^2)(1+\ve\Id\Gamma(\V{m})^2)^{-1},\quad\ve\in(0,1/E],\;E\ge1.
\end{equation}
and with $\vt:=\theta=1+\Id\Gamma(\omega)$ and $\delta=1$.
As a direct consequence of Lem.~\ref{lem-comm}
we may choose $E$ so large that $\const(p)(\|T_1(s)\|+\|\V{T}(s)\|^2)\le p/2$, 
for all $s\ge0$, where $\const(p)$ is the constant appearing in \eqref{spin81n}.
Then the sum of the first two lines on the right hand side of \eqref{spin81n} is less than or equal to
$$
-\frac{p}{2}\int_0^t\|\tilde\eta_{\ve,s}^{(N,M)}\|^{2p-2}
\|\Id\Gamma(\omega)^\eh\tilde\eta_{\ve,s}^{(N,M)}\|^2\Id s
+\frac{p}{2}\int_0^t\|\tilde\eta_{\ve,s}^{(N,M)}\|^{2p}\Id s,
$$
where
\begin{align*}
\tilde\eta_{\ve,t}^{(N,M)}&:=\Upsilon_{\ve}\WW{\vxi,t}{0,(N,M)}\,\eta.
\end{align*}
Using also the bound $\|T_2(s)\|\le\const$, which follows from \eqref{rb-vp1} and \eqref{nina4a} 
and is uniform in $\ve$, we see that \eqref{spin81n} implies, for all $t\in[0,\sup I)$,
\begin{align}\label{michi2000}
&f^\ve_{N,M}(t):=\|\tilde\eta_{\ve,t}^{(N,M)}\|^{2p}+\frac{p}{2}\int_0^t\|\tilde\eta_{\ve,s}^{(N,M)}\|^{2p-2}
\|\Id\Gamma(\omega)^\eh\tilde\eta_{\ve,s}^{(N,M)}\|^2\Id s
\\\nonumber
&\le f^\ve_{N,M}(0)+\frac{5p}{2}\int_0^t\|\tilde\eta_{\ve,s}^{(N,M)}\|^{2p}\Id s
+\const\int_0^t\|\tilde\eta_{\ve,s}^{(N-1,M-1)}\|^{2p}\Id s+2p\sup_{s\le t}|\sM_s|
\\\nonumber
&\quad+2p\int_0^t\|\tilde\eta_{\ve,s}^{(N,M)}\|^{2p-2}
\|\theta^{\nf{1}{4}}\|\big\|\wh{\V{T}}(s)\,
\theta^{\nf{1}{4}}\Upsilon_\ve^\eh \WW{\vxi,s}{0,(N,M)}\eta\big\|\,|\V{Y}_s|\Id s.
\end{align}
Here $\sM$ denotes the martingale defined in Lem.~\ref{lem-spin3} with $\Theta=\Upsilon_\ve$ and
$\vt=\theta$; recall that $\Id\V{X}_s=\Id\V{B}_s+\V{Y}_s\Id s$. Moreover, we abbreviate
$$
\wh{\V{T}}(s):=\theta^{-\nf{1}{4}}(\Ad_{\vp(\V{G}_{\V{X}_s})}\Upsilon_\ve)
\Upsilon_\ve^\mh\theta^{-\nf{1}{4}};
$$
then $\wh{\V{T}}(s)$ is bounded uniformly on $\Omega$ and in $\ve>0$ and $s\in I$, 
according to \eqref{nina4b}. Since the terms in the last two lines of \eqref{michi2000} are 
monotonically increasing in $t$ the estimate still holds true, if we replace $f^\ve_{N,M}(t)$ by 
$\sup_{s\le t}f^\ve_{N,M}(s)$ on the left hand side of \eqref{michi2000}. To bound the integral 
in the last line of \eqref{michi2000} we estimate
\begin{align}\nonumber
\|\theta^{\nf{1}{4}}\tilde\eta_{\ve,s}^{(N,M)}\|
&\le\|\tilde\eta_{\ve,s}^{(N,M)}\|^\eh\|\theta^{\nf{1}{2}}\tilde\eta_{\ve,s}^{(N,M)}\|^\eh,
\\\nonumber
\|\theta^{\nf{1}{4}}\Upsilon_\ve^\eh \WW{\vxi,s}{0,(N,M)}\eta\|
&\le\|\tilde\eta_{\ve,s}^{(N,M)}\|^\eh\|\theta\WW{\vxi,s}{0,(N,M)}\eta\|^{\nf{1}{4}}
\|\WW{\vxi,s}{0,(N,M)}\eta\|^{\nf{1}{4}},
\end{align}
and combine these two bounds with the geometric-arithmetic mean inequality 
$a^\eh b^{\nf{1}{4}}c^{\nf{1}{8}}d^{\nf{1}{8}}\le a/2+b/16+c/8+2d$ to get
\begin{align}\nonumber
&2p\|\wh{\V{T}}(s)\|\,\|\theta^{\nf{1}{4}}\tilde\eta_{\ve,s}^{(N,M)}\|
\big\|\theta^{\nf{1}{4}}\Upsilon_\ve^\eh
\WW{\vxi,s}{0,(N,M)}\eta\big\|\,|\V{Y}_s|
\\\label{david5}
&\le
\big(p(\cT-s)^{\delta-1}+p/8\big)\|\tilde\eta_{\ve,s}^{(N,M)}\|^2
+\frac{p}{8}\,\|\Id\Gamma(\omega)^{\nf{1}{2}}\tilde\eta_{\ve,s}^{(N,M)}\|^2
\\\nonumber
&\quad
+\frac{p}{4}\,(\cT-s)^{-4\delta}\|\theta\WW{\vxi,s}{0,(N,M)}\eta\|^2
+4p\|\wh{\V{T}}(s)\|^8(\cT-s)^{4}\|\WW{\vxi,s}{0,(N,M)}\eta\|^2|\V{Y}_s|^8,
\end{align}
where $\delta\in(0,\nf{1}{4p})$. To bound the expectation of $\sup_{s\le t}|\sM_s|$ in \eqref{michi2000}
we employ Lem.~\ref{lem-spin3} (with $\epsilon=1/8p$ in \eqref{mone1}). Putting these remarks
together and setting
\begin{align*}
\vr_{N,M}^\ve(t)&:=\EE\big[\sup_{s\le t}f^\ve_{N,M}(s)\big],
\end{align*}
we infer from \eqref{michi2000} that, for all $t\in[0,\sup I)$,
\begin{align}\nonumber
\vr_{N,M}^\ve(t)
&\le\vr_{N,M}^\ve(0)+\int_0^t\big(p(\cT-s)^{\delta-1}+\tfrac{21p}{8}+8p\const\|\V{T}\|_\infty^2\big)
\EE\big[\|\tilde\eta_{\ve,s}^{(N,M)}\|^{2p}\big]\Id s
\\
&\nonumber
+\const\int_0^t\EE\big[\|\tilde\eta_{\ve,s}^{(N-1,M-1)}\|^{2p}\big]\Id s
+\frac{1}{4}\EE\big[\sup_{s\le t}\|\tilde\eta_{\ve,s}^{(N,M)}\|^{2p}\big]
\\\nonumber
&+\big(8p\const\|\V{T}\|_\infty^2+\tfrac{p}{8}\big)
\int_0^t\EE\big[\|\tilde\eta_{\ve,s}^{(N,M)}\|^{2p-2}
\|\Id\Gamma(\omega)^{\nf{1}{2}}\tilde\eta_{\ve,s}^{(N,M)}\|^2\big]\Id s
\\\label{michi2001}
&+\frac{p}{4}\int_0^t
\EE\big[\|\tilde\eta_{\ve,s}^{(N,M)}\|^{2p-2}(\cT-s)^{-4\delta}\|\theta\WW{\vxi,s}{0,(N,M)}\eta\|^2\big]\Id s
\\\nonumber
&+4p\int_0^t\EE\big[\|\tilde\eta_{\ve,s}^{(N,M)}\|^{2p-2}
\|\wh{\V{T}}(s)\|^8(\cT-s)^{4}\|\WW{\vxi,s}{0,(N,M)}\eta\|^2|\V{Y}_s|^8\big]\Id s.
\end{align}
By enlarging $E\ge1$ further, if necessary, we may assume that 
$\|\V{T}\|_\infty^2:=\sup_{s\in I}\sup_\Omega\|\V{T}(s)\|^2\le1/8^2\const$.
Applying also H\"older's inequality ($\frac{2p-2}{2p}+\frac{1}{p}=1$) to the $(\Id\PP\Id s)$-integrals 
in the last two lines of \eqref{michi2001} and estimating 
$a^{\frac{p-1}{p}}b^{\frac{1}{p}}\le a+b/p$ after that we obtain
\begin{align}\nonumber
\vr_{N,M}^\ve(t)
&\le\vr_{N,M}^\ve(0)+\int_0^t\!\tilde{\beta}_{p,\cT}(s)\EE\big[\|\tilde\eta_{\ve,s}^{(N,M)}\|^{2p}\big]\Id s
+\const\!\int_0^t\!\EE\big[\|\tilde\eta_{\ve,s}^{(N-1,M-1)}\|^{2p}\big]\Id s
\\\nonumber
&+\frac{1}{4}\EE\big[\sup_{s\le t}\|\tilde\eta_{\ve,s}^{(N,M)}\|^{2p}\big]
+\frac{p}{4}\EE\Big[\int_0^t\|\tilde\eta_{\ve,s}^{(N,M)}\|^{2p-2}
\|\Id\Gamma(\omega)^{\nf{1}{2}}\tilde\eta_{\ve,s}^{(N,M)}\|^2\Id s\Big]
\\\nonumber
&+\frac{1}{4}\int_0^t(\cT-s)^{-4p\delta}\EE\big[\|\theta\WW{\vxi,s}{0,(N,M)}\eta\|^{2p}\big]\Id s
\\\label{david6}
&+4\|\wh{\V{T}}\|_\infty^{8p}\int_0^t
\EE\Big[(\cT-s)^{4p}\|\WW{\vxi,s}{0,(N,M)}\eta\|^{2p}|\V{Y}_s|^{8p}\Big]\Id s,
\end{align}
for all $t\in[0,\sup I)$, with
$$
\tilde{\beta}_{p,\cT}(s):=7p+p(\cT-s)^{\delta-1},\quad\text{so that}\quad\tilde{\beta}_{\cT}\in L^1([0,\cT]).
$$
Next, we observe that the sum of the two terms in the second line of \eqref{david6}
is $\le\frac{3}{4}\vr_{N,M}^\ve(t)$.
Applying also the bounds \eqref{bound-spin4} and \eqref{norm-WQ1} in the
third and fourth lines of \eqref{david6}, respectively,
we arrive at the following analog of \eqref{david3b},
\begin{align*}
\frac{1}{4}\vr_{N,M}^\ve(t)
&\le\vr_{N,M}^\ve(0)+\int_0^t\tilde{\beta}_{p,\cT}(s)\,\vr_{N,M}^\ve(s)\Id s
+\const\int_0^t\vr_{N-1,M-1}^\ve(s)\,\Id s
\\
&\quad+\const(p)\int_0^t\tilde{\alpha}_N^M(s)\,\Id s\,\EE[\|\theta\eta\|^{4p}]^\eh,\quad t\in[0,\sup I).
\end{align*}
Here we abbreviate, for integers $N\le M$, $0\le M$,
$$
\tilde{\alpha}_N^M(s):=\big\{(\cT-s)^{-4p\delta}c_{p,\V{Y},I}(s)
+e^{cs}(\cT-s)^{4p}\EE[|\V{Y}_s|^{16p}]^\eh\big\}\!\!
\sum_{\ell=0\vee N}^M\!\!\frac{(\const(p)s)^\ell}{\ell!},
$$
where $\const(p)>0$ is chosen bigger than the constant $c>0$ introduced in
the paragraph below \eqref{david3}.
Since by the choice of $\delta$ and by \eqref{hyp-Y2} the function in
the curly brackets $\{\cdots\}$ is integrable on $[0,\cT]$,
we may now conclude exactly as in the proof of Lem.~\ref{lem-spin4}.

Again the same proof works in the case $I=[0,\infty)$, if
all factors $(\cT-s)^a$ with some $a\in\RR$ are replaced by $1$.
\qed

 \begin{remark}{\rm \label{rem-gustav}
In the proofs of the next two lemmas we shall employ the following elementary
observation:

Let $\sK$ be a separable Hilbert space and let
$X,X^{(N)}$, $N\in\NN$, be $\sK$-valued processes on $I$ such that
$\sup_{s\le t}\|X^{(N)}_s-X_s\|\to0$ in $L^2(\PP)$, as $N\to\infty$, for all $t\in I$, and
$$
\EE\big[\sup_{s\le t}\|X^{(N)}_s-X_s^{(M)}\|^2\big]\le
\sum_{n=N+1}^Mc_n(t)\,,\quad 
0\le N<M<\infty,\;t\in I,
$$
where the $c_n:I\to(0,\infty)$ are monotonically increasing
such that $\{nc_n(t)\}\in\ell^2(\NN)$, $t\in I$.
Then, $\PP$-a.s., the limit
$\lim_{N\to\infty}X_\bullet^{(N)}=X_\bullet$ exists locally uniformly
on $I$.

In fact, a priori it is clear that $X_\bullet^{(N_\ell)}\to
X_\bullet$, $\PP$-a.s., along some subsequence.
An easy argument employing the monotone convergence theorem shows,
however, that $\sum_{N=1}^\infty\sup_{s\le
  t}\|X^{(N)}_s-X_s^{(N+1)}\|\in L^2(\PP)$, for every $t\in I$, which
readily implies
that, $\PP$-a.s., $\{X^{(N)}_\bullet\}_{N\in\NN}$ is a locally uniform
Cauchy sequence.
 }\end{remark}

Recall that we consider $\wh{\dom}$ as a Hilbert space equipped with the graph norm of
$M_1(\V{0})=\id_{\CC^L}\otimes(\tfrac{1}{2}\Id\Gamma(\V{m})^2+\Id\Gamma(\omega))$.

\begin{lemma}\label{lem-spin2000}
Assume that $\V{q}$ in \eqref{Ito-eq-X} is bounded, let $0\le N\le M\le \infty$, $p\in\NN$,
and let $\eta:\Omega\to\wh{\dom}$ be $\fF_0$-measurable 
with $\EE[\|\eta\|_{\wh{\dom}}^{4p}]<\infty$. Then the following holds:

\smallskip

\noindent
{\rm(1)} For $\PP$-a.e. $\vgamma\in\Omega$, we have 
$(\WW{\vxi,\bullet}{0,(N,M)}\eta)(\vgamma)\in C(I,\wh{\dom})$ and
$(\WW{\vxi,\bullet}{0,(0,N)}\eta)(\vgamma)\to(\WW{\vxi,\bullet}{0}\eta)(\vgamma)$, $N\to\infty$,
in $C(I,\wh{\dom})$. Furthermore,
\begin{align}\label{gustav0}
\EE\big[\sup_{s\le t}\|\WW{\vxi,s}{0,(N,M)}\eta\|_{\wh{\dom}}^{2p}\big]
&\le \hat{c}_{p,\V{Y},I}(t)\,\EE[\|\eta\|_{\wh{\dom}}^{4p}]^\eh
\sum_{n=N}^M\frac{(\const(p)t)^n}{n!}\,,\quad t\in I.
\end{align}
{\rm(2)} The integral process
$(\int_0^t\V{v}(\vxi,\V{X}_s)\WW{\vxi,s}{0,(N,M)}\eta\,\Id\V{B}_s)_{t\in I}$
is an $L^2$-martingale.
\end{lemma}

{\proof} 
(1) The bound \eqref{gustav0} follows by combining \eqref{bound-spin4} and \eqref{bound-spin-k}.
It shows that Rem.~\ref{rem-gustav} is applicable with $\sK=\wh{\dom}$
and $X^{(N)}=\WW{\vxi,s}{0,(0,N)}\eta$, whence, $\PP$-a.s. and locally uniformly on $I$,
we have $\WW{\vxi,\bullet}{0,(0,N)}\eta\to\WW{\vxi,\bullet}{0}\eta$ in $\wh{\dom}$.

Next, let $M^\ve:=(M_1(\V{0})+1)(\ve M_1(\V{0})+1)^{-1}$, $\ve>0$.
Then each $\FHR$-valued process $M^\ve\WW{\vxi}{0,(0,N)}\psi$ with $N\in\NN$,
$\ve>0$, and $\psi\in\FHR$ has continuous paths
outside some $\psi$-independent $\PP$-zero set by Lem.~\ref{lem-spin2}.
Therefore, each process $M^\ve\WW{\vxi}{0,(0,N)}\eta$ with $N\in\NN$ and
$\ve>0$ has continuous paths $\PP$-a.s.
By virtue of \eqref{gustav0} we may apply the dominated convergence
theorem to show that $\sup_{s\le t}\|(M^\ve-M_1(\V{0}))\WW{\vxi,s}{0,(0,N)}\eta\|\to0$,
$\ve\downarrow0$, in $L^2(\PP)$ and for all $t\in I$, which implies that, $\PP$-a.s.,
$M_1(\V{0})\WW{\vxi}{0,(0,N)}\eta$ has continuous paths as an
$\FHR$-valued process or,
in other words, $(\WW{\vxi,\bullet}{0,(0,N)}\eta)(\vgamma)\in C(I,\wh{\dom})$,
for $\PP$-a.e. $\vgamma$ and for every $N\in\NN$.
By the remark in the first paragraph above it now follows that 
$(\WW{\vxi,\bullet}{0}\eta)(\vgamma)\in C(I,\wh{\dom})$ and 
$(\WW{\vxi}{0,(0,N)}\eta)(\vgamma)\to(\WW{\vxi}{0}\eta)(\vgamma)$
in $C(I,\wh{\dom})$, for $\PP$-a.e. $\vgamma$.

(2): The bound \eqref{rb-vp1}, Hyp.~\ref{hyp-G}(2), and Hyp.~\ref{hyp-B} imply that
the components of $\V{v}(\vxi,\V{X})$ are continuous $\LO(\wh{\dom},\FHR)$-valued adapted
processes whose operator norms are uniformly  bounded by deterministic constants. 
Hence, the assertion is a consequence of Prop.~\ref{prop-mart} and \eqref{gustav0}.
\qed

\smallskip

{\it Proof of Thm.~\ref{thm-Ito-spin}.}
Apart from the bound \eqref{norm-W}, which is derived in the uniqueness proof in the next paragraph,
Part~(1) of Thm.~\ref{thm-Ito-spin} is an immediate consequence of Lem.~\ref{lem-spin2}.

To prove the uniqueness part of Thm.~\ref{thm-Ito-spin}(2),
let $X\in\mathsf{S}_I(\FHR)$ be such that its paths belong $\PP$-a.s to $C(I,\wh{\dom})$
and such that it $\PP$-a.s. solves \eqref{SDE-spin}.
Then a computation analogous to \eqref{spin79} with $\Theta=\id$ (again using the skew-symmetry
of the components of $i\V{v}(\vxi,\V{x})$) yields, $\PP$-a.s. for all $t\in[0,\sup I)$,
\begin{align*}
\|X_t\|^2&=\|\eta\|^2-\int_0^t2\SPb{X_s}{
\big(\Id\Gamma(\omega)-\vsigma\cdot\vp(\V{F}_{\V{X}_s})+V(\V{X}_s)\big)X_s}\Id s.
\end{align*}
Together with the bound \eqref{qfb-vp} this $\PP$-a.s. implies that
$$
\|X_t\|^2\le\|\eta\|^2+\int_0^t2\big(\Lambda(\V{X}_s)^2-V(\V{X}_s)\big)\|X_s\|^2\Id s,\quad t\in I,
$$
where $\Lambda$ is defined in the statement of Thm.~\ref{thm-Ito-spin}(1), thus
$$
\|X_t\|\le\|\eta\|e^{\int_0^t(\Lambda(\V{X}_s)^2-V(\V{X}_s))\Id s},\quad t\in I.
$$ 
This entails the desired uniqueness statement and also proves \eqref{norm-W}.

To prove the existence part of Thm.~\ref{thm-Ito-spin}(2) we assume that $\V{q}$ (in
\eqref{Ito-eq-X}) is bounded for a start.

Let $\eta:\Omega\to\wh{\dom}$ be $\fF_0$-measurable
with $\EE[\|\eta\|^8_{\wh{\dom}}]<\infty$ and let $\eta_\ell$, $\ell\in\NN$,
be $\CC^L\otimes\sC[\mathfrak{d}_C]$-valued $\fF_0$-measurable simple functions
such that $\|\eta_\ell-\eta\|_{\wh{\dom}}\to0$ in $L^8(\PP)$.
We already know that each $\eta_\ell$, $\ell\in\NN$, may be plugged into
the stochastic integral equation \eqref{spin78} (where $M<\infty$). 
Set $\tilde{\eta}_s^{(\ell)}:=\WW{\vxi,s}{0,(N,M)}(\eta_\ell-\eta)$, where $0\le N\le M\le \infty$.
In view of \eqref{gustav0} we then see that
$\sup_{s\le t}\|\tilde{\eta}_s^{(\ell)}\|_{\wh{\dom}}\to0$, $\ell\to\infty$,
in $L^4(\PP)$, for all $t\in I$.
Moreover, we have the following bounds, uniformly in $\V{x}\in\RR^\nu$,
\begin{align}\label{spin123}
\|\wh{H}_{\scal}^0(\vxi,\V{x})\phi\|+&
\|\V{v}(\vxi,\V{x})\phi\|\le\const(\vxi)\,\|\phi\|_{\wh{\dom}}\,,
\quad
\|\vsigma\cdot\vp(\V{F}_{\V{x}})\phi\|\le\const\,\|\phi\|_{\wh{\dom}}\,.
\end{align}
Combined with Lem.~\ref{lem-spin2000} and Prop.~\ref{prop-stoch-calc}(2) the first one permits to get
\begin{align}\nonumber
\EE\Big[\sup_{s\le t}\Big\|\int_0^s\V{v}(\vxi,\V{X}_r)\,\tilde{\eta}_r^{(\ell)}\,\Id\V{B}_r\Big\|^2\Big]
&\le\const\,\EE\Big[\int_0^t\|\V{v}(\vxi,\V{X}_s)\,\tilde{\eta}_s^{(\ell)}\|^2\Id s\Big]
\\\label{gustav1}
&\le\const't\,\EE\big[\sup_{s\le t}\|\tilde{\eta}_s^{(\ell)}\|_{\wh{\dom}}^2\big]
\xrightarrow{\;\;\ell\to\infty\;\;}0,
\end{align}
for every $t\in I$. Moreover, \eqref{hyp-Y2} and \eqref{def-Yt} imply that the right hand side of
\begin{align}\label{gustav2}
&\EE\Big[\sup_{s\le
  t}\Big\|\int_0^s\V{v}(\vxi,\V{X}_r)\,\tilde{\eta}_r^{(\ell)}\,\V{Y}_r\Id r\Big\|^2\Big]
\\\nonumber
&\le\const''\Big(\int_0^t(\cT-s)^{-\nf{2}{3}}\Id
s\Big)^{\nf{3}{2}}
\Big(\int_0^t(\cT-s)^{2}\EE[|\V{Y}_s|^4]\Id s\Big)^{\nf{1}{2}}
\EE\big[\sup_{s\le  t}\|\tilde{\eta}_s^{(\ell)}\|_{\wh{\dom}}^{4}\big]^{\nf{1}{2}}
\end{align}
goes to zero, for every $t\in I$, as well. (If $I=[0,\infty)$, 
replace $(\cT-s)^a$ by $1$. Notice that the constants in
\eqref{gustav1} and \eqref{gustav2} depend in particular on $\V{q}$.)
It is now clear that we may plug $\eta_\ell$ into \eqref{spin78} and
pass to the limit $\ell\to\infty$ in that equation, since each term converges in
$L^2(\PP)$, locally uniformly on $I$.
Hence, \eqref{spin78} is available for all $\fF_0$-measurable $\eta:\Omega\to\wh{\dom}$  
with $\EE[\|\eta\|^8_{\wh{\dom}}]<\infty$, at least when $M<\infty$.

To pass to the limit $M\to\infty$
in the so-obtained extension of \eqref{spin78}, we pick some $\fF_0$-measurable
$\eta:\Omega\to\wh{\dom}$ with $\EE[\|\eta\|^8_{\wh{\dom}}]<\infty$ and 
observe that Lem.~\ref{lem-spin2000}(1) and \eqref{spin123} imply the
$\PP$-a.s. existence of the following limit in $C(I,\FHR)$
(equipped with the topology of locally uniform convergence),
\begin{align*}
&\lim_{N\to\infty}\int_0^\bullet\big(\wh{H}_{\scal}^0(\vxi,\V{X}_s)\WW{\vxi,s}{0,(0,N)}
-\vsigma\cdot\vp(\V{F}_{\V{X}_s})\WW{\vxi,s}{0,(0,N-1)}\big)\,\eta\,\Id s
\\
&=\int_0^\bullet\wh{H}^0(\vxi,\V{X}_s)\,\WW{\vxi,s}{0}\eta\,\Id s\,.
\end{align*}
Employing \eqref{gustav1} and \eqref{gustav2},
but with $\tilde{\eta}^{(\ell)}$ replaced by $\WW{\vxi}{0,(N+1,M)}\eta$,
$N<M\le\infty$, and invoking \eqref{gustav0},
we further see that Rem.~\ref{rem-gustav} applies with $\sK=\FHR$
and $X_t^{(N)}=\int_0^t\V{v}(\vxi,\V{X}_s)\,\WW{\vxi,s}{0,(0,N)}\,\eta\,\Id\V{X}_s$.
This shows that, $\PP$-a.s.,
\begin{align*}
&\lim_{N\to\infty}
\int_0^\bullet\V{v}(\vxi,\V{X}_s)\,\WW{\vxi,s}{0,(0,N)}\,\eta\,\Id\V{X}_s
=\int_0^\bullet\V{v}(\vxi,\V{X}_s)\,\WW{\vxi,s}{0}\,\eta\,\Id\V{X}_s
\quad\text{in $C(I,\FHR)$.}
\end{align*}
Thus, we have solved \eqref{SDE-spin} with $V=0$, for bounded $\V{q}$, and for
$\eta$ in $L^8_{\wh{\dom}}(\PP)$,
\begin{equation}\label{harald}
\WW{\vxi,\bullet}{0}\eta
=\eta-i\!\int_0^\bullet\!\V{v}(\vxi,\V{X}_s)\,\WW{\vxi,s}{0}\,\eta\,\Id\V{X}_s
-\int_0^\bullet\wh{H}^0(\vxi,\V{X}_s)\,\WW{\vxi,s}{0}\eta\,\Id s,\;\;\,\PP\text{-a.s.}
\end{equation} 

Next, if $\eta:\Omega\to\wh{\dom}$ is an arbitrary $\fF_0$-measurable map
and $\V{q}:\Omega\to\RR^\nu$ is $\fF_0$-measurable but otherwise
arbitrary as well,
then we may apply the results proven so far with
$\V{q}_n:=1_{\{|\V{q}|+\|\eta\|_{\wh{\dom}}\le n\}}\V{q}$,
$\eta_n:=1_{\{|\V{q}|+\|\eta\|_{\wh{\dom}}\le n\}}\eta$, $n\in\NN$.
If $\WW{\vxi}{0}\eta$ is constructed by means of $\V{q}$
(which is possible according to Lem.~\ref{lem-spin2}),
then we use the pathwise
uniqueness property explained in Rem.~\ref{rem-pw-unique} and
the pathwise uniqueness property $\V{X}^{\V{q}}=\V{X}^{\V{q}_n}$,
$\PP$-a.s. on $\{\V{q}=\V{q}_n\}$, and Lem.~\ref{lem-spin2000}(1) to argue
that $\WW{\vxi}{0}\eta$ has $\PP$-a.s. continuous paths as a $\wh{\dom}$-valued process. 
(See also \eqref{pia1} below for a pathwise uniqueness statement slightly more
general than necessary.) 
Hence, the (stochastic) integrals in \eqref{harald} are well-defined
elements of $\mathsf{S}_I(\FHR)$, for general $\V{q}$ and $\eta$ as
well. Then the pathwise uniqueness property of Rem.~\ref{rem-pw-unique}
and the pathwise uniqueness of the latter (stochastic) integrals
imply that \eqref{harald} is satisfied $\PP$-a.s. on the union of all
sets $\{|\V{q}|+\|\eta\|_{\wh{\dom}}\le n\}$, $n\in\NN$.

To conclude it only remains to include the potential $V$, which can be done
by applying It\={o}'s formula to $\WW{\vxi,t}{V}\eta=e^{-\int_0^tV(\V{X}_s)\Id s}\WW{\vxi,t}{0}\eta$.
\qed


\section{Dependence on initial conditions}\label{sec-initial}

\noindent
In this section we shall deal with families of driving
processes  indexed by the initial condition in \eqref{Ito-eq-X}. Recall that in Hyp.~\ref{hyp-B}
we introduced the notation $\V{X}^{\V{q}}$ for the process solving the SDE
$\Id\V{X}_t=\Id\V{B}_t+\V{\beta}(t,\V{X}_t)\Id t$ with initial condition $\V{X}_0=\V{q}$. 

Obviously, all quantities $\iota$, $(\w{\tau}{t})_{t\in I}$, $u^V_{\vxi}$, $U^\pm$, 
$(U_{\tau,t}^-)_{t\in I}$, $K$, $(K_{\tau,t})_{t\in I}$, and $\WW{\vxi}{V}$ depend on the choice of
the driving process (and in particular of $\BB$).
Since we are now dealing with different choices
of the driving process at the same time, we explicitly refer to this dependence
in the notation by writing $Z[\V{X}^{\V{q}}]$, if $Z$ is
any of the above quantities constructed by means of $\V{X}^{\V{q}}$.

In the first lemma below and in its corollary
we consider constant initial conditions $\V{q}=\V{x}$ and study
the pathwise continuous dependence of the above processes on $\V{x}$.
In the second lemma we prove a weaker form of continuous dependence
for a more general class of initial conditions. Both lemmas serve as a
preparation for the study of a Markovian flow introduced in Sect.~\ref{sec-Markov}. 
As usual, the existence of the flow will be inferred from an interplay between these two types
of continuous dependences.

\begin{lemma}\label{lem-UKx}
For any Hilbert space $\sK$, let $\mathsf{C}_{\sK}^{(\nu)}$ denote the set of maps
$Z:I\times\RR^\nu\times\Omega\to\sK$, $(t,\V{x},\vgamma)\mapsto Z_t(\V{x},\vgamma)$,
for which we can find another map $Z^{\sharp}:I\times\RR^\nu\times\Omega\to\sK$
satisfying the following two conditions:

\smallskip

\noindent{\rm(1)}
For all $\V{x}\in\RR^\nu$, we find some $\PP$-zero set $N_{\V{x}}$ such that
$Z_t(\V{x},\vgamma)=Z^\sharp_t(\V{x},\vgamma)$, for all
$(t,\vgamma)\in I\times(\Omega\setminus N_{\V{x}})$.

\smallskip

\noindent{\rm(2)} For every $\vgamma\in\Omega$, the map
$I\times\RR^\nu\ni(t,\V{x})\mapsto Z_t^\sharp(\V{x},\vgamma)\in\sK$ is continuous. 

\smallskip

\noindent
If $V$ is continuous, then the following map belongs to 
$\mathsf{C}_{\HP_{+1}\oplus\HP\oplus\HP\oplus\CC}^{(\nu)}$,
$$
(t,\V{x},\vgamma)\longmapsto\big(K_{t}[\V{X}^{\V{x}}],U^-_{t}[\V{X}^{\V{x}}],
U_t^+[\V{X}^{\V{x}}],u_{\vxi,t}^V[\V{X}^{\V{x}}]\big)(\vgamma).
$$
\end{lemma}

{\proof} 
Let $p\ge2$. 
It is well-known (see, e.g., \cite[Thm.~4.37]{daPrZa2014}) that there exists $\const_{p}>0$, such that,
for all separable Hilbert spaces $\sK$ and (e.g.) all adapted, continuous $\LO(\RR^\nu,\sK)$-valued 
processes $\V{A}$ on $I$,
\begin{equation}\label{Lp-maj}
\EE\Big[\sup_{t\le\sigma}\Big\|\int_0^t\V{A}_s\,\Id\V{B}_s\Big\|^p\Big]
\le\const_{p}\,\sigma^{\frac{p-2}{2}}\,\EE\Big[\int_0^\sigma\|\V{A}_s\|^p\,\Id s\Big],\quad\sigma\in I.
\end{equation}
Let us further assume that $p>\nu$ is such that \eqref{Lip-EX} is available and
apply the previous inequality to
$$
K^0_{\bullet}[\V{X}^{\V{x}}]:=\int_0^\bullet\iota_s[\V{X}^{\V{x}}]\,
\V{G}_{\V{X}_s^{\V{x}}}\,{\Id}\V{B}_s\in\mathsf{S}_{I}(\HP_{+1}),\quad\V{x}\in\RR^\nu.
$$
Employing \eqref{Lp-maj} with
$\V{A}_s=\iota_s[\V{X}^{\V{x}}]\V{G}_{\V{X}^{\V{x}}_s}-\iota_s[\V{X}^{\V{y}}]\V{G}_{\V{X}_s^{\V{y}}}$
and observing that the derivative of $(\V{x},\V{y})\mapsto e^{-i\V{m}\cdot\V{x}}\V{G}_{\V{y}}$ 
is uniformly bounded on $\RR^\nu$ as a consequence of Hyp.~\ref{hyp-G}(2),
we deduce that, for some $L_0>0$ and all $\V{x},\V{y}\in\RR^\nu$,
\begin{align*}
\EE\big[&\sup_{t\le\sigma}\big\|K^0_{t}[\V{X}^{\V{x}}]-K^0_{t}[\V{X}^{\V{y}}]\big\|^p\big]
\\
&\le\const_{p,\nu}\,L_0^p\,\sigma^{\frac{p-2}{2}}\Big(\sigma\,|\V{x}-\V{y}|^p
+\int_0^\sigma\EE\big[|\V{X}_s^{\V{x}}-\V{X}_s^{\V{y}}|^p\big]\,\Id s\Big)
\\
&\le\const_{p,\nu}\,L_0^p(1+L(\sigma)^p)\sigma^{\nf{p}{2}}\,|\V{x}-\V{y}|^p,\quad\sigma\in I,
\end{align*}
where we applied \eqref{Lip-EX} in the last step. Since $p>\nu$, this estimate implies that
$(t,\V{x},\vgamma)\mapsto K^0_{t}[\V{X}^{\V{x}}](\vgamma)$
belongs to $\mathsf{C}_{\HP_{+1}}^{(\nu)}$ according to
the Kolmogorov-Neveu lemma; see \cite[Lem.~3 of \textsection36 and
Exercise~E.5 of Chap.~8]{Me1982}. Moreover, it is easy to check that
$(t,\V{x},\vgamma)\mapsto K_{t}[\V{X}^{\V{x}}](\vgamma)-K^0_{t}[\V{X}^{\V{x}}](\vgamma)$
is in $\mathsf{C}_{\HP_{+1}}^{(\nu)}$ as the latter processes are given by the 
Bochner-Lebesgue integrals
\begin{align}\label{gustav71}
\int_0^tj_se^{i\V{m}\cdot(\V{x}-\vXi_s(\V{x},\vgamma))}\big\{\V{G}_{\vXi_s(\V{x},\vgamma)}\cdot
\V{\beta}(s,\vXi_s(\V{x},\vgamma))+\breve{q}_{\vXi_s(\V{x},\vgamma)}\big\}\Id s,\;t\in I,
\end{align}
for all $\vgamma$ outside a $\V{x}$-dependent $\PP$-zero set.
In fact, for every $\vgamma\in\Omega$, the integrand in \eqref{gustav71} is continuous in 
$(s,\V{x})\in[0,\sup I)\times\RR^\nu$ 
as a consequence of Hyp.~\ref{hyp-G}(2) and Hyp.~\ref{hyp-B}(2a).
Hence, we may apply the dominated convergence theorem to verify continuity 
of the integrals \eqref{gustav71} as $(t,\V{x})$ varies in any compact subset of
$[0,\sup I)\times\RR^\nu$.
In the case $I=[0,\cT]$ we have to employ the following additional observation
to include the endpoint $\cT<\infty$: For every $r\in\NN$, \eqref{hyp-Y2} implies
\begin{align*}
&\EE\Big[\int_0^\cT\sup_{|\V{x}|\le r}|\V{\beta}(s,\vXi_s(\V{x},\cdot))|\Id s\Big]
\\
&\le\Big(\int_0^{\cT}(\cT-s)^{-\nf{2}{3}}\Id s\Big)^{\nf{3}{4}}\Big(\int_0^\cT(\cT-s)^{2}\EE\big[
\sup_{|\V{x}|\le r}|\V{\beta}(s,\V{X}_s^{\V{x}})|^4\big]\Id s\Big)^{\nf{1}{4}}<\infty.
\end{align*}
As a consequence, we find a $\PP$-zero set $\sN$ such that, for all $\vgamma\in\sN^c$ 
and $r\in\NN$, we may use a suitable multiple of 
$1+\sup_{|\V{x}|\le r}|\V{\beta}(s,\vXi_s(\V{x},\vgamma))|$ 
as an integrable majorant when we apply the dominated convergence theorem to show continuity 
of the integral \eqref{gustav71} as $(t,\V{x})$ varies in $I\times\{|\V{x}|\le r\}$.
The remaining assertions now follow from the fact that
$(t,\V{x},\vgamma)\mapsto K_{t}[\V{X}^{\V{x}}](\vgamma)$
is in $\mathsf{C}_{\HP_{+1}}^{(\nu)}$ in combination with \eqref{def-Uminus} and \eqref{def-u}.
\qed

\begin{corollary}\label{cor-willi}
Let $V\in C(\RR^\nu,\RR)$ and $0\le N\le M\le \infty$. Then 
\begin{equation}\label{willi1}
(t,\V{x},\vgamma)\mapsto\WW{\vxi,t}{V,(N,M)}[\V{X}^{\V{x}}](\vgamma)\,\psi
\quad\text{belongs to}\;\,\mathsf{C}_{\FHR}^{(\nu)},
\end{equation}
for all $\psi\in\FHR$. More precisely, there exist operators
$\WW{\vxi,t}{V,(N,M)}[\V{X}^{\V{x}}]^\sharp(\vgamma)\in\LO(\FHR)$,
$t\in I$, $\V{x}\in\RR^\nu$, $\vgamma\in\Omega$, such that
\begin{equation}\label{willi11}
\forall\,(t,\vgamma)\in I\times\Omega:\quad
\sup_{s\le
  t}\sup_{\V{x}\in\RR^\nu}\big\|\WW{\vxi,s}{V,(N,M)}[\V{X}^{\V{x}}]^\sharp(\vgamma)\big\|
\le\const_t\sum_{\ell=N}^M\frac{(\const\,t)^\ell}{\ell!},
\end{equation}
and such that, for every $\psi\in\FHR$, 
$(t,\V{x},\vgamma)\mapsto\WW{\vxi,t}{V,(N,M)}[\V{X}^{\V{x}}]^\sharp(\vgamma)\,\psi$
is a modification of the map in \eqref{willi1} fulfilling the
requirements (1) and (2) of Lem.~\ref{lem-UKx}.
\end{corollary}

{\proof} {\em Step 1.}
By definition, Hyp.~\ref{hyp-G}, and Lem.~\ref{lem-UKx}, 
after a suitable modification the maps
$(s,t,\V{x})\mapsto\w{s}{t}[\V{X}^{\V{x}}]\,\V{F}_{\V{X}_s^{\V{x}}}\in\HP_C^S$, and 
$(s,t,\V{x})\mapsto{U}_{s,t}^-[\V{X}^{\V{x}}]$ are $\PP$-a.s. 
jointly continuous on $\{s\le t\in I\}\times\RR^\nu$.
More precisely, one has to replace $\V{X}^{\V{x}}$ in $\iota$, $w$, and $\V{F}$ by its version
$(\vXi_{0,t}(\V{x},\cdot))_{t\in I}$ given by Hyp.~\ref{hyp-B}(2), and
$(t,\V{x},\vgamma)\mapsto K_t[\V{X}^{\V{x}}](\vgamma)$ 
should be replaced by a suitable version $K^\sharp$ as in Lem.~\ref{lem-UKx}.
Combining this observation with Rem.~\ref{rem-smooth}, 
Lem.~\ref{lem-UKx}, \eqref{def-W}, and \eqref{def-Qh}
we may verify by hand that \eqref{willi1} holds true,
provided that $\psi\in\CC^L\otimes\sC[\mathfrak{d}_C]$, $0\le N\le M<\infty$, and $V$ is
continuous. We let $\WW{\vxi,t}{V,(N,M)}[\V{X}^{\V{x}}]^\sharp$ denote
the random operators defined on the domain
$\CC^L\otimes\sC[\mathfrak{d}_C]$ by the same formulas as 
$\WW{\vxi,t}{V,(N,M)}[\V{X}^{\V{x}}]\!\!\upharpoonright_{\CC^L\otimes\sC[\mathfrak{d}_C]}$, 
but with $\V{X}^{\V{x}}$ and $K$ replaced by $(\vXi_{0,t}(\V{x},\cdot))_{t\in I}$ and $K^\sharp$,
respectively. (Recall that $u_{\vxi}^V$ and $U^\pm$ are defined by means of $K$ and $\iota$.)
Then we find a $(M,N)$-independent $\PP$-zero set $\sN\in\fF$ such that,
for all $(t,\V{x},\vgamma)\in I\times\QQ^\nu\times(\Omega\setminus\sN)$,
$$
\WW{\vxi,t}{V,(N,M)}[\V{X}^{\V{x}}]\!\!\upharpoonright_{\CC^L\otimes\sC[\mathfrak{d}_C]}\!\!(\vgamma)
=\WW{\vxi,t}{V,(N,M)}[\V{X}^{\V{x}}]^\sharp(\vgamma).
$$
Applying the bound \eqref{norm-WQ1} to each of the countable number of
processes in the previous equation and enlarging the $\PP$-zero set $\sN$,
if necessary, we conclude that, for all $0\le N\le M<\infty$, $t\in I$,
$\vgamma\in\Omega\setminus\sN$, we have
\begin{align}\label{okke}
\sup_{s\le t}\big\|\WW{\vxi,s}{V,(N,M)}[\V{X}^{\V{x}}]^\sharp(\vgamma)\,\wt{\psi}\big\|\le
\const_t\,\|\wt{\psi}\|\sum_{\ell=N}^M\frac{(\const\,t)^\ell}{\ell!},
\quad \wt{\psi}\in\CC^L\otimes\sC[\mathfrak{d}_C],
\end{align}
a priori for all $\V{x}\in\QQ^\nu$. By continuity of
$(t,\V{x})\mapsto\WW{\vxi,t}{V,(N,M)}[\V{X}^{\V{x}}]^\sharp(\vgamma)\,\wt{\psi}$,
the bound \eqref{okke} is, however, even available for all $\V{x}\in\RR^\nu$.
Finally, we re-define $\WW{\vxi}{V,(N,M)}[\V{X}^{\V{x}}]^\sharp(\vgamma):=\delta_{0,N}\id$, 
if $\vgamma\in\sN$,
so that \eqref{okke} is valid for all $(t,\V{x},\vgamma)\in I\times\RR^\nu\times\Omega$.

{\em Step 2.}
Let $M<\infty$, $\psi\in\FHR$, and $\psi_n\in\CC^L\otimes\sC[\mathfrak{d}_C]$, $n\in\NN$,
with $\psi_n\to\psi$. Then, by the construction of $\WW{\vxi}{V,(N,M)}[\V{X}^{\V{x}}]$ in 
Lem.~\ref{lem-spin2}, 
$\WW{\vxi}{V,(N,M)}[\V{X}^{\V{x}}]\psi_n\to\WW{\vxi}{V,(N,M)}[\V{X}^{\V{x}}]\psi$
on $I$ outside some $\V{x}$-dependent $\PP$-zero set
$\sN_{\V{x},N,M}'$, which neither depends on $\psi$ nor on the
approximating sequence $\{\psi_n\}$. Therefore, defining
$\WW{\vxi,t}{V,(N,M)}[\V{X}^{\V{x}}]^\sharp\psi
:=\lim_{n\to\infty}\WW{\vxi,t}{V,(N,M)}[\V{X}^{\V{x}}]^\sharp\psi_n$, $t\in I$,
on $\Omega$, we certainly have
$\WW{\vxi,t}{V,(N,M)}[\V{X}^{\V{x}}]^\sharp\psi=\WW{\vxi,t}{V,(N,M)}[\V{X}^{\V{x}}]\psi$, $t\in I$,
on $\Omega\setminus(\sN\cup\sN_{\V{x},N,M}')$, so that
$\WW{\vxi,t}{V,(N,M)}[\V{X}^{\V{x}}]^\sharp\psi$ satisfies the
requirement (1) in Lem.~\ref{lem-UKx}.
Notice that, by \eqref{okke}, the above definition of
$\WW{\vxi,t}{V,(N,M)}[\V{X}^{\V{x}}]^\sharp\psi$
does not depend on the approximating sequence $\{\psi_n\}$ and
that \eqref{okke} extends to all $\wt{\psi}\in\FHR$.
Moreover, \eqref{okke}, thus extended, implies that, on $\Omega$,
the convergence $\WW{\vxi,t}{V,(N,M)}[\V{X}^{\V{x}}]\,\psi_n\to
\WW{\vxi,t}{V,(N,M)}[\V{X}^{\V{x}}]^\sharp\,\psi$, $n\to\infty$,
is locally uniform in $(t,\V{x})\in I\times\RR^\nu$.
Employing the results of the first step, we deduce that
$(t,\V{x})\mapsto\WW{\vxi,t}{V,(N,M)}[\V{X}^{\V{x}}]^\sharp(\vgamma)\,\psi$ 
is continuous on $I\times\RR^\nu$. This proves \eqref{willi1} and \eqref{willi11} for finite $M$.

{\em Step 3.}
Employing \eqref{willi11} (with finite $M$) we further see that the limits
$\WW{\vxi,t}{V,(N,\infty)}[\V{X}^{\V{x}}]^\sharp(\vgamma)\psi
:=\lim_{M\to\infty}\WW{\vxi,t}{V,(N,M)}[\V{X}^{\V{x}}]^\sharp(\vgamma)\psi$
are locally uniform in $(t,\V{x})$, for all $\vgamma\in\Omega$ and $\psi\in\FHR$,
and that, by the construction of
$\WW{\vxi,t}{V,(N,M)}[\V{X}^{\V{x}}]$ in Lem.~\ref{lem-spin2} and the remarks in Step~2, 
$\WW{\vxi}{V,(N,\infty)}[\V{X}^{\V{x}}]^\sharp\psi=\WW{\vxi}{V,(N,\infty)}[\V{X}^{\V{x}}]\psi$
holds outside some $\psi$-independent $\PP$-zero set $\sN_{\V{x},N}''$.
This implies \eqref{willi1} and \eqref{willi11} also in the general case.
\qed

\begin{lemma}\label{lem-flow-st}
Assume that $V$ is continuous and bounded. Let $\V{q},\V{q}_n:\Omega\to\RR^\nu$, $n\in\NN$, 
all be bounded and $\fF_0$-measurable such that $\V{q}_n\to\V{q}$,
$\PP$-a.s., as $n\to\infty$, and $\sup_n\|\V{q}_n\|_\infty<\infty$. Moreover, let
$\eta,\eta_n:\Omega\to\FHR$, $n\in\NN$, all be bounded and $\fF_0$-measurable
such that $\EE[\|\eta-\eta_n\|^2]\to0$, as $n\to\infty$. Then
$$
\EE\big[\sup_{t\le\tau}\big\|\WW{\vxi,t}{V}[\V{X}^{\V{q}}]\,\eta
-\WW{\vxi,t}{V}[\V{X}^{\V{q}_n}]\,\eta_n\big\|^2\big]
\xrightarrow{\;\;n\to\infty\;\;}0,\qquad\tau\in I.
$$
\end{lemma}

{\proof} 
In the case $I=[0,\cT]$ we assume that $\tau<\cT$ to start with.

Since $\|\WW{\vxi,t}{V}[\V{X}^{\V{q}}]\|\le \const_{\tau}$, $t\in[0,\tau]$,
$\PP$-a.s., with a $\V{q}$-independent constant $\const_{\tau}$, we may
assume that $\eta_n=\eta$, $n\in\NN$. As we can approximate $\eta$ by the vectors 
$\wt{\eta}_\ell:=(1+\Id\Gamma(\V{m})^2/\ell+\Id\Gamma(\omega)/\ell)^{-1}\eta:
\Omega\to\wh{\dom}$, 
which satisfy $\EE[\|\eta-\wt{\eta}_\ell\|^2]\to0$, $\ell\to\infty$, by dominated convergence,
we may also assume that $\eta:\Omega\to\wh{\dom}$ such that
$\|\eta\|_{\wh{\dom}}$ is bounded on $\Omega$. Under these assumptions we define
$\psi_t^{(n)}:=\WW{\vxi,t}{V}[\V{X}^{\V{q}}]\,\eta-\WW{\vxi,t}{V}[\V{X}^{\V{q}_n}]\,\eta$,
so that $\psi_0^{(n)}=0$. Abbreviate 
$$
\V{v}_s:=\vxi-\Id\Gamma(\V{m})-\vp(\V{G}_{\V{X}_s^{\V{q}}}),
$$
and let $\V{v}_s^{(n)}$ be defined analogously with $\V{q}_n$ in place of $\V{q}$.
Applying Thm.~\ref{thm-Ito-spin} in combination with Ex.~\ref{ex-Ito-SP} and taking 
$\Re\SPn{\psi_s^{(n)}}{i\V{v}_s\,\psi_s^{(n)}}=0$
into account, we $\PP$-a.s. obtain after some brief computations, 
for all $n\in\NN$ and $t\in[0,\sup I)$,
\begin{align}\nonumber
&\|\psi_t^{(n)}\|^2
=-\int_0^t2\|\Id\Gamma(\omega)^\eh\psi_s^{(n)}\|^2\Id s
\\\nonumber
&+\int_0^t2\Re\SPb{\psi_s^{(n)}}{\big(\vsigma\cdot\vp(\V{F}_{\V{X}^{\V{q}}_s})
+\tfrac{i}{2}\vp(q_{\V{X}^{\V{q}}_s})\big)\,\psi_s^{(n)}}\,\Id s
\\\nonumber
&-\int_0^t2\Re\SPb{\psi_s^{(n)}}{V(\V{X}^{\V{q}})\WW{\vxi,s}{V}[\V{X}^{\V{q}}]\eta-
V(\V{X}^{\V{q}_n})\WW{\vxi,t}{V}[\V{X}^{\V{q}_n}]\eta}\,\Id s
\\\nonumber
&+\int_0^t\Re\SPb{(\V{v}_s-\V{v}_s^{(n)})\WW{\vxi,s}{V}[\V{X}^{\V{q}}]\eta}{
(\V{v}_s-\V{v}_s^{(n)})\WW{\vxi,s}{V}[\V{X}^{\V{q}_n}]\eta}\,\Id s
\\\nonumber
&+\int_0^t\Re\SPb{\WW{\vxi,s}{V}[\V{X}^{\V{q}}]\eta}{
[\V{v}_s,\V{v}_s-\V{v}_s^{(n)}]\,\WW{\vxi,s}{V}[\V{X}^{\V{q}_n}]\eta}\,\Id s
\\\nonumber
&+\int_0^t2\Re\SPb{\psi_s^{(n)}}{\big(\vsigma\cdot
\vp(\V{F}_{\V{X}^{\V{q}}_s}-\V{F}_{\V{X}^{\V{q}_n}_s})
+\tfrac{i}{2}\vp(q_{\V{X}^{\V{q}}_s}-q_{\V{X}^{\V{q}_n}_s})\big)\,
\WW{\vxi,s}{V}[\V{X}^{\V{q}_n}]\eta}\,\Id s
\\\nonumber
&-\int_0^t2\Re\SPb{\psi_s^{(n)}}{i(\V{v}_s-\V{v}_s^{(n)})
\,\WW{\vxi,s}{V}[\V{X}^{\V{q}_n}]\eta}\,\Id\V{B}_s
\\\nonumber
&-\int_0^t2\Re\SPb{\psi_s^{(n)}}{i(\V{v}_s-\V{v}_s^{(n)})
\,\WW{\vxi,s}{V}[\V{X}^{\V{q}_n}]\eta}\,\V{\beta}(s,\V{X}_s^{\V{q}})\,\Id s
\\\label{flora7}
&-\int_0^t2\Re\SPb{\psi_s^{(n)}}{i\V{v}_s^{(n)}\,\WW{\vxi,s}{V}[\V{X}^{\V{q}_n}]\eta}
\big(\V{\beta}(s,\V{X}_s^{\V{q}})-\V{\beta}(s,\V{X}_s^{\V{q}_n})\big)\Id s.
\end{align}
Next, we observe that
$\V{v}_s-\V{v}_s^{(n)}=\vp(\V{G}_{\V{X}^{\V{q}_n}_s}-\V{G}_{\V{X}^{\V{q}}_s})$ and
\begin{align*}
[\V{v}_s,\V{v}_s-\V{v}_s^{(n)}]
&=[\Id\Gamma(\V{m}),\vp(\V{G}_{\V{X}^{\V{q}}_s}-\V{G}_{\V{X}^{\V{q}_n}_s})]
\\
&=i\vp\big(i\V{m}\cdot(\V{G}_{\V{X}^{\V{q}_n}_s}-\V{G}_{\V{X}^{\V{q}}_s})\big),
\end{align*}
where the field operators on the right hand sides can be controlled by
means of \eqref{rb-vp1}; i.e., setting $\theta:=1+\Id\Gamma(\omega)$, we obtain
\begin{align*}
\|(\V{v}_s-\V{v}_s^{(n)})\theta^\mh\|+\|[\V{v}_s,\V{v}_s-\V{v}_s^{(n)}]\theta^\mh\|
\le\const\,\|\V{G}_{\V{X}^{\V{q}_n}_s}-\V{G}_{\V{X}^{\V{q}}_s}\|_{\mathfrak{k}^\nu}&\le\const',
\\
\|\vsigma\cdot\vp(\V{F}_{\V{X}^{\V{q}}_s}-\V{F}_{\V{X}^{\V{q}_n}_s})\theta^\mh\|
\le\const\,\|\V{F}_{\V{X}^{\V{q}_n}_s}-\V{F}_{\V{X}^{\V{q}}_s}\|_{\mathfrak{k}^S}&\le\const',
\\
\|\vp({q}_{\V{X}^{\V{q}}_s}-{q}_{\V{X}^{\V{q}_n}_s})\theta^\mh\|\le\const\,\max_{j=1,...,\nu}
\|\partial_{x_j}\V{G}_{\V{X}^{\V{q}_n}_s}-\partial_{x_j}\V{G}_{\V{X}^{\V{q}}_s}\|_{\mathfrak{k}^\nu}
&\le\const'.
\end{align*}
Moreover,
$\|\vsigma\cdot\vp(\V{F}_{\V{X}^{\V{q}}_s})\psi^{(n)}_s\|,\|\vp(q_{\V{X}_s^{\V{q}}})\psi^{(n)}_s\|
\le\const\,\|\theta^\eh\psi_s^{(n)}\|$;
here we observe that terms containing one factor 
$\|\theta^\eh\psi_s^{(n)}\|=(\|\Id\Gamma(\omega)^\eh\psi_s^{(n)}\|^2+\|\psi^{(n)}_s\|^2)^\eh$ 
can be controlled by the first integral in the first line of \eqref{flora7} via the bound
$2ab\le\ve a^2+b^2/\ve$.
Taking these remarks into account, writing
$\WW{\vxi,s}{V}[\V{X}^{\V{q}_n}]\eta=\WW{\vxi,s}{V}[\V{X}^{\V{q}}]\eta-\psi_s^{(n)}$,
and applying Cauchy-Schwarz inequalities we easily see that the sum of
all terms of the right hand side of \eqref{flora7} which appear in the
first six lines is bounded from above by
$$
\big(\text{Lines 1.--6. of RHS of \eqref{flora7}}\big)\le
\const\int_0^t\big(\|\psi_s^{(n)}\|^2+\alpha_n(s)
\,\|\theta^\eh\WW{\vxi,s}{V}[\V{X}^{\V{q}}]\eta\|^2\big)\Id s,
$$
for $t\in[0,\tau]$. Here the constant depends (inter alia) on the supremum norm
of $V$ which is bounded by assumption, and the random variables $\alpha_n(s)$ are defined by
\begin{align*}
\alpha_n(s)&:=\max_{\vk=1,2,4}\|\V{G}_{\V{X}^{\V{q}_n}_s}
-\V{G}_{\V{X}^{\V{q}}_s}\|_{\mathfrak{k}^\nu}^\vk
+\|\V{F}_{\V{X}^{\V{q}_n}_s}-\V{F}_{\V{X}^{\V{q}}_s}\|_{\mathfrak{k}^S}^2
\\
&\quad+\max_{j=1,...,\nu}
\|\partial_{x_j}\V{G}_{\V{X}^{\V{q}_n}_s}-\partial_{x_j}\V{G}_{\V{X}^{\V{q}}_s}\|_{\mathfrak{k}^\nu}^2
+|V(\V{X}^{\V{q}_n}_s)-V(\V{X}^{\V{q}}_s)|^2.
\end{align*}
To treat the martingale in the seventh line of \eqref{flora7}, let us
call it $\fM$, we apply the special case
$\EE[\sup_{t\le\tau}|\fM_t|]\le\const \EE[\llbracket \fM,\fM\rrbracket_\tau^\eh]$
of an inequality due to Davis; see, e.g., \cite[Thm.~3.28 in Chap.~3]{KaratzasShreve}.
Here we have, for every $\ve>0$,
\begin{align*}
\EE\big[\llbracket \fM\rrbracket_\tau^\eh\big]
&=2\EE\Big[\Big(\int_0^\tau\big(\Re\SPb{\psi_s^{(n)}}{i(\V{v}_s-\V{v}_s^{(n)})
\,\WW{\vxi,s}{V}[\V{X}^{\V{q}_n}]\eta}\big)^2\Id s\Big)^\eh\Big]
\\
&\le\ve\,\EE\big[\sup_{t\le\tau}\|\psi_t^{(n)}\|^2\big]
+\frac{1}{\ve}\,\EE\Big[\int_0^\tau\big\|(\V{v}_s-\V{v}_s^{(n)})
\,\WW{\vxi,s}{V}[\V{X}^{\V{q}_n}]\eta\big\|^2\Id s\Big].
\end{align*}
Furthermore (ignore the factors $(\cT-s)^a$ in the case $I=[0,\infty)$),
\begin{align*}
&2\EE\Big[\int_0^t\|\psi_s^{(n)}\|\,\big\|(\V{v}_s-\V{v}_s^{(n)})
\,\WW{\vxi,s}{V}[\V{X}^{\V{q}_n}]\eta\big\|\,|\V{\beta}(s,\V{X}_s^{\V{q}})|\,\Id s\Big]
\\
&\le\int_0^t(\cT-s)^{-\nf{3}{4}}\EE\big[\sup_{r\le s}\|\psi_r^{(n)}\|^2\big]\,\Id s
+\Big(\int_0^t(\cT-s)^2\EE[|\V{\beta}(s,\V{X}_s^{\V{q}})|^4]\Id s\Big)^\eh
\\
&\hspace{3.3cm}
\cdot\Big(\int_0^t\EE\big[\alpha_n(s)(\cT-s)^{\mh}
\|\theta^\eh\WW{\vxi,s}{V}[\V{X}^{\V{q}_n}]\eta\|^4\big]\Id s\Big)^\eh,
\end{align*}
and, likewise,
\begin{align*}
&2\EE\Big[\|\psi_s^{(n)}\|\,\big\|\V{v}_s^{(n)}\,\WW{\vxi,s}{V}[\V{X}^{\V{q}_n}]\eta\big\|\,
|\V{\beta}(s,\V{X}_s^{\V{q}})-\V{\beta}(s,\V{X}_s^{\V{q}_n})|\,\Id s\Big]
\\
&\le\int_0^t(\cT-s)^{-\nf{3}{4}}\EE\big[\sup_{r\le s}\|\psi_r^{(n)}\|^2\big]\,\Id s
\\
&\;\,
+\Big(\int_0^t\EE\big[\const(\cT-s)^\mh\|\WW{\vxi,s}{V}[\V{X}^{\V{q}_n}]\eta\|_{\wh{\dom}}^4\big]
\Id s\Big)^\eh\Big(\int_0^t(\cT-s)^2\EE[\alpha_n'(s)^4]\Id s\Big)^\eh,
\end{align*}
with $\alpha_n'(s):=|\V{\beta}(s,\V{X}_s^{\V{q}})-\V{\beta}(s,\V{X}_s^{\V{q}_n})|$.

Putting all the above remarks together, using that, by \eqref{gustav0},
$$
\max_{\vk=1,2,4}\:\sup_{\wt{\V{q}}=\V{q},\V{q}_1,\V{q}_2,...}
\EE\big[\sup_{s\le\tau}\|\WW{\vxi,s}{V}[\V{X}^{\wt{\V{q}}}]\eta\|_{\wh{\dom}}^{2\vk}\big]\le
\const(\tau)(1+\EE[\|\eta\|_{\wh{\dom}}^{16}]),
$$
and employing \eqref{hyp-Y2}, we readily arrive at the
following estimate for $\rho(t):=\EE[\sup_{r\le t}\|\psi_t^{(n)}\|^2]$,
\begin{align}\label{patrice}
\rho(t)&\le\const\!\int_0^t[1\vee(\cT-s)^\mh]\,\rho(s)\,\Id s+c_n(\tau),\quad t\in[0,\tau],
\\\nonumber
c_n(\tau)&:=\const(\tau,\eta)\Big\{\max_{a=\nf{1}{2},\nf{1}{4}}
\Big(\int_0^\tau[1\vee(\cT-s)^\mh]\,\EE[\alpha_n(s)^2]\Id s\Big)^a
\\\nonumber
&\qquad\qquad\qquad\qquad+\Big(\int_0^\tau\!(\cT-s)^2\EE[\alpha_n'(s)^4]\Id s\Big)^\eh\Big\}.
\end{align}
Since $\V{q}_n\to\V{q}$, $\PP$-a.s., Hyp.~\ref{hyp-B}(2a)\&(2c) imply that
$\V{X}_s^{\V{q}_n}\to\V{X}_s^{\V{q}}$, $s\in[0,\tau]$, $\PP$-a.s.,
whence, by  Hyp.~\ref{hyp-G} and the continuity of $V$ and $\V{\beta}$, 
$\alpha_n(s)\to0$ and $\alpha_n'(s)\to0$, for all $s\in[0,\tau]$, $\PP$-a.s.
By virtue of Hyp.~\ref{hyp-G}, Hyp.~\ref{hyp-B}(3) (with $\mathfrak{q}=|\V{q}|\vee(\sup_n|\V{q}_n|)$),
and the boundedness of $V$ we may thus apply the dominated convergence theorem to 
see that $c_n(\tau)\to0$, as $n\to\infty$. In the case $I=[0,\cT]$ we further observe that
$c_n(\cT)\to0$ and that the bound \eqref{patrice} holds true for $\tau=\cT$ as well.
We may now apply Gronwall's lemma to conclude.
\qed


\section{Stochastic flow, strong Markov property, and strong solutions}\label{sec-Markov}

\noindent
In this section we prove the existence of a Markovian flow associated
with our model, always assuming that the potential $V$ is continuous and bounded.
To start with we recall that the time-shifted stochastic basis
$\BB_\tau$, where $\tau\in[0,\sup I)$, together with the time-shifted
Brownian motion and the drift vector field given, respectively, by
\begin{align*}
{}^\tau\!\V{B}_t:=\V{B}_{\tau+t}-\V{B}_\tau,\quad\V{\beta}_\tau(t,\V{x}):=\V{\beta}(\tau+t,\V{x}),
\quad t\in I^\tau,\,\V{x}\in\RR^\nu,
\end{align*}
again satisfy the conditions imposed by Hyp.~\ref{hyp-B};
cf. \eqref{shifted-stoch-basis} for the definition of $\BB_\tau$ and $I^\tau$.
In accordance with our earlier conventions,
we let ${}^\tau\!\V{X}^{\V{q}}\in\mathsf{S}_{I^\tau}(\RR^\nu)$ denote
the solution of the SDE
$\Id\V{X}_t=\Id^\tau\!\V{B}_t+\V{\beta}_\tau(t,\V{X}_t)\Id t$
with $\fF_{\tau}$-measurable initial condition
$\V{q}:\Omega\to\RR^\nu$ and $\BB_\tau$ as underlying stochastic basis.
Then the corresponding operators
\begin{equation}\label{def-WW(q)}
\WW{\vxi,t}{V}[^\tau\!\V{X}^{\V{q}}]\in\LO(\FHR),\quad t\in I^\tau,\quad\PP\text{-a.s.},
\end{equation}
are defined by Thm.~\ref{thm-Ito-spin} applied with $\BB_\tau$ as underlying basis.
For later reference we note that the pathwise uniqueness property of $\WW{\vxi}{V}[\,\cdot\,]$
explained in Rem.~\ref{rem-pw-unique} implies, for any
two $\fF_\tau$-measurable $\V{q},\wt{\V{q}}:\Omega\to\RR^\nu$ and $A\in\fF_\tau$,
\begin{align}\label{pia1}
\text{$\V{q}=\wt{\V{q}}$ $\PP$-a.s. on $A$}\;\;\Rightarrow\;\;
\big(\forall\:t\in I^\tau\!\!:\:\WW{\vxi,t}{V}[^\tau\!\V{X}^{\V{q}}]
=\WW{\vxi,t}{V}[^\tau\!\V{X}^{\wt{\V{q}}}]\big)\;\text{$\PP$-a.s. on $A$.}
\end{align}
Moreover, if $\eta:\Omega\to\wh{\dom}$ is $\fF_\tau$-measurable, then,
according to Hyp.~\ref{hyp-B} and Thm.~\ref{thm-Ito-spin},
$(^\tau\!\V{X}^{\V{q}},\WW{\vxi}{V}[^\tau\!\V{X}^{\V{q}}]\,\eta)$
is, up to indistinguishability, the unique element of $\mathsf{S}_{I^\tau}(\RR^\nu\times\FHR)$
whose paths belong $\PP$-a.s. to $C(I^\tau,\RR^\nu\times\wh{\dom})$
and which solves the following initial value problem for a system of SDE's for $(\V{X},X)$,
\begin{align}\label{bX1}
\V{X}_\bullet&=\V{q}+{}^\tau\!{\V{B}}_{\bullet}
+\int_0^\bullet\V{\beta}_\tau(s,\V{X}_s)\,\Id s,
\\\label{bX2}
X_\bullet&=\eta-i\int_0^{\bullet}{\V{v}(\vxi,\V{X}_s)}\,X_s\,\Id\V{X}_s
-\int_0^{\bullet}\wh{H}^V(\vxi,\V{X}_s)\,X_s\,\Id s.
\end{align}

In what follows we shall also set ${}^T\!\V{X}^{\V{q}}:=\V{q}$ and
$\WW{\vxi,0}{V}[^T\!\V{X}^{\V{q}}]:=\id$, for every $\fF_T$-measurable $\V{q}:\Omega\to\RR^\nu$,
and $I^T:=\{0\}$.

\begin{lemma}\label{lem-flora}
Let $0\le\sigma\le\tau\in I$ and let $(\V{q},\eta):\Omega\to\RR^\nu\times\FHR$
be $\fF_\sigma$-measurable. Then we $\PP$-a.s. have
\begin{equation}\label{eq-flora}
\forall\,t\in I,\,t\ge\tau:\quad
\WW{\vxi,t-\sigma}{V}[^\sigma\!\V{X}^{\V{q}}]\,\eta
=\WW{\vxi,t-\tau}{V}[^\tau\!\V{X}^{{}^\sigma\!\V{X}_{\tau-\sigma}^{\V{q}}}]\,
\WW{\vxi,\tau-\sigma}{V}[^\sigma\!\V{X}^{\V{q}}]\,\eta.
\end{equation}
\end{lemma}

{\proof} 
If $\eta$ attains its values in $\wh{\dom}$, then
it is straightforward to infer the statement from the above remarks.
If $\eta$ is arbitary, we apply \eqref{eq-flora} first to the $\fF_\tau$-measurable random vectors
$\eta_n:=(1+\Id\Gamma(\V{m})^2/n+\Id\Gamma(\omega)/n)^{-1}\eta:\Omega\to\wh{\dom}$.
Then there is a $\PP$-zero set $N$ such that \eqref{eq-flora} holds with $\eta$ replaced by $\eta_n$
on $\Omega\setminus N$ and for all $n\in\NN$. By \eqref{def-WW(q)}
we may then pass to the limit $n\to\infty$ pointwise on $\Omega\setminus N$.
\qed

We summarize parts of our previous discussion in the following theorem.
With the results proven so far at hand its proof follows traditional lines:

\begin{theorem}[Existence of a stochastic flow]\label{thm-flow}
Assume that $V$ is continuous and bounded. Then there exists a family of maps
$\Lambda_{\tau,t}:\RR^\nu\times\FHR\times\Omega\to\RR^\nu\times\FHR$,
$0\le\tau\le t\in I$, satisfying the following:

\smallskip

\noindent{\rm(1)} For all $\tau\in[0,\sup I)$, $\phi\in\FHR$, and $\vgamma\in\Omega$,
the following two maps are continuous,
\begin{align*}
\RR^\nu\times\FHR\ni(\V{x},\psi)&\longmapsto
\Lambda_{\tau,\tau+\bullet}(\V{x},\psi,\vgamma)\in C(I^\tau,\RR^\nu\times\FHR),
\\
I^\tau\ni t&\longmapsto\Lambda_{\tau,\tau+t}(\,\cdot\,,\phi,\vgamma)\in
C(\RR^\nu,\RR^\nu\times\FHR).
\end{align*}

\smallskip

\noindent{\rm(2)} Let $\tau\in I$. Then
$\Lambda_{\tau,\tau}(\V{x},\psi,\vgamma)=(\V{x},\psi)$, for all
$(\V{x},\psi,\vgamma)\in\RR^\nu\times\FHR\times\Omega$.
If $(\V{q},\eta):\Omega\to\RR^\nu\times\FHR$ is $\fF_\tau$-measurable
with $\tau<\sup I$, then 
\begin{align}\label{def-flow}
\Lambda_{\tau,\tau+\bullet}(\V{q}(\vgamma),\eta(\vgamma),\vgamma)&=
\big({}^\tau\!\V{X}_\bullet^{\V{q}},\WW{\vxi,\bullet}{V}[^\tau\!\V{X}^{\V{q}}]\,\eta\big)
(\vgamma)\quad\text{on}\;\,I^\tau,
\end{align}
for $\PP$-a.e. $\vgamma$.
In particular, if $\eta$ attains its values in $\wh{\dom}$,
then $\Lambda_{\tau,\tau+\bullet}(\V{q}(\cdot),\eta(\cdot),\cdot)$
is, up to indistinguishability, the only element of $\mathsf{S}_{I^\tau}(\RR^\nu\times\FHR)$
whose paths belong $\PP$-a.s. to $C(I^\tau,\RR^\nu\times\wh{\dom})$
and which solves \eqref{bX1}\&\eqref{bX2}.

\smallskip

\noindent
{\rm(3)} For $0\le\sigma\le\tau\in I$, we find a $\PP$-zero set
$N_{\sigma,\tau}$
such that, for all
$(\V{x},\psi)\in\RR^\nu\times\FHR$,
\begin{align}\label{flow-eq}
\Lambda_{\sigma,t}(\V{x},\psi,\vgamma)
&=\Lambda_{\tau,t}(\Lambda_{\sigma,\tau}(\V{x},\psi,\vgamma),\vgamma),\qquad
\tau\le t\in I,
\;\vgamma\in\Omega\setminus N_{\sigma,\tau}.
\end{align}
{\rm(4)} For $0\le\tau\le t$, the map
$[\tau,t]\times\RR^\nu\times\FHR\times\Omega\ni(s,\V{x},\psi,\vgamma)\mapsto
\Lambda_{\tau,s}(\V{x},\psi,\vgamma)$ is
$\fB([\tau,t])\otimes\fB(\RR^\nu\times\FHR)\otimes\fF_{\tau,t}$-measurable,
where $\fF_{\tau,t}$ is the completion of the $\sigma$-algebra generated
by all increments $\V{B}_s-\V{B}_\tau$ with $s\in[\tau,t]$.
In particular, $\Lambda_{\tau,t}(\V{x},\psi,\cdot)$ is $\fF_\tau$-independent.
\end{theorem}

{\proof} 
If $(\V{q},\eta)=(\V{x},\psi)\in\RR^\nu\times\FHR$ is constant, then we define 
$$
\Lambda_{\tau,\tau+t}(\V{x},\psi,\vgamma):=
\big(\vXi_{\tau,\tau+t}(\V{x},\vgamma),\WW{\vxi,t}{V}[^\tau\!\V{X}^{\V{x}}]^\sharp\,\psi\big),
\quad\tau\in I,\;t\in I^\tau,\;\vgamma\in\Omega.
$$
Then $\Lambda$ satisfies (1) and \eqref{def-flow} (with $(\V{q},\eta)=(\V{x},\psi)$)
according to Hyp.~\ref{hyp-B}(2) and Cor.~\ref{cor-willi} (applied to the time-shifted data).

Next, let $A_1,\ldots,A_\ell$ be disjoint elements of $\fF_\tau$
whose union equals $\Omega$ and let $(\V{x}_j,\psi_j)\in\RR^\nu\times\FHR$, $j=1,\ldots,\ell$.
Then \eqref{pia1} implies that, $\PP$-a.s. on $I^\tau$,
\begin{align}\label{flow3}
\WW{\vxi,\bullet}{V}[^\tau\!\V{X}^{\hat{\V{q}}}]\hat{\eta}
&=\sum_{j=1}^\ell
1_{A_j}\,\WW{\vxi,\bullet}{V}[^\tau\!\V{X}^{\V{x}_j}]\psi_j,\;\;\text{where}
\;\;(\hat{\V{q}},\hat{\eta})=\sum_{j=1}^\ell(\V{x}_j,\psi_j)\,1_{A_j}.
\end{align}
Since, by the remarks in the first paragraph of this proof, the process on
the right hand side of the first identity in \eqref{flow3} and the second component of
$(\Lambda_{\tau,\tau+t}(\hat{\V{q}}(\cdot),\hat{\eta}(\cdot),\cdot))_{t\in I^\tau}$
are indistinguishable, we see that \eqref{def-flow} holds true, for simple
$\fF_\tau$-measurable functions $(\V{q},\eta)=(\hat{\V{q}},\hat{\eta})$ as in \eqref{flow3}.

Now, let $({\V{q}},{\eta}):\Omega\to\RR^\nu\times\FHR$
be $\fF_\tau$-measurable and bounded. Then there exist simple functions
$(\hat{\V{q}}_n,\hat{\eta}_n)$, $n\in\NN$, like the one in \eqref{flow3},
such that $\sup_n\|\hat{\V{q}}_n\|_\infty<\infty$, $\hat{\V{q}}_n\to{\V{q}}$, $\PP$-a.s., 
and $\EE[\|{\eta}-\hat{\eta}_n\|^2]\to0$, as $n\to\infty$.
By applying Lem.~\ref{lem-flow-st} to the time-shifted data, we may assume -- after passing to
a suitable subsequence if necessary -- that also
$\hat{\eta}_n\to{\eta}$ and $\WW{\vxi,t}{V}[^\tau\!\V{X}^{\hat{\V{q}}_n}]\hat{\eta}_n
\to\WW{\vxi,t}{V}[^\tau\!\V{X}^{\V{q}}]{\eta}$, $t\in I^\tau$,
on the complement of some $t$-independent $\PP$-zero set.
Since $(\V{x},\psi)\mapsto\Lambda_{\tau,t}(\V{x},\psi,\vgamma)$ is continuous,
we may thus pass to the limit $n\to\infty$ in
\begin{align*}
\Lambda_{\tau,t}(\hat{\V{q}}_n(\vgamma),\hat{\eta}_n(\vgamma),\vgamma)&=
\big({}^\tau\!\V{X}^{\hat{\V{q}}_n}_t,\WW{\vxi,t}{V}[^\tau\!\V{X}^{\hat{\V{q}}_n}]\,\hat{\eta}_n\big)
(\vgamma),
\end{align*}
for all $t\in I^\tau$ and $\vgamma$ outside another $t$-independent $\PP$-zero set.
This proves \eqref{def-flow} for bounded $(\V{q},\eta)$.
For general $(\V{q},\eta)$, we plug $\wt{\V{q}}_n:=1_{|\V{q}|\le n}\V{q}$ and
$\wt{\eta}_n:=1_{\|{\eta}\|\le n}\,{\eta}$, $n\in\NN$,
into \eqref{def-flow} which then holds outside a $\PP$-zero set $N_n$.
Then \eqref{pia1} permits to argue that both sides of the resulting identity converge pointwise
on $\Omega\setminus\cup_nN_n$, as $n\to\infty$, for every $t\ge\tau$.

Altogether we have now proved (1) and (2). The assertions of (4) follow from Part~(1),
Hyp.~\ref{hyp-B}(2),  and Lem.~\ref{lem-flora} together with \eqref{bX2} and \eqref{def-flow}.

To prove (3) we pick a countable dense subset, $\{(\V{x}_n,\psi_n):
\,n\in\NN\}$, of $\RR^\nu\times\FHR$. Then a straightforward
combination of Lem.~\ref{lem-flora} and \eqref{def-flow} shows that
the equality in \eqref{flow-eq} with $(\V{x},\psi)=(\V{x}_n,\psi_n)$
holds true, for all $t\in I$ with $t\ge\tau$ and $n\in\NN$, as long as $\vgamma$ 
does not belong to some $(n,\sigma,\tau)$-dependent $\PP$-zero set,
say $N_{\sigma,\tau}^{(n)}$. Taking the continuity of
$(\V{x},\psi)\mapsto\Lambda_{r,s}(\V{x},\psi,\vgamma)$ into account we conclude that
\eqref{flow-eq} is valid, for all $\V{x}\in\RR^\nu$, $\psi\in\FHR$, $\tau\le t\in I$, 
and $\vgamma\in\Omega\setminus\cup_n N_{\sigma,\tau}^{(n)}$.
\qed

In the next proposition $C_b(\RR^\nu\times\FHR,\sK)$ is the set of
bounded and continuous maps from $\RR^\nu\times\FHR$ into some Hilbert
space $\sK$.

\begin{proposition}[Feller and Markov properties]\label{prop-Markov}
Assume that $V$ is continuous and bounded.
Let $\sK$ be a Hilbert space. For $0\le\tau\le t\in I$ and every
bounded Borel-measurable function
$f:\RR^\nu\times\FHR\to\sK$, we define
\begin{equation}\label{def-Ptaut}
(P_{\tau,t}f)(\V{x},\psi):=\int_{\Omega}f(\Lambda_{\tau,t}(\V{x},\psi,\vgamma))\,\Id\PP(\vgamma),\quad
\V{x}\in\RR^\nu,\,\psi\in\FHR.
\end{equation}
Then the family $(P_{\tau,t})_{\tau\le t\in I}$ enjoys the Feller property, i.e.
$P_{\tau,t}$ maps the set $C_b(\RR^\nu\times\FHR,\sK)$ into itself.
In fact, for every $f\in C_b(\RR^\nu\times\FHR,\sK)$, the following map is continuous,
\begin{equation}\label{Markov0}
I^\tau\times\RR^\nu\times\FHR\ni(t,\V{x},\psi)\longmapsto(P_{\tau,\tau+t}f)(\V{x},\psi)\in\sK.
\end{equation}
Furthermore, if $0\le\sigma\le\tau\le t\in I$, if $f$ is a
real-valued bounded Borel function or $f\in C_b(\RR^\nu\times\FHR,\sK)$, and if
$(\V{q},\eta):\Omega\to\RR^\nu\times\FHR$
is $\fF_\sigma$-measurable, then we have, for $\PP$-a.e. $\vgamma$, 
\begin{equation}\label{Markov1}
\big(\EE^{\fF_\tau}\big[f(\Lambda_{\sigma,t}[\V{q},\eta])\big]\big)(\vgamma)
=(P_{\tau,t}f)\big(\Lambda_{\sigma,\tau}(\V{q}(\vgamma),\eta(\vgamma),\vgamma)\big),
\end{equation}
where $\Lambda_{r,s}[\V{q},\eta]$ denotes the random variable
$\Omega\ni\vgamma\mapsto\Lambda_{r,s}(\V{q}(\vgamma),\eta(\vgamma),\vgamma)$.
\end{proposition}

{\proof} 
The Feller property and the continuity of \eqref{Markov0} follow from Thm.~\ref{thm-flow}(1) and the
dominated convergence theorem.

To prove the Markov property \eqref{Markov1} we argue similarly as in, e.g., 
\cite[Thm.~9.14]{daPrZa2014}. There it is also explained why, without loss of generality, 
we may assume $f$ to be continuous in the case $\sK=\RR$ as well.
On account of \eqref{flow-eq} it suffices to show that
\begin{equation}\label{Markov2}
\EE^{\fF_\tau}\big[f(\Lambda_{\tau,t}[\wt{\V{q}},\wt{\eta}])\big]
=(P_{\tau,t}f)(\wt{\V{q}},\wt{\eta}),\quad\text{$\PP$-a.s.},
\end{equation}
holds, for all $\sigma(\Lambda_{\sigma,\tau}[\V{q},\eta])$-measurable maps
$(\wt{\V{q}},\wt{\eta}):\Omega\to\RR^\nu\times\FHR$ and in particular for 
$\Lambda_{\sigma,\tau}[\V{q},\eta]$ itself.
If $(\wt{\V{q}},\wt{\eta})$ is $\PP$-a.s. constant equal to some $(\V{x},\psi)\in\RR^\nu\times\FHR$, 
then, according to the second assertion of Thm.~\ref{thm-flow}(4),
we may replace the conditional expectation $\EE^{\fF_\tau}$  by $\EE$ on the
left hand side of \eqref{Markov2} which then reduces to the definition of $P_{\tau,t}$.

Next, let $A_1,\ldots,A_\ell$ be disjoint Borel subsets of $\RR^\nu\times\FHR$
whose union equals $\RR^\nu\times\FHR$ and 
set $\chi_j:=1_{A_j}(\Lambda_{\sigma,\tau}[\V{q},\eta])$. Then, of course,
\begin{align}\label{Markov3}
\Lambda_{\tau,t}[\hat{\V{q}},\hat{\eta}]
&=\sum_{j=1}^\ell\Lambda_{\tau,t}(\V{x}_j,\psi_j,\cdot)\,\chi_j\,,\quad\text{where}
\;\;(\hat{\V{q}},\hat{\eta})=\sum_{j=1}^\ell(\V{x}_j,\psi_j)\,\chi_j,
\end{align}
with constant $(\V{x}_j,\psi_j)\in\RR^\nu\times\FHR$, for $j=1,\ldots,\ell$.
Since $\Lambda_{\tau,t}(\V{x}_j,\psi_j,\cdot)=:\Lambda_{\tau,t}[\V{x}_j,\psi]$ is $\fF_\tau$-independent
and $\chi_j$ is $\fF_\tau$-measurable it follows that
\begin{align}\label{Markov4}
\EE^{\fF_\tau}\big[f(\Lambda_{\tau,t}[\hat{\V{q}},\hat{\eta}])\big]
&=\sum_{j=1}^\ell\EE\big[f(\Lambda_{\tau,t}[\V{x}_j,\psi_j])\big]\,\chi_j
=(P_{\tau,t}f)(\hat{\V{q}},\hat{\eta}),\;\;\PP\text{-a.s.},
\end{align}
with $\hat{\V{q}},\hat{\eta}$ as in \eqref{Markov3}.
For general $(\wt{\V{q}},\wt{\eta})$, we construct simple functions $(\hat{\V{q}}_n,\hat{\eta}_n)$,
$n\in\NN$, as the one in \eqref{Markov3} such that
$(\hat{\V{q}}_n,\hat{\eta}_n)\to(\wt{\V{q}},\wt{\eta})$, $\PP$-a.s.,
plug them into \eqref{Markov4}, and pass to the limit $n\to\infty$ using the continuity of $f$,
$P_{\tau,t}f$, and $(\V{x},\psi)\mapsto\Lambda_{\tau,t}(\V{x},\psi,\vgamma)$.
\qed 

\begin{corollary}\label{Chapman-Kolmogorov}
Assume that $V$ is continuous and bounded,
let $\sK$ be a Hilbert space, and let 
$f:\RR^\nu\times\FHR\to\sK$ be bounded and Borel measurable. Then 
\begin{equation}\label{ChKoEq}
P_{\sigma,t}f=P_{\sigma,\tau}P_{\tau,t}f\quad\text{on
$\RR^\nu\times\FHR$, if $\,0\le\sigma\le\tau\le t\in I$.}
\end{equation}
\end{corollary}

{\proof} 
The asserted identity follows from \eqref{Markov1} and $\EE\EE^{\fF_\tau}=\EE$.
\qed

By standard procedures we may finally infer the strong Markov property
of the flow $(\Lambda_{\tau,t})_{\tau\le t}$ from Prop.~\ref{prop-Markov}. 
In order to state it precisely in the next theorem we denote the law of the process
$(\Lambda_{s,s+t}(\V{x},\psi,\cdot))_{t\ge0}$, where $s\ge0$ and 
$(\V{x},\psi)\in\RR^\nu\times\FHR$ is deterministic, by
$$
\PP^{s,(\V{x},\psi)}:=\PP\circ\big(\Lambda_{s,s+\bullet}[\V{x},\psi]\big)^{-1}.
$$
Here we consider only the case $I=[0,\infty)$ and in particular the above formula
defines a measure on the Borel subsets of
$C([0,\infty),\RR^\nu\times\FHR)$; the corresponding expectation is
denoted by $\EE^{s,(\V{x},\psi)}[\,\cdot\,]$.

If $\tau:\Omega\to[0,\infty]$ is a stopping time, then $\fF_\tau$ denotes as usual the $\sigma$-algebra 
consisting of all events $A\in\fF$ such that $\{\tau\le t\}\cap A\in\fF_t$, for every $t\in I$. 
Moreover,
$\Lambda_{s,\tau+\bullet}[\V{q},\eta]:\Omega\to C([0,\infty),\RR^\nu\times\FHR)$
is the path map assigning the path
$[0,\infty)\ni t\mapsto\Lambda_{s,\tau(\vgamma)+t}(\V{q}(\vgamma),\eta(\vgamma),\vgamma)$
to $\vgamma\in\Omega$.

\begin{theorem}[Strong Markov property]\label{thm-str-Markov}
Assume that $V$ is bounded and continuous.
Consider the case $I=[0,\infty)$, let $s\in [0,\infty)$, 
and let $\tau\ge s$ be a stopping time. Furthermore, suppose that
$(\V{q},\eta):\Omega\to\RR^\nu\times\FHR$ is $\fF_s$-measurable
and that $f:C([0,\infty),\RR^\nu\times\FHR)\to[0,\infty)$ is Borel-measurable.
Then we have, for $\PP$-a.e. $\vgamma\in\{\tau<\infty\}$,
\begin{equation}\label{str-Markov}
\big(\EE^{\fF_\tau}\big[f(\Lambda_{s,\tau+\bullet}[\V{q},\eta])\big]\big)(\vgamma)
=\EE^{\tau(\vgamma),\Lambda_{s,\tau(\vgamma)}(\V{q}(\vgamma),\eta(\vgamma),\vgamma)}[f].
\end{equation}
\end{theorem}

{\proof} 
With Prop.~\ref{prop-Markov} at hand we may -- for the most part literally -- follow the
exposition in \cite[pp. 250--252]{daPrZa2014}; here the continuity of \eqref{Markov0} is used.
\qed

Next, we formulate a Blagove\v{s}\v{c}ensky-Freidlin type theorem. To this end we let
$\Omega_{\mathrm{W}}:=C(I,\RR^\nu)$ denote the Wiener space, $\fF^{\mathrm{W}}$ 
the completion of the corresponding Borel $\sigma$-algebra with respect to the Wiener measure 
$\PP_{\mathrm{W}}$, and $\fF_t^{\mathrm{W}}$ the completion
of the $\sigma$-algebra $\sigma(\pr_s:\,0\le s\le t)$ generated
by the evaluation maps $\pr_t(\vgamma):=\vgamma(t)$, $t\in I$, $\vgamma\in\Omega_{\mathrm{W}}$.
(Then $(\fF_t^{\mathrm{W}})_{t\in I}$ is known to be right continuous.)

\begin{theorem}[Strong solutions]\label{thm-str-sol}
Assume that $V$ is bounded and continuous.
Let $(\Lambda_{\tau,t}^{\mathrm{W}})_{\tau\le t\in I}$ denote the stochastic
flow constructed in Thm.~\ref{thm-flow} for the special choices
$\BB=(\Omega_{\mathrm{W}},\fF^{\mathrm{W}},(\fF_t^{\mathrm{W}})_{t\in I},\PP_{\mathrm{W}})$ and
$\V{B}=\pr$. Then $(\Lambda_{0,t}^{\mathrm{W}})_{t\in I}$ is a
strong solution of \eqref{bX1}\&\eqref{bX2} in the sense that,
for any stochastic basis $(\Omega,\fF,(\fF_t)_{t\in I},\PP)$ and Brownian
motion $\V{B}$ as in Hyp.~\ref{hyp-B}, and for any $\fF_0$-measurable
$(\V{q},\eta):\Omega\to\RR^\nu\times\FHR$, the up to indistinguishability unique solution of
\eqref{bX1}\&\eqref{bX2} is given, for $\PP$-a.e. $\vgamma$, by the
following formula 
\begin{equation}\label{eq-str-sol}
\big(\V{X}^{\V{q}}_t,\WW{\vxi,t}{V}[\V{X}^{\V{q}}]\eta\big)(\vgamma)
=\Lambda_{0,t}^{\mathrm{W}}(\V{q}(\vgamma),\eta(\vgamma),\V{B}_\bullet(\vgamma)),\quad t\in I.
\end{equation}
\end{theorem}

{\proof} 
First, let $(\V{q},\eta)=(\V{x},\psi)\in\RR^\nu\times\FHR$ be constant with $\psi\in\wh{\dom}$. Then
\eqref{eq-str-sol} follows from the uniqueness statement of Thm.~\ref{thm-Ito-spin} and a
transformation argument applied to the system of
SDEs solved by the process $(\Lambda_{0,t}^{\mathrm{W}}[\V{x},\psi])_{t\in I}$; 
cf. \cite[Satz~6.26 and Lem.~6.27]{HackenbrochThalmaier1994} for details.
(We apply Ex.~\ref{ex-Ito-limprob} to transform the stochastic integral in that system.) 
Employing the continuity of 
$(\V{x},\psi)\mapsto\Lambda_{0,\bullet}^{\mathrm{W}}(\V{x},\psi,\vgamma)$,
we then extend the result to general $(\V{q},\eta)$ by the 
approximation procedure already used in the first part of the proof of Thm.~\ref{thm-flow}.
\qed

\begin{corollary}\label{cor-P-time-hom}
Assume that $V$ is continuous and bounded,
that $I=[0,\infty)$, and that the vector field $\V{\beta}$ appearing in Hyp.~\ref{hyp-B}
is time-independent, i.e. $\V{\beta}\in C(\RR^\nu,\RR^\nu)$.
Then the flow $(\Lambda_{\tau,t})_{\tau\le t}$ is stationary, i.e.
$P_{\tau,t}f=P_{0,t-\tau}f$, for all $0\le\tau\le t$ and $f$ as in Cor.~\ref{Chapman-Kolmogorov}.
\end{corollary}

{\proof} 
If $\V{\beta}$ does not depend explicitly on $t$, then the whole
system \eqref{bX1}\&\eqref{bX2} is autonomous. If the initial condition is constant,
$(\V{q},\eta)=(\V{x},\psi)\in\RR^\nu\times\FHR$,
it follows that its solution corresponding to $(\BB,\V{B})$ and its solution
corresponding to the time-shifted data
$(\BB_\tau,\V{B}_{\tau+\bullet})$ are obtained by inserting $\V{B}$
and $\V{B}_{\tau+\bullet}$, respectively, into the strong solution
$(\Lambda_{0,t}^{\mathrm{W}}(\V{x},\psi,\cdot))_{t\ge0}$.
Now the result follows from \eqref{def-Ptaut} and the fact that $\V{B}$
and $\V{B}_{\tau+\bullet}$ have the same law.
\qed


\section{Symmetric semi-groups}\label{sec-C0}

\noindent
In our verifications of the Feynman-Kac formulas in Sect.~\ref{sec-ext} we shall employ the 
Hille-Yosida theorem on generators of strongly continuous semi-groups of bounded self-adjoint 
operators. For this purpose we shall show in the present section that the expressions on the 
``probabilistic'' side of the Feynman-Kac formulas define such symmetric semi-groups.

{\em In the whole section we fix some $t\in I$, $t>0$.} To study the symmetry
we shall consider certain reversed processes running backwards from
$t$. To start with we denote the reverse of the driving process $(\V{X}_\tau)_{\tau\in[0,t]}$ (fulfilling
Hyp.~\ref{hyp-B}) by $\bar{\V{X}}$ and the associated stochastic basis by $\bar{\BB}$. That is,
\begin{align}\label{def-barX}
\bar{\V{X}}_\tau&:=\V{X}_{t-\tau},\;\tau\in[0,t],\qquad
\bar{\BB}:=(\Omega,\fF,(\bar{\fF}_\tau)_{\tau\in[0,t]},\PP).
\end{align}
Here the filtration $(\bar{\fF}_\tau)_{\tau\in[0,t]}$ is defined as follows:
For every $\tau\in[0,t]$, set $\fG_\tau:=\sigma(\V{X}_{t-\tau};\V{B}_{t-s}-\V{B}_t:\,s\in[0,\tau])$
and let $\fH_\tau$ denote the smallest $\sigma$-algebra containing $\fG_\tau$ and all
$\PP$-zero sets. Set $\fH_{t+\ve}:=\fH_t$, for all $\ve>0$.
Then it follows easily from Hyp.~\ref{hyp-B}(2) that $(\fH_\tau)_{\tau\ge0}$ is a filtration, and
we define $\bar{\fF}_\tau:=\bigcap_{\ve>0}\fH_{\tau+\ve}$.
By construction, $\bar{\BB}$ satisfies the usual assumptions and 
$\bar{\V{X}}$ is adapted to $\bar{\BB}$. 
Under certain assumptions on the drift vector field $\V{\beta}$
appearing in Hyp.~\ref{hyp-B} and the law of $\V{X}_\tau$,
$\tau\in(0,t]$, it is possible to guarantee that $\bar{\V{X}}$ is again
a diffusion process and in particular a continuous semi-martingale with respect to $\bar{\BB}$; see
\cite{HaussmannPardoux1986,Pardoux-LNM1204} and Rem.~\ref{rem-rev-proc} below.
In the first two lemmas of this section we content ourselves, 
however, to work with the somewhat implicit assumption that $\bar{\V{X}}$ again fulfills 
Hyp.~\ref{hyp-B} (together with the new basis $\bar{\BB}$, of course). We verify this postulate
only in the two main examples of interest in the present paper,
namely Brownian motion and Brownian bridges.

Since all quantities 
$\iota$, $(\w{\tau}{t})_{t\in I}$, $u^V_{\vxi}$, $U^\pm$, $(U_{\tau,t}^-)_{t\in I}$,
$K$, $(K_{\tau,t})_{t\in I}$, and $\WW{\vxi}{V}$ depend on the choice of
$\V{X}$, and since we are again dealing with different choices
of the driving process at the same time, we again refer to this dependence
in the notation by writing $Z[\V{X}]$ or $Z[\bar{\V{X}}]$, if $Z$ is
any of the above processes constructed by means of $\V{X}$ or $\bar{\V{X}}$, respectively.

In Eq. \eqref{cassandra00} below and its proof we extend the conjugation $C$ of Hyp.~\ref{hyp-G}
trivially to $\HP_{+1}\cong L^2(\RR,\Id k_0)\otimes\HP$; for short we shall again
write $C$ instead of $\id\otimes C$. Under this convention we have, for instance,
\begin{equation}\label{comm-Cj}
j_sC=Cj_{-s},\qquad s\in \RR.
\end{equation}

\begin{lemma}\label{horror} 
Assume that $\bar{\V{X}}$ is a continuous semi-martingale on $[0,t]$
with respect to $\bar{\BB}$ satisfying Hyp.~\ref{hyp-B}.
Then there exists a $\PP$-zero set $N$ such that the following identities
hold on $\Omega\setminus N$, for all $\tau\in[0,t]$,
\begin{align}\label{cassandra00}
{{K}}_{\tau,t}[\bar{\V{X}}]&=-C\,e^{i\V{m}\cdot(\V{X}_t-\V{X}_0)+ik_0t}\,{K}_{t-\tau}[\V{X}],
\\\label{cassandra1}
{U}_{\tau,t}^{-}[\bar{\V{X}}]&=-U_{t-\tau}^{+}[\V{X}],\qquad
U_\tau^+[\bar{\V{X}}]=-U_{t-\tau,t}^-[\V{X}],
\\\label{cassandra-u}
{u}^{V}_{\vxi,t}[\bar{\V{X}}]&=\ol{u}^V_{\vxi,t}[\V{X}]={u}^V_{-\vxi,t}[\V{X}].
\end{align}
\end{lemma}

{\proof}  
Plugging $\V{X}$ and $\bar{\V{X}}$ into the formula \eqref{ulk} for the sum 
$\Sigma_{\tau,t}^n$ and employing the identity (see \eqref{hyp-sym2}, \eqref{def-jt}, 
and \eqref{comm-Cj})
\begin{align*}
j_{s}\,e^{-i\V{m}\cdot(\V{X}_{t-s}-\V{X}_{t-0})}C=
Ce^{i\V{m}\cdot(\V{X}_t-\V{X}_0)+ik_0t}\,j_{t-s}\,e^{-i\V{m}\cdot(\V{X}_{t-s}-\V{X}_0)}
\end{align*}
it is straightforward to verify that
\begin{align*}
{\Sigma}_{\tau,t}^n[\bar{\V{X}}]
=-Ce^{i\V{m}\cdot(\V{X}_t-{\V{X}}_0)+ik_0t}\,{\Sigma}_{0,t-\tau}^{n}[\V{X}].
\end{align*}
By the assumption on $\bar{\V{X}}$,
the approximation formula \eqref{approx-KR} applies to {\em both} $\V{X}$
and $\bar{\V{X}}$ and shows that \eqref{cassandra00} holds
$\PP$-a.s., for all $\tau\in[0,t]\cap\QQ$.
By continuity (see Lem.~\ref{lem-U-K}(5)) we then see that \eqref{cassandra00} even holds
for arbitrary $\tau\in[0,t]$, outside a $\tau$-independent $\PP$-zero set.
Multiplying \eqref{cassandra00} with 
$\iota_\tau^*[\bar{\V{X}}]=j_\tau^*e^{i\V{m}\cdot(\V{X}_{t-\tau}-\V{X}_t)}$ and using 
$j_\tau^*Ce^{ik_0t}=Cj_{t-\tau}^*$, we further obtain, $\PP$-a.s. for all $0\le\tau\le t$,
$$
{U}^-_{\tau,t}[\bar{\V{X}}]=\iota_\tau^*[\bar{\V{X}}]\,{K}_{\tau,t}[\bar{\V{X}}]
=-C\iota_{t-\tau}^*[\V{X}]{K}_{t-\tau}[\V{X}]=-C\,U^+_{t-\tau}[\V{X}],
$$ 
which yields the first identity in \eqref{cassandra1}, if take Lem.~\ref{lem-U-K}(4) into account.
The second identity in \eqref{cassandra1} follows from
\begin{align*}
U^+_\tau[\bar{\V{X}}]&=\iota^*_\tau[\bar{\V{X}}]\big(K_{0,t}[\bar{\V{X}}]-K_{\tau,t}[\bar{\V{X}}]\big)
\\
&=-C\iota_{t-\tau}^*[\V{X}]\big(K_t[\V{X}]-K_{t-\tau}[\V{X}]\big)=-CU_{t-\tau,t}^-[\V{X}].
\end{align*}
Finally, \eqref{cassandra00} implies $\|{K}_t[\bar{\V{X}}]\|=\|K_t[\V{X}]\|$ 
which permits to get \eqref{cassandra-u}.
\qed

Next, we study the influence of the time-reversal of the driving process on $\WW{\vxi}{V}$.
This is easily done starting from the convenient formulas in Rem.~\ref{rem-for-W-exp-vec}.
Again, we indicate the dependence on the driving process
of the processes appearing in Rem.~\ref{rem-for-W-exp-vec} 
by adding the extra variables $[\V{X}]$ or $[\bar{\V{X}}]$ to the corresponding symbols.

\begin{lemma}\label{lem-sym-Qgh}
Assume that the time-reversed data $(\bar{\V{X}},\bar{\BB})$ fulfills Hyp.~\ref{hyp-B} as well.
Then the following two relations,
\begin{align*}
Q_{t}(g,h)[\bar{\V{X}}]=Q_t(h,g)[{\V{X}}]^*,\;\;\;
\SPn{\zeta(g)}{\WW{\vxi,t}{V}[\bar{\V{X}}]\,\zeta(h)}
=\SPn{\WW{\vxi,t}{V}[\V{X}]\,\zeta(g)}{\zeta(h)},
\end{align*}
hold true outside a $\PP$-zero set which does not depend on $g,h\in\mathfrak{d}_C$, and
\begin{equation}\label{sym-WW}
\WW{\vxi,t}{V}[\bar{\V{X}}]=\WW{\vxi,t}{V}[\V{X}]^*,\quad\PP\text{-a.s.}
\end{equation}
\end{lemma}

{\proof} 
We consider the various terms in the formula \eqref{spin2} for $\sQ^{(n)}_{t}(g,h,t_{[n]})[\bar{\V{X}}]$:
The relations $\w{r}{s}[\bar{\V{X}}]\,\V{F}_{\bar{\V{X}}_r}=\olw{t-s}{t-r}[\V{X}]\,\V{F}_{\V{X}_{t-r}}$
obviously hold true on $\Omega$, for $0\le r\le s\le t$. 
In view of \eqref{spin2} this together with \eqref{cassandra1} shows that
\begin{equation*}
\sQ^{(n)}_{t}(g,h,t_1,\ldots,t_n)[\bar{\V{X}}]=\sQ^{(n)}_t(h,g,t-t_n,\ldots,t-t_1)[\V{X}]^*
\end{equation*}
outside a $\PP$-zero set which neither depends on $(t_1,\ldots,t_n)\in t\simplex_n$ nor $g,h$.
Combining this with \eqref{spin2a} and substituting
$t_1':=t-t_n$, $\ldots\;$, $t_n'=t-t_1$ in the integrals over $t\simplex_n$,
$n\in\NN$, we obtain the first asserted identity.
Taking also \eqref{cassandra1}, \eqref{cassandra-u}, and \eqref{spin2bW} into account we
arrive at the second one. Since $\zeta(g)$ and $\zeta(h)$ can be chosen from total subset of $\sF$
and since $\WW{\vxi,t}{V}[\V{X}]$ and $\WW{\vxi,t}{V}[\bar{\V{X}}]$ are $\PP$-a.s. bounded
we also obtain the relation \eqref{sym-WW}.
\qed 

Before we discuss our main examples we quote a special case of a result from
\cite{HaussmannPardoux1986,Pardoux-LNM1204}:

 \begin{remark}{\rm \label{rem-rev-proc}
Suppose that, for all $\tau\in(0,t]$, the law of $\V{X}_\tau$ is
absolutely continuous with respect to the Lebesgue measure and assume (for simplicity)
that the corresponding density, $d_\tau:\RR^\nu\to[0,\infty)$, 
is strictly positive and continuously differentiable. Set $d_0:=1$.
Assume further that the vector field $\V{\beta}(\tau,\cdot)$ is globally Lipschitz
continuous, uniformly in $\tau\in[0,t]$. Then 
\begin{align}\label{for-barB}
\bar{\V{B}}_\tau&:=\V{B}_{t-\tau}-\V{B}_t-\int_{t-\tau}^t(\nabla\ln d_{s})(\V{X}_s)\,\Id s,
\quad \tau\in[0,t],
\end{align}
defines a $\bar{\BB}$-Brownian motion $\bar{\V{B}}$ on $[0,t]$ and it is elementary to check that
\begin{align}\label{rev-eq}
\bar{\V{X}}_\tau&=\bar{\V{X}}_0+\int_0^\tau\bar{\V{\beta}}(s,\bar{\V{X}}_s)\,\Id s
+\bar{\V{B}}_\tau,\quad \tau\in[0,t),
\\\label{def-barbeta}
\bar{\V{\beta}}(s,\cdot)&:=-\V{\beta}(t-s,\cdot)+\nabla\ln d_{t-s},\quad s\in[0,t).
\end{align}
 }\end{remark}

\begin{example}{\rm \label{ex-Bx-rev}
Assume that $\V{X}=\V{B}^{\V{x}}$ is a translated Brownian motion, where
\begin{equation}\label{def-Bx}
\V{B}^{\V{x}}:=\V{x}+\V{B},\quad \V{x}\in\RR^\nu.
\end{equation} 
The density of $\V{B}_\tau^{\V{x}}$ is given by the Gaussian $p_\tau(\V{x},\cdot)$ 
defined in \eqref{gaussian}. From Rem.~\ref{rem-rev-proc} we infer the existence of
a $\bar{\BB}$-Brownian motion, $\bar{\V{B}}$, on $[0,t]$ such that
$\bar{\V{X}}=(\V{B}^{\V{x}}_{t-\tau})_{\tau\in[0,t]}$ 
is a solution with initial condition ${\V{q}}=\bar{\V{X}}_0=\V{B}_t^{\V{x}}$ of
\begin{align}\label{sde-bridge-retro}
\V{b}_\tau={\V{q}}+\int_0^\tau\frac{\V{x}-\V{b}_s}{t-s}\,\Id s+\bar{\V{B}}_\tau,
\quad\tau\in[0,t),\quad\V{b}_t=\V{x}.
\end{align}
This is the SDE for a Brownian bridge from $\V{q}$ to $\V{x}$ in time $t$.
 }\end{example}

Since the drift vector field in the SDE for a Brownian bridge is singular at the end point,
the results of \cite{Pardoux-LNM1204} do not apply directly to reversed Brownian bridges. 
One can, however, adapt
the arguments of \cite{Pardoux-LNM1204} to verify the following lemma.
For the reader's convenience we present a detailed proof of it in App.~\ref{app-bridge}.

\begin{lemma}\label{lem-rev-bridge}
Let $\V{b}^{t;\V{x},\V{y}}$ denote the Brownian bridge from $\V{x}$ to $\V{y}$ in time $t$ defined as 
the, up to indistinguishability, unique solution of the SDE
\begin{align}\label{sde-bridge}
\V{b}_\tau=\V{x}+\int_0^\tau\frac{\V{y}-\V{b}_s}{t-s}\,\Id s+\V{B}_\tau,\quad\tau\in[0,t),
\end{align}
which has a limit at $t$, $\PP$-a.s., namely $\V{b}^{t;\V{x},\V{y}}_{t}:=\V{y}$. 
Define $(\fH_s)_{s\ge0}$ and $(\bar{\fF}_s)_{s\in[0,t]}$ as in the beginning of this section
with $\V{X}=\V{b}^{t;\V{x},\V{y}}$. Then
\begin{align}\label{def-barB-bridget}
\hat{\V{B}}_s&:=
\V{b}_{t-s}^{t;\V{x},\V{y}}-\V{y}+\int_{t-s}^{t}\frac{\V{b}^{t;\V{x},\V{y}}_r-\V{x}}{r}\Id r,
\quad s\in[0,t),
\end{align}
defines a Brownian motion with respect to $(\fH_s)_{s\in[0,t)}$. Its unique extension to a martingale
on $[0,t]$ with respect to $(\fH_s)_{s\in[0,t]}$, henceforth again denoted by $\hat{\V{B}}$, is even
a $(\bar{\fF}_s)_{s\in[0,t]}$-Brownian motion.
Furthermore, we $\PP$-a.s. have
\begin{align}\label{bridge-rev}
\V{b}^{t;\V{x},\V{y}}_{t-s}&=\V{y}+\hat{\V{B}}_s
+\int_0^s\frac{\V{x}-\V{b}^{t;\V{x},\V{y}}_{t-r}}{t-r}\Id r,\quad s\in[0,t).
\end{align}
\end{lemma}

Hence, if $\V{X}=\V{b}^{t;\V{x},\V{y}}$ is a Brownian bridge, then
$\bar{\V{X}}$ is a semi-martingale realization of the Brownian bridge from $\V{y}$ to $\V{x}$
in time $t$ with respect to the $\bar{\BB}$-Brownian motion $\hat{\V{B}}$. In the situation of the
previous lemma we thus write
\begin{equation}\label{def-hatb}
\hat{\V{b}}\!\!\phantom{s}^{t;\V{y},\V{x}}_{\bullet}:=\V{b}^{t;\V{x},\V{y}}_{t-\bullet}.
\end{equation}

\begin{example}{\rm (1)
As a consequence of Lem.~\ref{lem-rev-bridge}, \eqref{sym-WW}, and \eqref{def-hatb},
\begin{align}\label{kernel-sym}
\WW{\vxi,t}{V}[\V{b}^{t;\V{x},\V{y}}]^*=\WW{\vxi,t}{V}[\hat{\V{b}}\!\!\phantom{s}^{t;\V{y},\V{x}}],
\quad\PP\text{-a.s.}
\end{align}
(2) Here we continue Ex.~\ref{ex-Bx-rev}, i.e., we again set 
$\bar{\V{X}}_\bullet=\V{B}^{\V{x}}_{t-\bullet}$ and define the stochastic basis $\bar{\BB}$
as in \eqref{def-barX} with $\V{X}=\V{B}^{\V{x}}$. 
Furthermore, we assume that $V$ is bounded and continuous.
If $(\bar{\Lambda}_{s,\tau}^{t;\V{x}})_{s\le\tau\le t}$ denotes the
stochastic flow constructed in Thm.~\ref{thm-flow} in the case where
\eqref{sde-bridge-retro} is substituted for \eqref{bX1} and the stochastic basis
$\bar{\BB}$ is used, then Part~(1) of the present example $\PP$-a.s. implies 
\begin{align}\label{richard-1}
\bar{\Lambda}_{0,\tau}^{t;\V{x}}[\V{y},\phi]&=
\big(\hat{\V{b}}\!\!\phantom{s}_\tau^{t;\V{y},\V{x}},
\WW{\vxi,\tau}{V}[\hat{\V{b}}\!\!\phantom{s}^{t;\V{y},\V{x}}]\phi\big),
\quad\tau\in[0,t],\;\V{y}\in\RR^\nu,\;\phi\in\FHR.
\end{align}
Let $\Psi:\RR^\nu\to\FHR$ be Borel measurable in what follows.
Since $\bar{\V{X}}_0=\V{B}_t^{\V{x}}$ and $\Psi(\V{B}_t^{\V{x}})$
are $\bar{\fF}_0$-measurable, we $\PP$-a.s. arrive at the formulas
\begin{align}\label{richard0}
\big(\bar{\V{X}}_t,\WW{\vxi,t}{V}[\V{B}^{\V{x}}]^*\Psi(\V{B}_t^{\V{x}})\big)
=\big(\bar{\V{X}}_t,\WW{\vxi,t}{V}[\bar{\V{X}}]\,\Psi({\V{B}}_t^{\V{x}})\big)
=\bar{\Lambda}_{0,t}^{t;\V{x}}[\V{B}_t^{\V{x}},\Psi(\V{B}_t^{\V{x}})].
\end{align}
Next, let $\sK$ be a separable Hilbert space and let $f:\RR^\nu\times\FHR\to\sK$
be bounded and Borel-measurable. 
Using the $\bar{\fF}_0$-measurability of $\V{B}_t^{\V{x}}$ and $\Psi(\V{B}_t^{\V{x}})$ as well as the Markov property \eqref{Markov1} in the second step, we further observe that
\begin{align}\nonumber
\EE\big[f(\bar{\Lambda}_{0,t}^{t;\V{x}}[\V{B}_t^{\V{x}},\Psi(\V{B}_t^{\V{x}})])\big]
&=\EE\big[\EE^{\bar{\fF}_0}[f(\bar{\Lambda}_{0,t}^{t;\V{x}}[\V{B}_t^{\V{x}},\Psi(\V{B}_t^{\V{x}})])]\big]
\\\label{richard1}
&=\EE\big[(\bar{P}_{0,t}^{t;\V{x}}f)(\V{B}_t^{\V{x}},\Psi(\V{B}_t^{\V{x}}))\big],
\end{align}
where $\bar{P}_{s,\tau}^{t;\V{x}}$ is the transition operator associated with
$(\bar{\Lambda}_{s,\tau}^{t;\V{x}})_{s\le\tau\le t}$ according to \eqref{def-Ptaut}.
Since $\V{B}_t^{\V{x}}$ is $p_t(\V{x},\cdot)$-distributed we may re-write \eqref{richard1} as
\begin{align}\label{richard2}
\EE\big[f(\bar{\Lambda}_{0,t}^{t;\V{x}}[\V{B}_t^{\V{x}},\Psi(\V{B}_t^{\V{x}})])\big]
&=\int_{\RR^\nu}(\bar{P}_{0,t}^{t;\V{x}}f)(\V{y},\Psi({\V{y}}))\,p_t(\V{x},\V{y})\,\Id\V{y}.
\end{align}
Now, assume in addition that $\Psi$ is bounded and choose $f(\V{x},\psi):=1_{\|\psi\|<R}\psi$, 
with some $R>e^{\|\Lambda\|_\infty^2t+\|V\|_\infty t}\|\Psi\|_\infty$;
recall \eqref{norm-W}.
Then \eqref{def-Ptaut}, \eqref{richard-1}, 
\eqref{richard0}, and \eqref{richard2} in combination yield
\begin{align}\label{richard3}
\EE\big[\WW{\vxi,t}{V}[\V{B}^{\V{x}}]^*\Psi(\V{B}_t^{\V{x}})\big]
=\int_{\RR^\nu}\EE\big[\WW{\vxi,t}{V}[\hat{\V{b}}\!\!\phantom{s}^{t;\V{y},\V{x}}]\,\Psi(\V{y})\big]
\,p_t(\V{x},\V{y})\,\Id\V{y}.
\end{align}
On account of \eqref{norm-W} and the boundedness of $V$, it is easy to extend the relation
\eqref{richard3} to, e.g., all $\Psi\in L^p(\RR^\nu,\FHR)$ with $p\in[1,\infty]$. 
 }\end{example}

Next, we introduce some abbreviations for quantities appearing on the 
``probabilistic'' side of the Feynman-Kac formulas we shall derive in Sect.~\ref{sec-ext}.

\begin{definition}[{\bf Feynman-Kac operators}]\label{def-FKO} 
(1) Let $\V{m}=\V{0}$, $t>0$, and $\V{x},\V{y}\in\RR^\nu$ be such that 
$V(\V{b}_\bullet^{t;\V{y},\V{x}})\in L^1([0,t])$, $\PP$-a.s.,
and $e^{-\int_0^tV(\V{b}_s^{t;\V{y},\V{x}})\Id s}$ is $\PP$-integrable. Then we define
\begin{align}\label{def-T(x,y)}
T^V_t(\V{x},\V{y}):=p_t(\V{x},\V{y})\,\EE\big[\WW{\V{0},t}{V}[{\V{b}}^{t;\V{y},\V{x}}]\big].
\end{align}
(2) Let $t\ge0$, $\V{x}\in\RR^\nu$, and $\Psi\in\HR$ such that
$V(\V{B}_\bullet^{\V{x}})\in L_\loc^1([0,\infty))$, $\PP$-a.s., and
$e^{-\int_0^tV(\V{B}_s^{\V{x}})\Id s}\|\Psi(\V{B}_t^{\V{x}})\|_{\FHR}$ is $\PP$-integrable.
Then we set
\begin{align*}
T^V_t\Psi(\V{x})&:=\EE\big[\WW{\V{0},t}{V}[\V{B}^{\V{x}}]^*\,\Psi(\V{B}_t^{\V{x}})\big].
\end{align*}
(3) If $\V{G}$ and $\V{F}$ are $\V{x}$-independent, then we define, 
for all $t\ge0$ and $\vxi\in\RR^\nu$,
\begin{align}\label{def-whTxi}
\wh{T}_t(\vxi)\psi&:=\EE\big[\WW{\vxi,t}{0}[\V{B}]^*\psi\big]=\EE\big[\WW{\vxi,t}{0}[\V{B}]\psi\big],
\quad\psi\in\FHR.
\end{align}
\end{definition}

 \begin{remark}{\rm \label{rem-FKBL}
(1) The Bochner-Lebesgue integrability of
the operator-valued map $\WW{\V{0},t}{V}[\V{b}^{t;\V{y},\V{x}}]:\Omega\to\LO(\FHR)$ in 
\eqref{def-T(x,y)} is  (and in particular its measurability and the fact it is $\PP$-almost 
separably valued) follows from \eqref{norm-W}, Prop.~\ref{prop-meas}, Bochner's theorem,
and the additional assumption $\V{m}=\V{0}$.

\smallskip

\noindent(2) Assume that $V$ is continuous and $\V{m}=\V{0}$.
Then we actually know that we can modify the processes $\WW{\V{0}}{V}[\V{b}^{t;\V{y},\V{x}}]$,
$(\V{x},\V{y})\in\RR^{2\nu}$, in such a way that the map
$[0,t]\times\RR^{2\nu}\times\Omega\ni(s,\V{x},\V{y},\vgamma)\mapsto
\WW{\V{0},s}{V}[\V{b}^{t;\V{y},\V{x}}](\vgamma)\in\LO(\FHR)$ becomes
$\fB([0,t]\times\RR^{2\nu})\otimes\fF$-$\fB(\LO(\FHR))$-measurable
with a separable image,
$\WW{\V{0},s}{V}[\V{b}^{t;\V{y},\V{x}}]:\Omega\to\LO(\FHR)$
is $\fF_s$-$\fB(\LO(\FHR))$-measurable for all $(s,\V{x},\V{y})\in[0,t]\times\RR^{2\nu}$,
and $(0,t]\times\RR^{2\nu}\ni(s,\V{x},\V{y})\mapsto
\WW{\V{0},s}{V}[\V{b}^{t;\V{y},\V{x}}](\vgamma)\in\LO(\FHR)$
is continuous for all $\vgamma\in\Omega$.

This claim follows from the solution formula \eqref{sol-for-BB} for the Brownian bridge, 
Prop.~\ref{prop-app-cont}, and an obvious analogue
of Lem.~\ref{lem-UKx} (where $(\V{x},\V{y})$ will adopt the role of $\V{x}$).

If $V$ is bounded and continuous, then we may conclude that 
$\RR^{2\nu}\ni(\V{x},\V{y})\mapsto T_t^V(\V{x},\V{y})$ is operator norm continuous.

\smallskip

\noindent(3)
If, in addition to our standing hypotheses, we assume that $|\V{m}|\le c\omega$, for some $c>0$,
then \eqref{norm-W} and Prop.~\ref{prop-meas} imply that $\WW{\vxi,t}{0}[\V{B}]$ 
in \eqref{def-whTxi} is Bochner-Lebesgue integrable and
$\wh{T}_t(\vxi)=\EE[\WW{\vxi,t}{0}[\V{B}]]=\EE[\WW{\vxi,t}{0}[\V{B}]^*]$.
 }\end{remark}

\begin{lemma}\label{lem-antonio}
Assume that $V$ is continuous and bounded and
let $\eta:\Omega\to\FHR$ be $\fF_0$-measurable and
square-integrable. Then there exist $c,c_V>0$ such that, for all $t\ge0$,
\begin{equation}\label{antonio0}
\sup_{\V{x}\in\RR^\nu}\EE\big[\sup_{s\le t}\|(1+M_1(\vxi))^\mh
(\WW{\vxi,s}{V}[\V{B}^{\V{x}}]-\id)\eta\|^2\big]\le cte^{c_Vt}\EE[\|\eta\|^2].
\end{equation}
\end{lemma}

{\proof}
We abbreviate $\Theta:=1+M_1(\vxi)$ and
 $\psi_t:=(\WW{\vxi,t}{V}[\V{B}^{\V{x}}]-\id)\eta$, so that $\psi_0=0$
and $\|\psi_t\|\le2e^{c_Vt}\|\eta\|$, $t\ge0$, $\PP$-a.s.
We assume without loss of generality that $\eta$ maps $\Omega$ into $\wh{\dom}$.
(Otherwise approximate $\eta$ by the vectors $(1+M_1(\vxi)/n)^{-1}\eta$, $n\in\NN$.)
By Thm.~\ref{thm-Ito-spin} and the It\={o} formula in Ex.~\ref{ex-Ito-SP}, we $\PP$-a.s. have
\begin{align*}
\|\Theta^\mh\psi_t\|^2
&=-2\int_0^t\SPb{\psi_s}{\Theta^{-1}\wh{H}^V(\vxi,\V{B}_s^{\V{x}})
\WW{\vxi,s}{V}[\V{B}^{\V{x}}]\eta}\Id s
\\
&\quad
+\int_0^t\big\|\Theta^\mh\V{v}(\vxi,\V{B}_s^{\V{x}})\WW{\vxi,s}{V}[\V{B}^{\V{x}}]\eta\big\|^2\Id s
\\
&\quad-2\int_0^t\Re\SPb{\Theta^\mh\psi_s}{i\Theta^\mh\V{v}(\vxi,\V{B}_s^{\V{x}})
\WW{\vxi,s}{V}[\V{B}^{\V{x}}]\eta}\Id\V{B}_s,\quad t\ge0.
\end{align*}
In view of Prop.~\ref{prop-mart}, \eqref{norm-W}, and the bound
$\sup_{\V{x}}\|\V{v}(\vxi,\V{x})\Theta^\mh\|<\infty$ 
(recall \eqref{rb-a}, \eqref{rb-ad}, and Hyp.~\ref{hyp-G}),
the stochastic integral in the third line, call it $\sM$, is a martingale.
Employing \eqref{norm-W} and \eqref{ub-whH}
to estimate the integrals in the first and second line
and using Davis' inequality $\EE[\sup_{s\le t}|\sM_s|]\le\const\,\EE[\llbracket\sM,\sM\rrbracket_t^\eh]$,
$t\ge0$, we readily arrive at the asserted bound.
\qed

\begin{lemma}\label{lem-C0-fiber}
Assume that $\V{G}$ and $\V{F}$ are $\V{x}$-independent and let
$\vxi\in\RR^\nu$. Then the family $(\wh{T}_t(\vxi))_{t\ge0}$ defines
a strongly continuous, bounded, and self-adjoint semi-group on $\FHR$. 
\end{lemma}

{\proof}
The (locally uniform) boundedness and self-adjointness are obvious from \eqref{norm-W}
and \eqref{def-whTxi}. Having observed these facts, it only remains to show that
$\wh{T}_t(\vxi)\psi\to\psi$, $t\downarrow0$, for all $\psi\in\fdom(M_1(\vxi))$. 
For such $\psi$, we have, however,
\begin{align*}
&\big\|(\wh{T}_t(\vxi)-\id)\psi\big\|
=\sup_{{\|\phi\|=1}}\big|\SPb{\phi}{\EE\big[(\WW{\vxi,t}{0}[\V{B}]^*-\id)\psi\big]}\big|
\\
&\le\sup_{{\|\phi\|=1}}\EE\big[\big\|(1+M_1(\vxi))^\mh
(\WW{\vxi,t}{0}[\V{B}]-\id)\phi\big\|^2\big]^\eh\|(1+M_1(\vxi))^\eh\psi\|,
\end{align*}
and we conclude by applying \eqref{antonio0}.

\begin{lemma}\label{lem-C0}
Let $V\in C_b(\RR^\nu,\RR)$. Then 
\begin{equation}\label{T(x,y)sym}
T_t^V(\V{y},\V{x})=T_t^V(\V{x},\V{y})^*\in\LO(\FHR),
\end{equation}
for all $\V{x},\V{y}\in\RR^\nu$ and $t>0$.
Furthermore, $(T^V_t)_{t\ge0}$ is a strongly continuous one-parameter
semi-group of bounded self-adjoint operators on the Hilbert space
\begin{align}\label{total}
\HR:=L^2\big(\RR^\nu,\FHR\big)=\int_{\RR^\nu}^\oplus\FHR\,\Id \V{x}.
\end{align}
Morever,
\begin{align}\label{for-kernel}
T^V_t\Psi(\V{x})&=\int_{\RR^\nu}T_t^V(\V{x},\V{y})\,\Psi(\V{y})\,\Id\V{y},
\quad \Psi\in\HR,\;\V{x}\in\RR^\nu.
\end{align} 
\end{lemma}

{\proof} 
Under the stated assumptions the (locally uniform) boundedness of $(T^V_t)_{t\ge0}$
is obvious from \eqref{norm-W}. (Observe that the function defined by
$f(\V{x}):=\EE[\|\Psi(\V{B}_t^{\V{x}})\|]=(e^{t\Delta/2}\|\Psi(\cdot)\|_{\FHR})(\V{x})$, $\V{x}\in\RR^\nu$,
belongs to $L^2(\RR^\nu)$ with $\|f\|\le\|\Psi\|$, if $\Psi\in\HR$.)
In particular, $T_t^V$ is well-defined on all of $\HR$.

On account of Thm.~\ref{thm-str-sol} the definition of $T_t^V(\V{x},\V{y})$ does
not depend on the choice of the Brownian motion used to construct (via \eqref{sde-bridge}) the
Brownian bridge ${\V{b}}^{t;\V{y},\V{x}}$ in \eqref{def-T(x,y)}. In
particular, it may be replaced by the bridge $\hat{\V{b}}\!\!\phantom{s}^{t;\V{y},\V{x}}$
defined in \eqref{def-hatb}. Then \eqref{kernel-sym} implies
\eqref{T(x,y)sym} and the formula \eqref{for-kernel} is
nothing else than \eqref{richard3}. Of course, \eqref{T(x,y)sym} and
\eqref{for-kernel} imply that $T_t^V$ is bounded and selfadjoint.

The semi-group property follows from a special case of \eqref{ChKoEq}:
Consider the bounded Borel function 
$f(\V{x},\psi):=\SPn{\psi}{\Phi(\V{x})}\,1_{\|\psi\|\le R}$, $(\V{x},\psi)\in\RR^\nu\times\FHR$,
where $\Phi\in C(\RR^\nu,\FHR)\cap L^2(\RR^\nu,\FHR)$ is bounded and $R>0$
is chosen so large that
$\|\WW{\vxi,\tau}{V}[\V{B}^{\V{x}}]\|\le R$ $\PP$-a.s., for all $\V{x}$ and $\tau\in[0,s+t]$.
If $\tau\le s+t$ and $\|\psi\|\le1$, we then have
$$
P_\tau f(\V{x},\psi)
=\EE[\SPn{\WW{\V{0},\tau}{V}[\V{B}^{\V{x}}]\,\psi}{\Phi(\V{B}_\tau^{\V{x}})}]
=\SPn{\psi}{T_\tau^V\Phi(\V{x})},
$$
where $T_\tau^V\Phi\in C(\RR^\nu,\FHR)\cap L^2(\RR^\nu,\FHR)$ is again bounded. Consequently, 
$$
\SPn{\psi}{T_s^VT_t^V\Phi(\V{x})}
=P_sP_tf(\V{x},\psi)=P_{s+t}(\V{x},\psi)=\SPn{\psi}{T_{s+t}^V\Phi(\V{x})},
$$
if $\|\psi\|\le1$. Since each $T_\tau^V$ is bounded, this entails $T_s^VT_t^V=T_{s+t}^V$.
Choosing $\V{x}=\V{0}$ we may also conclude that 
$\wh{T}_s(\vxi)\wh{T}_t(\vxi)=\wh{T}_{s+t}(\vxi)$ in the translation invariant case.

To prove the strong continuity of $(T^V_t)_{t\ge0}$, we set 
$\Theta:=1+\Id\Gamma(\omega)+\Id\Gamma(\V{m})^2/2$.
Thanks to its by now proven locally uniform boundedness 
and semi-group property, it suffices to show that $T^V_t\Psi\to\Psi$, as $t\downarrow0$,
for all $\Psi\in\HR$ with $\|\Theta^\eh\Psi(\cdot)\|_{\FHR}\in L^2(\RR^\nu)$, 
which clearly form a dense subset of $\HR$.  
Then, for such $\Psi$, Lem.~\ref{lem-antonio} implies
\begin{align*}
&\|(T_t^V-\id)\Psi\|^2
=\int_{\RR^\nu}\big\|\EE\big[(\WW{\V{0},t}{V,*}[\V{B}^{\V{x}}]-\id)
\,\Psi(\V{B}_t^{\V{x}})\big]\big\|_{\FHR}^2\Id\V{x}
\\
&=\int_{\RR^\nu}\sup_{{\phi\in\FHR\atop\|\phi\|=1}}\big|\SPb{\phi}{
\EE\big[(\WW{\V{0},t}{V}[\V{B}^{\V{x}}]^*-\id)
\,\Psi(\V{B}_t^{\V{x}})\big]}\big|^2\Id\V{x}
\\
&=\int_{\RR^\nu}\sup_{{\phi\in\FHR\atop\|\phi\|=1}}\big|\EE\big[\SPb{\Theta^\mh
(\WW{\V{0},t}{V}[\V{B}^{\V{x}}]-\id)\phi}{\Theta^\eh
\Psi(\V{B}_t^{\V{x}})}\big]\big|^2\Id\V{x}
\\
&\le\sup_{{\V{y}\in\RR^\nu\atop s\le t}}\sup_{{\phi\in\FHR\atop\|\phi\|=1}}
\EE\big[\|\Theta^\mh
(\WW{\V{0},s}{V}[\V{B}^{\V{y}}]-\id)\phi\|^2\big]\int_{\RR^\nu}\EE\big[\|\Theta^\eh
\Psi(\V{B}_t^{\V{x}})\|^2\big]\Id\V{x}
\\
&\le cte^{c_Vt}\big\|e^{t\Delta/2}\|\Theta^\eh\Psi(\cdot)\|^2\big\|_{L^1(\RR^\nu)}
\le cte^{c_Vt}\big\|\|\Theta^\eh\Psi(\cdot)\|^2\big\|_{L^1(\RR^\nu)},
\end{align*}
where the last $L^1$-norm equals $\|\Theta^\eh\Psi\|^2$.
\qed


\section{Feynman-Kac formulas}\label{sec-ext}

This final section is devoted to the Feynman-Kac formulas. 
With the results proven in the previous sections at hand
all proofs given here are essentially standard and they are repeated
only for the convenience of the reader. We start with the fiber Hamiltonian. 

\begin{theorem}[Feynman-Kac formula: fiber case]\label{thm-FH}
Assume that $\V{G}$ and $\V{F}$ are $\V{x}$-independent and 
let $\vxi\in\RR^{\nu}$ and $t\ge0$. Then 
\begin{align}\label{FK-fiber1}
e^{-t\wh{H}(\vxi)}\psi=\EE\big[\WW{\vxi,t}{0}[\V{B}]^*\psi\big]
=\EE\big[\WW{\vxi,t}{0}[\V{B}]\psi\big],\quad\psi\in\FHR.
\end{align}
If $|\V{m}|\le c\omega$, for some $c>0$, then the Feynman-Kac formula can also be written
in terms of a $\LO(\FHR)$-valued Bochner-Lebesgue integral,
\begin{align}\label{FK-fiber2}
e^{-t\wh{H}(\vxi)}=\EE\big[\WW{\vxi,t}{0}[\V{B}]\big].
\end{align}
\end{theorem}

{\proof} 
Since $(\wh{T}_t(\vxi))_{t\ge0}$ is a symmetric $C_0$-group, by the Hille-Yosida theorem, 
it has a unique self-adjoint generator, say $\wh{K}(\vxi)$.
Let $\psi\in\wh{\dom}$. Then 
\begin{align}\label{integ-fib}
\wh{T}_t(\vxi)\psi=\psi+\int^t_0\wh{T}_s(\vxi)\wh{H}(\vxi)\psi\Id s,
\end{align}
by Thm.~\ref{thm-Ito-spin}(2), where we used Lem.~\ref{lem-spin2000}(2)
to exploit the fact that expectations of $L^2$-martingales starting from zero vanish. 
Since $t\mapsto \wh{T}_t(\vxi)\psi$ is continuous at $t=0$, \eqref{integ-fib} implies
$\lim_{t\downarrow0}t^{-1}(\wh{T}_t(\vxi)\psi-\psi)=\wh{H}(\vxi)\psi$.
We deduce that $\wh{\dom}\subset\dom(\wh{K}(\vxi))$ and
$\wh{K}(\vxi)=\wh{H}(\vxi)$ on $\wh{\dom}$.
Since $\wh{H}(\vxi)$ is essentially self-adjoint on $\wh{\dom}$ by Prop.~\ref{prop-esa},
it follows that $\wh{K}(\vxi)=\wh{H}(\vxi)$ and, hence, $e^{-t\wh{H}(\vxi)}=\wh{T}_t(\vxi)$,
i.e., \eqref{FK-fiber1} is valid. Then \eqref{FK-fiber2} follows from Rem.~\ref{rem-FKBL}.
\qed

Next, we treat the total Hamiltonian which is acting in the Hilbert space $\HR$
defined in \eqref{total}. The proofs given here are essentially standard and they are repeated
only for the convenience of the reader.

If $V$ is bounded, then it is well-known that the formula \eqref{def-HV-intro} with
$\Psi\in\sD_0$, where
\begin{equation}\label{def-sD0}
\sD_0:=\mathrm{span}_\CC\big\{f\psi\in\HR\big|\,f\in C_0^\infty(\RR^\nu),\,
\psi\in{\CC^L}\otimes\sC[\mathfrak{d}_C]\big\},
\end{equation} 
defines an essentially self-adjoint operator with domain $\sD_0$;
see \cite{HaslerHerbst2008} for a simple analytic proof and \cite{Hiroshima2000esa,Hiroshima2002}. 
We denote the self-adjoint closure of this operator again by the symbol $H^V$.
Furthermore, if $V=V_{+}-V_-$ is a decomposition of $V$ into measurable functions 
$V_{\pm}:\RR^{\nu}\to[0,\infty)$ with $V_\pm\in L^1_{\loc}(\RR^{\nu})$, and
if the densely defined symmetric sesqui-linear form in $\HR$ given by
\begin{align}\label{form-HV}
\sD_0^2\ni(\Psi_1,\Psi_2)\longmapsto \SPn{\Psi_1}{H^0\,\Psi_2}+\SPn{{V}_+^\eh\Psi_1}{
{V}_+^\eh\Psi_2}-\SPn{{V}_-^\eh\Psi_1}{{V}_-^\eh\Psi_2},
\end{align}
is semi-bounded from below and closable, then we denote the self-adjoint operator 
associated with its closure by $H^V$ as well. This is consistent with the above definition
for bounded $V$.

For instance, the form \eqref{form-HV}
is semi-bounded and closable if $V_-=0$. Hence, by the KLMN-theorem, it is still
semi-bounded and closable with $\fdom(H^{V})=\fdom(H^{V_+})$, if ${V}_-$ is $H^{V_+}$-form 
bounded with relative form bound $<1$. If \eqref{form-HV} is semi-bounded and 
$V\in L_\loc^2(\RR^\nu)$, then it is closable as well and $H^V$ is the Friedrichs extension
of $(H^0+{V})\!\!\upharpoonright_{\sD_0}$.
Likewise, whenever the densely defined symmetric sesqui-linear form in $L^2(\RR^{\nu})$ given by
\begin{align*}
&C_0^\infty(\RR^{\nu})^2\ni(f_1,f_2)\longmapsto \SPn{f_1}{-\tfrac{1}{2}\Delta\,f_2}
+\SPn{V_+^\eh f_1}{V_+^\eh f_2}-\SPn{V_-^\eh f_1}{V_-^\eh f_2},
\end{align*}
is semi-bounded from below and closable, then we denote the self-adjoint Schr\"odinger operator 
associated with its closure by $S^V$. 

Finally, we note that $\WW{\V{0}}{V}[\V{B}^{\V{x}}]$ (resp. $\WW{\V{0}}{V}[\V{b}^{t;\V{x},\V{y}}]$) 
are always well-defined for a.e. $\V{x}$ (resp. a.e. $(\V{x},\V{y})\in \RR^{\nu}\times \RR^{\nu}$, for a 
given $t>0$) under our standing hypothesis that $V$ be locally integrable.
The latter follows from observing that, for a given $\tilde{V}\in L^1_{\loc}(\RR^{\nu})$, one has
\begin{align}
\PP\{\tilde{V}(\V{B}^{\V{x}}_{\bullet})\in L^1_{\loc}([0,\infty))\}=1,
\quad\PP\{\tilde{V}(\V{b}^{t;\V{x},\V{y}}_{\bullet})\in L^1([0,t])\}=1,\label{inter}
\end{align}
for a.e. $\V{x}$ and for all $t>0$ and a.e. $(\V{x}, \V{y})$, respectively.
Here the first equality has been noted in Lem.~2 of \cite{FarisSimon1975}.
The second one then follows from standard properties of the law of $\V{b}^{t;\V{x},\V{y}}$. 

\begin{proposition}\label{prop-Ito-H}
Let $V$ be bounded, $\V{m}=\V{0}$, $\chi\in C_0^\infty(\RR^\nu,\RR)$, $\phi\in\wh{\dom}$,
and $\V{x}\in\RR^\nu$. Then
\begin{align}\label{christian1}
(T^V_t(\chi\phi))(\V{x})-\chi(\V{x})\phi+\int_0^t(T_s^VH^V(\chi\phi))(\V{x})\Id s=0,\quad t\ge0.
\end{align}
\end{proposition}

{\proof}
Let $\V{x}\in\RR^\nu$.
For every $\psi\in\wh{\dom}$, we infer from It\={o}'s formula that, $\PP$-a.s.,
\begin{align*}
\SPb{&\phi}{\chi(\V{B}_t^{\V{x}})\WW{\V{0},t}{V}[\V{B}^{\V{x}}]\psi}
=\SPn{\phi}{\chi(\V{x})\psi}
\\
&+\int_0^t\SPb{\phi}{\big(\nabla\chi(\V{B}_s^{\V{x}})+i\chi(\V{B}_s^{\V{x}})
\vp(\V{G}_{\V{B}^{\V{x}}_s})\big)\WW{\V{0},s}{V}[\V{B}^{\V{x}}]\psi}\Id \V{B}_s\:
\\
&+\int_0^t\SPb{\phi}{\big(\tfrac{1}{2}\Delta\chi(\V{B}_s^{\V{x}})
+i\nabla\chi(\V{B}_s^{\V{x}})\cdot\vp(\V{G}_{\V{B}^{\V{x}}_s})-\wh{H}^V(\V{0},\V{B}_s^{\V{x}})\big)
\WW{\V{0},s}{V}[\V{B}^{\V{x}}]\psi}\Id s,
\end{align*}
for all $t\ge0$.  Applying \eqref{rb-vp1}, Hyp.~\ref{hyp-G}, and \eqref{norm-W}), we next observe that,
for every $t\ge0$, the expression
$\sup_{s\le t}|\SPn{(\nabla\chi(\V{B}_s^{\V{x}})-i\chi(\V{B}_s^{\V{x}})
\vp(\V{G}_{\V{B}^{\V{x}}_s}))\phi}{\WW{\V{0},s}{V}[\V{B}^{\V{x}}]\psi}|$ is bounded by
some deterministic constant. In view of Prop.~\ref{prop-mart} this shows that the stochastic integral in 
the second line above is a martingale. Taking the expectation, re-arranging some terms,
and using \eqref{def-HV-intro}, we
thus arrive at $\SPn{L_t(\V{x})}{\psi}=0$, $\psi\in\wh{\dom}$, where $L_t(\V{x})$ denotes the
vector on the left hand side of \eqref{christian1}.
\qed

\begin{theorem}[Feynman-Kac formula]\label{thm-PF} 
Assume that $V$ admits a decomposition $V=V_{+}-V_-$ into measurable functions 
$V_{\pm}:\RR^{\nu}\to[0,\infty)$ such that $V_+\in L^1_{\loc}(\RR^{\nu})$ and $V_-$ is 
$S^{V_+}$-form bounded with relative form bound $b\le1$. Then 
${\Psi}\in\fdom(H^{V_+})$ implies $\|{\Psi}(\cdot)\|_{\FHR}\in\fdom(S^{V_+})$ with
\begin{align}\label{dia1}
\big\|(S^{V_+})^\eh\|{\Psi}(\cdot)\|_{\FHR}\big\|
\le\|(H^{V_+}+\const)^\eh{\Psi}\|.
\end{align}
In particular, ${V}_-$ is $H^{V_+}$-form bounded with form bound $\le b$ and the
form \eqref{form-HV} is semi-bounded. If the form \eqref{form-HV} is closable as well, 
then the following 
Feynman-Kac formulas are valid, for all $t>0$, $\Psi\in\HR$, and a.e. $\V{x}\in\RR^{\nu}$, 
\begin{align}\nonumber
(e^{-tH^V}\!\Psi)(\V{x})&=\EE\big[\WW{\V{0},t}{V}[\V{B}^{\V{x}}]^*\Psi(\V{B}_t^{\V{x}})\big]
\\
&=\int_{\RR^{\nu}}p_t(\V{x},\V{y})\EE\big[\WW{\V{0},t}{V}[\V{b}^{t;\V{y},\V{x}}]\,\big]
\Psi(\V{y})\Id \V{y}.\label{feyn}
\end{align}
Here the integral $\EE\big[\WW{\V{0},t}{V}[\V{b}^{t;\V{y},\V{x}}]\big]\in\LO(\FHR)$ is well-defined 
in the Bochner-Lebesgue sense, for all $t>0$ and a.e. $(\V{x}, \V{y})\in\RR^{\nu}\times\RR^{\nu}$.
\end{theorem}

{\proof}  
Notice that, by definition, $S^{V_+}$-form boundedness of $V_-$ includes the second relation in
$C_0^\infty(\RR^\nu)\subset\fdom(S^{V_+})\subset\fdom(V_-)$, which entails 
$V_-\in L_\loc^1(\RR^\nu)$.

For $V$ \emph{bounded and continuous}, 
the proof of the first equation in \eqref{feyn} parallels the one of Thm.~\ref{thm-FH},
with Prop.~\ref{prop-Ito-H} applied instead of \eqref{integ-fib}.
Moreover, we employ the fact (recalled above) that $H^V$ is essentially self-adjoint on
$\sD_0$, i.e., on the complex linear hull of vectors $\chi\phi$ as considered in Prop.~\ref{prop-Ito-H}. 
The disintegration formula in the second line of \eqref{feyn} 
is precisely the content of \eqref{for-kernel}, for bounded continuous $V$.

Next, we record a simple fact that will be used implicitly in the approximation arguments below: 
For any measurable $N\subset \RR^{\nu}$ with Lebesgue measure zero, one has
\begin{equation}\label{nullnummer}
\int^t_0 \PP\{\V{B}^{\V{x}}_s\in N\}\,\Id s=0=\int^t_0 \PP\{\V{b}^{t;\V{x},\V{y}}_s\in N\}\,\Id s.
\end{equation}
The relations in \eqref{nullnummer} ensure that, if $\tilde{V}$ and $\tilde{V}_n$, $n\in\NN$,
belong to $L^1_\loc(\RR^\nu,\RR)$, and if $\tilde{V}_n(\V{z})\to \tilde{V}(\V{z})$
and $\tilde{V}_n(\V{z})\le\tilde{V}(\V{z})$, for a.e. $\V{z}$, then, for fixed $t\in I$, 
$\int_0^t\tilde{V}_n(\V{X}_s)\Id s\to\int_0^t\tilde{V}(\V{X}_s)\Id s$, $\PP$-a.s.,
where $\V{X}$ is $\V{B}^{\V{x}}$ or $\V{b}^{t;\V{x},\V{y}}$ with
$\V{x}$ and $\V{y}$ satisfying \eqref{inter}.

Let us now extend \eqref{feyn} to the case when \emph{$V$ is bounded}: 
Then we can use Friedrichs mollifiers to construct a sequence
of smooth potentials  $V_n$ with $|V_n(\V{x})|\le\|V\|_{\infty}$ 
and $V_n(\V{x})\to V(\V{x})$, $n\to\infty$, for a.e. $\V{x}$. 
Clearly, $H^{V_n}\to H^V$, $n\to\infty$, in the strong resolvent
sense and, in particular, $e^{-tH^{V_n}}\to e^{-tH^V}$ strongly, for
every $t>0$. Let $\Psi:\V{x}\mapsto\Psi(\V{x})$ be in $\HR$ and fix some $t>0$.
Then we find integers $0<n_1<n_2<\ldots$ such that
$(e^{-t H^{V_{n_j}}}\Psi)(\V{x})\to(e^{-t H^V}\Psi)(\V{x})$,
$j\to\infty$, in $\FHR$ and for a.e. $\V{x}$.
By the validity of \eqref{feyn} for bounded continuous potentials, 
it now suffices to show that, for a.e. $\V{x}$, one has
\begin{align}\label{mono2}
\EE\big[\WW{\V{0},t}{V_n}[\V{B}^{\V{x}}]^*\,\Psi(\V{B}_t^{\V{x}})\big]
&\to\EE\big[\WW{\V{0},t}{V}[\V{B}^{\V{x}}]^*\,\Psi(\V{B}_t^{\V{x}})\big],
\\\nonumber
\int_{\RR^{\nu}}p_t(\V{x},\V{y})\EE\big[\WW{\V{0},t}{V_n}[\V{b}^{t;\V{y},\V{x}}]\,\big]\Psi(\V{y})\Id \V{y}
&\to\int_{\RR^{\nu}}p_t(\V{x},\V{y})\EE\big[\WW{\V{0},t}{V}[\V{b}^{t;\V{y},\V{x}}]\,\big] 
\Psi(\V{y})\Id \V{y},
\end{align}
as $n\to\infty$. This follows, however, readily by dominated convergence, 
as \eqref{norm-W} gives us the uniform bounds
\begin{align}\label{wiebke}
\|\WW{\V{0},t}{V_n}[\V{B}^{\V{x}}]^*\|  \vee \|\WW{\V{0},t}{V_n}[\V{b}^{t;\V{y},\V{x}}]\,\|
\le e^{(\const -\inf V) t},\>\>\text{ $\PP$-a.s.}
\end{align}

Next, we extend \eqref{feyn} to the case when $V$ is \emph{bounded from below}. 
Since the sequence of bounded potentials given by $V_n:=n\wedge V$, $n\in\NN$,
is monotonically increasing, the corresponding Hamiltonians $H^{V_n}$ again
converge to $H^V$ in strong resolvent sense (see \cite[Thm.~7.10]{Faris1975}), and it suffices to 
verify both limit relations in (\ref{mono2}). This follows, however, again by dominated convergence
since we again have the bounds \eqref{wiebke}. Then, in order to see \eqref{dia1}, we can employ 
\eqref{norm-W}, \eqref{feyn}, and a standard scalar Feynman-Kac formula to get
\begin{align*}
\|(e^{-tH^{V_+}}\Psi)(\V{x})\| 
\le\EE\big[e^{\const t-\int_0^t V_+(\V{B}_s^{\V{x}})\Id s}\|\Psi(\V{B}_t^{\V{x}})\|\big]
=\big(e^{-t(S^{V_+}- \const)}\|\Psi(\cdot)\|\big)(\V{x}),
\end{align*}
for a.e. $\V{x}$, which entails
$$
\int_{\RR^\nu}\|\Psi(\V{x})\|\big(\|\Psi(\cdot)\|-e^{-t(S^{V_+}-\const)}\|\Psi(\cdot)\|\big)(\V{x})\,\Id\V{x}\le
\SPn{\Psi}{\Psi-e^{-tH^{V_+}}\Psi},
$$
for all $t>0$ and $\Psi\in\HR$. Dividing by $t>0$, passing to the limit $t\downarrow0$, and
invoking the spectral calculus we see that
${\Psi}\in\fdom(H^{V_+})$ implies $\|{\Psi}(\cdot)\|\in\fdom(S^{V_+})$ and \eqref{dia1} is proven.

Finally, we consider \emph{general $V$ as in the statement} and assume that
the form \eqref{form-HV} is closable. Then the sequence
$V_n:=(-n)\vee V$, $n\in\NN$, is monotonically decreasing and we again know that
$H^{V_n}$ converges to $H^V$ in strong resolvent sense; see \cite[Thm.~7.9]{Faris1975}. 
It remains to prove the two convergences in \eqref{mono2}. Since $b\le1$,
we also know that there exists a semi-bounded self-adjoint operator $S_\infty$ in $L^2(\RR^\nu)$
such that $S^{V_n}$ converges to $S_\infty$ in strong resolvent sense \cite[Thm.~S.16]{ReedSimonI}.
Fix $t>0$ and $\Psi\in\HR$ in what follows. Again we use \eqref{norm-W} to get
\begin{align}
\|\WW{\V{0},t}{V_n}[\V{B}^{\V{x}}]^*\|
&\le e^{\const t - \int^t_0 V(\V{B}^{\V{x}}_s)\Id s },\quad\;\text{ $\PP$-a.s.,}\label{maj0} \\
\|\WW{\V{0},t}{V_n}[\V{b}^{t;\V{y},\V{x}}]\,\|
&\le e^{\const t- \int^t_0 V(\V{b}^{t;\V{y},\V{x}}_s)\Id s },\>\>\text{ $\PP$-a.s.,}\label{maj}
\end{align}
for $\V{x}$ and $(\V{x},\V{y})$ as in \eqref{inter}, respectively. On the other hand we know that
\begin{align*}
\EE\big[e^{-\int_0^tV^{(n)}(\V{B}_s^{\V{x}})\Id s}\|\Psi(\V{B}_t^{\V{x}})\|\big]
=(e^{-tS^{V_n}}\|\Psi(\cdot)\|)(\V{x})\to (e^{-tS_\infty}\|\Psi(\cdot)\|)(\V{x}),
\end{align*}
for a.e. $\V{x}$. Thus, $e^{-\int_0^tV(\V{B}_s^{\V{x}})\Id s}\|\Psi(\V{B}_t^{\V{x}})\|\in L^1(\PP)$,
for a.e. $\V{x}$, by monotone convergence. (Here we argue similarly as in \cite{Voigt1986}.)
Hence, the first limit relation in
\eqref{mono2} follows, for a.e. $\V{x}$, from the dominated convergence theorem and \eqref{maj0}, 
using $e^{\const t-\int_0^tV(\V{B}_s^{\V{x}})\Id s}\|\Psi(\V{B}_t^{\V{x}})\|$ as a 
$\PP$-integrable majorant. Analogously, in order to prove the second relation in \eqref{mono2}, 
we can use dominated convergence and \eqref{maj}, noting that, for a.e. $\V{x}$,
$$
\int_{\RR^{\nu}}p_t(\V{x},\V{y})
\EE\big[ e^{-\int^t_0V(\V{b}^{t;\V{y},\V{x}}_s) \Id s}\,\big] \|\Psi(\V{y})\|\Id \V{y}
=\EE\big[e^{-\int_0^tV(\V{B}_s^{\V{x}})\Id s}\|\Psi(\V{B}_t^{\V{x}})\|\big]<\infty.
$$
The latter relation also implies that, for a.e. $(\V{x},\V{y})\in\RR^\nu\times\RR^\nu$, 
one has $ e^{-\int^t_0 V(\V{b}^{t;\V{y},\V{x}}_s) \Id s}\in L^1(\PP)$. This completes the proof.
\qed

 \begin{remark}{\rm 
In the scalar case, i.e. if $\V{F}=\V{0}$, the bound \eqref{dia1} holds
true with $\const=0$. This follows immediately from \eqref{norm-W} and the proof of \eqref{dia1}. 
In this case, \eqref{dia1} is one example of a diamagnetic inequality;
see, e.g., \cite{KMS2013,LHB2011} and the references given therein for other versions and alternative derivations of diamagnetic inequalities for quantized vector potentials.
 }\end{remark}


\appendix

\section{Examples}\label{app-examples}

\subsection{Non-relativistic quantum electrodynamics}

\begin{example}{\rm \label{ex-G} 
In all items below we choose $\cM=\RR^3\times\{1,2\}$, equipped with the product of
the Lebesgue and counting measures, i.e.,
$\HP=L^2(\RR^3\times\{1,2\})$.

\smallskip 

\noindent(1)
In the {\em standard model of NRQED} for {\em one} electron interacting with
the electromagnetic radiation field with sharp ultra-violet cut-off one chooses $\nu=3$, 
$\omega(\V{k},j)=|\V{k}|$, for $(\V{k},j)\in\RR^3\times\{1,2\}$, $\V{m}=\V{0}$, and $\V{G}$ is given by
\begin{align*}
\V{G}_{\V{x}}^\Lambda(\V{k},j)
:=(\alpha/2)^\eh(2\pi)^{-\nf{3}{2}}|\V{k}|^\mh\,\chi_\Lambda(\V{k})\,e^{-i\V{k}\cdot\V{x}}\veps(\V{k},j),
\quad\text{a.e. $(\V{k},j)$,}
\end{align*}
where $\alpha>0$ and $\chi_\Lambda$ is the characteristic function
of a ball of radius $\Lambda>0$ about the origin in $\RR^3$.
The vectors $|\V{k}|^{-1}\V{k}$, $\veps(\V{k},1)$, and $\veps(\V{k},2)$
form a.e. an oriented orthonormal basis of $\RR^3$, so that the Coulomb gauge condition
$\Div_{\V{x}}\V{G}_{\V{x}}^\Lambda=0$ is satisfied in $\HP$.
If the electron spin is neglected, then one chooses
$L=1$ and $\V{F}=\V{0}$. To include the electron spin one takes
$L=2$, $S=3$, $\sigma_1$, $\sigma_2$, and $\sigma_3$ are 
the $2$\texttimes$2$-Pauli-spin matrices, and for $\V{F}$ one chooses 
$$
\V{F}_{\V{x}}^\Lambda(\V{k},j):=-\tfrac{i}{2}\V{k}\times\V{G}_{\V{x}}^\Lambda(\V{k},j),\quad
\V{x}\in\RR^3,\;\text{a.e.}\;(\V{k},j).
$$
Applying a suitable unitary transformation to the total Hamiltonian,
if necessary, one may always assume that the polarization vectors are given by
$$
\veps(\V{k},1)=|\V{e}\times\V{k}|^{-1}\V{e}\times\V{k},\quad
\veps(\V{k},2)=|\V{k}|^{-1}\V{k}\times\veps(\V{k},1),\quad\text{a.e.}\;\V{k},
$$
where $\V{e}$ is some unit vector in $\RR^3$. Then a suitable conjugation is given
by $(Cf)(\V{k},j):=(-1)^{j}\ol{f(-\V{k},j)}$, for a.e. $(\V{k},j)$ and $f\in\HP$.

\smallskip

\noindent(2)
To cover the standard model of NRQED for $N\in\NN$ electrons we choose $\nu=3N$,
write $\ul{\V{x}}=(\V{x}_1,\ldots,\V{x}_N)\in(\RR^3)^N$ instead of $\V{x}$, and set
$$
\V{G}_{\ul{\V{x}}}^{\Lambda,N}(\V{k},j):=
\big(\V{G}_{\V{x}_1}^\Lambda(\V{k},j),\ldots,\V{G}_{\V{x}_N}^\Lambda(\V{k},j)\big)\in(\CC^3)^N.
$$
If spin is neglected, then we again set $L=1$ and $\V{F}=\V{0}$.
To include spin, we choose $L=2^N$, so that ${\CC^L}=(\CC^2)^{\otimes_N}$, $S=3N$, and
$$
\sigma_{3\ell+j}:=\id_{\CC^2}^{\otimes_\ell}\otimes\sigma_j\otimes\id_{\CC^2}^{\otimes_{N-\ell-1}},
\qquad\ell=0,\ldots,N-1,\;j=1,2,3,
$$
with the Pauli matrices $\sigma_1$, $\sigma_2$, and $\sigma_3$, as well as
$$
\V{F}_{\ul{\V{x}}}^{\Lambda,N}(\V{k},j):=
\big(\V{F}_{\V{x}_1}^\Lambda(\V{k},j),\ldots,\V{F}_{\V{x}_N}^\Lambda(\V{k},j)\big)\in(\CC^3)^N.
$$
(3) In the {standard model of NRQED} for $N$ electrons in the electrostatic potential of
$K\in\NN$ nuclei with atomic numbers $\sZ=(Z_1,\ldots,Z_K)\in(0,\infty)^K$ located at 
the sites $\sR=(\V{R}_1,\ldots,\V{R}_K)\in(\RR^3)^K$, the potential $V$ is given by
the Coulomb interaction potential,
\begin{align*}
V^N_{\mathscr{R},\mathscr{Z}}(\V{x}_1,\ldots,\V{x}_N)
&:=-\sum_{i=1}^N\sum_{\vk=1}^K\frac{\alpha\,Z_\vk}{|\V{x}_i-\V{R}_\vk|}
+\sum_{1\le i<j\le N}\frac{\alpha}{|\V{x}_i-\V{x}_j|}.
\end{align*}
It is infinitesimally Laplace-bounded \cite{Kato1951}. 
The corresponding total Hamiltonain acts in 
$\HR=L^2((\RR^{3})^N,(\CC^{2})^{\otimes_N}\otimes \sF)$ and attains the form
\begin{align*}
H_{\mathscr{R},\mathscr{Z}}^{\Lambda,N}
&:=\sum_{\ell=1}^N\big\{\tfrac{1}{2}(-i\nabla_{\V{x}_\ell}-\vp(\V{G}_{\V{x}_\ell}^\Lambda))^2
-\vsigma^{(\ell)}\cdot\vp(\V{F}_{\V{x}_\ell}^\Lambda)\big\}+\Id\Gamma(\omega)+
{V}^N_{\mathscr{R},\mathscr{Z}},
\end{align*}
where $\vsigma^{(\ell)}:=(\sigma_{3\ell-2},\sigma_{3\ell-1},\sigma_{3\ell})$.
Here we abuse notation: all terms in the previous formula have to be considered as operators
in $\HR$ in the canonical way; see, e.g., \cite{HaslerHerbst2008,LHB2011} for careful discussions.
According to the Pauli principle the physical Hamiltonian is actually given by the restriction of
$H_{\mathscr{R},\mathscr{Z}}^{\Lambda,N}$ to the reducing subspace of functions which are 
anti-symmetric under simultaneous permutations of the $N$ position-spin degrees of freedom.
By the permutation symmetry of the Hamiltonian the Feynman-Kac formula
for the restricted, physical Hamiltonian is, however, the same as for the non-restricted one.

\smallskip

\noindent(4) {\em Fiber decompositions in the translation-invariant case.}
Consider again the situation in Part~(1) of this example. Let $H^0$ be the corresponding
total Hamiltonian for one electron interacting with the quantized photon field and with
a vanishing electrostatic potential.
Then it turns out that $H^0$ is unitarily equivalent to a direct integral,
$\int_{\RR^3}^\oplus\wh{H}(\vxi)\,\Id\vxi$,
of fiber Hamiltonians attached to the total momenta $\vxi\in\RR^3$ of the system,
\begin{align}\label{fibHPF}
\wh{H}(\vxi)&=\tfrac{1}{2}(\vxi-\Id\Gamma(\V{m})-\vp(\V{G}^\Lambda_{\V{0}}))^2
-\vsigma\cdot\vp(\V{F}^\Lambda_{\V{0}})+\Id\Gamma(\omega).
\end{align}
In \eqref{fibHPF} we have $\V{m}(\V{k},j)=\V{k}$.
The transformation is achieved by applying first
$\int_{\RR^3}^\oplus e^{i\V{x}\cdot\Id\Gamma(\V{m})}\Id\V{x}$ and then a 
($\CC^2\otimes\sF$-valued) Fourier transform acting on the $\V{x}$-variables;
recall that
$e^{i\V{x}\cdot\Id\Gamma(\V{m})}\vp(e^{-i\V{m}\cdot\V{x}}f)e^{-i\V{x}\cdot\Id\Gamma(\V{m})}=\vp(f)$.
 }\end{example}


\subsection{The Nelson model}

\begin{example}{\rm \label{ex-Nelson}
Let $L=S=1$, $\sigma_1=-1$, and $\V{G}=\V{0}$.
Then $\V{F}_{\V{x}}$ has only one component which we denote by $F_{\V{x}}$.
With the usual abuse of notation,
the total Hamiltonian then attains the general form of the {\em Nelson Hamiltonian},
\begin{align*}
H^V_\NE&:=-\tfrac{1}{2}\Delta +\vp(F_{\V{x}})+\Id\Gamma(\omega)+{V}.
\end{align*}
The easiest way to treat Nelson's model is to adapt the proof of Thm.~\ref{thm-Ito1} by 
replacing $-i\vp(q)$ by $\vp(F)$ in the computations.
To illustrate the involved formulas of Def.~\ref{defn-TOE}
we shall, however, demonstrate how they simplify in the above  situation:
Of course, $\V{G}=\V{0}$ entails $K_t=U_t^{\pm}=U_{s,t}^-=0$.
Recalling also that $\w{s}{t}=\iota_t^*\iota_s$, if $s\le t$, we see that 
the quantity defined in \eqref{spin2} satisfies
\begin{align}\nonumber
\sQ_t^{(n)}(g,h;t_{[n]})
&=(-1)^n\!\!\!\!\sum_{{\cA\cup\cB\cup\cC=[n]\atop\#\cC\in2\NN_0}}
\sum_{{\cC=\cup\{c_p,c_p'\}}}
\Big(\prod_{p=1}^{\#\cC/2}\frac{1}{2}\SPn{\iota_{t_{c_p'}}F_{\V{X}_{t_{c_p'}}}}{
\iota_{t_{c_p}}F_{\V{X}_{t_{c_p}}}}\Big)\times
\\\label{Edward1}
&\qquad\qquad\;\;\times
\Big(\prod_{a\in\cA}-i\SPn{\iota_tg}{\iota_{t_a}{F}_{\V{X}_{t_a}}}\Big)
\prod_{b\in\cB}i\SPn{\iota_{t_b}{F}_{\V{X}_{t_b}}}{\iota_0h}.
\end{align}
Here we dropped the condition $c_p<c_p'$ in the partitions of $\cC$, i.e. we sum
over all possibilities to partition $\cC$ into {\em ordered} pairs;
thus the new factors $\frac{1}{2}$ appearing in \eqref{Edward1}.
In doing so we exploited that the scalar products in the first line of \eqref{Edward1} are real.
Written in this way the sum on the right hand side of \eqref{Edward1}
becomes permutation symmetric as a function of $t_1,\ldots,t_n$.
Instead of integrating it over the simplex $t\simplex_n$,
we may just as well integrate it over the cube $[0,t]^n$ and multiply the result by $1/n!$. Therefore,
\begin{align*}
&\int_{t\simplex_n}\!\!\!\sQ_t^{(n)}(g,h;t_{[n]})\Id t_{[n]}
\\
&\quad=\frac{(-1)^n}{n!}\!\!\sum_{{\cA\cup\cB\cup\cC=[n]\atop\#\cC\in2\NN_0}}
\frac{(\#\cC)!}{(\#\cC/2)!}\Big(\frac{\|K^{\NE}_t\|^2}{2}\Big)^{\frac{\#\cC}{2}}
\SPn{ig}{U^{\NE,+}_t}^{\#\cA}\SPn{U^{\NE,-}_t}{ih}^{\#\cB},
\end{align*}
where the analogs of the basic processes for Nelson's model are given by
\begin{align*}
K^{\NE}_t&:=\int_0^t\iota_sF_{\V{X}_s}\,\Id s,\quad U^{\NE,+}_t:=\iota_t^*K^{\NE}_t,\quad
U^{\NE,-}_t:=\iota_0^*K^{\NE}_t,
\end{align*}
i.e. only by Bochner-Lebesgue integrals. A little combinatorics reveals that
\begin{align*}
\sum_{n=0}^\infty&\int_{t\simplex_n}\sQ_t^{(n)}(g,h;t_{[n]})\,\Id t_{[n]}
\\
&=\sum_{n=0}^\infty\frac{1}{n!}\sum_{P=0}^{\lfloor\nf{n}{2}\rfloor}\frac{(-1)^{n-2P}n!}{(n-2P)!P!}
\Big(\frac{\|K^{\NE}_t\|^2}{2}\Big)^P
\big(\SPn{ig}{U^{\NE,+}_t}+\SPn{U^{\NE,-}_t}{ih}\big)^{n-2P}
\\
&=\sum_{P=0}^\infty\frac{1}{P!}\Big(\frac{\|K^{\NE}_t\|^2}{2}\Big)^P
\sum_{\ell=0}^\infty\frac{(-1)^\ell}{\ell!}
\big(\SPn{ig}{U^{\NE,+}_t}+\SPn{U^{\NE,-}_t}{ih}\big)^{\ell}.
\end{align*}
Combining this formula with \eqref{def-W} and \eqref{spin2b} we arrive at
\begin{align*}
\lim_{M\to\infty}\SPn{\zeta(g)}{\WW{\vxi,t}{V,(0,M)}\,\zeta(h)}
&=e^{-u^{\NE,V}_{-\vxi,t}+\SPn{g}{\w{0}{t}h}+i\SPn{g}{U^{\NE,+}_t}-i\SPn{U^{\NE,-}_t}{h}}
\\
&=\SPn{\zeta(g)}{W^{\NE,V}_{\vxi,t}\zeta(h)},
\end{align*}
where (observe the flipped sign of the first term in comparison to \eqref{def-u})
$$
u_{\vxi,\bullet}^{\NE,V}:=-\frac{1}{2}\,\|K_{\bullet}^{\NE}\|^2+\int_0^\bullet
V(\V{X}_s)\,\Id s-i\vxi\cdot(\V{X}_\bullet-\V{X}_0),
$$
and where (the exponentials converge strongly on the normed space $\sC[\mathfrak{d}_C]$)
\begin{align*}
W^{\NE,V}_{\vxi,\bullet}\psi
&:=e^{-u^{\NE,V}_{-\vxi,\bullet}}\exp\{-\ad(U^{\NE,+}_\bullet)\}\Gamma(\w{0}{\bullet})
\exp\{-a(U^{\NE,-}_\bullet)\}\psi,\quad\psi\in\sC[\mathfrak{d}_C].
\end{align*}
Applying \eqref{APC-fact} we see that
$$
W^{\NE,V}_{\vxi,t}=\Gamma(\iota_t^*)e^{-\vp(K^{\NE}_t)}
\Gamma(\iota_0)e^{i\vxi\cdot(\V{X}_t-\V{X}_0)-\int_0^tV(\V{X}_s)\Id s}
\quad\text{on $\sC[\mathfrak{d}_C]$,}
$$ 
which is the formula appearing, e.g., in \cite{LHB2011}.
 }\end{example}


\section{Self-adjointness of fiber Hamiltonians}\label{app-esa}

\noindent
The following short proof of Prop.~\ref{prop-esa} combines three observations: 
a first one by M.~K\"onenberg (see \cite{KMS2013}) who noticed that, by putting an artificial, large 
constant in front of $\Id\Gamma(\omega)$ (instead of assuming weak 
coupling), one obtains a manifestly self-adjoint and surprisingly useful comparison operator.
The second one is borrowed from \cite{HaslerHerbst2008} where a double commutator analog to 
the one in \eqref{dcHH} appears. The third ingredient is the following result \cite[Thm.~2.b)]{Wuest1972}: 

\begin{theorem}\label{thm-Wuest}
If $A$ is a self-adjoint operator in some Hibert space $\sK$, $B$ is symmetric in $\sK$ 
and $A$-bounded, and $A+tB$ is closed, for all $t\in[0,1]$, then $A+B$ is self-adjoint.
\end{theorem}

The following proof can {\em mutatis mutandis} also be used for the total Hamiltonian.

\smallskip

{\it Proof of Prop.~\ref{prop-esa}.}
{\em Step 1.}
Starting from \eqref{def-fib-Ham-spin} and the representation of the scalar Hamiltonian
in the second and third lines of \eqref{def-fib-Ham-scalar}
the bounds \eqref{rb-whH} and \eqref{ub-whH} follow, for sufficiently
large $a\ge1$, from \eqref{rb-vp1}, \eqref{rb-vp2},
and brief and elementary estimations using
$$
\|(1+\Id\Gamma(\omega))^\eh(\vxi-\Id\Gamma(\V{m}))\psi\|^2
\le\|(\vxi-\Id\Gamma(\V{m}))^2\psi\|\,\|(1+\Id\Gamma(\omega))\psi\|.
$$ 
By virtue of the Kato-Rellich theorem the bound \eqref{rb-whH} shows that 
$T:=\wh{H}^0(\vxi,\V{x})+(a-1)\,\Id\Gamma(\omega)$
is closed (resp. self-adjoint if $q_{\V{x}}=0$) on $\wh{\dom}$ and that every core of ${M}_a(\vxi)$
is a core of $T$; in fact, $T-{M}_a(\vxi)=\wh{H}^0(\vxi,\V{x})-{M}_{1}(\vxi)$. We further have
\begin{equation}\label{fred}
a\,\|\Id\Gamma(\omega)\,\psi\|
\le\|{M}_a(\vxi)\,\psi\|\le2\,\|T\psi\|+\const\,\|\psi\|,
\end{equation}
and, hence, $\|\wh{H}^0(\vxi,\V{x})\,\psi\|\le3\|T\,\psi\|+\const\,\|\psi\|$, for all $\psi\in\wh{\dom}$.
Since $\CC^L\otimes\sC[\mathfrak{d}_C]$ is a core of $T$, this implies 
$\wh{H}(\vxi,\V{x})\subset\ol{\wh{H}(\vxi,\V{x})\!\!\upharpoonright_{\CC^L\otimes\sC[\mathfrak{d}_C]}}$.

Abbreviate $\V{v}:=\vxi-\Id\Gamma(\V{m})-\vp(\V{G}_{\V{x}})$ and assume that $\omega$ is bounded 
for the moment. Then \eqref{CCR}, \eqref{comm-dGamma-aad}, and \eqref{qfb-vp} yield
\begin{align}\label{dcHH}
2\Re\SPb{\Id\Gamma(\omega)\,\psi}{&\V{v}^2\,\psi}
=2\SPn{\V{v}\,\psi}{\Id\Gamma(\omega)\,\V{v}\,\psi}
+\SPb{\psi}{\big[\V{v},[\V{v},\Id\Gamma(\omega)]\big]\,\psi}
\\\nonumber
&\ge-2\,\|\omega^\eh\V{G}_{\V{x}}\|^2\,\|\psi\|^2+\SPn{\psi}{\vp(\omega\,\V{m}\cdot\V{G}_{\V{x}})\,\psi}
\\\nonumber
&\ge
-\big(2\,\|\omega^\eh\V{G}_{\V{x}}\|^2+\|\omega^\eh\V{m}\cdot\V{G}_{\V{x}}\|^2\big)\,\|\psi\|^2
-\SPn{\psi}{\Id\Gamma(\omega)\,\psi},
\end{align}
for all $\psi\in\CC^L\otimes\sC[\mathfrak{d}_C]$.
Returning to our general assumptions on $\omega$ we apply \eqref{dcHH} with
$\omega\wedge n$ instead of $\omega$, for every $n\in\NN$, and pass to the limit
$n\to\infty$ on the left hand side and in the last line.
(Notice that Hyp.~\ref{hyp-G} does not imply that $\omega\,\V{m}\cdot\V{G}_{\V{x}}\in\HP$.)
In combination with \eqref{rb-vp1} the so-obtained estimate entails
\begin{align*}
\|T\psi\|^2&\le2\big(\|\tfrac{1}{2}\V{v}^2\,\psi\|^2
+2a\,\Re\SPn{\tfrac{1}{2}\V{v}^2\,\psi}{\Id\Gamma(\omega)\,\psi}
+a^2\,\|\Id\Gamma(\omega)\,\psi\|^2\big)
\\
&\quad
+\const\,\|(\Id\Gamma(\omega)+1)^\eh\,\psi\|^2
\\
&\le2a^2\,\|(\tfrac{1}{2}\V{v}^2+\Id\Gamma(\omega))\,\psi\|^2
+\const'\SPn{\psi}{(\Id\Gamma(\omega)+1)\,\psi}
\\
&\le4a^2\,\|\wh{H}^0(\vxi,\V{x})\,\psi\|^2+\const''\SPn{\psi}{(\Id\Gamma(\omega)+1)\,\psi},
\end{align*}
for all $\psi\in\CC^L\otimes\sC[\mathfrak{d}_C]$. Since, by \eqref{fred},
$\SPn{\psi}{\Id\Gamma(\omega)\,\psi}\le\ve\,\|T\,\psi\|^2+\const(\ve)\,\|\psi\|^2$, 
we obtain $\|T\psi\|\le\const(a\|\wh{H}^0(\vxi,\V{x})\psi\|+\|\psi\|)$, for all
$\psi\in\CC^L\otimes\sC[\mathfrak{d}_C]$. Together with the above remarks this implies that
$\wh{H}(\vxi,\V{x})=\ol{\wh{H}(\vxi,\V{x})\!\!\upharpoonright_{\CC^L\otimes\sC[\mathfrak{d}_C]}}$ 
and that the graph norms of $T$ and $\wh{H}(\vxi,\V{x})$ are equivalent. 

{\em Step 2.} Assume that $q_{\V{x}}=0$ in the rest of this proof.
To conclude that $\wh{H}(\vxi,\V{x})$ is selfadjoint in this case we apply Thm.~\ref{thm-Wuest}
with $A=T$ and $B=(1-a)\Id\Gamma(\omega)$. In fact, we then have
$$
A+tB=\tfrac{1}{2}(\vxi-\Id\Gamma(\V{m})-\vp(\V{G}_{\V{x}}))^2
-\vsigma\cdot\vp(\V{F}_{\V{x}})+\Id\Gamma(\omega_t)
$$  
on $\wh{\dom}$, where $\omega_t:=(1-t)a\omega+t\omega$, $t\in[0,1]$. In particular, 
$A+tB$ is closed by Step~1, since the tuple $(\omega_t,\V{m},\V{G},\V{F})$
satisfies Hyp.~\ref{hyp-G}, for every $t\in[0,1]$. 
\qed


\section{Commutator estimates}\label{app-comm}

\noindent
In the next lemma we prove a number of commutator estimates which have
been used in Sect.~\ref{sec-weights}.

\begin{lemma}\label{lem-comm}
Define $\theta_\ve$ by \eqref{def-Xi}, $\Upsilon_\ve$ by \eqref{def-Upsilon},
and let $\theta:=\theta_0=1+\Id\Gamma(\omega)$.
Then the following bounds hold true, for all 
$E\ge1$, $\ve\in(0,1/E]$, $\alpha\in[\nf{1}{2},1]$, and $f\in\mathfrak{k}$,
\begin{align}\label{nina1}
\|\theta_\ve^\mh\,\Ad_{\vp(f)}\theta_\ve\|
=\|(\Ad_{\vp(f)}\theta_\ve)\,\theta_\ve^\mh\|
&\le\const\,\|\omega^\eh(1+\omega)^\eh f\|_{\HP},
\\\label{nina2}
\|\Ad_{\vp(f)}^2\theta_\ve\|&\le\const\,\|\omega^\eh f\|_{\HP}^2,
\\\label{nina3}
\|\theta_\ve^{-1}\,(\Ad_{\vp(f)}\,\theta_\ve^2)\,\theta_\ve^{-1}\|
&\le\const\,\|\omega^\eh(1+\omega)^\eh f\|_{\HP}^2,
\\\label{nina4}
\|(\Ad_{\vp(f)}\Upsilon_\ve)\Upsilon_\ve^{-\alpha}\theta^\mh\|
&\le\const\,E^{\eh-\alpha}\,\|f\|_{\mathfrak{k}},
\\\label{nina4a}
\|\theta^\mh(\Ad_{\vp(f)}\Upsilon_\ve)\Upsilon_\ve^{-\alpha}\|
&\le\const\,E^{\eh-\alpha}\,\|f\|_{\mathfrak{k}},
\\\label{nina4b}
\|\theta^{-\nf{1}{4}}(\Ad_{\vp(f)}\Upsilon_\ve)\Upsilon_\ve^{-\alpha}\theta^{-\nf{1}{4}}\|
&\le\const\,E^{\eh-\alpha}\,\|f\|_{\mathfrak{k}},
\\\label{nina5}
\|\theta^\mh\Upsilon_\ve^{-\alpha}\Ad_{\vp(f)}\Upsilon_\ve\|
&\le\const\,E^{\eh-\alpha}\,\|f\|_{\mathfrak{k}},
\\\label{nina6}
\|\theta^\mh\Re[\Upsilon_\ve^{-1}(\Ad_{\vp(f)}^2\Upsilon_\ve)]\theta^\mh\|
&\le\const\,E^{\mh}\,\|f\|_{\mathfrak{k}}^2,
\\\label{nina7}
\|\theta^\mh\Upsilon_\ve^{-1}(\Ad_{\vp(f)}^2\Upsilon_\ve^2)\Upsilon_\ve^{-1}\theta^\mh\|
&\le\const\,E^{\mh}\,\|f\|_{\mathfrak{k}}^2.
\end{align}
\end{lemma}

{\proof} 
We remark that all algebraic identities between various
combinations of operators below are valid on the dense domain $\sC[\mathfrak{d}_C]$.
All norms have to be read as norms of operators which are densely
defined and bounded on $\sC[\mathfrak{d}_C]$.

First, we observe that, if $\Theta$ denotes one of the weights
$\theta_\ve$ or $\Upsilon_\ve$, then
\begin{align*}
\Theta^{-1}\,(\Ad_{\vp(f)}\,\Theta^2)\,\Theta^{-1}
&=2\{\Theta^{-1}\,(\Ad_{\vp(f)}\Theta)\}\{(\Ad_{\vp(f)}\Theta)\,\Theta^{-1}\}
\\
&\quad
+\Theta^{-1}\,\Ad_{\vp(f)}^2\Theta+(\Ad_{\vp(f)}^2\Theta)\,\Theta^{-1},
\end{align*}
so that \eqref{nina3} follows from \eqref{nina1} and \eqref{nina2}
and \eqref{nina7} from \eqref{nina4}, \eqref{nina5}, and \eqref{nina6}. Writing
$$
\Theta_\ve=(1+\Id\Gamma(\omega))(1+\ve\Id\Gamma(\omega))^{-1}=
\ve^{-1}\big(\id-(1-\ve E)(1+\ve\Id\Gamma(\omega))^{-1}\big)
$$
and applying \eqref{for-vp}, \eqref{CR-Segal3}, and 
$\Ad_s(TT')=T\Ad_sT'+(\Ad_ST)T'$ as well as $\Ad_ST^{-1}=-T^{-1}(\Ad_ST)\,T^{-1}$, 
repeatedly we obtain
\begin{align}\label{conny1}
\Ad_{\vp(f)}\Theta_\ve
&=(\ve E-1)(1+\ve\Id\Gamma(\omega))^{-1}\,i\vp(i\omega f)
(1+\ve\Id\Gamma(\omega))^{-1},
\\\nonumber
\Ad_{\vp(f)}^2\Theta_\ve
&=2\ve\,(1-\ve E)(1+\ve\Id\Gamma(\omega))^{-1}\,\big(\vp(i\omega f)
(1+\ve\Id\Gamma(\omega))^{-1}\big)^2
\\\label{conny2}
&\quad+(1-\ve E)\,\|\omega^\eh f\|^2\,(1+\ve\Id\Gamma(\omega))^{-2}.
\end{align}
As a consequence of \eqref{rb-a} we have
$\ve^\eh\|a(\omega f)(1+\ve\Id\Gamma(\omega))^\mh\|\le\|\omega^\eh f\|$,
which together with \eqref{for-vp} and \eqref{conny2} implies \eqref{nina2}.
From \eqref{rb-vp1} and \eqref{conny1} we readily infer that \eqref{nina1} is satisfied.

Likewise, by writing 
$$
\Upsilon_\ve=(E+\Id\Gamma(m_j)^2)(1+\ve\Id\Gamma(m_j)^2)^{-1}
=\ve^{-1}\big(\id-(1-\ve E)\,R_\ve\big)
$$
with $R_\ve:=(1+\ve\Id\Gamma(m_j)^2)^{-1}$, we deduce that
\begin{align*}
\Ad_{\vp(f)}\Upsilon_\ve&=(1-\ve E)\,R_\ve\,\{\Ad_{\vp(f)}(\Id\Gamma(m_j)^2)\}\,R_\ve,
\end{align*}
where, on account of \eqref{CR-Segal3},
\begin{align*}
\Ad_{\vp(f)}(\Id\Gamma(m_j)^2)
&=2i\Id\Gamma(m_j)\,\vp(im_jf)+\vp(m_j^2f)
\\
&=2i\vp(im_jf)\,\Id\Gamma(m_j)-\vp(m_j^2f).
\end{align*}
Consequently, for $\alpha\in[\nf{1}{2},1]$, $\gamma\in[0,1]$, and $\beta:=1-\gamma$,
\begin{align*}
\big\|&\theta^{-\nf{\beta}{2}}(\Ad_{\vp(f)}\Upsilon_\ve)\Upsilon_\ve^{-\alpha}
\theta^{-\nf{\gamma}{2}}\big\|
\\&
\le|1-\ve E|\,\|\theta^{-\nf{\beta}{2}}\vp(m_j^2f)\,\theta^{-\nf{\gamma}{2}}\|\,
\|(E+\Id\Gamma(m_j)^2)^{-\alpha}\|
\\
&\quad+2|1-\ve E|\,\big\|\Id\Gamma(m_j)(E+\Id\Gamma(m_j)^2)^{-\alpha}\big\|\,
\|\theta^{-\nf{\beta}{2}}\vp(im_jf)\,\theta^{-\nf{\gamma}{2}}\|,
\end{align*}
which proves \eqref{nina4}, \eqref{nina4a}, and \eqref{nina4b}. Here we use that
\eqref{rb-vp1} implies the bounds
\begin{equation}\label{nina1999}
\|\theta^{-\nf{1}{4}}\vp(g)\,\theta^{-\nf{1}{4}}\|,\|\theta^{-\nf{1}{4}}\vp(ig)\,\theta^{-\nf{1}{4}}\|
\le2^\eh\|(1+\omega^{-1})^\eh g\|.
\end{equation}
Using the above identities for a single commutator we further find
\begin{align*}
\Ad_{\vp(f)}^2\Upsilon_\ve
&=2(1-\ve E)R_\ve\,2i\vp(im_jf)\,\frac{\ve\Id\Gamma(m_j)^2}{1+\ve\Id\Gamma(m_j)^2}
\,2i\vp(im_jf)\,R_\ve
\\
&\quad+8(1-\ve E)\,\Re\Big[
R_\ve\,2i\vp(im_jf)\,\frac{\ve\Id\Gamma(m_j)}{1+\ve\Id\Gamma(m_j)^2}\,\vp(m_j^2f)\, R_\ve\Big]
\\
&\quad-2\ve(1-\ve E)R_\ve\,\vp(m_j^2f)R_\ve\,\vp(m_j^2f)\, R_\ve
\\
&\quad
-2(1-\ve E)\,\big(R_\ve\,\vp(im_jf)^2\,R_\ve+\|\,|m_j|^\eh f\|^2\,R_\ve\,\Id\Gamma(m_j)\,R_\ve\big).
\end{align*}
Now, we multiply the previous identity both from the left and from the
right with $\theta^\mh=(1+\Id\Gamma(\omega))^\mh$. By \eqref{rb-vp1} the latter operators
control all unbounded fields. Multiplying the previous identity
in addition with $\Upsilon_{\ve}^{\mh}$ from the left or from the right
we can also control the operator $\Id\Gamma(m_j)$ in the last line,
where no power of $\ve$ can be employed to control it by means of the
resolvents $R_\ve$. From these remarks we readily infer \eqref{nina6}.
\qed


\section{Admissibility of Brownian bridges}\label{app-bridge}

\noindent
In this appendix we verify that the semi-martingale realizations of Brownian bridges satisfy
the technical condition \eqref{hyp-Y2} of Hyp.~\ref{hyp-B}. After that we also present a detailed
proof of Lem.~\ref{lem-rev-bridge} on time-reversals of Brownian bridges.

In all what follows, $\V{y}\in\RR^\nu$ and $\V{q}:\Omega\to\RR^\nu$ is $\fF_0$-measurable
such that $\EE[|\V{q}|^n]<\infty$, for all $n\in\NN$. Recall that the 
(up to indistinguishability unique) solution of 
$\V{b}_\bullet=\V{q}+\V{B}_\bullet+\int_0^\bullet\frac{\V{y}-\V{b}_s}{\cT-s}\,\Id s$ 
is explicitly given by
\begin{align}\label{sol-for-BB}
&\V{b}_t^{\mathcal{T};\V{q},\V{y}}:=\begin{cases}
\; \frac{t}{\mathcal{T}}\V{y}+\frac{\cT-t}{\cT}\V{q}+\V{B}_t
-(\mathcal{T}-t)\int^t_0\frac{\V{B}_s}{(\mathcal{T}-s)^2}\Id s,
\>\> \text{if $0\le t<\mathcal{T}$,}\\
\;\V{y},\>\> \text{if $t=\mathcal{T}$.}\end{cases}
\end{align}

\begin{lemma}\label{lem-bridget2000}
The drift vector field in the SDE 
$\V{b}_\bullet^{\cT;\V{q},\V{y}}=\V{B}_\bullet^{\V{q}}+\int_0^\bullet\V{Y}_s\Id s$ 
satisfied by the process in \eqref{sol-for-BB} can $\PP$-a.s. be written as
\begin{align}
\V{Y}_t&:=\frac{\V{y}-\V{b}_t^{\cT;\V{q},\V{y}}}{\cT-t}=
\frac{\V{y}-\V{q}}{\cT}\,-\int_0^t\frac{1}{\cT-s}\,\Id\V{B}_s,\quad 0\le t<\cT.\label{drift1}
\end{align}
\end{lemma}

{\proof} 
Plugging the formula \eqref{sol-for-BB} for
$\V{b}_t^{\cT;\V{q},\V{y}}$ into the expression in the middle in \eqref{drift1}
and taking the following consequence of It\={o}'s formula into account,
\begin{align*}
-\frac{\V{B}_t}{\cT-t}+\int_0^t\frac{\V{B}_s}{(\cT-s)^2}\,\Id s
&=-\int_0^t\frac{1}{\cT-s}\,\Id\V{B}_s,\quad 0\le t<\cT,\;\;\text{$\PP$-a.s.,}
\end{align*}
we arrive at the formula on the right hand side of \eqref{drift1}.
\qed 

\begin{lemma}\label{lem-bridge}
For all $\cT>0$, $p\in\NN$, and $t\in[0,\cT)$,
\begin{align}\label{eq-drift}
\EE\big[|\V{Y}_t|^{2p}\big]&=
\sum_{\ell=0}^p\frac{(2p-2+\nu)!!}{(2(p-\ell)-2+\nu)!!}
{p\choose\ell}\frac{\EE\big[|\V{q}-\V{y}|^{2(p-\ell)}\big]}{\cT^{2(p-\ell)}}
\,\frac{t^\ell}{\cT^\ell(\cT-t)^\ell},
\end{align}
where $(2j)!!:=2^j j!$, $(2j+1)!!=(2j+1)!/(2j)!!$,
and in particular
\begin{align}\label{bd-drift}
\int_0^\cT(\cT-s)^{\vk}\EE\big[|\V{Y}_s|^{2\vk}\big]\,\Id s
\le\const(\nu,\vk,\cT)\,\EE\big[(1+|\V{q}-\V{y}|)^{2\vk}\big], \quad \vk>0.
\end{align}
\end{lemma}

{\proof} 
By \eqref{drift1} and It\={o}'s formula
(ignore the last integral, if $p=1$)
\begin{align*}
|\V{Y}_t|^{2p}
&=\frac{|\V{q}-\V{y}|^{2p}}{\cT^{2p}}-2p\int_0^t|\V{Y}_s|^{2(p-1)}\frac{\V{Y}_s}{\cT-s}\,\Id\V{B}_s
\\
&\;\;\,
+p\,\nu\int_0^t|\V{Y}_s|^{2(p-1)}\,\frac{\Id s}{(\cT-s)^2}
+\frac{p(p-1)}{2}\int_0^t|\V{Y}_s|^{2(p-2)}\,\frac{4|\V{Y}_s|^2}{(\cT-s)^2}\,\Id s,
\end{align*}
for all $t\in[0,\cT)$, $\PP$-a.s., and therefore,
\begin{align*}
\EE\big[|\V{Y}_t|^{2p}\big]
&=\frac{\EE\big[|\V{q}-\V{y}|^{2p}\big]}{\cT^{2p}}
+(2p-2+\nu)p\int_0^t\EE\big[|\V{Y}_s|^{2(p-1)}\big]\,\frac{\Id s}{(\cT-s)^2}.
\end{align*}
Iterating this we find
\begin{align*}
\EE\big[|\V{Y}_t|^{2p}\big]
&=\sum_{\ell=0}^p\frac{(2p-2+\nu)!!\,p!}{(2(p-\ell)-2+\nu)!!}
\,\frac{\EE\big[|\V{q}-\V{y}|^{2(p-\ell)}\big]}{(p-\ell)!\,\cT^{2(p-\ell)}}
\int_{t\simplex_\ell}\prod_{j=1}^\ell\frac{\Id t_j}{(\cT-t_j)^2},
\end{align*}
which is \eqref{eq-drift} since the integral over $t\simplex_n$ equals $(\int_0^t\Id s/(\cT-s)^2)^\ell/\ell!$.
\qed 

As announced above, we shall now work out the details on the time-reversal of a Brownian
bridge:

\smallskip

{\it Proof of Lem.~\ref{lem-rev-bridge}.}
In this proof the letter $\cT$ plays the role of the letter $t$ in the statement of 
Lem.~\ref{lem-rev-bridge}, i.e., we consider the bridge $\V{b}^{\cT;\V{x},\V{y}}$ reversed at $\cT$.

Let $\fN\subset\fF$ be the set of $\PP$-zero sets. Recall that, for every $t\in[0,\cT]$, we defined
$\fH_t$ to be the smallest $\sigma$-algebra containing $\fN$ and the $\sigma$-algebra
$\sigma(\V{b}^{\cT;\V{x},\V{y}}_{\cT-t};\,\V{B}_{s}-\V{B}_{\cT}:\,\cT-t\le s\le\cT)
=\sigma(\V{b}^{\cT;\V{x},\V{y}}_{\cT-t};\,\V{B}_{s}-\V{B}_{r}:\,\cT-t\le r\le s\le\cT)$.

{\em Step 1.} We claim that $(\fH_t)_{t\in[0,\cT]}$ is a filtration and that $\V{b}^{\cT;\V{x},\V{y}}_{\cT-s}$
is $\fH_t$-measurable, for all $0\le s\le t\le\cT$.

Of course, the second assertion implies the first.
Since $\V{b}^{\cT;\V{x},\V{y}}_{\cT}=\V{y}$, $\PP$-a.s., and $\sigma(\fN)=\fH_0\subset\fH_t$,
$t\in(0,\cT]$, we see that $\V{b}^{\cT;\V{x},\V{y}}_{\cT}$ is $\fH_t$-measurable, for all $t\in[0,\cT]$.
Let $0<s<t\le\cT$. Then, up to indistinguishability, $(\V{b}_{\cT-t+s}^{\cT;\V{x},\V{y}})_{s\in[0,t]}$ is the
unique semi-martingale with respect to $\BB_{\cT-t}$ on $[0,t]$ which $\PP$-a.s. solves
\begin{align*}
\V{X}_\bullet&=\V{b}_{\cT-t}^{\cT;\V{x},\V{y}}+\V{B}_{\cT-t+\bullet}-\V{B}_{\cT-t}+
\int_0^\bullet\frac{\V{y}-\V{X}_r}{t-r}\Id r\;\;\text{on $[0,t)$},\quad\V{X}_t=\V{y}.
\end{align*}
The standard solution theory for SDE thus implies that, 
for every $\ve\in(0,s)$, the random variable $\V{b}_{\cT-s}^{\cT;\V{x},\V{y}}$ is 
measurable with respect to the smallest $\sigma$-algebra containing $\fN$ and
$\sigma(\V{b}^{\cT;\V{x},\V{y}}_{\cT-t};\,\V{B}_{r}-\V{B}_{\cT-t}:\,\cT-t\le r\le\cT-s+\ve)$.
In particular, $\V{b}_{\cT-s}^{\cT;\V{x},\V{y}}$ is $\fH_t$-measurable.

{\em Step 2.} Next, we claim that \eqref{def-barB-bridget}
defines a continuous martingale with respect to $(\fH_t)_{t\in[0,\cT)}$ starting at zero. 

Obviously, all paths of $\hat{\V{B}}$ are continuous on $[0,\cT)$ and,
by Step 1 and \eqref{def-barB-bridget}, $\hat{\V{B}}$ is adapted to $(\fH_t)_{t\in[0,\cT)}$.
Using $\EE[\V{b}_t^{\cT;\V{x},\V{y}}]=\frac{t}{\cT}\V{y}+\frac{\cT-t}{\cT}\V{x}$, $t\in[0,\cT]$,
which is obvious from \eqref{sol-for-BB},
we read off from \eqref{def-barB-bridget} that $\hat{\V{B}}_t$ is integrable
and $\EE[\hat{\V{B}}_t]=\V{0}$, for all $t\in[0,\cT]$.
Of course, $\hat{\V{B}}_0=\V{0}$, $\PP$-a.s., and
$$
\EE^{\fH_0}[\hat{\V{B}}_t]=\EE^{\sigma(\fN)}[\hat{\V{B}}_t]
=\EE[\hat{\V{B}}_t]=\V{0},\quad t\in(0,\cT).
$$
Let $0<s<t<\cT$. Taking the SDE solved by $\V{b}^{\cT;\V{x},\V{y}}$ into account we see that
\begin{align}\nonumber
\hat{\V{B}}_t-\hat{\V{B}}_s
&=\V{B}_{\cT-t}-\V{B}_{\cT-s}-\int_{\cT-t}^{\cT-s}
\Big(\frac{\V{x}-\V{b}^{\cT;\V{x},\V{y}}_u}{u}+\frac{\V{y}-\V{b}^{\cT;\V{x},\V{y}}_u}{\cT-u}\Big)\Id u
\\\label{bridget88}
&=\V{B}_{\cT-t}-\V{B}_{\cT-s}-\int_{\cT-t}^{\cT-s}\nabla\ln d_u^{\cT;\V{x},\V{y}}(\V{b}^{\cT;\V{x},\V{y}}_u)
\Id u,\quad\text{$\PP$-a.s.,}
\end{align}
where 
\begin{equation}\label{bridge-law}
d_u^{\cT;\V{x},\V{y}}:=p_u(\V{x},\cdot)p_{\cT-u}(\V{y},\cdot)\big/p_{\cT}(\V{x},\V{y}),\quad u\in(0,\cT).
\end{equation}
We may now employ the arguments of \cite[\textsection4]{Pardoux-LNM1204} to show that
$\EE^{\sigma(\V{b}_{\cT-s}^{\cT;\V{x},\V{y}})}[\hat{\V{B}}_t-\hat{\V{B}}_s]=\V{0}$, $\PP$-a.s.;
see Lem.~\ref{lem-Pardoux} below. For all $\cT-s\le r\le u\le\cT$, we further know that
$\sigma(\V{B}_u-\V{B}_r)$ and $\sigma(\V{b}_{\cT-s}^{\cT;\V{x},\V{y}};\hat{\V{B}}_t-\hat{\V{B}}_s)$
are independent since $\V{b}_{\cT-s}^{\cT;\V{x},\V{y}}$ and $\hat{\V{B}}_t-\hat{\V{B}}_s$  
are $\fF_{\cT-s}$-measurable while $\V{B}_u-\V{B}_r$ is $\fF_{\cT-s}$-independent.
For trivial reasons, $\sigma(\fN)$ and 
$\sigma(\V{b}_{\cT-s}^{\cT;\V{x},\V{y}};\hat{\V{B}}_t-\hat{\V{B}}_s)$ are independent as well.
These remarks entail 
$\EE^{\fH_s}[\hat{\V{B}}_t-\hat{\V{B}}_s]=\EE^{\sigma(\V{b}_{\cT-s}^{\cT;\V{x},\V{y}})}
[\hat{\V{B}}_t-\hat{\V{B}}_s]=\V{0}$, $\PP$-a.s.
Since $\hat{\V{B}}_s$ is $\fH_s$-measurable, we arrive at
$$
\EE^{\fH_{s}}[\hat{\V{B}}_t]=\hat{\V{B}}_s,\quad\text{$\PP$-a.s.}
$$
Hence, $\hat{\V{B}}$ is a continuous martingale with respect to $(\fH_t)_{t\in[0,\cT)}$
starting $\PP$-a.s. at zero.

{\em Step 3.} Invoking a martingale convergence theorem, we see that $(\hat{\V{B}}_t)_{t\in[0,\cT)}$
has a unique extension to a continuous $(\fH_t)_{t\in[0,\cT]}$-martingale (starting at zero), 
again denoted by $\hat{\V{B}}$.
Furhermore, a glance at \eqref{def-barB-bridget} reveals that
the quadratic variation process of $\hat{\V{B}}$ is $(t\id)_{t\in[0,\cT]}$.
In view of L\'{e}vy's characterization we now see that $\hat{\V{B}}$ is a Brownian motion
with respect to $(\fH_t)_{t\in[0,\cT]}$ and, hence, also with respect to its standard extension 
$(\bar{\fF}_t)_{t\in[0,\cT]}$; see \cite[p.~219]{HackenbrochThalmaier1994}.

{\em Step 4.} Substituting $u(s):=\cT-s$ pathwise in \eqref{def-barB-bridget} we obtain 
\eqref{bridge-rev}. By Step~1, $(\V{b}^{\cT;\V{x},\V{y}}_{\cT-t})_{t\in[0,\cT]}$ is
adapted to $(\bar{\fF}_t)_{t\in[0,\cT]}$ and we conclude.
\qed

\begin{lemma}\label{lem-Pardoux}
Let $0<s<t<\cT$ and $f:\RR^\nu\to\RR$ be bounded and Borel measurable. 
With $d^{\cT;\V{x},\V{y}}$ defined by \eqref{bridge-law}, we then have
\begin{align*}
\EE\big[f(\V{b}_t^{\cT;\V{x},\V{y}})\big]&=\int_{\RR^\nu}d_t^{\cT;\V{x},\V{y}}(\V{z})f(\V{z})\Id \V{z},
\\
\EE\big[f(\V{b}_{\cT-s}^{\cT;\V{x},\V{y}})(\V{B}_{\cT-s}-\V{B}_{\cT-t})\big]
&=\EE\Big[f(\V{b}_{\cT-s}^{\cT;\V{x},\V{y}})\int_{\cT-t}^{\cT-s}
\nabla\ln d_u^{\cT;\V{x},\V{y}}(\V{b}^{\cT;\V{x},\V{y}}_u)\Id u\Big].
\end{align*}
\end{lemma}

{\proof}
We shall show the second asserted identity with $0<\cT-t<\cT-s<\cT$ replaced by $0<s<t<\cT$. 
By an approximation argument, we may actually assume $f$ to be continuous with compact support, 
which we do from now on.

For fixed $\cT>0$ and $\V{y}\in\RR^\nu$, we set
\begin{align*}
\vr_{s,t}(\V{z},\V{a})&:=p_{t-s}(\V{z},\V{a})p_{\cT-t}(\V{a},\V{y})\big/p_{\cT-s}(\V{z},\V{y}),
\quad\V{a},\V{z}\in\RR^\nu,\,0\le s<t<\cT.
\end{align*}
Then
\begin{align*}
\Big(\partial_s+\tfrac{1}{2}\Delta_{\V{z}}+\frac{\V{y}-\V{z}}{\cT-s}\cdot\nabla_{\V{z}}\Big)
\vr_{s,t}(\V{z},\V{a})=0,
\end{align*}
for all $\V{a},\V{z}\in\RR^\nu$ and $0\le s<t<\cT$. We set
\begin{align*}
(\pi_{s,t}f)(\V{z}):=\int_{\RR^\nu}\vr_{s,t}(\V{z},\V{a})f(\V{a})\Id\V{a},\quad
\V{z}\in\RR^\nu,\,0\le s<t<\cT.
\end{align*}
Since $f$ is bounded, it is then also clear that
$(s,\V{z})\mapsto(\pi_{s,t}f)(\V{z})$ belongs to $C^2([0,t)\times\RR^\nu)$ with
\begin{align*}
\Big(\partial_s+\tfrac{1}{2}\Delta_{\V{z}}+\frac{\V{y}-\V{z}}{\cT-s}\cdot\nabla_{\V{z}}\Big)
(\pi_{s,t}f)(\V{z})&=0,\quad\V{z}\in\RR^\nu,\,0\le s<t<\cT.
\end{align*}
Hence, It\={o}'s formula (applied with respect to the time-shifted basis $\BB_s$) $\PP$-a.s. entails,
for all $0\le s\le r<t$,
\begin{align}\label{emcy-1}
(\pi_{r,t}f)(\V{b}_r^{\cT;\V{x},\V{y}})-(\pi_{s,t}f)(\V{b}_s^{\cT;\V{x},\V{y}})
&=\int_s^r\nabla(\pi_{u,t}f)(\V{b}_u^{\cT;\V{x},\V{y}})\Id\V{B}_u.
\end{align}
Since $f\in C_0(\RR^\nu)$, 
we further know that $(s,\V{z})\mapsto\pi_{s,t}f$ has a unique bounded and continuous extension onto 
$[0,t]\times\RR^\nu$ with $\pi_{t,t}f(\V{z})=f(\V{z})$, $\V{z}\in\RR^\nu$.
The function $(s,\V{z})\mapsto\nabla(\pi_{s,t}f)(\V{z})$ is bounded on every set $[0,r]\times\RR^\nu$
with $r\in[0,t)$.
Let $F:\Omega\to\RR$ be bounded and $\fF_s$-measurable.
Then the dominated convergence theorem and \eqref{emcy-1} yield
\begin{align*}
\EE\big[F\big(f(\V{b}_t^{\cT;\V{x},\V{y}})-(\pi_{s,t}f)(\V{b}_s^{\cT;\V{x},\V{y}})\big)\big]
=\lim_{r\uparrow t}\EE\Big[\int_s^rF\,\nabla(\pi_{u,t}f)(\V{b}_u^{\cT;\V{x},\V{y}})\Id\V{B}_u\Big]=0.
\end{align*}
This proves the following relation,
\begin{align}\label{emcy2015}
\EE^{\fF_s}[f(\V{b}_t^{\cT;\V{x},\V{y}})]=(\pi_{s,t}f)(\V{b}_s^{\cT;\V{x},\V{y}}),\;\text{$\PP$-a.s.,}
\quad0\le s<t.
\end{align}
In particular, $\EE[f(\V{b}_t^{\cT;\V{x},\V{y}})]=\EE[\EE^{\fF_0}[f(\V{b}_t^{\cT;\V{x},\V{y}})]]
=(\pi_{0,t}f)(\V{x})$, which is the first asserted identity.
Applying \eqref{emcy-1} first and It\={o}'s formula with respect to $\BB_s$ 
afterwards, we $\PP$-a.s. obtain
\begin{align}\label{emcy}
&(\pi_{r,t}f)(\V{b}_r^{\cT;\V{x},\V{y}})({B}_{\ell,r}-{B}_{\ell,s})
=\int_s^r\partial_\ell (\pi_{u,t}f)(\V{b}_u^{\cT;\V{x},\V{y}})\Id u
\\\nonumber
&\quad+\int_s^r({B}_{\ell,u}-{B}_{\ell,s})\nabla(\pi_{u,t}f)(\V{b}_u^{\cT;\V{x},\V{y}})\Id\V{B}_u
+\int_s^r(\pi_{u,t}f)(\V{b}_u^{\cT;\V{x},\V{y}})\Id B_{\ell,u},
\end{align}
for all $r\in[s,t)$.
The dominated convergence theorem and \eqref{emcy} now imply
\begin{align}\nonumber
\EE\big[f(\V{b}_{t}^{\cT;\V{x},\V{y}})(\V{B}_{t}-\V{B}_{s})\big]&=\lim_{r\uparrow t}
\EE\big[(\pi_{r,t}f)(\V{b}_r^{\cT;\V{x},\V{y}})(\V{B}_{r}-\V{B}_{s})\big]
\\\label{emcy2}
&=\lim_{r\uparrow t}\int_s^r\EE\big[\nabla(\pi_{u,t}f)(\V{b}_u^{\cT;\V{x},\V{y}})\big]\Id u.
\end{align}
Here we further infer from the first asserted identity (extended to bounded measurable $f$)
and from \eqref{emcy2015} that
\begin{align}\nonumber
\int_s^r\EE\big[&\nabla(p_{u,t}f)(\V{b}_u^{\cT;\V{x},\V{y}})\big]\Id u
=\int_s^r\int_{\RR^\nu}\vr_{0,u}(\V{x},\V{z})\nabla(p_{u,t}f)(\V{z})\Id\V{z}\Id u
\\\nonumber
&=-\int_s^r\int_{\RR^\nu}(\nabla_{\V{z}}\vr_{0,u})(\V{x},\V{z})(p_{u,t}f)(\V{z})\Id\V{z}\Id u
\\\nonumber
&\xrightarrow{\;\,r\uparrow t\;\,}-\int_s^t\int_{\RR^\nu}(\vr_{0,u})(\V{x},\V{z})(p_{u,t}f)(\V{z})
(\nabla\ln d_u^{\cT;\V{x},\V{y}})(\V{z})\Id\V{z}\Id u
\\\nonumber
&=-\int_s^t\EE\big[(p_{u,t}f)(\V{b}_u^{\cT;\V{x},\V{y}})
(\nabla\ln d_u^{\cT;\V{x},\V{y}})(\V{b}_u^{\cT;\V{x},\V{y}})\big]\Id u
\\\nonumber
&=-\int_s^t\EE\big[\EE^{\fF_u}[f(\V{b}_t^{\cT;\V{x},\V{y}})]
(\nabla\ln d_u^{\cT;\V{x},\V{y}})(\V{b}_u^{\cT;\V{x},\V{y}})\big]\Id u
\\\nonumber
&=-\int_s^t\EE\big[f(\V{b}_t^{\cT;\V{x},\V{y}})
(\nabla\ln d_u^{\cT;\V{x},\V{y}})(\V{b}_u^{\cT;\V{x},\V{y}})\big]\Id u
\\\label{emcy3}
&=-\EE\Big[f(\V{b}_t^{\cT;\V{x},\V{y}})\int_s^t
(\nabla\ln d_u^{\cT;\V{x},\V{y}})(\V{b}_u^{\cT;\V{x},\V{y}})\Id u\Big].
\end{align}
Combinig \eqref{emcy2} and \eqref{emcy3} we arrive at the second asserted identity.
\qed


\section{On time-ordered integration of a stochastic integral}\label{app-cont-tn}

\noindent
After the application of the stochastic calculus in Sect.~\ref{sec-alg} we obtain the relation 
\eqref{spin-vera} on the complement of a $\PP$-zero set which depends {\em inter alia} on the 
parameters $t_{[n]}=(t_1,\dots,t_n)$. Hence, it is clear a priori that, $\PP$-a.s., \eqref{spin-vera} is 
available for all rational $t_{[n]}\in I\simplex_n\cap\QQ^n$ at the same time, where
$I\simplex_n:=\{0\le t_1\le\dots\le t_n\in I\}\subset\RR^{n}$.
To obtain \eqref{spin-vera} for all $t_{[n]}\in I\simplex$ on the complement of
one fixed $\PP$-zero set, we shall exploit the continuity in $t_{[n]}$ of the various terms in 
\eqref{spin-vera}. To this end we have
to show in particular that the stochastic integrals in \eqref{spin-vera} posses modifications which 
define a process that is jointly continuous in $(t,t_{[n]})$. This is essentially what is done in the proof 
of the first of the two following lemmas. In the second one we justify the use of the stochastic Fubini
theorem in the proof of Lem.~\ref{lem-spin1} at the end of Sect.~\ref{sec-alg}.
In this appendix the results of
Sects.~\ref{sec-proc} and~\ref{sec-Ito} may be used without 
producing logical inconsistences, and the vectors $g,h\in\mathfrak{d}_C$ are fixed. 

\begin{lemma}\label{lem-cont-tn}
On the complement of some $(t,t_{[n]})$-independent $\PP$-zero set,
the stochastic integral formula \eqref{spin-vera} holds true, for all $t\in[0,\sup I)$, $n\in\NN$, 
and $t_{[n]}\in t\simplex_n$.
\end{lemma}

{\proof}
{\em Step 1.}
Employing \eqref{CCR}, \eqref{exp-vec1},  \eqref{exp-vec2}, \eqref{def-W}, and \eqref{def-Qh}
we first observe that the integrand
$\SPn{\zeta(g)}{i\V{v}(\vxi,\V{X}_\tau)\,\sQ_{\tau}^{(n)}(h;t_{[n]})\W{\vxi,\tau}{0}\zeta(h)}$
of the stochastic integral in \eqref{spin-vera} is a linear combination of terms of the form
\begin{align}\nonumber
\V{L}_\tau[t_{[n]}]&:=\IN{\cC}\,\ADg{\cA}{\tau}\,\Ah{\cB}
\\\label{def-VLt}
&\qquad\cdot
\SPn{i\V{m}g+\V{G}_{\V{X}_\tau}}{\w{t_d}{\tau}F_{\alpha_d,\V{X}_{t_d}}}^\vk
\SPn{\zeta(g)}{\W{\vxi,\tau}{0}\zeta(h)},
\end{align}
with disjoint (and possibly empty) subsets $\cA,\cB,\cC,\{d\}\subset[n]$
and $\vk\in\{0,1\}$. As a consequence, if we define
\begin{align}\nonumber
\sI_t[t_{[n]}]&:=\int_{0}^t1_{\tau> t_n}\V{L}_\tau[t_{[n]}]\,\Id\V{B}_\tau\,,
\quad t\in I,\;t_{[n]}\in I\simplex_n,
\end{align}
then it suffices to verify the following:

\smallskip

{\em Claim.}
There exists a $\fB(I\simplex_n)\otimes\fF$-measurable map
$\sJ^\sharp:(t_{[n]},t,\vgamma)\mapsto\sI_t^\sharp[t_{[n]}](\vgamma)\in\RR$
such that, for each fixed $(t_{[n]},t)\in I\simplex_n\times[0,\sup I)$,
we $\PP$-a.s. have $\sI_t^\sharp[t_{[n]}]=\sI_t[t_{[n]}]$, and
such that, for all $\vgamma\in\Omega$, the map
$I\simplex_n\times[0,\sup I)\ni(t_{[n]},t)\mapsto\sI_t^\sharp[t_{[n]}](\vgamma)$ is continuous. 

\smallskip

{\em Step 2.}
To begin with we argue that 
we may additionally assume that $\V{X}=\V{X}^{\V{q}}$,
for some {\em bounded} $\fF_0$-measurable $\V{q}:\Omega\to\RR^\nu$,
so that \eqref{hyp-Y2} is available.
For, if $\V{X}_0=\V{q}$ is unbounded, then we can set $\V{q}_m:=1_{|\V{q}|\le m}\V{q}$, $m\in\NN$,
and verify the claim in Step 1 for each $\V{X}^{\V{q}_m}$. After that we invoke the pathwise
uniqueness property $\V{X}_\bullet^{\V{q}}=\V{X}_\bullet^{\V{q}_m}$, $\PP$-a.s. on $\{|\V{q}|\le m\}$,
which entails $u_{\vxi,\bullet}^0[\V{X}^{\V{q}}]=u_{\vxi,\bullet}^0[\V{X}^{\V{q}_m}]$,
$U_\bullet^+[\V{X}^{\V{q}}]=U_\bullet^+[\V{X}^{\V{q}_m}]$,
and $U_{s,\bullet}^-[\V{X}^{\V{q}}]=U_{s,\bullet}^-[\V{X}^{\V{q}_m}]$,
$\PP$-a.s. on $\{|\V{q}|\le m\}$, for each $s\in I$.
(Here we use the notation explained in the second paragraph of Sect.~\ref{sec-initial}.)
 
So let $\V{q}$ be bounded.
Then the claim in Step 1 follows from the Kolmogoroff-Neveu lemma
(see, e.g., \cite[Satz~2.11${}'$]{HackenbrochThalmaier1994} or 
\cite[Exercise~E.4 of Chap.~8]{Me1982}) as soon as we can find ($n$-dependent) $p,\ve>0$ 
and some function $c:I\to(0,\infty)$ such that
\begin{align*}
\EE\Big[\sup_{\tau\le\sigma}\|\sI_\tau[t_{{n}}]-\sI_\tau[s_{[n]}]\|^p\Big]
\le c(\sigma)\,|t_{[n]}-s_{[n]}|^{n+\ve},\quad\sigma\in[0,\sup I),
\end{align*}
for all $t_{[n]},s_{[n]}\in I\simplex_n$ with $|t_{[n]}-s_{[n]}|\le1$.
To this end we shall prove that
\begin{align}\label{evanthia1}
\EE\Big[\Big(\int_0^\sigma\big|1_{\tau>t_n}\V{L}_\tau[t_{{n}}]-
1_{\tau>s_n}\V{L}_\tau[s_{[n]}\big|^2\Id\tau\Big)^{\nf{p}{2}}\Big]
\le c_{n,p}(\sigma)\,|t_{[n]}-s_{[n]}|^{\frac{p-2}{2}},
\end{align}
for all $\sigma,t_{[n]},s_{[n]}$ as above and for all $p\ge2$.

{\em Step 3.} First, we derive suitable bounds on the scalar products
whose products define $\V{L}_\tau[t_{[n]}]$; recall \eqref{def-ADg} and \eqref{def-Ah}.
In fact, by Hyp.~\ref{hyp-G} the terms 
$$
\V{a}^{(\ell)}_{s,t}:=\SPb{i\V{m}\,g+\V{G}_{\V{X}_t}}{\w{s}{t}{F}_{\ell,\V{X}_{s}}}
\;\;\,\text{and}\;\;\,
\V{a}^{(S+\ell)}_{s,t}:=\SPn{g}{\w{s}{t}{F}_{\ell,\V{X}_{s}}},\quad\ell=1,\ldots,S,
$$
are bounded on $\Omega$, uniformly in $0\le s\le t\in I$. 
Moreover, it is straightforward to infer the following bounds from Hyp.~\ref{hyp-G},
\begin{align*}
|\V{a}^{(\ell)}_{s,t}-\V{a}^{(\ell)}_{\tilde{s},t}|&\le\const
\big(|s-\tilde{s}|+|\V{X}_s-\V{X}_{\tilde{s}}|\big),\quad s,\tilde{s}\le t\in I,
\;\ell=1,\ldots,2S,
\end{align*}
on $\Omega$ with a $t$-independent constant $\const>0$, 
where \eqref{Ito-eq-X}  $\PP$-a.s. implies
\begin{align}\label{evanthia2}
|\V{X}_s-\V{X}_{\tilde{s}}|
&\le|\V{B}_s-\V{B}_{\tilde{s}}|+\int_{\tilde{s}}^s|\V{\beta}(\tau,\V{X}_\tau)|\,\Id\tau,
\quad 0\le\tilde{s}\le s<\sup I,
\end{align}
Taking \eqref{hyp-Y2} into account we deduce that, for all $p\ge2$
and $\sigma\in[0,\sup I)$,
\begin{align}\nonumber
&\EE\Big[\int_0^\sigma|\V{a}^{(\ell)}_{s,\tau}-\V{a}^{(\ell)}_{\tilde{s},\tau}|^p\Id\tau\Big]
\\\nonumber
&\le\const(p)\Big(|s-\tilde{s}|^p+\EE\big[|\V{B}_s-\V{B}_{\tilde{s}}|^p\big]
+|s-\tilde{s}|^{p-1}\int_0^\sigma\EE\big[|\V{\beta}(\tau,\V{X}_\tau)|^p\big]\,\Id\tau\Big)
\\\label{evanthia3}
&\le\const'(p,\sigma)\,|s-\tilde{s}|^{\nf{p}{2}},\quad s,\tilde{s}\in[0,\sigma],\;
|s-\tilde{s}|\le1.
\end{align}
Furthermore, in view of \eqref{def-ulK} and \eqref{def-Uminus} the scalar products 
$$
\V{a}_{s,t}^{(\ell)}:=\SPn{U_{s,t}^-}{{F}_{\ell,\V{X}_s}},\quad\ell=2S+1,\ldots,3S,
$$
satisfy, for all $p\ge 2$, $\sigma\in[0,\sup I)$, and $s\in[0,\sigma]$,
\begin{align}\nonumber
&\EE[\sup_{t\le\sigma}|\V{a}_{s,t}^{(\ell)}|^p]
\\\nonumber
&\le\const^p\,\EE\Big[\sup_{t\le\sigma}
\Big\|\int_0^t1_{r>s}\iota_r\V{G}_{\V{X}_r}\Id\V{B}_r
+\int_0^t1_{r>s}\iota_r(\V{G}_{\V{X}_r}\cdot\V{\beta}(r,\V{X}_r)+\breve{q}_{\V{X}_r})\Id r\Big\|^p\Big]
\\\nonumber
&\le\const'(p)\,\sigma^{\frac{p-2}{2}}
\EE\Big[\int_0^\sigma\|\V{G}_{\V{X}_r}\|^p\Id r\Big]
+\const'(p)\,\sigma^{p-1}\int_0^\sigma\EE\big[1+|\V{\beta}(r,\V{X}_r)|^p\big]\,\Id r
\\\label{evanthia5}
&\le\const''(p,\sigma),\quad\ell=2S+1,\ldots,3S.
\end{align}
For all $p\ge2$ and $\tilde{s}\le s\le\sigma<\sup I$ with $|s-\tilde{s}|\le1$, we likewise have  
\begin{align}\nonumber
&\EE\big[\sup_{0\le\tau\le\sigma}\|U_{s,\tau}^--U_{\tilde{s},\tau}^-\|^p\big]
\\\nonumber
&=\EE\Big[\sup_{0\le\tau\le\sigma}
\Big\|\int_0^\tau\!\!1_{\tilde{s}< r\le s}\iota_r\V{G}_{\V{X}_r}\Id\V{B}_r
+\int_0^\tau\!\!1_{\tilde{s}< r\le s}\iota_r(\V{G}_{\V{X}_r}\V{\beta}(r,\V{X}_r)
+\breve{q}_{\V{X}_r})\Id r\Big\|^p\Big]
\\\nonumber
&\le
\const(p)\EE\Big[\Big(\int_0^\sigma\!\!1_{\tilde{s}<r\le s}\|\V{G}_{\V{X}_r}\|^2\Id r\Big)^{\nf{p}{2}}\Big]
+\const(p)|s-\tilde{s}|^{p-1}\!\int_{0}^\sigma\!\!\EE\big[1+|\V{\beta}(r,\V{X}_r)|^p\big]\Id r
\\\label{evanthia4}
&\le\const'(p,\sigma)|s-\tilde{s}|^{\nf{p}{2}}.
\end{align}
Together with the global Lipschitz continuity of $\V{x}\mapsto\V{F}_{\V{x}}$,
\eqref{evanthia2}, and an estimate analog to \eqref{evanthia3}, 
the bound \eqref{evanthia4} implies
\begin{align}\label{evanthia4b}
\EE\big[\sup_{0\le\tau\le\sigma}|\V{a}_{s,\tau}^{(\ell)}-\V{a}_{\tilde{s},\tau}^{(\ell)}|^p\big]
\le\const(p,\sigma)\,|s-\tilde{s}|^{\nf{p}{2}},
\quad\ell=2S+1,\ldots,3S,
\end{align}
under the above conditions on $s$, $\tilde{s}$, $\sigma$, and $p$.

Let us finally consider the $\tau$-independent terms in  \eqref{def-VLt}. It is clear that 
\begin{align*} 
{a}_{r,s}^{(j,\ell)}:=\SPn{F_{j,\V{X}_s}}{\w{r}{s}F_{\ell,\V{X}_r}}\quad
\text{and}\quad
\V{a}^{(3S+1)}_{s,\tau}:=\V{a}^{(3S+1)}_s:=\SPn{\V{F}_{\V{X}_s}}{\w{0}{s}h},
\end{align*}
are bounded on $\Omega$ uniformly in $r,s\in I$ (and $\tau$, of course).
Thanks to the above discussion it is also clear that $\V{a}^{(3S+1)}_{s,\tau}$
satisfies a bound analog to \eqref{evanthia3} and that
\begin{align}\label{evanthia6}
\EE\big[|a_{r,s}^{(j,\ell)}-a_{\tilde{r},\tilde{s}}^{(j,\ell)}|^p\big]
&\le\const(p,\sigma)\big(|r-\tilde{r}|+|s-\tilde{s}|\big)^{\nf{p}{2}},
\end{align}
for all $r,\tilde{r},s,\tilde{s}\in[0,\sigma]$ with $|r-\tilde{r}|\le1$ and $|s-\tilde{s}|\le1$.
Finally, setting $\V{a}^{(3S+2)}_{s,\tau}:=\V{a}^{(3S+2)}_s:=\SPn{\V{F}_{\V{X}_s}}{U^+_s}$, we
get $\EE[|\V{a}^{(3S+2)}_s|^p]\le\const(p,\sigma)$ and a bound analog to \eqref{evanthia4b}.

{\em Step 4.} Next, we derive the bound \eqref{evanthia1} assuming
that $s_n\le t_n\le\sigma<\sup I$ with $|t_n-s_n|\le1$ without loss of generality:
Notice that $L_\tau[t_{[n]}]$ is the product of $m\le n$ scalar products
which are either uniformly bounded or can be estimated as in \eqref{evanthia5},
whence
\begin{align}\nonumber
&\EE\Big[\Big(\int_{s_n}^{t_n}\big(|\V{L}_\tau[t_{[n]}]|
+|\V{L}_\tau[s_{[n]}]|\big)^2\Id\tau\Big)^{\nf{p}{2}}\Big]
\\\nonumber
&\le\const(n,p)\,|t_n-s_n|^{\frac{p-2}{2}}
\!\!\!\!\sup_{{s\le\sigma\atop j=1,...,3S+2}}\!\int_{s_n}^{t_n}\EE\big[
1+|\V{a}^{(j)}_{s,\tau}|^{np}\big]\Id\tau
\le\const'(n,p,\sigma)\,|t_n-s_n|^{\frac{p-2}{2}}.
\end{align}
Furthermore, representing the difference $\V{L}_\tau[t_{[n]}]-\V{L}_\tau[s_{[n]}]$
as a telescopic sum and using the bound \eqref{norm-W-scalar}, we readily deduce that
\begin{align*}
&\EE\Big[\Big(\int_{t_n}^{\sigma}\big|\V{L}_\tau[t_{[n]}]
-\V{L}_\tau[s_{[n]}]\big|^2\Id\tau\Big)^{\nf{p}{2}}\Big]
\le\sigma^{\frac{p-2}{2}}\EE\Big[\int_0^\sigma\big|\V{L}_\tau[t_{[n]}]-\V{L}_\tau[s_{[n]}]\big|^p\Id\tau\Big]
\\
&\le\const\!\max_{{1\le j,k\le3S+2}}\EE\Big[\int_0^\sigma\Big(\sum_{m=1}^n
|\V{a}^{(j)}_{t_m,\tau}-\V{a}^{(j)}_{s_m,\tau}|\prod_{{\ell=1\atop\ell\not=m}}^n
\big(1+|\V{a}_{s_\ell,\tau}^{(k)}|+|\V{a}_{t_\ell,\tau}^{(k)}|\big)\Big)^p
\Id\tau\Big]
\\
&\quad
+\const\!\max_{{1\le k\le 3S+2\atop1\le j,\ell\le S}}\EE\Big[\int_0^\sigma\Big(
\sum_{{a,b=1\atop a<b}}^n|a_{t_a,t_b}^{(j,\ell)}-a_{s_a,s_b}^{(j,\ell)}|
\prod_{{c=1\atop c\not=a,b}}^n
\big(1+|\V{a}_{s_c,\tau}^{(k)}|+|\V{a}_{t_c,\tau}^{(k)}|\big)\Big)^p\Id\tau\Big]
\\
&\le\const'\!\!\!\max_{{1\le j,k\le3S+2\atop m=1,...,n}}
\EE\Big[\int_0^\sigma|\V{a}^{(j)}_{t_m,\tau}-\V{a}^{(j)}_{s_m,\tau}|^{pn}\Id\tau\Big]^{\nf{1}{n}}
\!\!\!\sup_{s\in[0,\sigma]}
\EE\Big[\int_0^\sigma\big(1+|\V{a}_{s,\tau}^{(k)}|^{pn}\big)\Id\tau\Big]^{\frac{n-1}{n}}
\\
&\quad
+\const'\!\!\!\max_{{{1\le k\le3S+2\atop1\le a<b\le n}\atop1\le j,\ell\le S}}\!\!\EE\big[
|a^{(j,\ell)}_{t_a,t_b}-a^{(j,\ell)}_{s_a,s_b}|^{np}\big]^{\nf{1}{n}}\!\!
\sup_{s\in[0,\sigma]}\EE\Big[\int_0^\sigma
\big(1+|\V{a}_{s,\tau}^{(k)}|^{np}\big)\Id\tau\Big]^{\frac{n-2}{n}}
\\
&\le\const''\,|t_{[n]}-s_{[n]}|^{\nf{p}{2}}.
\end{align*}
Here the constants $\const,\const',\const''>0$ depend on $g$, $h$, $n$, $p$, and $\sigma$.
Altogether this proves \eqref{evanthia1}, where $|t_{[n]}-s_{[n]}|\le1$.

{\em Conclusion.} A priori we know that \eqref{spin-vera} is valid, for all $t\in[t_n,\sup I)$
and all rational $t_{[n]}\in I\simplex_n\cap\QQ^n$, ouside some $(t,t_{[n]})$-independent
$\PP$-zero set. The above steps show, however, that the stochastic integral
appearing in \eqref{spin-vera} has a suitable modification which is jointly continuous in $(t,t_{[n]})$.
Using Hyp.~\ref{hyp-B}(2), \eqref{Qh-Qgh}, and Rem.~\ref{rem-sec6}(2) it is straightforward to 
see that all remaining terms on both sides of \eqref{spin-vera} have continuous modifications as well. 
Hence, we can extend \eqref{spin-vera} by continuity to all $t_{[n]}\in I\simplex_n$ and 
$t_n\le t<\sup I$ such that it holds outside of one fixed $\PP$-zero set. 
\qed 

\begin{lemma}\label{lem-stoch-Fub}
The following relation holds $\PP$-a.s., for all $t\in[0,\sup I)$, 
\begin{align*}
&\int_{I^n}\int_{0}^t1_{\tau\simplex_n}(t_{[n]})
\SPb{\zeta(g)}{i\V{v}(\vxi,\V{X}_\tau)\,
\sQ_\tau^{(n)}(h;t_{[n]})\,\W{\vxi,\tau}{0}\zeta(h)}\,\Id\V{B}_\tau\,\Id t_{[n]}
\\
&=\int_{0}^t\int_{I^n}1_{\tau\simplex_n}(t_{[n]})
\SPb{\zeta(g)}{i\V{v}(\vxi,\V{X}_\tau)\,
\sQ_\tau^{(n)}(h;t_{[n]})\,\W{\vxi,\tau}{0}\zeta(h)}\,\Id t_{[n]}\,\Id\V{B}_\tau.
\end{align*}
\end{lemma}

{\proof}
Since both sides of the asserted identity are continuous in $t$ (according to Lem.~\ref{lem-cont-tn}),
it suffices to prove it ($\PP$-a.s.) for some fixed $t$. So, let $t\in[0,\sup I)$ in what follows.
By the remark in the very beginning of the proof of Lem.~\ref{lem-cont-tn}, we then have to show that,
$\PP$-a.s.,
\begin{align}\label{guido1}
\int_{I^n}\int_{0}^t&1_{\tau\simplex_n}(t_{[n]})\V{L}_\tau[t_{[n]}]\,\Id\V{B}_\tau\,\Id t_{[n]}
=\int_{0}^t\int_{I^n}1_{\tau\simplex_n}(t_{[n]})\V{L}_\tau[t_{[n]}]\,\Id t_{[n]}\,\Id\V{B}_\tau,
\end{align}
where $\V{L}_\tau[t_{[n]}]$ is given by \eqref{def-VLt}.
Invoking the pathwise uniqueness properties discussed in Step~2 of the proof of 
Lem.~\ref{lem-cont-tn} and the pathwise uniqueness property of stochastic integrals with respect to
Brownian motion, we may again argue that it suffices to prove \eqref{guido1} under the additional
assumption that the initial condition $\V{q}$ in the SDE solved by $\V{X}=\V{X}^{\V{q}}$ be bounded.
In order to justify the application of the stochastic Fubini
theorem it then suffices (see, e.g., \cite[Rem.~4.35]{daPrZa2014}) to check that
\begin{equation}\label{guido2}
\int_{t\simplex_n}\!\!\Big(\EE\Big[\int_{t_n}^t\big|\V{L}_\tau[t_{[n]}]\big|^2\Id\tau\Big]\Big)^\eh
\Id t_{[n]}<\infty.
\end{equation}
Since $\V{q}$ is assumed to be bounded we know, however, from the arguments
in the proof of Lem.~\ref{lem-cont-tn} that, for all $t\in[0,\sup I)$,
\begin{align*}
\EE\Big[\int_{t_n}^t\big|\V{L}_\tau[t_{[n]}]\big|^2\Id\tau\Big]
&\le\const(n)\sup_{{s\le t\atop j=1,\ldots,3S+2}}\int_{t_n}^t
\EE\big[1+|\V{a}_{s,\tau}^{(j)}|^{np}\big]<\infty,
\end{align*}
where we use the notation introduced in Step~3 of the proof of Lem.~\ref{lem-cont-tn}.
Clearly, this implies \eqref{guido2} and we conclude.
\qed


\section{Measurability of the operator-valued map $\WW{\vxi,t}{0}$}\label{app-meas}

\noindent
Recall that a measurable map from a measurable space into a Banach space equipped with its
Borel $\sigma$-algebra can be (a.e.) approximated by measurable simple functions, 
if and only if its range is (a.e.) separable. 
In particular, it is not possible to define its Bochner-Lebesgue integral, if its range is
not (a.e.) separable. Since $\LO(\FHR)$ is a non-separable Banach space,
we shall therefore prove the following two propositions in this appendix:
 
\begin{proposition}\label{prop-meas}
Let $\vxi\in\RR^\nu$ and assume, in addition to our standing hypotheses, that $|\V{m}|\le c\omega$, 
for some $c>0$. Then, after a suitable modification, the operator-valued map 
$\WW{\vxi}{V}:I\times\Omega\to\LO(\FHR)$
has a separable image and defines an adapted $\LO(\FHR)$-valued process whose paths are continuous on $I\setminus\{0\}$. In particular, it is predictable.
\end{proposition}

\begin{proposition}\label{prop-app-cont}
Let $\vxi\in\RR^\nu$ and let $\sT$ be a locally compact metric space.
Assume that $V$ is continuous and that $|\V{m}|\le c\omega$, for some $c>0$.
Assume further that the driving process depends parametrically on $x\in\sT$, which we indicate by
writing $\V{X}^x$, such that $I\times\sT\ni(t,x)\mapsto\V{X}_t^x(\vgamma)$ is continuous,
for all $\vgamma\in\Omega$. Finally, assume that the basic processes can and have been chosen
such that 
$$
I^2\times\sT\ni(\tau,t,x)\longmapsto
(u_{-\vxi,t}^V[\V{X}^x],U^+_t[\V{X}^x],U_{\tau,t}^-[\V{X}^x])(\vgamma)\in\CC\oplus\mathfrak{k}^2
$$
is continuous, for all $\vgamma\in\Omega$. Then we can modify each process 
$\WW{\vxi}{V}[\V{X}^x]$, $x\in\sT$, such that $(t,x,\vgamma)\mapsto\WW{\vxi}{V}[\V{X}^x](\vgamma)$
is measurable from $I\times\sT\times\Omega$ to $\LO(\FHR)$ with a separable image,
$\WW{\vxi,t}{V}[\V{X}^x]:\Omega\to\LO(\FHR)$ is $\fF_t$-$\fB(\LO(\FHR))$-measurable, for
all $(t,x)\in I\times\sT$, and 
$(I\setminus\{0\})\times\sT\ni(t,x)\mapsto\WW{\vxi,t}{V}[\V{X}^x](\vgamma)$ 
is operator norm continuous, for all $\vgamma\in\Omega$.
\end{proposition}

 \begin{remark}{\rm 
(1) Note that, in the trivial case where $\V{m}$, $\V{G}$, and $\V{F}$ are all equal to zero,
we have $\WW{\V{0},t}{0}=e^{-t\Id\Gamma(\omega)}$, which is not continuous at $t=0$
with respect to the operator norm.

\smallskip

\noindent(2) Employing the bounds derived in Lem.~\ref{lem-meas} below,
we can actually verify, without using Thm.~\ref{thm-Ito-spin}, that the series of time-ordered 
integrals \eqref{limit-WW} converges with respect to the operator norm pointwise on $\Omega$.
The bounds on the norm of $\WW{\vxi,t}{V}$ thus obtained are, however, not $\PP$-integrable
in general and way too rough in order to discuss the SDE \eqref{SDE-spin}.
 }\end{remark}

To prove the above propositions we shall employ the bound
\begin{align}\label{michi0}
\|\ad(f_1)\dots\ad(f_m)\psi\|
&\le2^{\frac{m}{2}}(m!)^\eh\Big(\prod_{j=1}^m\|f_j\|_{\omega}\Big)
\Big(\sum_{\ell=0}^m\frac{1}{\ell!}\big\|\Id\Gamma(\omega)^{\frac{\ell}{2}}\psi\big\|^2\Big)^{\frac{1}{2}}
\end{align}
with $\|f\|_\omega^2:=\|f\|^2+\|\omega^\mh f\|^2$ and where $f,f_1,\ldots,f_m$ and $\psi$ are
such that the norms on the right hand sides are well-defined. We leave the proof of \eqref{michi0}
as an exercise to the reader.

For general information on analytic maps from one Hilbert space into another, like the one 
appearing in the next lemma, we refer again to \cite[\textsection III.3.3]{HillePhillips1957}.

\begin{lemma}\label{lem-meas}
Let $t>0$ and $m\in\NN_0$. Then the map $F_{m,t}:\mathfrak{k}^{m+1}\to\LO(\FHR)$,
\begin{align}\label{michi-1}
F_{m,t}(f_1,\ldots,f_m,g)
&:=\sum_{n=0}^\infty\frac{1}{n!} 
\ad(f_m)\ldots\ad(f_1)\ad(g)^ne^{-t\Id\Gamma(\omega)},
\end{align}
is well-defined, analytic on $\mathfrak{k}^{m+1}$, 
and satisfies
\begin{align}\label{michi-2}
\|F_{m,t}(f_1,\ldots,f_m,g)\|&\le(m!)^\eh\Big(\prod_{j=1}^m2T^{\nf{1}{2}}\|f_j\|_{\omega}\Big)
s\big((2T)^{\nf{1}{2}}\|g\|_\omega\big),
\end{align}
where $T:=1\vee(1/2t)$ and $s(z):=\sum_{n=0}^\infty(n!)^\mh z^n$, $z\in\CC$.
If $\ell\in\NN$ and $f_{1},\ldots,f_{m+\ell},g\in\mathfrak{k}$, then
$\Ran(F_{m,t}(f_1,\ldots,f_m,g))\subset\dom(\ad(f_{m+\ell})\dots\ad(f_{m+1}))$
and 
\begin{align}\label{michi-3}
\ad(f_{m+\ell})\dots\ad(f_{m+1})F_{m,t}(f_1,\ldots,f_m,g)
=F_{m+\ell,t}(f_1,\ldots,f_{m+\ell},g).
\end{align}
In particular, we may write 
\begin{equation*}
F_{m,t}(f_1,\ldots,f_m,g)=\ad(f_m)\ldots\ad(f_1)\exp\{\ad(g)\}e^{-t\Id\Gamma(\omega)}
\end{equation*}
with $\exp\{\ad(g)\}e^{-t\Id\Gamma(\omega)}:=F_{0,t}(g)$. 
For every $s>0$, we finally have
\begin{equation}\label{michi777}
F_{m,t+s}(f_1,\ldots,f_m,g)=F_{m,t}(f_1,\ldots,f_m,g)e^{-s\Id\Gamma(\omega)}.
\end{equation}
\end{lemma}

{\proof}
Let $t>0$. It follows immediately from \eqref{michi0} that, for all $\ell\in\NN_0$, the multi-linear map 
$\mathfrak{k}^{\ell}\ni(h_1,\ldots,h_\ell)\mapsto
\ad(h_1)\ldots\ad(h_\ell)e^{-t\Id\Gamma(\omega)}\in\LO(\sF)$
is bounded and, in particular, analytic. Therefore, to show analyticity of $F_{m,t}$,
it suffices to show that the series in \eqref{michi-1}
converges uniformly on every bounded subset of $\mathfrak{k}^{m+1}$. 
Applying \eqref{michi0} we obtain, for all $\phi\in\bigcap_{\ell\in\NN}\dom(\Id\Gamma(\omega)^{\ell})$,
\begin{align*}
&\frac{1}{n!}\big\|\ad(f_1)\ldots\ad(f_m)\ad(g)^n\phi\big\|
\\
&\le(2T)^{\frac{m}{2}}\Big(\prod_{j=1}^m\|f_j\|_{\omega}\Big)
\frac{((m+n)!)^{\frac{1}{2}}}{n!}
(2T)^{\frac{n}{2}}\|g\|_\omega^n\Big(\sum_{\ell=0}^{m+n}\frac{T^{-\ell}}{\ell!}
\SPb{\phi}{\Id\Gamma(\omega)^\ell\phi}\Big)^{\frac{1}{2}}
\\
&\le(2T^\eh)^m(m!)^\eh\Big(\prod_{j=1}^m\|f_j\|_{\omega}\Big)
\frac{(2T^\eh\|g\|_\omega)^n}{(n!)^\eh}\Big(\sum_{\ell=0}^{\infty}\frac{T^{-\ell}}{\ell!}
\SPb{\phi}{\Id\Gamma(\omega)^\ell\phi}\Big)^{\frac{1}{2}}.
\end{align*}
Here we used the bound $\frac{(m+n)!}{m!n!}<2^{m+n}$ in the second step. Since 
$$
\sum_{\ell=0}^{\infty}\frac{T^{-\ell}}{\ell!}
\SPb{e^{-t\Id\Gamma(\omega)}\psi}{\Id\Gamma(\omega)^\ell e^{-t\Id\Gamma(\omega)}\psi}
=\big\|e^{-(t-1/2T)\Id\Gamma(\omega)}\psi\big\|^2\le\|\psi\|^2,
$$
for all $\psi\in\sF$, this implies
\begin{align*}
\frac{1}{n!}\big\|\ad(f_1)\ldots\ad(f_m)\ad(g)^n&e^{-t\Id\Gamma(\omega)}\big\|
\\
&\le(2T^{\frac{1}{2}})^m(m!)^{\frac{1}{2}}\Big(\prod_{j=1}^m\|f_j\|_{\omega}\Big)
\frac{(2T^{\frac{1}{2}}\|g\|_\omega)^n}{(n!)^{\frac{1}{2}}}.
\end{align*}
Therefore, the series in \eqref{michi-1} converges absolutely in operator norm,
uniformly on every bounded subset of $\mathfrak{k}^{m+1}$, and we also obtain \eqref{michi-2}.
The relation \eqref{michi-3} follows inductively from the fact that $\ad(f)$ is closed, 
for every $f\in\HP$,
and \eqref{michi777} is obvious from the fact that right multiplication with 
$e^{-s\Id\Gamma(\omega)}$ is continuous on $\LO(\sF)$.
\qed 

\begin{corollary}\label{cor-meas1}
Let $r,s,\tau>0$ and $m\in\NN_0$. Then, for all $f_1,\ldots,f_m,g\in\mathfrak{k}$, the operator
$G_{m,s}(f_1,\ldots,f_m,g)$ defined on the dense domain $\sC[\mathfrak{d}_C]$ by
\begin{align*}
G_{m,s}(f_1,\ldots,f_m,g)\psi:=e^{-s\Id\Gamma(\omega)}\exp\{a(g)\}a(f_1)\ldots a(f_m)\psi,
\quad\psi\in\sC[\mathfrak{d}_C],
\end{align*}
is bounded and its unique extension to an element of $\LO(\sF)$ is given by
\begin{align*}
\ol{G_{m,s}(f_1,\ldots,f_m,g)}=F_{m,s}(f_1,\ldots,f_m,g)^*.
\end{align*}
If $n\in\NN_0$ and $|\V{m}|\le c\omega$, for some $c>0$, then the map 
$D_{r,s,\tau}^{(m,n)}:\CC\times[0,\infty)\times\RR^\nu\times\mathfrak{k}^{m+n+2}\to\LO(\sF)$ 
defined by
\begin{align}\label{michi9}
D_{r,s,\tau}^{(m,n)}(&a,t,\V{x},f_1,\ldots,f_m,\tilde{f}_1,\ldots,\tilde{f}_n,g,\tilde{g})
\\\nonumber
&:= aF_{m,r}(f_1,\ldots,f_m,g)\Gamma(e^{-(\tau+t)\omega+i\V{m}\cdot\V{x}})
F_{n,s}(\tilde{f}_1,\ldots,\tilde{f}_n,\tilde{g})^*
\end{align}
is uniformly continuous on every bounded subset of 
$\CC\times[0,\infty)\times\RR^\nu\times\mathfrak{k}^{m+n+2}$ and has a separable image.
Moreover, $D_{r,s,\tau}^{(m,n)}=D_{\tilde{r},\tilde{s},\tilde{\tau}}^{(m,n)}$, 
for all $\tilde{r},\tilde{s},\tilde{\tau}>0$ satisfying $\tilde{r}+\tilde{s}+\tilde{\tau}=r+s+\tau$.
\end{corollary}

{\proof}
The first assertion follows from Lem.~\ref{lem-meas}, and the continuity of the map \eqref{michi9} 
follows from Lem.~\ref{lem-meas} and the bound
\begin{align*}
\big\|\Gamma(&e^{-(\tau+t)\omega+i\V{m}\cdot\V{x}})
-\Gamma(e^{-(\tau+u)\omega+i\V{m}\cdot\V{y}})\big\|
\\
&\le\big\|(\id-e^{(t-u)\Id\Gamma(\omega)+i(\V{x}-\V{y})\cdot\Id\Gamma(\V{m})})
e^{-(t+\tau)\Id\Gamma(\omega)}\big\|\;
\\
&\le(u-t)\big\|\Id\Gamma(\omega)e^{-\tau\Id\Gamma(\omega)}\big\|
+|\V{x}-\V{y}|\big\|\Id\Gamma(|\V{m}|)e^{-\tau\Id\Gamma(\omega)}\big\|
\\
&\le\big(u-t+c|\V{x}-\V{y}|\big)\big\|\Id\Gamma(\omega)e^{-\tau\Id\Gamma(\omega)}\big\|,
\end{align*}
for all $\V{x},\V{y}\in\RR^\nu$ and $u>t>0$. The map \eqref{michi9} has a separable image because
it is continuous and its domain 
$\CC\times[0,\infty)\times\RR^\nu\times\mathfrak{k}^{m+n+2}$ is separable.
The relation $D_{r,s,\tau}^{(m,n)}=D_{\tilde{r},\tilde{s},\tilde{\tau}}^{(m,n)}$ is a consequence of
\eqref{michi777}.
\qed

\begin{corollary}\label{cor-meas2}
Let $\sT$ be a locally compact metric space, let $\sK$ be a separable Hilbert space,
and let $\mathsf{T}_{\sK}$ be the set of measurable maps
$X:I\times\sT\times\Omega\to\sK$, $(t,x,\vgamma)\mapsto X_t^x(\vgamma)$,
such that $X^x$ is an adapted process, for every $x\in\sT$, and
$I\times\sT\ni(t,x)\mapsto X_t^x(\vgamma)$ is continuous, for all $\vgamma\in\Omega$.

Let $r,s,\tau>0$, $\ell,m,n\in\NN_0$, $\tilde{\V{X}}\in\mathsf{T}_{\RR^\nu}$,
$Z_1,\ldots,Z_m,\wt{Z}_1,\ldots,\wt{Z}_n,Y,\wt{Y}\in\mathsf{T}_{\mathfrak{k}}$, and
$h:I^\ell\times\sT\times\Omega\to\CC$, $(t_{[\ell]},x,\vgamma)\mapsto h_{t_{[\ell]}}^x(\vgamma)$ 
be measurable such that its restriction to 
$[0,t]^\ell\times\sT\times\Omega$ is $\fB([0,t]^\ell)\otimes\fB(\sT)\otimes\fF_t$-measurable, 
for every $t\in I$, and
such that $I^\ell\times\sT\ni(t_{[\ell]},x)\mapsto h_{t_{[\ell]}}^x(\vgamma)$ 
is continuous, for all $\vgamma\in\Omega$. 
For all $(t_{[\ell+m+n]},\rho_{[3]},t,x)\in\sG:=I^{\ell+m+n+3}\times[0,\infty)\times\sT$, 
define a function $\Omega\to\LO(\sF)$ by
\begin{align}\label{michi10}
&B_{t,\rho_{[3]}}^x(t_{[\ell+m+n]},\cdot)
\\\nonumber
&:=D_{r,s,\tau}^{(m,n)}\big(h_{t_{[\ell]}}^x,t,\tilde{\V{X}}_{\rho_1}^x,Z_{1,t_{\ell+1}}^x,
...,Z_{m,t_{\ell+m}}^x,\wt{Z}_{1,t_{\ell+m+1}}^x,...,\wt{Z}_{n,t_{\ell+m+n}}^x,Y_{\rho_2}^x,
\wt{Y}_{\rho_3}^x\big).
\end{align}
Then $B:\sG\times\Omega\to\LO(\sF)$ is measurable,
it has a separable image, its restriction to
$[0,t]^{\ell+m+n+3}\times[0,\infty)\times\sT\times\Omega\to\LO(\sF)$ is 
$\fB([0,t]^{\ell+m+n+3}\times[0,\infty)\times\sT)\otimes\fF_t$-$\fB(\LO(\sF))$-measurable,
and the map $(t_{[\ell+m+n]},\rho_{[3]},t,x)\mapsto B_{t,\rho_{[3]}}^x(t_{[\ell+m+n]},\vgamma)$
is continuous on $\sG$, for all $\vgamma$. 
Furthermore, the $\LO(\sF)$-valued Bochner-Lebesgue integrals in
\begin{align}\label{michi11}
J(\tilde{t},\rho_{[3]},t,x,\vgamma):=\int_{\tilde{t}\triangle_{m+n+\ell}}
B_{t,\rho_{[3]}}^x(t_{[\ell+m+n]},\vgamma)\Id t_{[\ell+m+n]},\quad\tilde{t}\in I,
\end{align}
are well-defined, the map $J:\sG':=I^4\times[0,\infty)\times \sT\times\Omega\to\LO(\sF)$ is 
measurable with a separable image and its restriction to
$[0,t]^4\times[0,\infty)\times\sT\times\Omega\to\LO(\sF)$ is
$\fB([0,t]^4\times[0,\infty)\times\sT)\otimes\fF_t$-$\fB(\LO(\FHR))$-measurable,
for every $t\in I$. Finally, for all $\vgamma\in\Omega$, the map
$(\tilde{t},\rho_{[3]},t,x)\mapsto J(\tilde{t},\rho_{[3]},t,x,\vgamma)$ is continuous on $\sG'$.
\end{corollary}

{\proof}
The measurability properties of $B$ are clear 
by definition and Cor.~\ref{cor-meas1}, since $B$ is the composition of two maps
which are measurable in the appropriate sense. 
(Here we use that $\otimes_{i=1}^n\fB(\mathfrak{k})=\fB(\mathfrak{k}^n)$
which follows from the separability of $\mathfrak{k}$.)
Since the image of $B$ is contained in the image of \eqref{michi9}, it is separable. 
Cor.~\ref{cor-meas1} also shows that,
at each fixed $\vgamma$, $B$ can be written as a composition of two continuous maps.
In particular, the integral in \eqref{michi11} is a (well-defined) Bochner-Lebesgue 
integral of a continuous function over a compact simplex. The measurability properties of $J$
thus follow from a standard result in integration theory and the image of $J$ is contained in any 
closed separable subspace of $\LO(\sF)$ containing the image of $B$. 
The continuity of $J$ follows from the dominated convergence theorem and the local 
compactness of $\sG$.
\qed

 \begin{remark}{\rm \label{rem-for-W-meas}
Let $t>0$ and pick arbitrary $r,s,\tau>0$ with $r+s+\tau<t$. Then the following statements hold
true on all of $\Omega$:

\smallskip

\noindent
(1) In view of \eqref{APC-fact} and \eqref{def-W} we have the following factorization,
\begin{align*}
\W{\vxi,t}{V}\psi
&=e^{-u_{-\vxi,t}^V}\exp\{i\ad(U_t^+)\}\,\Gamma(\w{0}{t})\exp\{ia(U_t^-)\}\psi,\quad\psi\in\sC[\HP].
\end{align*}
Thus, $\W{\vxi,t}{V}=D_{r,s,\tau}^{(0,0)}\big(e^{-u_{-\vxi,t}^V},t-\tau,\V{X}_t-\V{X}_0,U_t^+,U_t^-\big)$ with
$D_{r,s,\tau}^{(0,0)}$ as in \eqref{michi9}.

\smallskip

\noindent(2)
Let $n\in\NN$. Then $\WW{\vxi,t}{V,(n)}$ can be written as a linear combination 
(with coefficients in $\LO(\CC^L)$) of $\LO(\sF)$-valued Bochner-Lebesgue integrals,
\begin{align}\label{michi72}
\WW{\vxi,t}{V,(n)}=\sum_{\alpha\in[S]^n}
\sigma_{\alpha_n}\dots\sigma_{\alpha_1}\!\!\!\!\sum_{{\cA\cup\cA'\cup\cB\cup\cB'\cup\cC=[n]\atop\#\cC\in2\NN_0}}\int_{t\simplex_n}\!\!D_{r,s,\tau}^{(\#\cA,\#\cB)}\big(\aleph(t,t_{[n]})\big)\Id t_{[n]},
\end{align}
where the argument of the integrand is given by
{\small
\begin{align*}
&\aleph(t,t_{[n]})
\\
&:=\Big(h_{t,t_{\cA'\cup\cB'\cup\cC}},t-\tau,
\V{X}_t-\V{X}_0,\big\{\w{t_a}{t}{F}_{\alpha_a,\V{X}_{t_a}}\big\}_{a\in\cA},
\big\{\olw{0}{t_b}{F}_{\alpha_b,\V{X}_{t_b}}\big\}_{b\in\cB},iU_t^+,iU_t^-\Big),
\\
&h_{t,t_{\cA'\cup\cB'\cup\cC}}
\\
&:=\IN{\cC}e^{-u_{-\vxi,t}^V}
\Big({\prod_{a'\in\cA'}}\{i\SPn{U_{t_{a'},t}^{-}}{{F}_{\alpha_{a'},\V{X}_{t_{a'}}}}\}\Big)
{\prod_{b'\in\cB'}}\{i\SPn{{F}_{\alpha_{b'},\V{X}_{t_{b'}}}}{U_{t_b}^{+}}\}.
\end{align*}
}
(3) We may compute the adjoint of $\WW{\vxi,t}{V,(n)}$ by replacing the integrand in \eqref{michi72} by 
its adjoint. Hence, in combination with \eqref{michi9} we obtain a fairly detailed formula for 
$\WW{\vxi,t}{V*}=\sum_{n=0}^\infty\WW{\vxi,t}{V,(n)*}$ in terms of the basic processes.
 }\end{remark}

{\it Proof of Prop.~\ref{prop-app-cont}.}
Since $\WW{\vxi,0}{V,(n)}=\delta_{0,n}\id$ on $\Omega$, the $\fF_0$-measurability 
of $\WW{\vxi,0}{V}$ is trivial. Thus,
for every $n\in\NN_0$, the statement of the proposition with $\WW{\vxi}{V}$ replaced by
$\WW{\vxi}{V,(n)}$ follows immediately from
Cor.~\ref{cor-meas2} in combination with the formulas of Rem.~\ref{rem-for-W-meas}.
Combining this result with the bound \eqref{norm-WQ1}, we conclude that, $\PP$-a.s.,
the convergence $\WW{\vxi,t}{V}[\V{X}^x]=\lim_{N\to\infty}\WW{\vxi,t}{V,(0,N)}[\V{X}^x]$ 
in $\LO(\FHR)$ is locally uniform in $(t,x)\in I\times\sT$.
Since each measure space $(\Omega,\fF_t,\PP)$ with $t\in I$ is complete, this proves the proposition.
\qed 
 
\smallskip

{\it Proof of Prop.~\ref{prop-meas}.}
Prop.~\ref{prop-meas} is proved in the same way as Prop.~\ref{prop-app-cont}.
\qed


\newpage
\section{General notation and list of symbols}\label{app-notation}

\noindent
$s\wedge t:=\min\{s,t\}$ and $s\vee t:=\max\{s,t\}$, for $s,t\in\RR$.

\noindent
$1_A$ is the characteristic function of a set $A$. 

\smallskip

\noindent{\bf Vectors and vector spaces}

\smallskip

\noindent
$\dom(\cdot)$ denotes the domain of linear operators,
and $\fdom(\cdot)$ the quadratic form domain of suitable linear operators.  
$\LO(\sK_1,\sK_2)$ is the space of bounded linear operators between two normed linear spaces 
$\sK_1$, $\sK_2$; $\LO(\sK_1):=\LO(\sK_1,\sK_1)$. 

\noindent
$x^{\otimes_n}$ denotes the $n$-fold tensor product of a vector $x$ with itself.

\medskip

\begin{tabular}{ll}
$\HP=L^2(\cM,\fA,\mu)$ \slash $\mathfrak{k}$, $\mathfrak{d}$ \slash $\HP_{+1}$, 
$\mathfrak{k}_{+1}$&
\eqref{def-one-boson-space}\slash Hyp.~\ref{hyp-G} \slash Sect.~\ref{sec-proc}\\
$\sF=\Gamma_{\mathrm{s}}(\HP)$ \slash $\FHR=\CC^L\otimes\sF$ \slash $\HR$&
\eqref{def-Fockspace} \slash \eqref{def-FHR} \slash \eqref{total}\\
$\zeta(h)$ \slash $\sE[\mathfrak{v}]$, $\sC[\mathfrak{v}]$&\eqref{def-exp-vec} \slash \eqref{def-E}\\
$\wh{\dom}$ \slash $\sD_0$&\eqref{def-whD-intro} \slash \eqref{def-sD0}\\
$\HP_C$ \slash $\mathfrak{k}_C$, $\mathfrak{d}_C$ \slash $\sF_C$&Hyp.~\ref{hyp-G}
\slash \eqref{def-dCkC} \slash \eqref{conj-F}\\
\end{tabular}

\medskip

\noindent{\bf Quantities determining the model, operators}

\medskip

\begin{tabular}{ll}
$\sW(f,U)$, $\sW(f)$, $\Gamma(U)$&Subsect.~\ref{ssec-Fock}\\
$\vp(f)$, $\Id\Gamma(T)$, $\ad(f)$, $a(f)$&Subsect.~\ref{ssec-Fock}\\
$\omega$, $\V{m}$, $\V{G}$, $\V{F}$, $C$, $\vsigma$, $\nu$, $L$, $S$&
Hyp.~\ref{hyp-G} \& preceding paragraphs\\
$q$, $\breve{q}$&\eqref{sym-q}\\
$\wh{H}^V(\vxi,\V{x})$, $\wh{H}_{\scal}^V(\vxi,\V{x})$, $\wh{H}(\vxi)$, $\V{v}(\vxi,\V{x})$&
Def.~\ref{defn-gen-Ham}\\
$M$ \slash $M_a(\vxi)$&\eqref{def-whD-intro} \slash \eqref{def-Maxi}\\
$V$ \slash $H^V$ &Hyp.~\ref{hyp-V} \slash \eqref{def-HV-intro} and Sect.~\ref{sec-ext}\\
$p_t$ \slash $\wh{T}_t(\vxi)$, $T_t^V$, $T_t^V(\V{x},\V{y})$&\eqref{gaussian} 
\slash Def.~\ref{def-FKO}
\end{tabular}

\medskip

\noindent{\bf Measure theoretic and probabilistic objects, processes}

\smallskip

\noindent
$\fB(\sT)$ denotes the Borel $\sigma$-algebra of a topological space $\sT$.

\noindent
$\lambda^\nu$ is the $\nu$-dimensional Lebesgue-Borel measure and $\lambda:=\lambda^1$.

\medskip

\begin{tabular}{ll}
$I$, $\cT$, $\BB=(\Omega,\fF,(\fF_t)_{t\in I},\PP)$, $\EE$, $\EE^{\fH}$ &
Beginning of Subsect.~\ref{sec-prel}\\
$I^s$, $\BB_s$&\eqref{shifted-stoch-basis}\\
$\V{B}$, $\V{X}$, $\V{X}^{\V{q}}$, ${}^s\!\V{X}^{\V{q}}$, $\V{\beta}$, $\vXi$, $\fF_{s,t}$
&Hyp.~\ref{hyp-B} \slash \eqref{def-Yt}\\
$\V{Y}$ \slash $\V{B}^{\V{x}}:=\V{x}+\V{B}$&\eqref{def-Yt} \slash \eqref{def-Bx}\\
$\V{b}^{\cT;\V{x},\V{y}} \slash \smash{\hat{\V{b}}}^{\cT;\V{y},\V{x}}$&
Lem.~\ref{lem-rev-bridge} \& App.~\ref{app-bridge} \slash \eqref{def-hatb}\\
$\mathsf{S}_I(\sK)$&Beginning of Subsect.~\ref{sec-prel}\\
$\llbracket \cdot\,,\, \cdot\cdot\rrbracket$
&Rem.~\ref{cons}\\
$j_t$ \slash $\iota_t$ \slash $\w{\tau}{t}$, $\olw{\tau}{t}$&\eqref{def-jt} \slash \eqref{def-iota} 
\slash \eqref{def-w}\\
$u_{\vxi}^V$, $U^\pm$, $(U_{\tau,t}^-)_{t\in I}$, $K_{\tau,t}$, $K_t$&Def.~\ref{defn-basic-proc}\\
$\W{\vxi}{V}$ \slash $\WW{\vxi}{V}$ \slash $\WW{\vxi}{V,(n)}$, $\WW{\vxi}{V,(N,M)}$&\eqref{W-Hi} 
\slash Thm.~\ref{thm-Ito-spin} \slash Def.~\ref{defn-TOE}\\
$t\simplex_n$, $\AD{\cA}{}$, $\A{\cB}$, $\IN{\cC}$&Def.~\ref{defn-TOE}\\
$\ADg{\cA}{}$, $\Ah{\cB}$, $\sQ_\tau^{(n)}$&Rem.~\ref{rem-sec6}\\
$\Lambda_{s,t}(\V{x},\psi)$ \slash $\Lambda_{s,t}[\V{q},\eta]$, $P_{s,t}$&
Thm.~\ref{thm-flow} \slash Prop.~\ref{prop-Markov}\\
$\bar{\V{X}}$, $\bar{\fF}_\tau$, $\bar{\BB}$ &\eqref{def-barX}
\end{tabular}

\medskip

\noindent
The meaning and use of an additional argument $[\V{X}]$, $[\V{B}^{\V{x}}]$, etc., of a process,
e.g., $U^\pm[\V{X}]$ or $\WW{\vxi,t}{V}[\V{b}^{t:\V{y},\V{x}}]$, 
is explained in the beginning of Sect.~\ref{sec-initial}.



\vspace{0.5cm}

\noindent{\bf Acknowledgements.}
BG has been financially supported by the SFB 647: Raum-Zeit-Materie.
OM has been partially supported by the Lundbeck Foundation, the Villum Foundation, and by the 
European Research Council under the European
Community's Seventh Framework Program (FP7/2007--2013)/ERC grant
agreement 202859. 



${}$
\vspace{2cm}

{\small
\noindent
{\sc Batu G\"uneysu}\\
Institut f\"ur Mathematik, Humboldt
Universit\"at zu Berlin\\ Rudower Chaussee 25,
D-12489 Berlin, Germany\\
{\tt gueneysu@math.hu-berlin.de}           

\vspace{0.6cm}

\noindent
{\sc Oliver Matte $\cdot$ Jacob Schach  M{\o}ller}\\
Institut for Matematik, Aarhus Universitet\\
Ny Munkegade 118, DK-8000 Aarhus C, Denmark\\
{\tt matte@math.au.dk} $\cdot$ {\tt jacob@math.au.dk}
}

\end{document}